\documentclass[twocolumn]{aastex62}

\newcommand{\RNum}[1]{\uppercase\expandafter{\romannumeral #1\relax}}
\newcommand{\bef}{\begin{figure}}      
\newcommand{\eef}{\end{figure}}      
\newcommand{\bea}{\begin{eqnarray}}    
\newcommand{\eea}{\end{eqnarray}}      
\newcommand{\be}{\begin{equation}}      
\newcommand{\ee}{\end{equation}}  
\newcommand\HI{$\textrm{H}\scriptstyle\mathrm{I}$}
\newcommand\HII{$\textrm{H}\scriptstyle\mathrm{II}$}

\usepackage{lineno}
\graphicspath{{./}{figures/}}

\received{???}
\revised{???}
\accepted{???}
\shorttitle{Breaking the degeneracy between warps and radial flows  in external galaxies}
\shortauthors{Sylos Labini et al.}
\usepackage{amsmath}

\begin{document}

\title{Breaking the degeneracy between warps and radial flows  in external galaxies}
\author{Francesco  Sylos Labini}
\affil{Centro  Ricerche Enrico Fermi, Via Pansiperna 89a, 00184 Rome, Italy}
\affil{Istituto Nazionale Fisica Nucleare, Unit\`a Roma 1, Dipartimento di Fisica, Universit\'a di Roma ``Sapienza'', 00185 Rome, Italy}
\author{Giordano De Marzo}
\affil{Centro  Ricerche Enrico Fermi, Via Pansiperna 89a, 00184 Rome, Italy}
\affil{Dipartimento di Fisica, Sapienza, Universit\'a di  Roma,  I-00185, Roma, Italia}
\affil{Complexity Science Hub Vienna, Josefstaedter Strasse 39, 1080, Vienna, Austria}
\affil{University of Konstanz, Universit\"{a}tstra$\beta$e 10, 78457 Konstanz, Germany}
\author{Matteo Straccamore}
\affil{Centro  Ricerche Enrico Fermi, Via Pansiperna 89a, 00184 Rome, Italy}
\affil{Sony CSL - Rome, Joint Initiative CREF-SONY, Centro Ricerche Enrico Fermi, Via Panisperna 89/A, 00184, Rome, Italy}

\correspondingauthor{FSL}
\email{sylos@cref.it}

\begin{abstract}
Observations of the line-of-sight component of emitter velocities in galaxies are valuable for reconstructing their 2D velocity fields, albeit requiring certain assumptions. A common one is that radial flows can be neglected in the outer regions of galaxies, while their geometry can be deformed by a warp. A specular approach assumes that galactic discs are flat but allows for the presence of radial flows. This approach enables the reconstruction of 2D velocity maps that encompass both the transversal and radial velocity fields. Through the study of velocity fields in toy disc models, we find that the presence of warps is manifested as a dipolar correlation between the two velocity components obtained by assuming a flat disc. This shows that the analysis of angular velocity anisotropies provides an effective tool for breaking the degeneracy between warps and radial flows. We have applied these findings to the analysis of velocity fields of the galaxies from the THINGS sample and M33. Many of these galaxies exhibit such a dipolar correlation, indicating the presence of warps. However, we have found that the warp alone cannot explain all variations in the velocity field, suggesting that intrinsic perturbations are common. Furthermore, we have observed that the spatial distribution of the line-of-sight velocity dispersion may correlate with both velocity components providing independent evidence of non-trivial velocity fields. These findings offer a robust approach to reconstructing the velocity fields of galaxies, allowing us to distinguish between the presence of warps and complex velocity structures assessing their relative amplitude.
\end{abstract}

\section{Introduction}
\label{intro}

The characterization of velocity fields in external galaxies is crucial for understanding their dynamics and evolution. One way to measure the line-of-sight  (LOS)  velocity field across a galactic disc is by using stellar or gas emission.   High-resolution measurements of the LOS velocity field of the stellar component and of the ionized gas from \HII\ regions  provide maps of the inner disc (see, e.g., \cite{Erroz-Ferrer_etal_2015a}), while the distribution of gas, particularly traced by the 21cm line of \HI\ (neutral hydrogen), allows mapping of the external regions as well. In the past, observations of \HI\ were limited by low resolution. However, nowadays, there are high spectral and spatial resolution surveys available for \HI\ emission in nearby galaxies, such as the observations leading to the THINGS galaxy sample conducted by \cite{Walter_etal_2008}. 

The two-dimensional  LOS velocity field resulting from these observations must necessarily be analyzed by making certain assumptions about the geometry of the galactic disc and/or the nature of the velocity field. Indeed, the simplest model of a galactic disc, where matter moves on strictly co-planar circular orbits around the center, is rarely observed in the data. On one hand, there are perturbations due to galactic structures such as bars, spiral arms, satellites, etc. that give rise to non-circular motions. On the other hand, the disc may not be flat but may have a warped geometry, which {  should }  be more pronounced in the outermost regions {  where tidal forces are possibly larger. Indeed, warps have been directly observed  in the outer regions of edge-on disc galaxies \citep{Sancisi_1976, Reshetnikov+Combes_1998, Schwarzkopf+Dettmar_2001, Garcia-Ruiz_etal_2002, Sanchez-Saavedra_etal_2003}, typically starting at the optical radius. The warp angle, defined as the angle between the outermost detected point and the mean position of the plane of symmetry, typically does not exceed  $ 10^\circ$ \citep{Garcia-Ruiz_etal_2002, Sanchez-Saavedra_etal_2003, Reshetnikov_etal_2016, Peters_etal_2017}. }

{  In general, both deviations from planarity and circular orbits {  can be }  present simultaneously, creating a situation that makes it difficult to reconstruct the actual velocity field  of the 2D map obtained from the observations of the one component LOS velocity. In particular, one of the main limitations to the reconstruction of a galaxy's velocity field is that non-circular motions, especially radial motions, are challenging to disentangle from a non-flat geometry of a disc. In other words, there is an intrinsic degeneracy between the geometry of the disc and  radial motions. This degeneracy arises from the fact that the different methods used to reconstruct the 2D velocity field from the LOS velocity can lead to confusion between radial flows and warps.}

The projection of a disc galaxy on the sky is characterized by, at least, two orientation angles. The first, the inclination angle $i$, determines the  overall inclination of the disc with respect to the observer. The second,  the position angle (P.A.), is the  angle of the galactic major axis with respect to the North galactic pole. If the disc is warped, then one of these angles, or more commonly both, are not constant across the galactic disc, but depends on the distance from the galactic center. For this reason,  kinematics of galaxies are usually studied by analyzing the two-dimensional (2D) LOS  field using the  tilted-ring model (TRM) introduced by \cite{Warner_etal_1973,Rogstad_etal_1974}. This method allows the inclination angle $i$ and/or the P.A.  depend on radius, {  so that the disc's geometry instead of being flat is warped}. 

More specifically, the TRM assumes that a galaxy can be approximated by a sequence of rings, where each ring is characterized, at least, by its own P.A., inclination angle, and rotational velocity (other free parameters, such as the center of the ring, can be indeed introduced in the analysis of the TRM). {   Real radial motions in the TRM framework  appear  as  a change of the orientation angles  with the distance from the galactic center so that warps can suppress the detection of real radial motions (see, e.g., \cite{Schmidt_etal_2016} and references therein). That is, the TRM method may not be able to accurately reconstruct radial motions that are instead interpreted as distortions of the galactic disk. 

In its simplest implementation, the TRM assumes radial velocities to be zero. This assumption is, however, not a necessary condition in the framework of the TRM method and it is possible to consider cases in which they are different from zero.  Methods able to reconstruct radial and non-axisymmetric motions in galaxies have relied on a combination of Fourier decomposition with the TRM by considering that non-axisymmetric distortions to the planar flow can always be described by a harmonic analysis \citep{Franx_etal_1994,Schoenmakers_etal_1997,Wong_etal_2004,Simon_etal_2005,Chemin_etal_2006,Gentile_etal_2007,Trachternach_etal_2008}.  For instance,  \cite{Schoenmakers_etal_1997} introduced a tool to handle mildly non-circular flows. This method, which is based on epicycle theory, is  valid for small departures from circular orbits and can only fit mildly elliptical streamlines whereas is may give misleading results if the observed non-circular motions are not small compared to the circular orbital speed.  

In general, the problem with methods based on the TRM analysis lies in the fact that the initial fit using the TRM may suppress real radial motions by interpreting them as spurious warps in the galactic geometry. For this reason, while radial flows have been found by using the TRM already two decades ago \citep{Fraternali_etal_2001}, it is still unclear whether their full extent has been correctly  measured.

In their recent study, \cite{DiTeodoro+Peek_2021} investigated a large sample of galaxies by incorporating radial velocities directly into the TRM  fit. They employed a three-step fitting method to isolate the radial motion. Initially, they fitted the warped geometry of the disk and the circular motion based on the kinematic major axis. Subsequently, they focused on fitting the radial motion using the regions along the kinematic minor axis. This approach effectively attributed the radial distortion of the kinematic major axis solely to the presence of a warp, while disregarding distortions caused by radial flows.
However, it remains uncertain whether this method captures all radial velocities adequately due to the degeneracy with warps. Specifically, it may underestimate the radial motion on the disk, as it explicitly assigns the distortion of the kinematic major axis to the warp alone. This issue has been noted in previous researches  as well \citep{Schmidt_etal_2016}.
In a recent publication, \cite{Wang+Lilly_2023} developed a statistical test to determine whether warps or radial inflows dominate  the overall population of galaxies. Their approach relies on breaking the symmetry inherent in requiring the radial flow to be inflow rather than outflow. They found that systematic inflow effects on the 2D velocity field bear similarities to those caused by geometric warps, resulting in twisted distortions of both the kinematic major and minor axes. This similarity presents a challenge in differentiating between the two phenomena in practical scenarios. Indeed, warps in the P.A. inherently result in a rigid rotation of the velocity field. On the other hand, the effects of radial flows depend on whether they are isotropic and have a constant amplitude. Therefore, the degeneracy between radial flows in velocity fields and warps cannot be generally disentangled using the  TRM.

To summarize, distinguishing between warps and radial flows remains a challenging problem, with potential confusion between the two. Despite numerous attempts, resolving this degeneracy continues to pose a challenge. This is a significant open question as understanding the properties of radial flows, as well as estimating local gradients in the velocity field, is crucial for assessing the stability and dynamics of galactic discs.
It is important to emphasize that methods based on the TRM are  designed to measure  velocity profiles averaged over rings, and do not inherently enable the detection of angular velocity anisotropies. However, as discussed hereafter, the method presented in our paper offers a unique capability to precisely differentiate between warps and radial flows by analyzing angular velocity anisotropies.

Recently, the velocity ring model (VRM) was introduced \citep{SylosLabini_etal_2023}. This model shares the assumption of a flat galactic disc with other existing methods \cite{Barnes+Sellwood_2003,Spekkens+Sellwood_2007,Sellwood+Sanchez_2010,Sellwood_etal_2021}. However, the VRM offers the ability to characterize non-axisymmetric and heterogeneous velocity fields of any form. Within this framework, both radial and transverse velocity components can exhibit arbitrarily complex spatial anisotropic patterns. The VRM method is able to reconstruct coarse-grained two-dimensional maps, allowing for the investigation of spatial anisotropies in angular sectors at various distances from the galactic center. { \cite{SylosLabini_etal_2023} have applied the VRM reconstruction method to a sub-sample of the  THINGS galaxy sample \citep{Walter_etal_2008} whereas 
\cite{SylosLabini_etal_2025} have studied a sub-sample of the LITTLE THINGS galaxy sample \citep{Hunter_etal_2012}.}

 While this approach is particularly well-suited for examining optical discs, where warps are uncommon, its applicability to peripheral regions is uncertain. This is because, as mentioned above, a warp's signal may be mistakenly interpreted as a radial velocity flow when analyzing the LOS velocity map using the VRM. In this work, we address this limitation by discussing how the spatial velocity anisotropy patterns in the coarse-grained two-dimensional maps obtained through the VRM method can be used to characterize warps. By carefully examining these velocity anisotropy patterns, it becomes possible to differentiate between the effects of warps and radial velocity flows within the VRM framework. This provides a means to accurately identify and characterize the presence of warps, even in peripheral regions of galaxies.

 More specifically, in this work we present  evidence that examining the spatial properties of velocity anisotropies, obtained in the VRM  by assuming a flat disc, offers a distinct and definitive approach to detecting the potential presence of a warp. By analyzing these velocity anisotropies, we can differentiate between the effects of warps and other intrinsic velocity perturbations. Importantly, our analysis enables us to assess the amplitude of intrinsic velocity perturbations, regardless of whether a warp is actually present.  The key result is that if the disc is warped, a clear  dipolar correlation between the two velocity components is observed when assuming a flat disc and using the VRM to reconstruct the velocity field. This finding is significant because if the disc is intrinsically warped, applying VRM, which assumes a flat disc, allows us to break the degeneracy between geometric deformations of the disc and radial motions. 

In light of this result, we re-analyzed the fields of galaxies from the THINGS sample, as previously considered in \cite{SylosLabini_etal_2023}, and that of M33 \citep{Corbelli_Schneider_1997,Corbelli_etal_2014}.
Additionally, we explored the correlation between the velocity field reconstructed by VRM and the velocity dispersion field  (i.e., the second moment of the velocity) that is also available for the galaxies considered. This approach allows us to assess not only whether a galaxy is warped but also it permits  to characterize the properties of  intrinsic perturbations to the velocity field as well as their correlation with the velocity dispersion field.

The paper is organized as follows: In Section \ref{sect:methods}, we briefly review the main features of VRM and introduce the estimators of the correlations between the velocity fields and the velocity dispersion field. Then, in Section \ref{sect:warps}, we present results obtained by considering toy galactic disc models that illustrate how a warp corresponds to specific spatial correlations between the radial and transversal velocity fields obtained by the VRM method. The analysis of a few illustrative examples of  galaxies is presented in Section \ref{sect:things}, while in the Appendix there are reported the analyses for all the remaining galaxies in our sample. Finally, Section \ref{sect:discussion} contains our conclusions.



\section{Methods} 
\label{sect:methods} 

After having briefly reviewed  the main characteristics of the VRM, we discuss the estimators  of the spatial correlations between the various fields. 

\subsection{The Velocity Ring Model} 
\label{sect:vrm} 

{  The VRM is a method that,   by assuming that galactic disc is flat, allows for the reconstruction of transverse and radial velocity component 2D maps from observations of the LOS velocity map of an external galaxy \citep{SylosLabini_etal_2023}.} Let us briefly recall its essential features.  The map of an external galaxy consists of the angular coordinates $(r,\phi)$ in the projected image of the galaxy onto the plane of the sky, along with the LOS  component of the velocity $v_{\text{los}}(r,\phi)$. These can be transformed in
\be
\label{eq:vlos} 
v_{\text{los}}  = \left[v_t  \cos(\theta) + v_r  \sin(\theta) \right] \sin(i)  \;, 
\ee
where the transversal and radial velocity components, $v_t=v_t(R, \theta)$ and $v_r=v_r(R, \theta)$, refer to the velocity of the galaxy in the plane of the galaxy, where $R$ and $\theta$ are the polar coordinates  in the plane of the galaxy and $i$ is the inclination angle (the other orientation angle, the P.A., enters into the transformation from  $(r,\phi)$ to $(R, \theta$)). In the intrinsic coordinates of the galaxy, assuming a flat disc, it is possible to divide the disc into $N_r$ rings, all having the same inclination angle and P.A. The velocity field within each ring can be decomposed into a radial component $v_r = v_r(R)$ and a transverse component $v_t = v_t(R)$. By inverting Eq. \ref{eq:vlos}, the VRM provides the values of $v_r(R)$ and $v_t(R)$, { the profiles of the two velocity components averaged over a ring, in each ring.}
 
{  It is possible to extend the VRM to consider non-axisymmetric motions induced by galactic structures such as the bar, spiral arms, and others.  This is done by splitting each ring into $N_a$ arcs, each characterized by a different radial and transversal velocity. In this way we introduce also a dependency on the angular coordinate $\theta$, meaning that $v_t=v_t(R, \theta)$, $v_r=v_r(R, \theta)$, again under the assumption that the inclination angle and  the P.A. are constant across the disc.  The VRM divides the galaxy image into a grid of $N_{\text{cells}}$ cells, determined by the product of the number of rings and the number of arcs, i.e. $N_{\text{cells}}=N_r \times N_a$ \footnote{  
Hereafter we use $N_r=50$ rings and $N_a=32$ arcs. Whereas we have studied the effect of resolution by varying the number of arcs as $N_a=1,2,4,8,16$ we do not discuss these maps and we refer the interested reader to  \cite{SylosLabini_etal_2023} for a more detailed discussion of the method and for its testing against a number of toy galaxy models as well as for the results of the galaxies of the THINGS sample.} . Within the $i^{th}$ cell, the VRM provides the transverse velocity component, $v_t^i$, and the radial velocity component, $v_r^i$. 

Note that the VRM does not provide uncertainties for the velocity components in each cell. To control the effect of changing the resolution of the coarse-grained reconstructed map, one approach is to vary the number of arcs while keeping the number of rings fixed. By doing so, it is possible to ensure that the velocity field in a particular angular region, which is covered by at least one cell, yields consistent velocity components values regardless of the number of arcs. This consistency indicates that the maps of transverse and radial velocities have converged to the same spatial distribution of anisotropies.}

From a quantitative perspective, it is possible to analyze the convergence of several moments (see \cite{SylosLabini_etal_2023} for details) that measure the velocity field in a given region  by varying the resolution of the coarse-grained map. This involves, for instance, adjusting the number of arcs while keeping the number of rings fixed. The convergence of these moments, i.e., the fact that the reconstructed velocity field does not depend on the resolution used by the VRM, is crucial for assessing its reliability. It ensures that the noise introduced by the VRM reconstruction, unavoidably present,  does not surpass the signal. Tests discussed in \cite{SylosLabini_etal_2023} indicate that the moments generally converge when the velocity fluctuations field is sufficiently smooth. However, in certain cases, the VRM may struggle to accurately reconstruct a complex input field, particularly when the perturbation field exhibits rapid  changes in localized patches. Consequently, it is not possible to straightforwardly estimate the amplitude of velocity perturbations, e.g. the radial and transversal dispersion in function of radius, as the reconstruction method can in some cases significantly impact the results.
{ We refer the interested reader to \citep{SylosLabini_etal_2023} for a detailed discussion of the performance of the VRM reconstruction method, based on tests made on artificial disk models.}

The effects introduced by geometric deformations such as a galactic warp have distinct characteristics and can generate a genuine signal rather than noise. These deformations result in a slowly varying velocity field that{, as we discuss in detail below,} can be effectively identified by the VRM. In the case of a geometric deformation like a warp, the moments of the velocity field still converge as long as the noise in the measurements is smaller than the signal induced by the deformation. This means that if there is no convergence when varying the resolution, it implies that the noise in the measurements is larger than the signal, and the presence of a warp or other geometric deformation cannot be confirmed. On the other hand, if  convergence is observed, it suggests that a geometric deformation such as a warp may be present. However, it is important to note that convergence alone does not provide definitive proof of a warp. As we are going to discuss in details in what follows, other factors and observations need to be considered to establish the presence and characteristics of a warp in a galactic system.

A crucial aspect when comparing the results of the TRM and the  VRM is to consider the probability density function of their respective residuals. This refers to the difference between the observed velocity field and the reconstructed velocity field by each model. In the analysis presented in \cite{SylosLabini_etal_2023}, it is shown that when the number of free parameters in both models is the same, there are no statistically significant differences between the two methods. Thus, relying solely on the statistical analysis of the residuals cannot yield a definitive conclusion regarding the superiority of one method over the other or provide conclusive insights into the properties of the velocity field or the geometry of the disk. To gain a more comprehensive understanding of the velocity field and the underlying disk geometry, the analysis developed in this work proves to be highly valuable.


\subsection{Cross correlation}
\label{sect:crosscorr} 

Once we have reconstructed the tangential $v_t(R,\theta)$ and radial $v_r(R,\theta)$ velocity fields, we cross-correlate them among each other and, in the case of real galaxies, 
with the velocity dispersion field $\sigma(R,\theta)$ (see below for details).  To quantify the correlation between $v_r(R,\theta)$, $v_t(R,\theta)$ and $\sigma(R,\theta)$ we consider the Pearson correlation  coefficient. This, for two generic fields $f(R,\theta)$ and  $g(R,\theta)$ (where $f,g= v_t, v_t, \sigma$) can be defined    as
(see, e.g., \cite{Numerical_Recepies})
\be 
\label{eq:corrcoff} 
r_{f g}   = \frac{1}{N_{\text{cells}}-1} \sum_{i=1}^{N_{\text{cells}}} r_{f g}^i(R,\theta)
\ee
where 
\be
\label{eq:rxy} 
 r_{f g }^i(R,\theta) = \left( \frac{f^i(R,\theta)-\overline{f}}{s_{f}} \right)\left( \frac{g^i(R,\theta)-\overline{g}}{s_{g}} \right)
\ee
and we denoted with $f^i(R,\theta)$ the value of the field in the $i^{th}$ cell with $i=1,...,N_{\text{cells}}$. In addition, we have defined 
\be
\label{eq:ave} 
\overline{f} =  \frac{1}{N_{\text{cells}}} \sum_{i=1}^{N_{\text{cells}}} f^i(R,\theta)
\ee
 (the same for $\overline{g}$) and 
 \be
\label{eq:sigma}
s_{f} = \sqrt{ \frac{1}{N_{\text{cells}}-1} \sum_{i=1}^{N_{\text{cells}}}  (f^i(R,\theta) - \overline{f})^2} \:, 
\ee
(the same for $s_{g}$).

We also employ Spearman correlation coefficients (see, e.g., \cite{Numerical_Recepies}), which are defined as the Pearson correlation coefficients between the rank variables. In other words, instead of using the actual values of the variables 
$f^i(R,\theta)$ and $g^i(R,\theta)$, the correlations are computed based on their ranks. Through extensive testing using both toy models and real galaxies, we have chosen to present results obtained with the Spearman correlation coefficients. While the results obtained with both correlation coefficients are generally similar, the advantage of using the Spearman coefficient is that it is not influenced by factors such as the wide variation in velocity dispersion observed in real galaxies. This prevents biases in the estimation of averages and variances towards larger values.

While the overall values of the correlation coefficients, as given by Eq. \ref{eq:corrcoff}, provide a measure of the strength and direction of the correlations between the velocity components and among them and the velocity dispersion, their significance lies in their spatial distribution across the galactic discs. The presence of spatial correlations indicates that the relationships between these quantities extend beyond localized regions and encompass a larger portion of the galactic disc. For this reason, we specifically focused on analyzing the spatial distribution maps of the correlation coefficients $r_{v_t v_r}(R,\theta)$, $r_{\sigma v_r}(R,\theta)$, and $r_{\sigma v_t}(R,\theta)$, as these maps have the potential to reveal spatial correlations within the galactic discs. It is through these correlations that we can gain insights into the presence of warps or of complex velocity structures and their interplay. This information is crucial in understanding the underlying dynamics and physical processes operating within the galaxies.


\section{Effects of warps on the velocity ring model: tests with toy models} 
\label{sect:warps} 

In this section we illustrate the procedure that allows us to conclude whether or not a warp is present in the outer regions of a disc. To this aim, we generate artificial disc models with properties defined by the radial behaviors of the averaged velocity profiles  $v_t(R)$ and $v_r(R)$ and of the orientation angles $i(R)$ and $\phi_0(R)$. Each toy disc model is then projected onto the plane of the sky to obtain a LOS velocity map $v_{los}(r,\phi)$ that mimics real observations. The properties of this map are then measured using both the VRM and TRM methods. We provide a few controlled and simple examples that illustrate that by analyzing the results obtained from both the VRM and TRM methods, it becomes possible to disentangle the geometrical effects arising from the warp and the radial-dependent features of the velocity field.

The reconstruction process with the VRM enables us to obtain coarse-grained two-dimensional velocity maps that encompass both the transverse and radial velocity fields. However, due to the incorrect assumption of a flat disc encoded by the VRM, the geometric deformations associated with the warp introduce artifacts in both velocity component maps. We will denote these artifacts as extrinsic velocity anisotropies to differentiate them from intrinsic ones, which arise from real velocity perturbations in the disc. More importantly, we show that a warp introduces a specific spatial correlation between these two velocity components, and that such a spatial correlation serves as an indicator of the warp's existence. By measuring such correlation one can both asses the presence of the warp and recover the underlying intrinsic radial velocity profiles of the disc. Indeed, we also show that if the disc is flat and characterized by an isotropic radial flow such spatial correlation  vanishes. On the other hand, in that case, by using the  TRM, the isotropic radial flow appears as a geometric deformation, highlighting the intrinsic degeneracy between warps and radial flows in the context of the  TRM.

\subsection{Analysis of a simple toy warped disc model} 

\begin{figure*}
\gridline{\fig{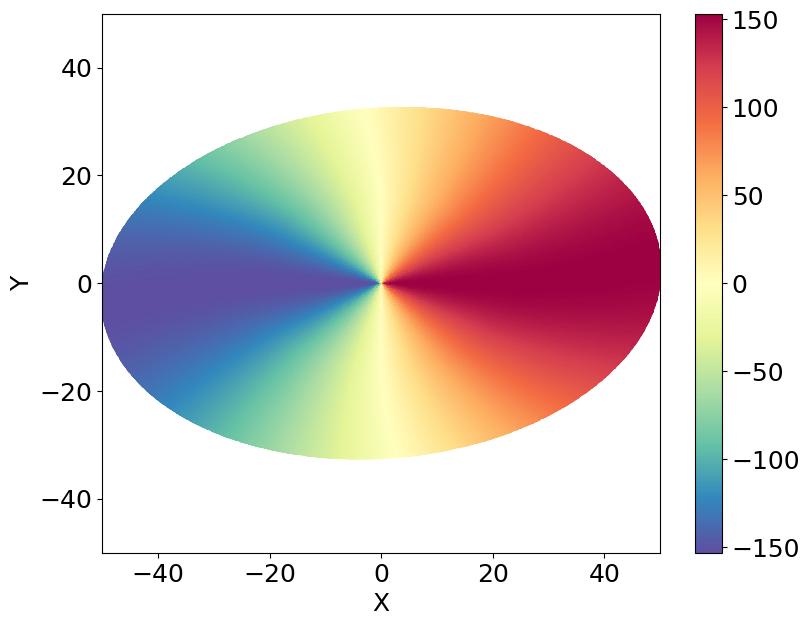}{0.3\textwidth}{(a)}
              \fig{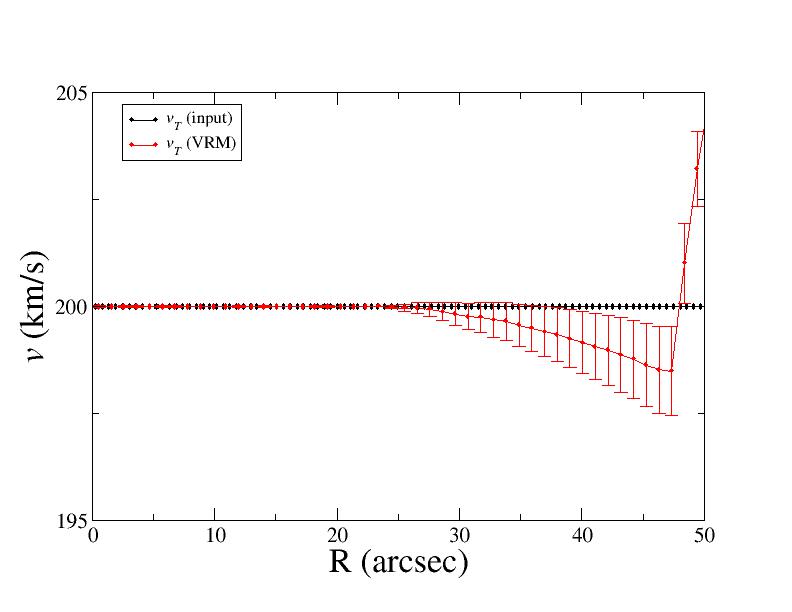}{0.3\textwidth}{(b)}
               \fig{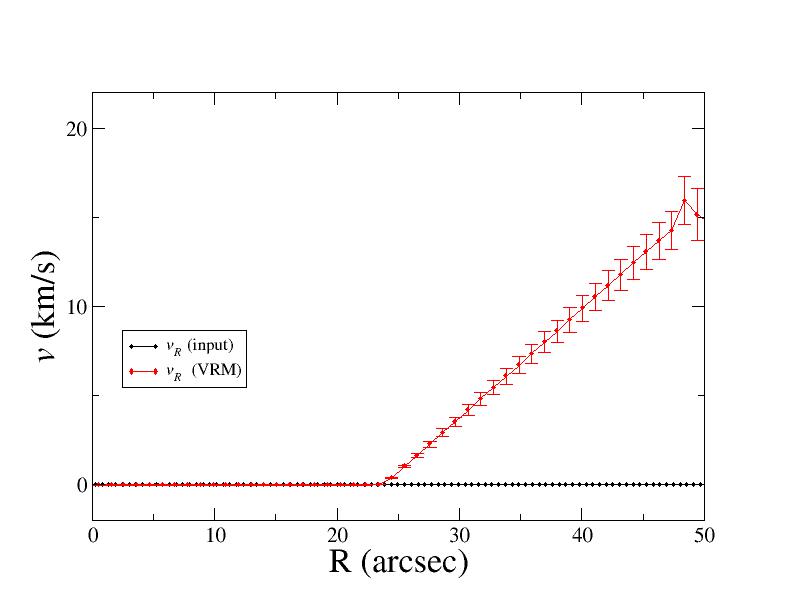}{0.3\textwidth}{(c)}}
\gridline{\fig{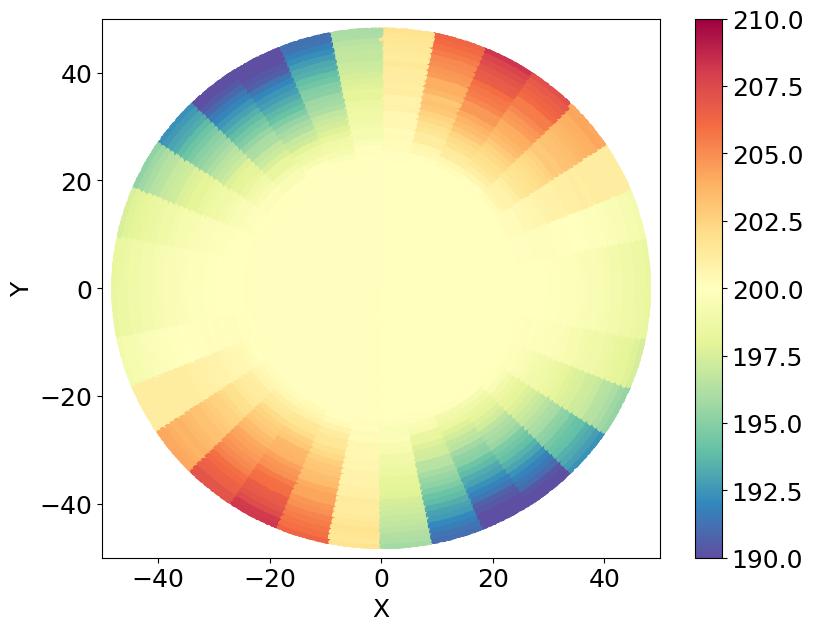}{0.3\textwidth}{(d)}
              \fig{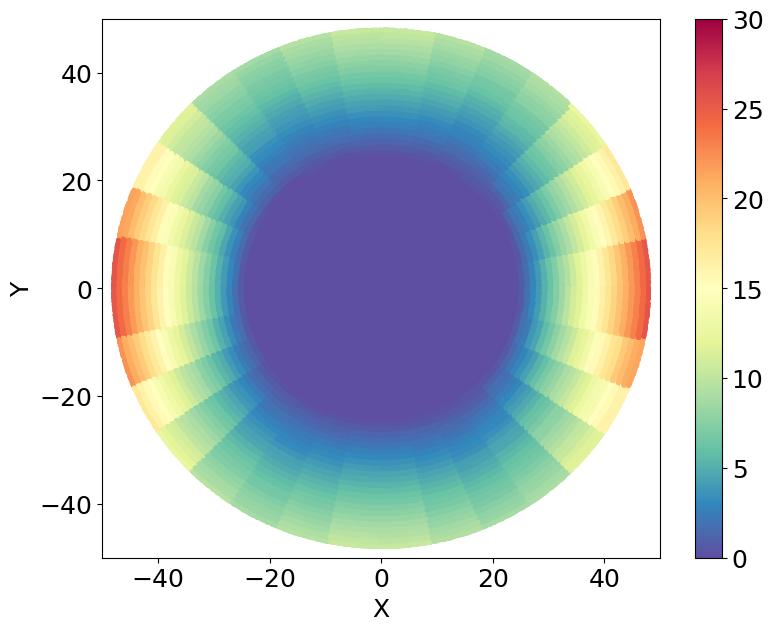}{0.3\textwidth}{(e)}
              \fig{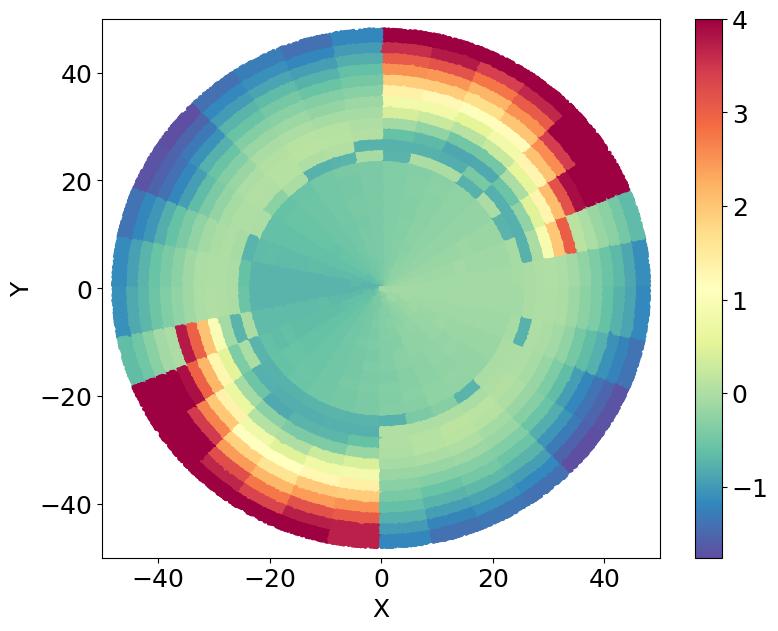}{0.3\textwidth}{(f)}}
\gridline{\fig{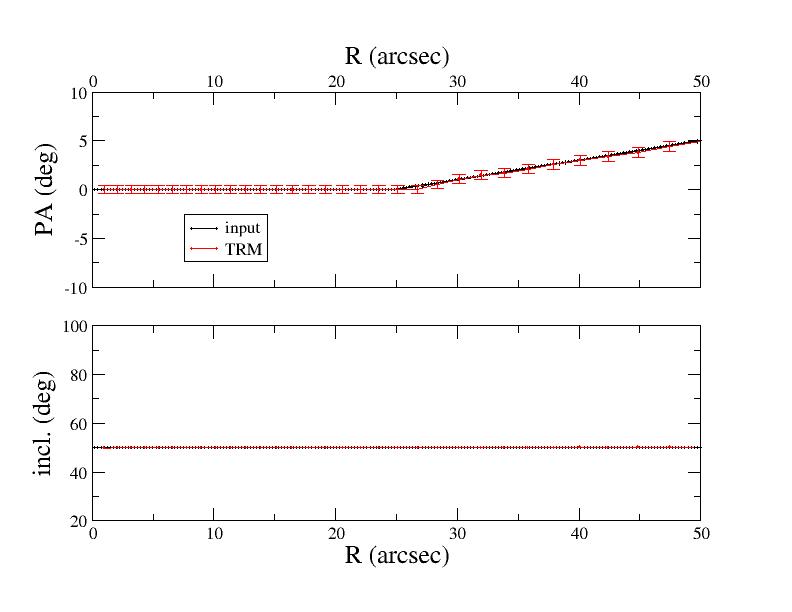}{0.3\textwidth}{(g)} 
              \fig{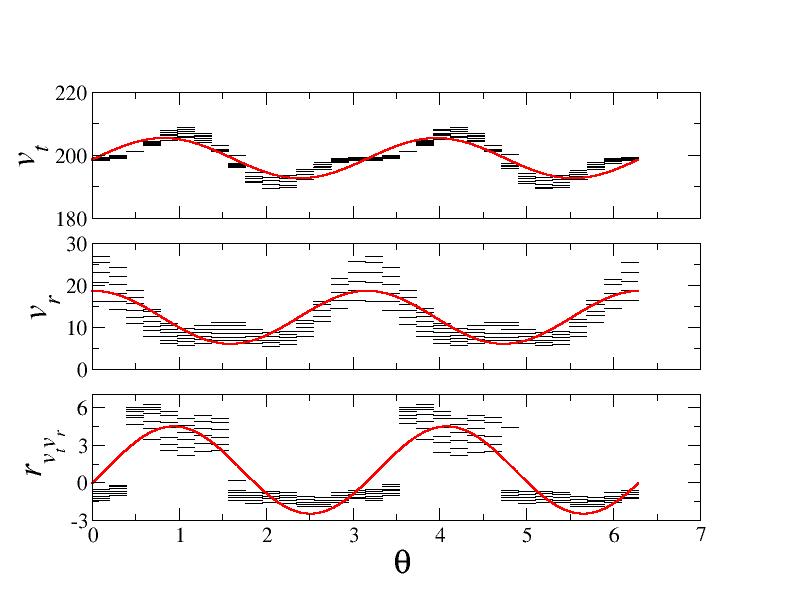}{0.3\textwidth}{(h)}}
\caption{Results for a toy disc model; the panels respectively show: 
(a)  the LOS velocity map projected on the plane of the sky $v_{los}(r,\phi)$ that, mimicking real observations, represent the input data of the analysis;
{ (b)  the VRM-reconstructed  velocity transversal profile (VRM) in function of the distance from the center;
(c)  the same as (b) but for the radial component; }
(d)  the  transversal velocity $v_t(R,\theta)$ map in the plane of the toy disc model reconstructed using the VRM method  with $N_r=50$ and $N_a=32$;
(e)  the  same for the  radial velocity $v_t(R,\theta)$ map in the plane of the toy disc model ;
(f)   the rank correlation coefficient map $r_{v_t v_r} (R,\theta)$ in the plane of the toy disc model ;
(g)  the P.A. (upper panel) and inclination angle (bottom panel) of the input model (these are given by the TRM); 
(h) {   the velocity transversal component, the velocity radial component  and rank velocity correlation coefficient in the outer part (i.e., for $r > 0.75 R_{max}=50'') $ of the warped disc in function of the polar angle $\theta$. }
} 
\label{Mod_1} 
\end{figure*}

We first  consider the simple toy disc model depicted in Figure \ref{Mod_1} as an illustrative example. This model is constructed as follows: within a disk of radius 50", the transverse velocity field  remains constant at 200 km s$^{-1}$, while the radial velocity field is null. Additionally, the disk is warped in such a way that the P.A. linearly changes by $\Delta$ P.A. = $5^\circ$ between a radius of  25" and a radius of 50", while, for simplicity, the inclination angle remains constant. The observer is positioned far enough from the disk to view it with an inclination of $i=50^\circ$. 
The map of the LOS velocity  in the plane of the sky $v_{los}(r,\phi)$  is shown in panel (a) of  Fig.\ref{Mod_1}: this map represents the input data which mimics observations. The input orientation angles  used to generate the warped disk are reported in panel (g) of  Fig.\ref{Mod_1}:  these {coincide with the orientation angles measured by TRM given that we have assumed that  $v_r=0$.

The VRM correctly estimates the velocity components profiles (see  panels (b) and  (c) of  Fig.\ref{Mod_1}) only  in the region where the disk is flat, i.e., $v_t=200$ km s$^{-1}$  and $v_r=0$ km s$^{-1}$  for $R \le 25$".  In the outer disc region, i.e., for 25"$<R<$50", both $v_t$ and $v_r$ show  variations  that, locally, can be as large as 30 km s$^{-1}$ , even though on average they do not exceed $\sim 20$ km s$^{-1}$. Such variations are artifacts resulting from the flat disk assumption encoded in the VRM method, which is at odds with the properties of the given toy disk model.   

The radial and transversal velocity coarse-grained maps reconstructed by the VRM (panels (d) and (e) of Figure \ref{Mod_1}) show angular anisotropies that smoothly vary with the angular coordinates $\theta$: these are also artifacts due to the warp as such variations were not encoded in the input velocity field of the toy disc model. Actually, from a simple visual inspection we can conclude that  these maps present  dipolar oscillations in function of the polar angle $\theta$, although with different phase for $v_r$ and $v_t$. The same kind of dipolar oscillation is present in the map of the correlation coefficient  $r_{v_t v_r}(R,\theta)$ (panel (f) of Fig. \ref{Mod_1}). 
More quantitatively,  we find that in the outermost regions of the disk, i.e., 35" $<R<$ 50", $v_t, v_r$ and $r_{v_t v_r}$ { show a  dipolar modulation of the type }
\be
\label{eq:oscil} 
f(\theta) = f_0 + A \sin( \omega \theta + \theta_0) 
\ee
where $\omega=2$ corresponds to  the dipolar oscillation's frequency, $\theta_0$ is the phase,  $A$ is its amplitude and $f_0$ is a plateau. In the panel (h) of Fig. \ref{Mod_1}  we have plotted the best fits to  $v_t, v_r$ and $r_{v_t v_r}$  with Eq.\ref{eq:oscil} by considering $f_0, A, \theta_0$ free parameters while the frequency has been taken fixed   and equal to  $\omega=2$. It is important to note that the function in Eq.\ref{eq:oscil} is not intended to capture the entire behavior of $v_r$, $v_t$, and $r_{v_t v_r}$. Its purpose is solely to identify the periodicity of the peaks, if any is present in the data.  
As reference,  black lines in panel (g) of Fig. \ref{Mod_1} show the  behavior in function of the polar angle $\theta$ of the velocity transversal and radial components  and of the rank velocity correlation coefficient in the outer part of a non-warped disc: in this case the dipolar modulation obviously is not present. 

 From this simple test we can conclude that the  presence of a warp in the disk has  a significant impact on the transverse and radial velocity components reconstructed by the VRM. Even though the VRM fails in reconstructing the correct input properties of the toy model the analysis of the rank correlation coefficient allows one to conclude that a warp is indeed present in the system. 
Indeed, on  the one hand, the VRM rings averaged profiles of $v_t(R)$ and $v_r(R)$ present artifacts due to the presence of a warp. On the other hand, the  angular dependence of $v_t(R, \theta)$, $v_r(R, \theta)$, and $r_{v_t v_r}(R, \theta)$ allows us to conclude that a warp is indeed present in the system. This is the key information provided by such analysis: only if we can unambiguously conclude that a warp is present are we justified in using the TRM to estimate the velocity profiles (even though, as discussed below, this situation does not necessarily imply that the results of the TRM are correct). Otherwise, reliable estimations of the profiles are given by the VRM. In the case of the model considered in Fig.\ref{Mod_1} the analysis of angular maps allows us to conclude that only in the inner part of the system the VRM provide with the correct information while in the outer regions its results are affected by the warp. 
{ It is worth mentioning that the sign and amplitude of the rank correlation coefficient is related to the shape of the warp, i.e., to the deformation introduced by the radial dependence of the two orientation angles. On the other hand, the sign of the radial velocity component is physically associated with whether there is an inflow or outflow of matter.
}

The toy disc model discussed in Fig. \ref{Mod_1} is very simple and encodes the key properties of our method: more complex toy disc models are presented { below when we discuss  the analysis of individual galaxies.} Fig. \ref{fig:sigma_PA} illustrates the  radial velocity dispersion  in  rings for different values of the variation of the P.A  while keeping the inclination angle fixed. This trend gives a useful order of magnitude of the velocity variation. However,  if there is also a variation in the inclination angle in addition to the P.A., the relationship is no longer approximately linear in $R$ as it is in this case but in the outermost rings, and requires further investigation.  This is the motivation, that we explore in the next section, to generate toy models of the observed galaxies with the more complex geometric deformation resulting from the TRM analysis of each object.
\begin{figure}  
\includegraphics[width=7.5cm,angle=0]{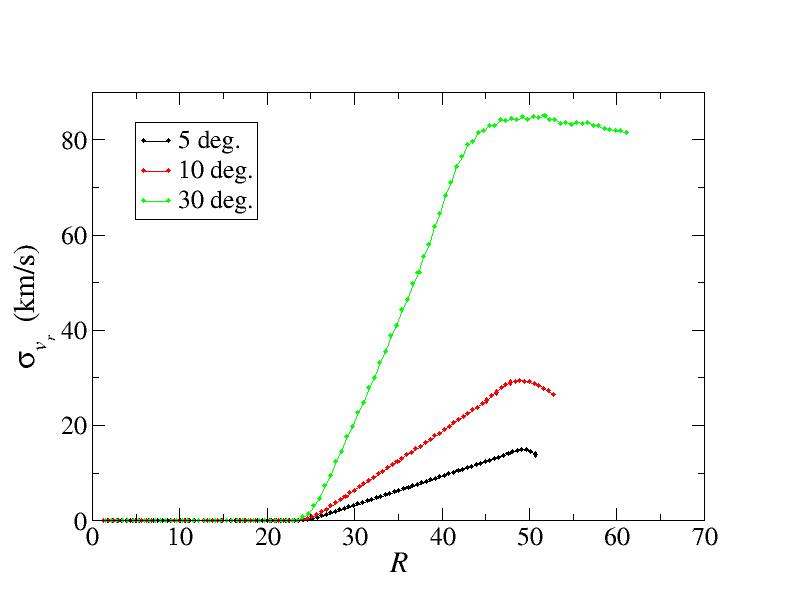}
\caption{Radial velocity dispersion in function of the radius of the ring and for toy models with different variation of the P.A. (see labels)} 
\label{fig:sigma_PA} 
\end{figure}


Let us now consider a toy disc model without a warp but with an isotropic radial flow for  $R>30''$ , with an amplitude that linearly increases as a function of radial distance. 
Such radial flow corresponds to an overall expansion (or contraction if directed inward) of the disk, which has little physical justification. In this case (see Fig.\ref{ToyMod2}), the  TRM finds a  P.A. that depends on the radial distance instead of being constant as the input values. This artifact is due to the effect of the non-zero radial motions, which are (incorrectly)  interpreted by the  TRM as indicative of a warped disc. The inclination angle is correctly reconstructed, as is the rotational velocity. On the other hand, in this case the VRM accurately reconstructs the mean radial profiles  of  $v_r$ and $v_t$. In addition, the analysis of the coarse-grained VRM-reconstructed $v_r$ and $v_t$ maps shows that their polar dependence  in the outer shells does not show any dipolar signal. The same occurs for the velocity rank correlation coefficient.  Thus in this case we can conclude that the TRM fails to detect the correct properties of the input system because of its underlying assumption that radial motions are null.  Instead, the absence of dipolar correlations in the velocity rank correlation coefficient  $r_{v_t v_r}(R, \theta)$ allows us to conclude that the disk is not warped and the VRM correctly estimate the intrinsic properties of the toy disk model.
\begin{figure*}
\gridline{\fig{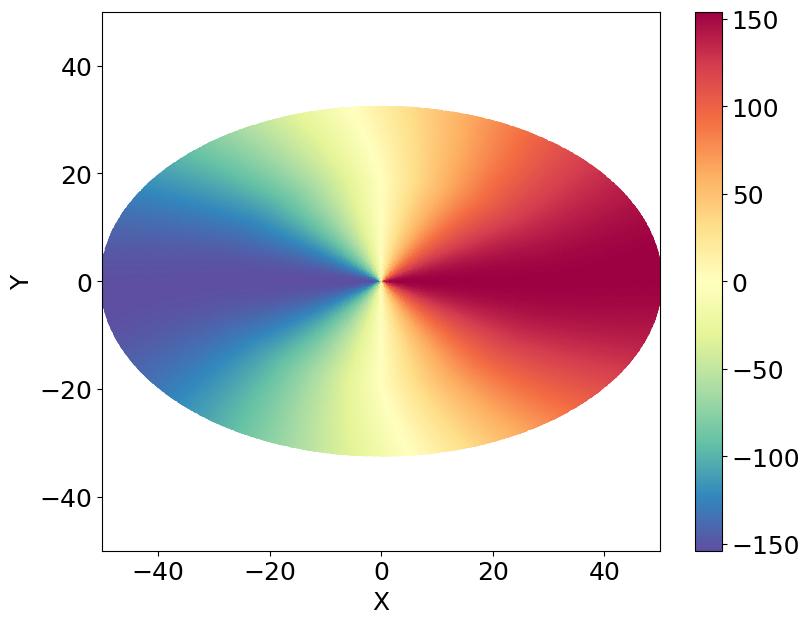}{0.3\textwidth}{(a)}
              \fig{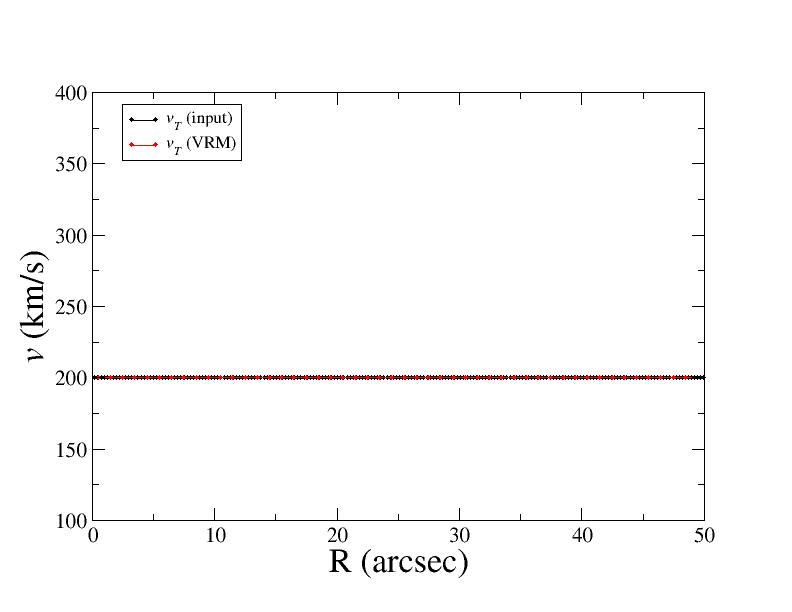}{0.3\textwidth}{(b)}
               \fig{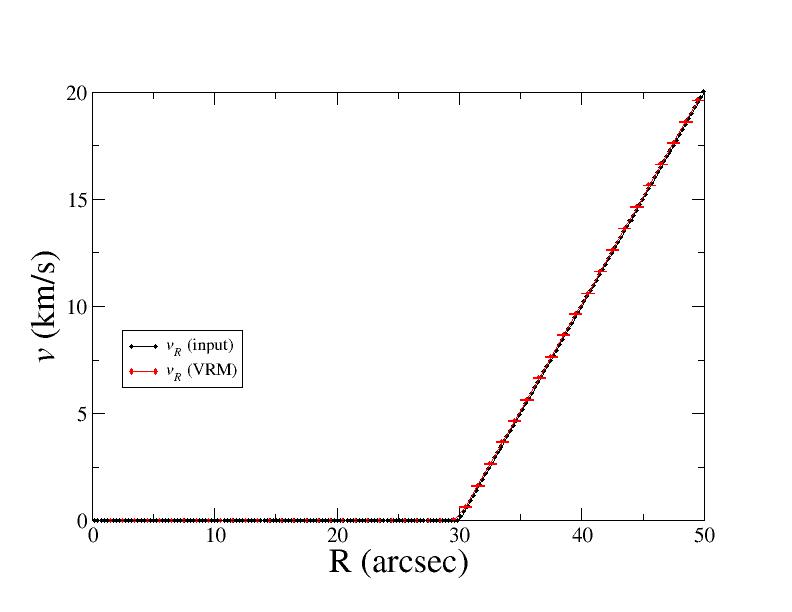}{0.3\textwidth}{(c)}}
\gridline{\fig{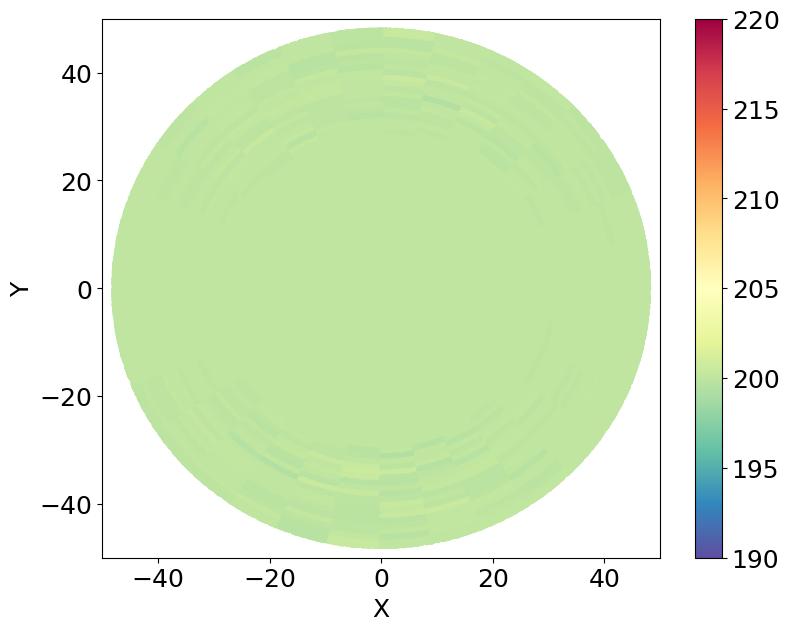}{0.3\textwidth}{(d)}
              \fig{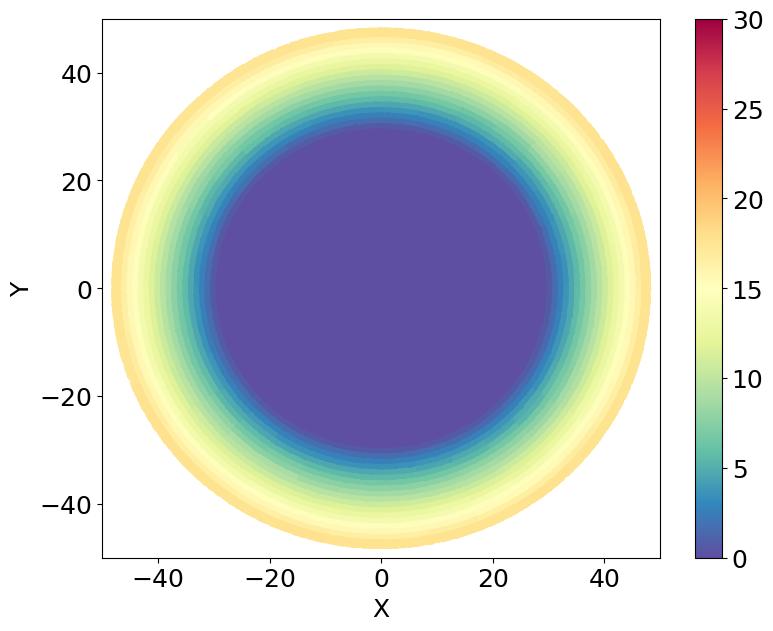}{0.3\textwidth}{(e)}
              \fig{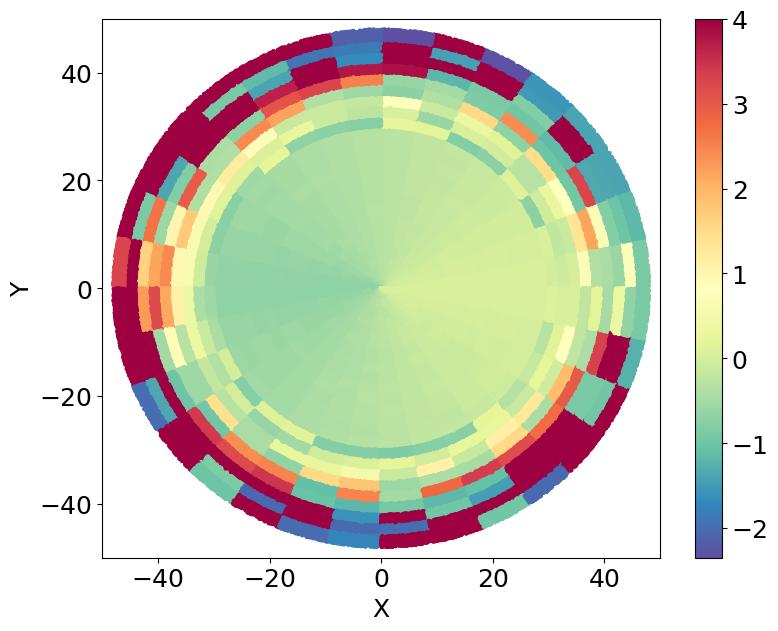}{0.3\textwidth}{(f)}}
\gridline{\fig{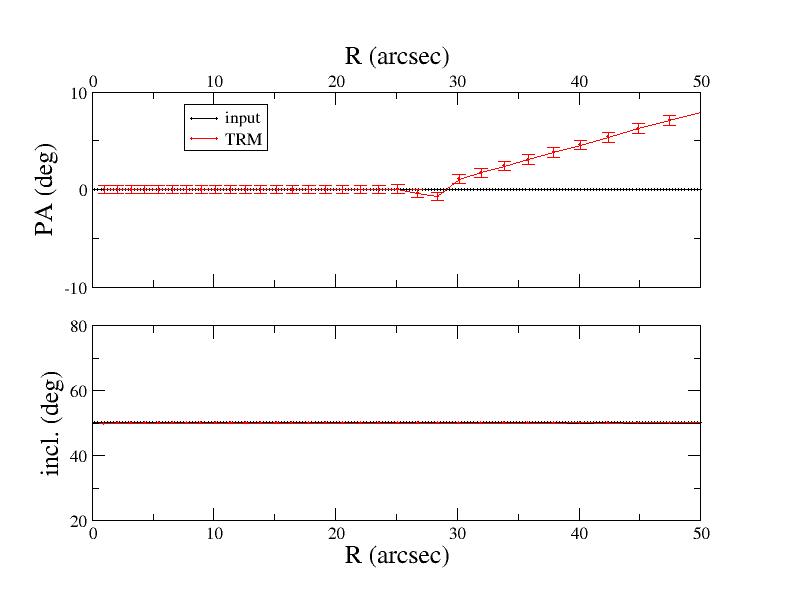}{0.3\textwidth}{(g)} 
              \fig{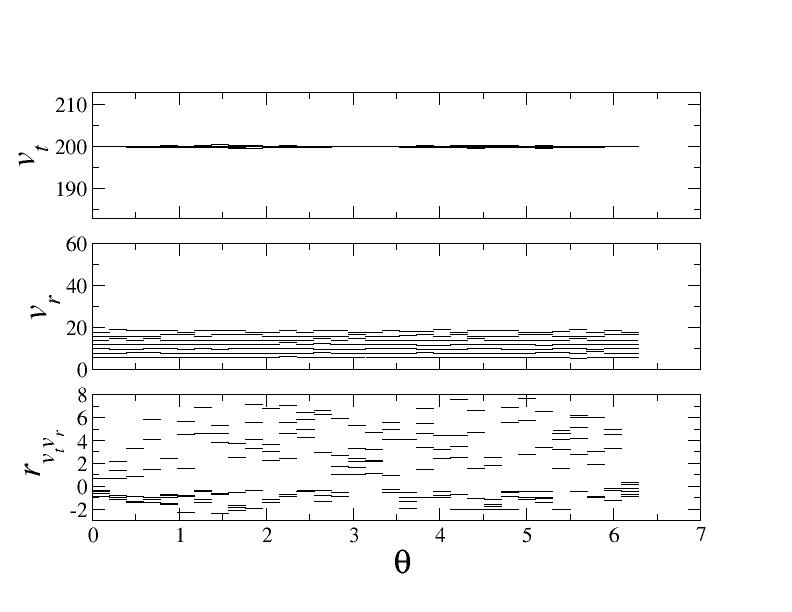}{0.3\textwidth}{(h)}}
\caption{ As Fig.\ref{Mod_1}  but for a toy disc model without warp but with an isotropic radial dependent velocity field} 
\label{ToyMod2}
\end{figure*}


{ We then } consider a toy warped-disc model with an isotropic, radially dependent velocity field:   the warp is the same as that in the case shown in Fig.\ref{Mod_1} and the  isotropic radial flow like the disc model shown in Fig.\ref{ToyMod2}. The TRM and VRM display the following behaviors: 
\begin{itemize} 

\item For the effect of the isotropic radial flow the TRM finds a P.A. that differs from the input value (while the inclination angle is the same --- see panel (g) of Fig.\ref{ToyMod3}). 
For this reason, the rotational velocity measured by the  TRM coincides with the input value, whereas the TRM clearly fails to estimate the radial velocity as it assumes that is null. 

\item  The VRM finds the correct values of the mean $v_t$ and $v_r$ in the inner regions where no warp is present. However, in the outer regions, both the VRM-estimated $v_r$ and $v_t$ differ from the input values because of the presence of the warp (see panels (b,c) of Fig.\ref{ToyMod3}).  Note that $v_t$ differs by $\sim$ 3 \% w.r.t. to the input value while the $v_r$ up to $\sim 50\%$.  

\item The warp is unambiguously identified  by considering  the coarse-grained VRM-reconstructed $v_r$ and $v_t$ maps, as well as by the velocity rank correlation coefficient, all of which display a dipolar modulation (see panels (d,e,f) of Fig.\ref{ToyMod3}).
\end{itemize} 

\begin{figure*}
\gridline{\fig{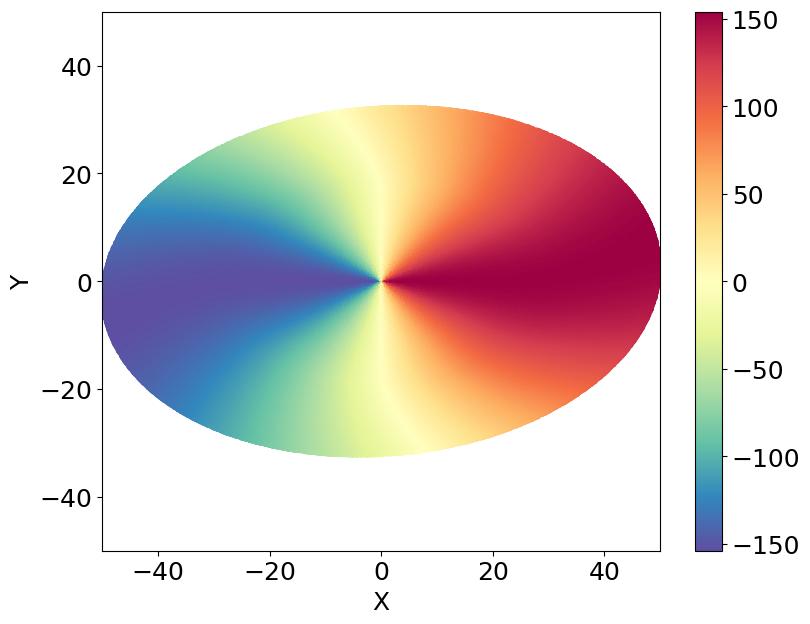}{0.3\textwidth}{(a)}
              \fig{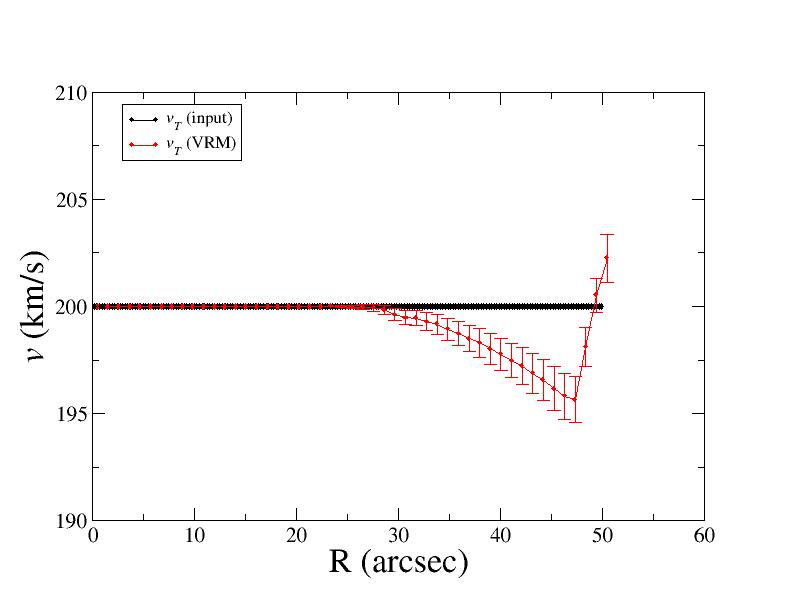}{0.3\textwidth}{(b)}
               \fig{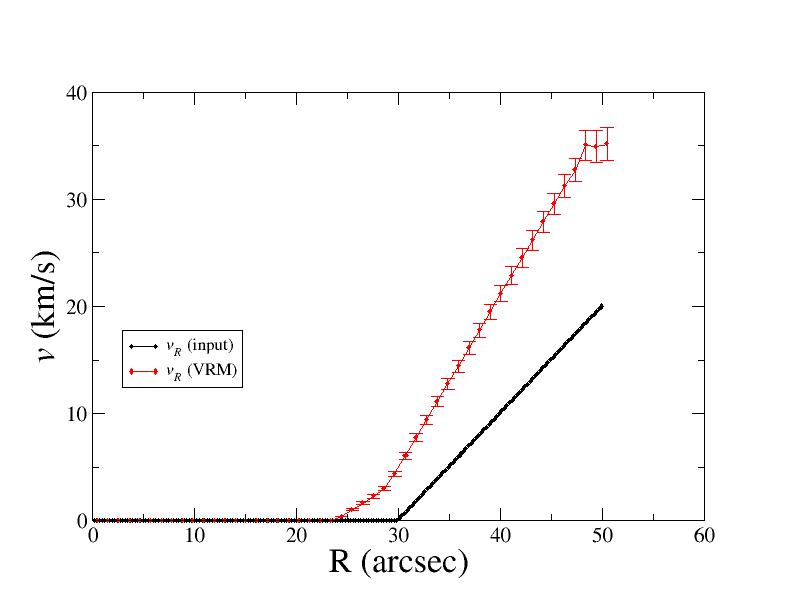}{0.3\textwidth}{(c)}}
\gridline{\fig{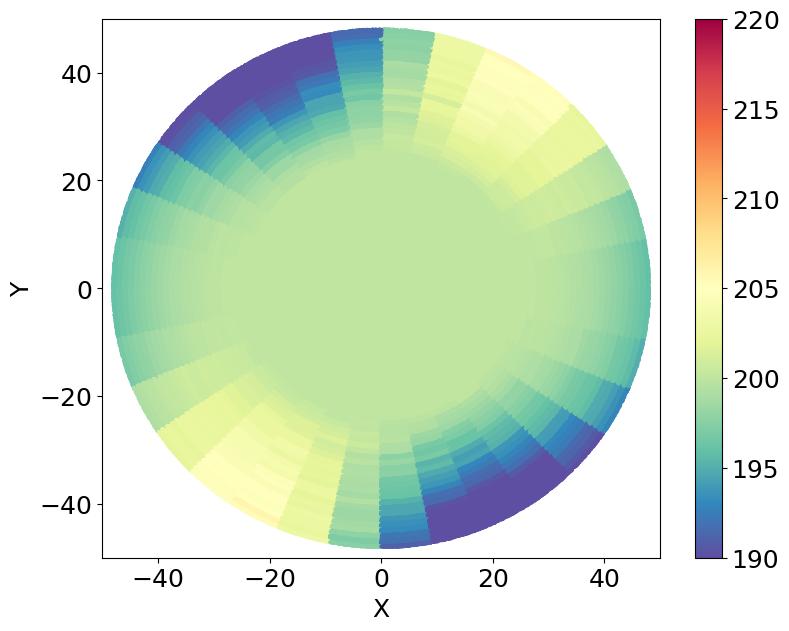}{0.3\textwidth}{(d)}
              \fig{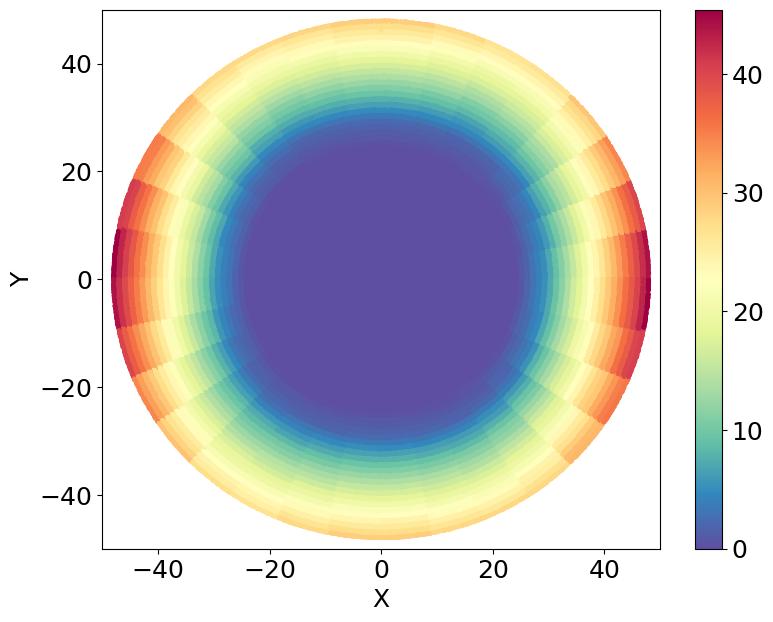}{0.3\textwidth}{(e)}
              \fig{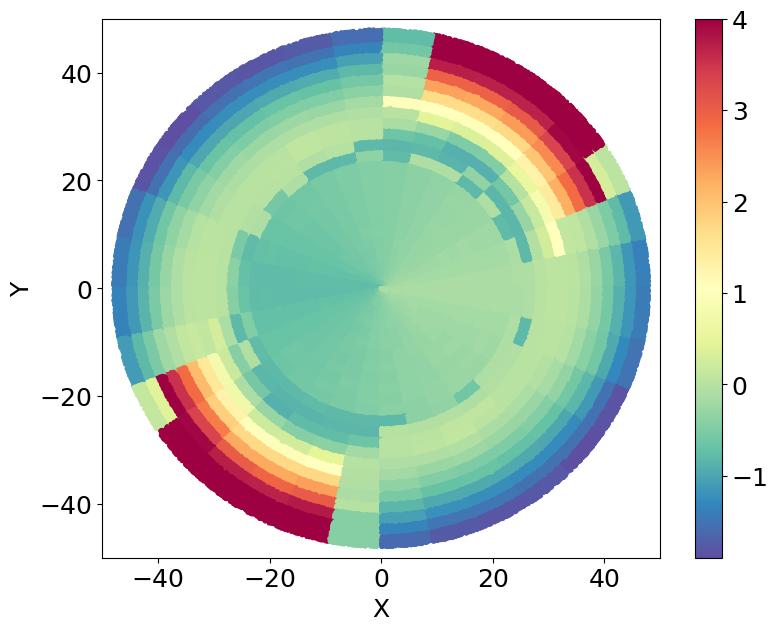}{0.3\textwidth}{(f)}}
\gridline{\fig{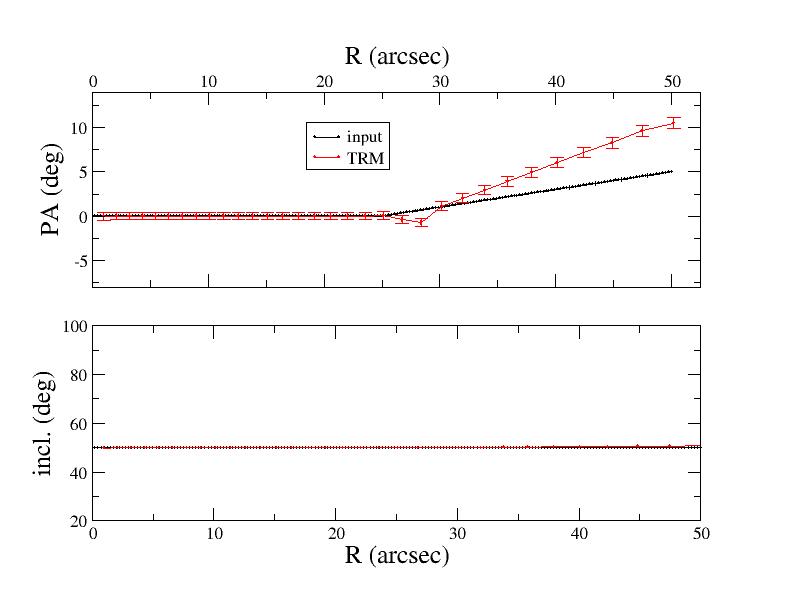}{0.3\textwidth}{(g)} 
              \fig{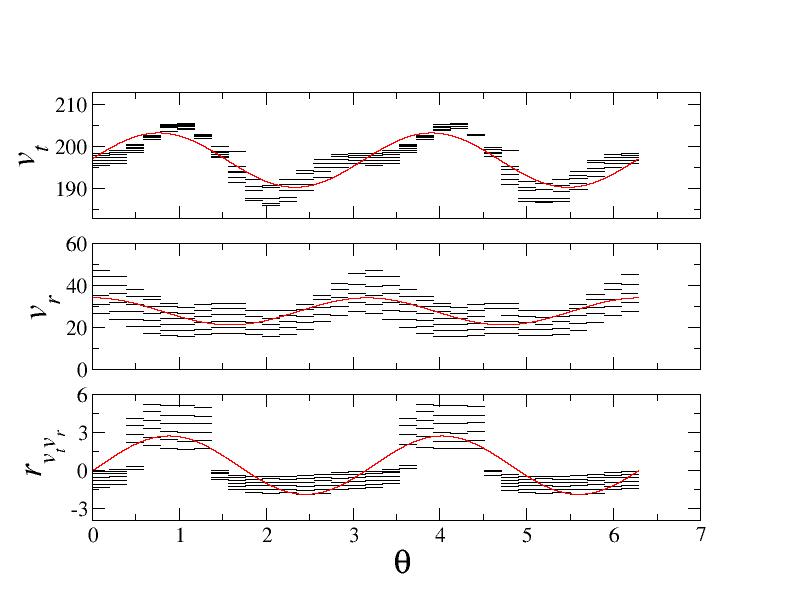}{0.3\textwidth}{(h)}}
\caption{ As Fig.\ref{Mod_1}  but for a toy warped disc model with an isotropic radial dependent velocity field} 
\label{ToyMod3}
\end{figure*}
\begin{figure}  
\includegraphics[width=7.5cm,angle=0]{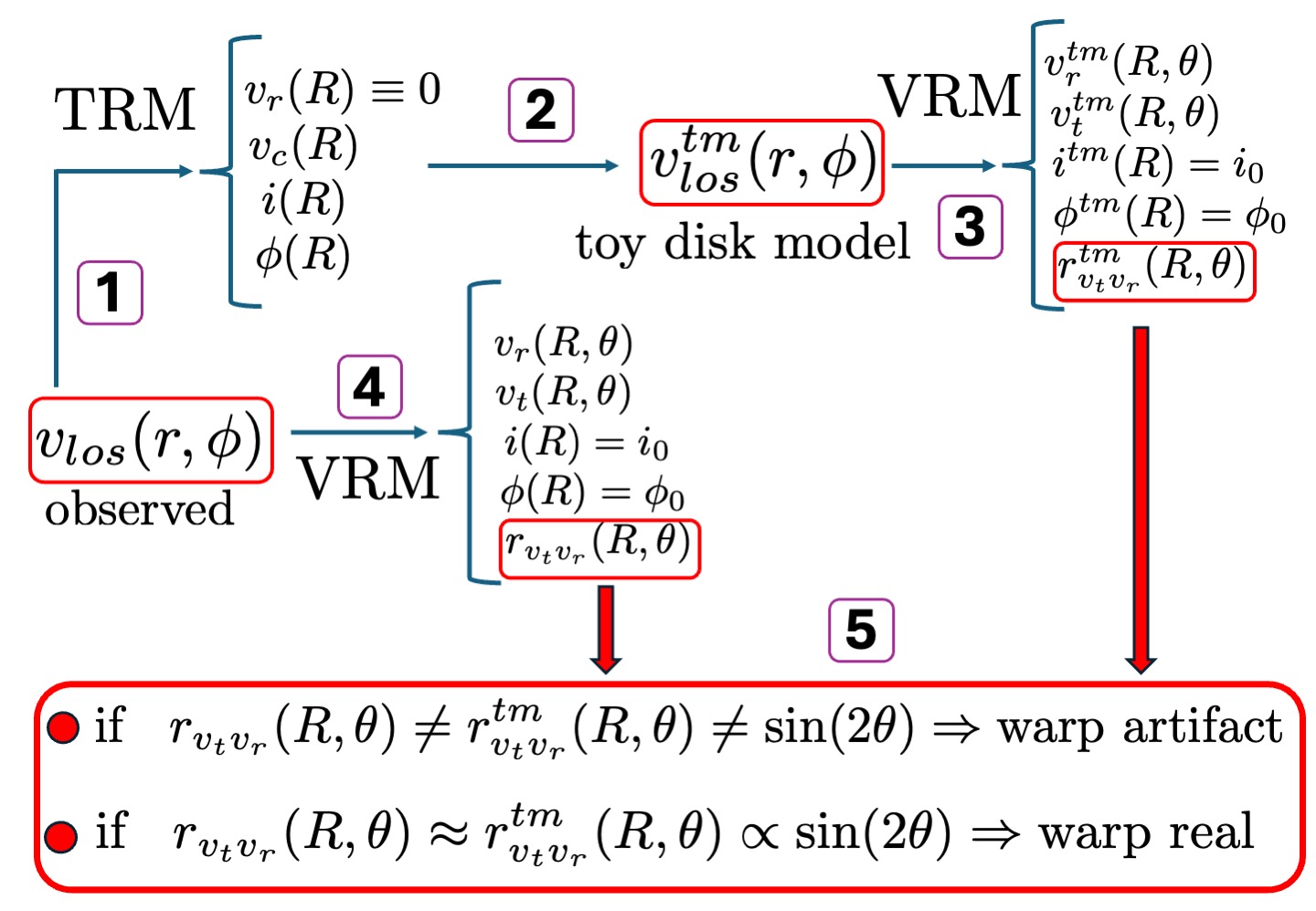}
\caption{
 Schematic representation of the procedure to identify whether a warp is present or not.
} 
\label{fig:Flux1} 
\end{figure}

In summary, while both the VRM and TRM fail to reconstruct the input velocity component profiles in the outer regions of the system --- where both warp and isotropic flow are present --- the most distinct signal detected is the dipolar modulation of the velocity rank correlation coefficient. This result implies that any non-dipolar angular anisotropy observed in the VRM-reconstructed, coarse-grained velocity component maps is unambiguously an intrinsic feature of the system's velocity field.

For this reason, the dipolar modulation of the velocity rank correlation coefficient should serve as a reliable marker for the presence of a warp, as it is unlikely that such modulation would arise solely from intrinsic velocity field fluctuations. However, in real galaxies, the velocity field --- beyond any warp --- may also exhibit intrinsic velocity anisotropies with amplitudes comparable to or even exceeding those caused by the warp (within the VRM framework). Thus, for real galaxies, detecting a dipolar modulation based only on velocity components may be challenging. Estimating the amplitude and location of intrinsic velocity anisotropies is essential for accurately characterizing a galaxy's velocity field. Velocity anisotropies often correlate with spatial structures, such as spiral arms, satellites, and other features, and thus provide valuable kinematic information crucial for the dynamical modeling of the system. This correlation allows us to better understand the galaxy's structural and dynamical properties, as well as the formation processes underlying these features.


{   
Finally, we consider a toy disk model without warp  in which the radial velocity is constructed to exhibit a dipolar anisotropy with maximum velocity in the direction along the $x$-axis. 
 Specifically, the radial velocity component grows linearly with radial distance and, for polar angles $\theta$ (in the plane of the disk) within $-20^\circ \leq \theta \leq 20^\circ$ and $160^\circ \leq \theta \leq 200^\circ$, it is characterized by a dipolar anisotropy. 
 The  transverse velocity is on average constant with $v_t=200$ km/s 
but it has random fluctuations with a relative amplitude of $10\%$. 

 Fig.~\ref{Mod_1_Anisotropic_Model} presents the VRM-reconstructed radial velocity field at different resolutions. The number of rings is fixed at $N_r=20$, while the number of arcs varies as $N_a=16, 32, 64$. It is evident that for $N_a=16$, the resolution is insufficient to accurately capture the dipolar anisotropy of the radial velocity field. However, for both $N_a=32$ and $N_a=64$, the dipolar structure in the radial velocity map is clearly identified, even though, in the latter case, the noise is higher due to the smaller cell size. The similarity between these two cases indicates the convergence of the reconstruction method.  Furthermore, the rank velocity correlation does not show any detectable dipolar anisotropy. It is worth noting that the transverse velocity map exhibits a low-amplitude symmetric structure, which is an artifact of the reconstruction method. However, this artifact does not leave any imprint on the rank velocity correlation coefficient. 
 
 Thus, if the radial velocity exhibits a genuine dipolar component, the VRM is capable of accurately reconstructing it. However, in this case, the rank correlation coefficient does not reveal any significant correlation between the radial and transverse velocity components (panel (l) of Fig.~\ref{Mod_1_Anisotropic_Model}). This lack of correlation is a clear indicator of the absence of a warp. While a dipolar modulation in a single velocity component is a necessary but not sufficient condition to infer the presence of a warp, a dipolar modulation in the rank correlation coefficient provides unequivocal evidence that a warp is indeed present.

It is important to note that, in principle, a disk could exhibit both radial and transverse velocity components with dipolar modulations that are correlated. However, such a scenario would represent an extremely contrived and physically implausible situation.
}
%
\begin{figure*}
\gridline{\fig{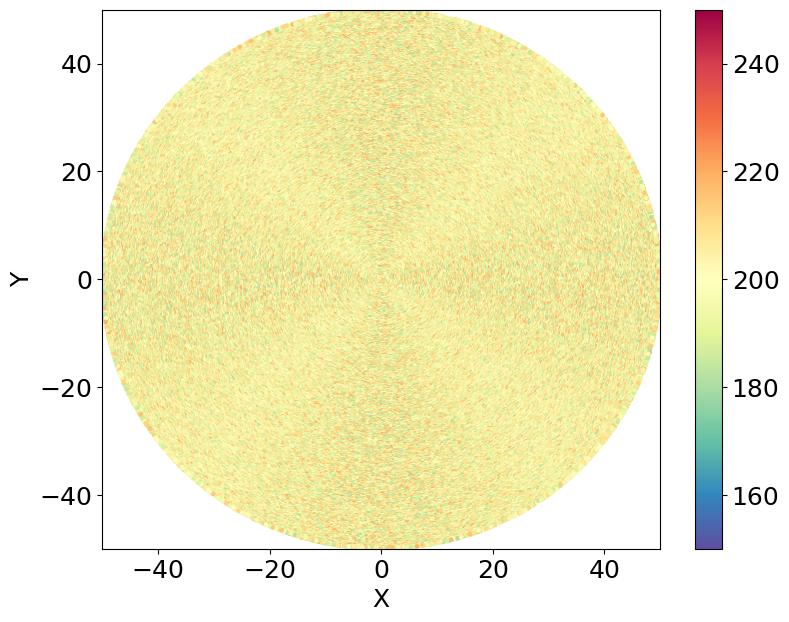}{0.24\textwidth}{(a)}
              \fig{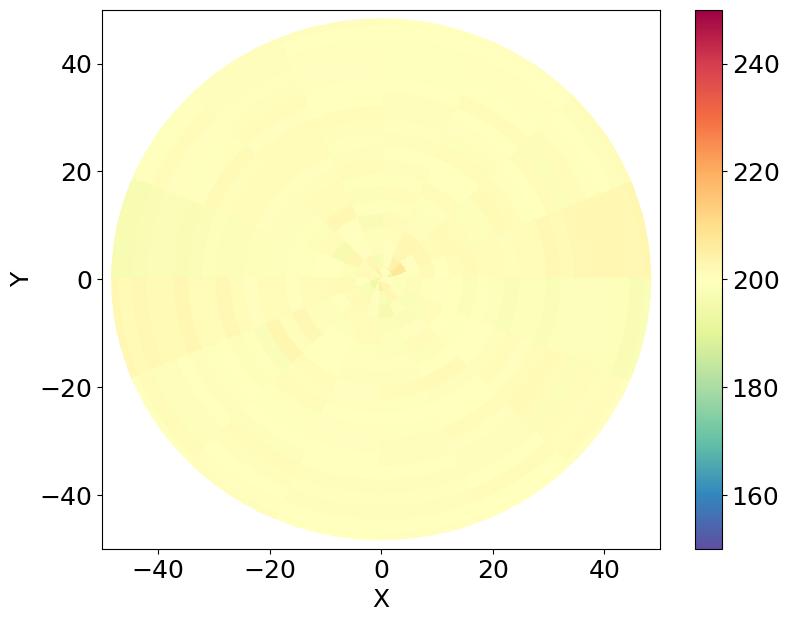}{0.24\textwidth}{(b)}              
              \fig{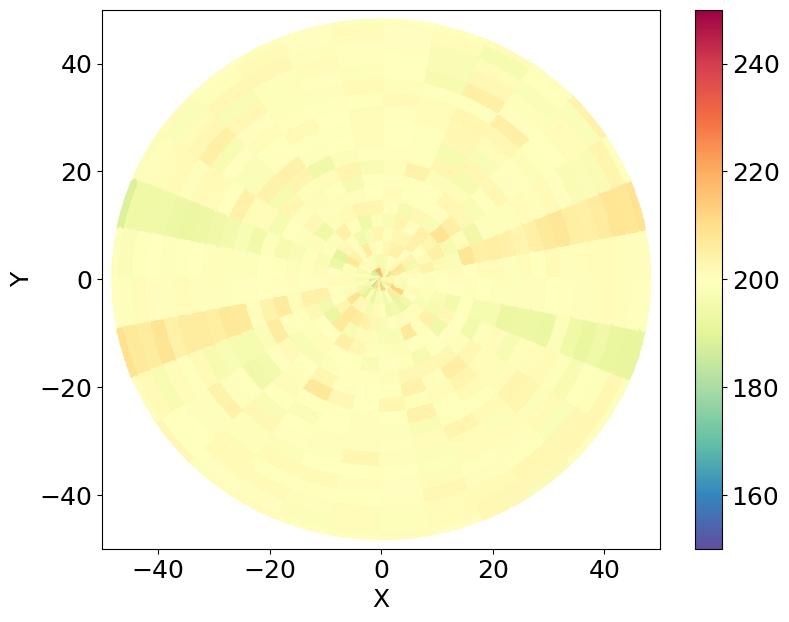}{0.24\textwidth}{(c)}
              \fig{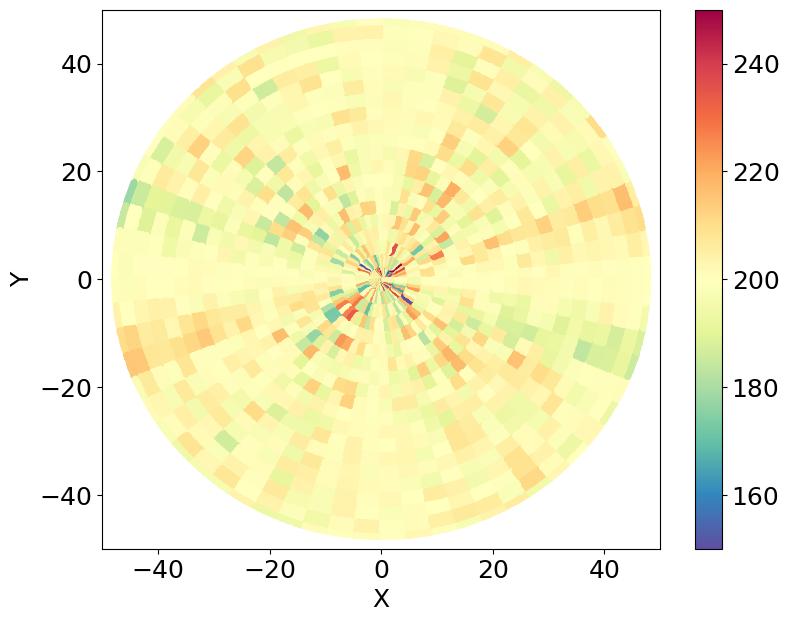}{0.24\textwidth}{(d)}
              }
\gridline{\fig{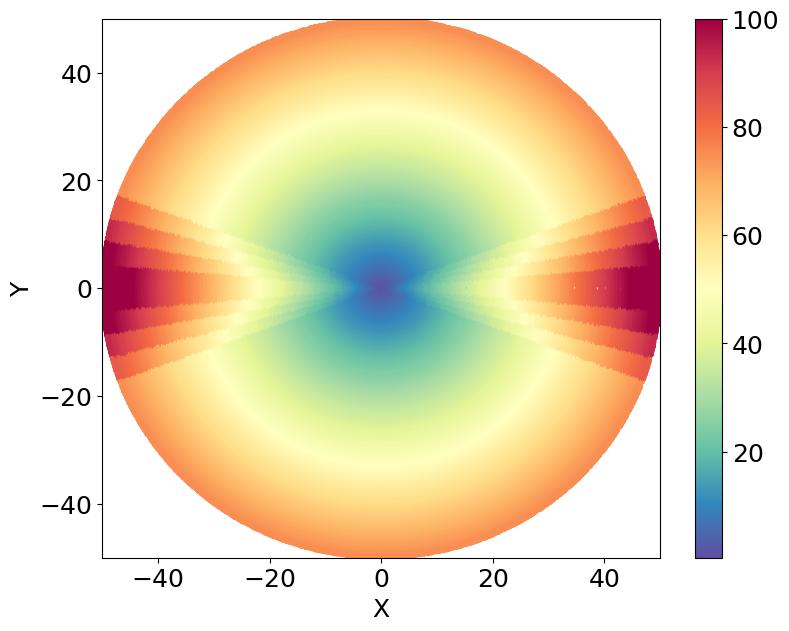}{0.24\textwidth}{(e)}
              \fig{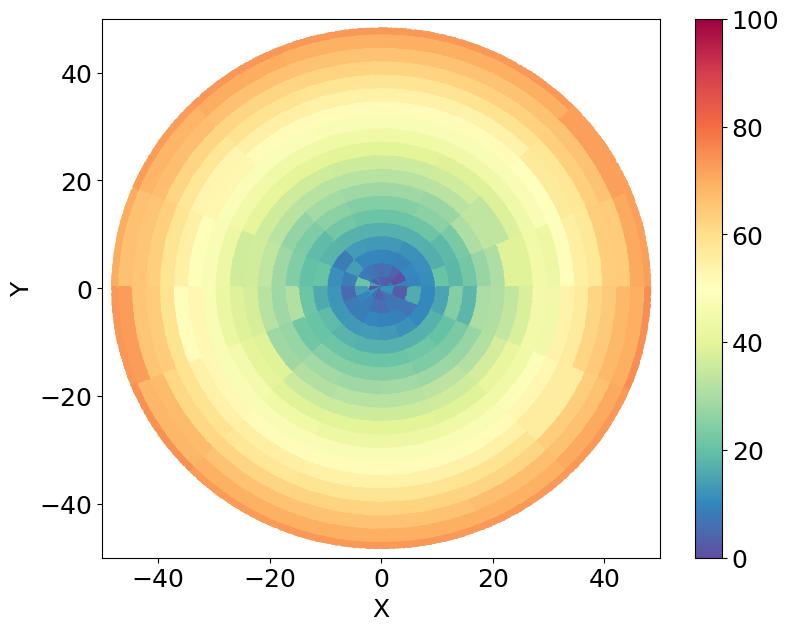}{0.24\textwidth}{(f)}              
              \fig{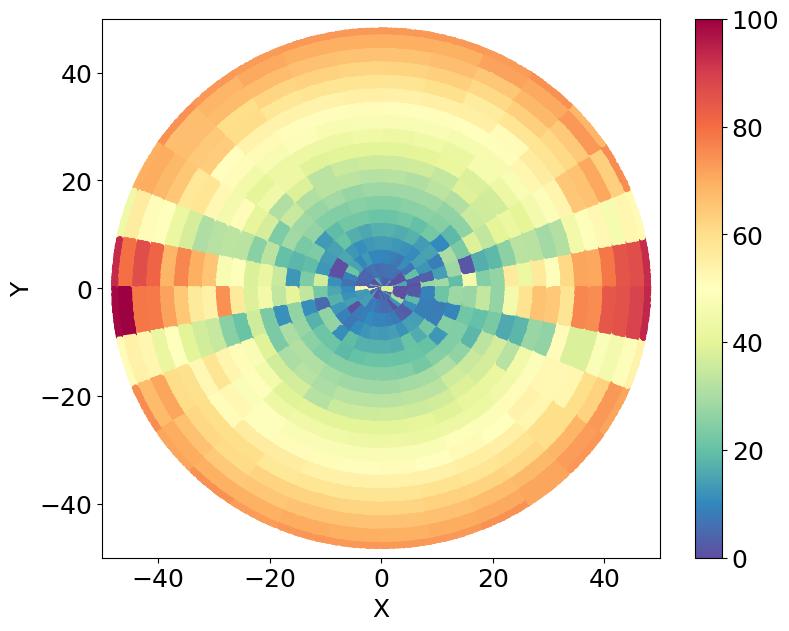}{0.24\textwidth}{(g)}
              \fig{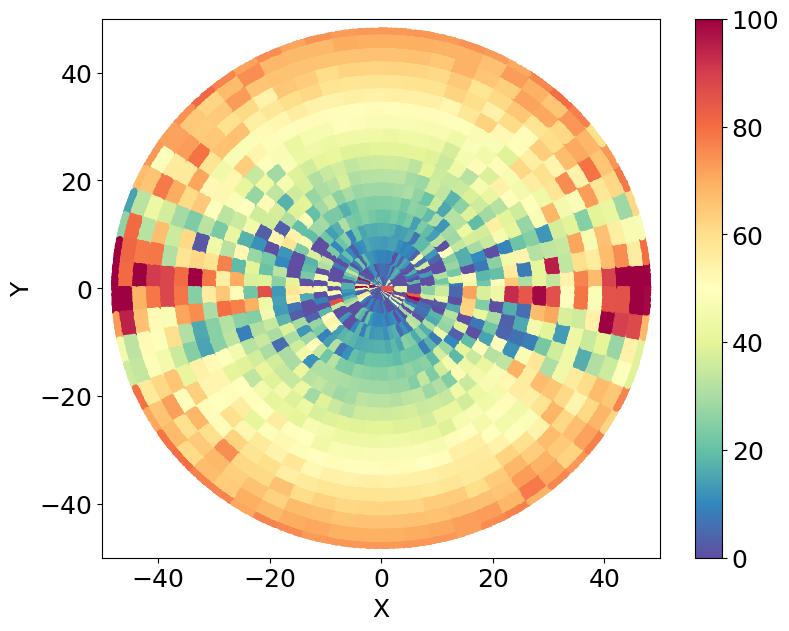}{0.24\textwidth}{(h)}
              }
\gridline{\fig{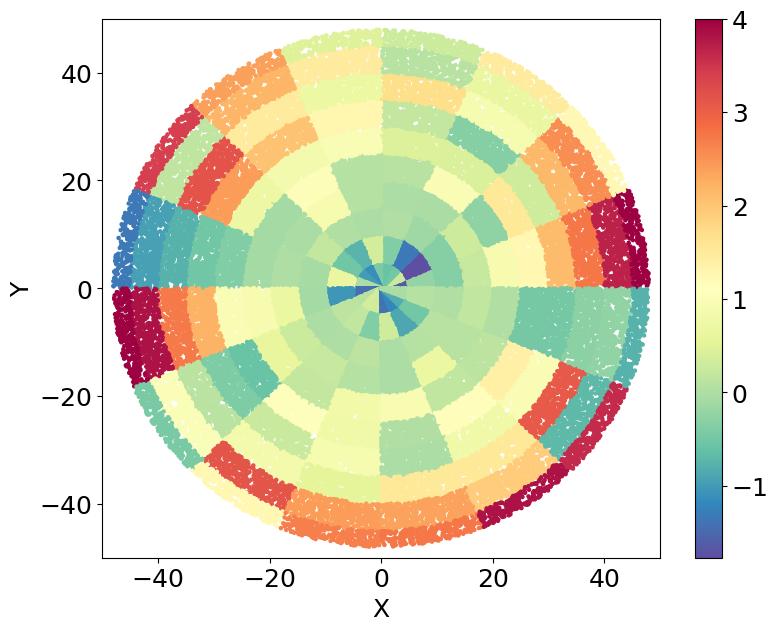}{0.24\textwidth}{(k)}
              \fig{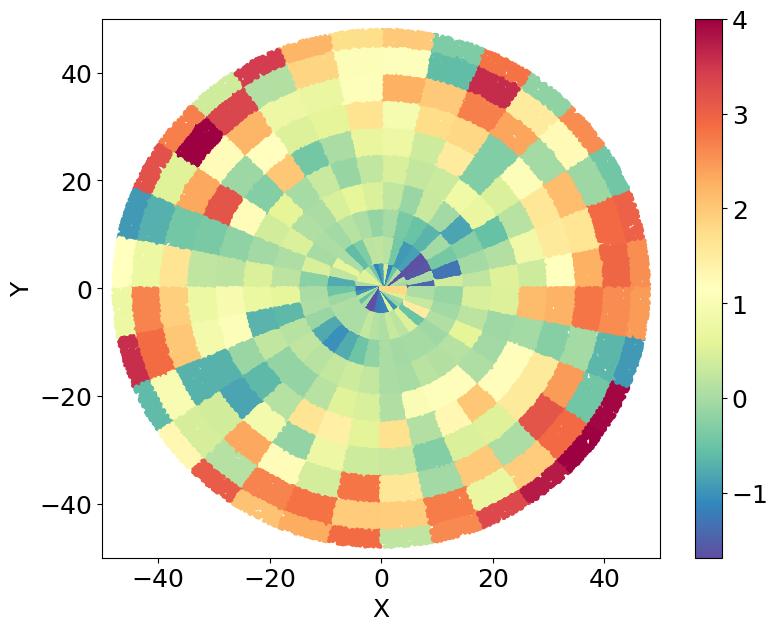}{0.24\textwidth}{(i)}              
              \fig{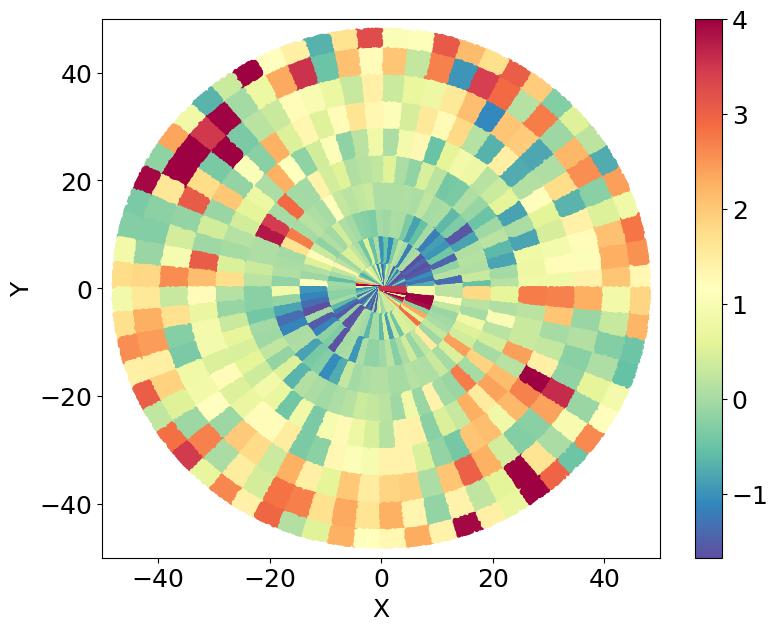}{0.24\textwidth}{(j)}
              \fig{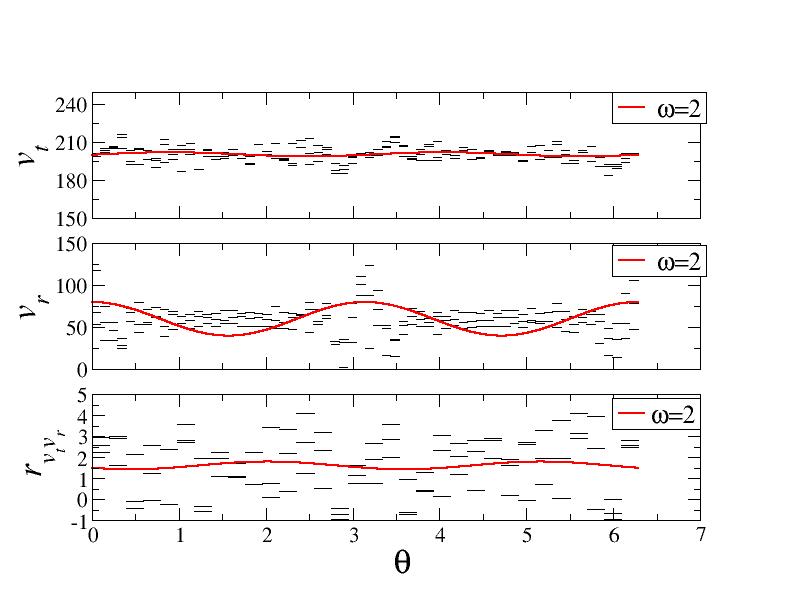}{0.26\textwidth}{(l)}
              }
\caption{    
Toy  flat disc model with an anisotropic (dipolar) radial dependent velocity field
(a) input transversal  velocity field; 
(b) VRM reconstructed transversal velocity field with resolution $N_r=20$ and $N_a=16$; 
(c)  as (b) but with resolution $N_r=20$ and $N_a=32$; 
(d)  as (b) but with resolution $N_r=20$ and $N_a=64$; 
(e) input radial  velocity field; 
(f) VRM reconstructed radial velocity field with resolution $N_r=20$ and $N_a=16$; 
(g)  as (f) but with resolution $N_r=20$ and $N_a=32$; 
(h)  as (f) but with resolution $N_r=20$ and $N_a=64$; 
(k) rank  velocity correlation  with resolution $N_r=20$ and $N_a=16$; 
(i)  as (k) but with resolution $N_r=20$ and $N_a=32$; 
(l)  as (k) but with resolution $N_r=20$ and $N_a=64$; 
(f) the velocity transversal component, the velocity radial component  and rank velocity correlation coefficient in the outer part of the warped disc in function of the polar angle $\theta$. } 
\label{Mod_1_Anisotropic_Model}
\end{figure*}


In summary, we have implemented the following procedure to identify whether a warp is present (see Fig.\ref{fig:Flux1}):}

\begin{itemize}
\item    
(1) We apply  the TRM to the observed LOS velocity map $v_{\text{los}}(r, \phi)$ so that we can  estimate the circular velocity $v_c(R)$, the inclination angle $i(R)$, and the P.A. $\phi_0(R)$. 
\item  
(2)  Using the derived parameters  $i(R)$,  $\phi_0(R)$ and $v_c(R)$ as input, we construct a disc galaxy model. The toy model disc is intentionally warped to match the observed deformation identified using the TRM. Additionally, the model assumes an isotropic circular velocity field and negligible radial motions within the disc. By employing this model, we generate a synthetic LOS velocity map $v_{\text{los}}^{tm}(r, \phi)$ as it would be seen from an observer placed far enough with a global inclination angle equal to the observed one.    This will be used as a null model to test the results of the VRM.
\item   
(3) From the LOS velocity map of the toy disc galaxy, $v_{\text{los}}^{tm}(r, \phi)$, we can derive its transverse velocity component, $v_t^{tm}(R, \theta)$, and radial velocity component, $v_r^{tm}(R,\theta)$, using the VRM. We can then measure the rank correlation coefficient, $r_{v_t v_r}^{tm}(R,\theta)$, between the two velocity components. 
\item  (4) From the observed galaxy LOS velocity map, $v_{\text{los}}(r, \phi)$, we can derive the transverse velocity component, $v_t(R, \theta)$, and the radial velocity component, $v_r(R, \theta)$, using the VRM. Furthermore, we can measure the rank correlation coefficient, $r_{v_t v_r}(R, \theta)$.
\item (5)     If the rank correlation coefficient  $r_{v_t v_r}(R,\theta)$   closely resembles the dipolar modulation  exhibited by $r_{v_t v_r}^{tm}(R,\theta)$ of the toy model then the observed disc is  warped, validating  the inclination angle   $i(R)$ and P.A. $\phi_0(R)$ by the TRM as faithful representations of the system's properties. 
To quantitatively establish the relation $r_{v_t v_r}(R,\theta) \approx r_{v_t v_r}^{tm}(R,\theta)$,  we compute the Spearman correlation coefficient $\cal{C}$    between these quantities at fixed  $(\theta,R)$ over the entire range   $\theta \in [0,2\pi]$. If $\cal{C}$$ \approx 0$, the warp can be excluded; whereas if $\cal{C} $$ \ge 0.2 $  the warp is confirmed. We have place the threshold to this relatively low value because from our tests we have noticed that in many cases the correlations holds only for the peaks of the dipolar signal.
\end{itemize}

In summary, the concept behind this procedure is to employ, for each galaxy, its corresponding toy model constructed with the velocity components profiles and geometric deformation detected by the TRM, as a null-hypothesis test. In this way we can identify the features of the velocity maps that cannot be solely explained as a warp. These features, therefore, correspond to the extrinsic velocity perturbations. 



\section{Analysis of the THINGS galaxies} 
\label{sect:things}

In this section, our primary focus is to investigate, for the galaxies in our sample,  the spatial maps of the correlation coefficient between the radial and transversal velocity components determined using VRM with arcs, which assumes a flat disc hypothesis. Our main objective is to determine if there is a distinguishable signature of a warp in these maps that, as discussed in the previous section, corresponds to a dipolar modulation in the rank velocity correlation coefficient of the type  $r_{v_t v_r}(R, \theta) \sim \sin(2 \theta)$. In that case even the two velocity components $v_t$ and $v_r$  should have  the same kind of dipolar oscillation: these are the extrinsic anisotropies that arise in the VRM reconstructed velocity field because of the inconsistent assumption that the disc is flat, while in reality, it is warped.

It is important to consider that intrinsic anisotropies of the velocity field, stemming from actual angular perturbations of the two velocity components, can contribute along with the extrinsic anisotropies. The detectability of these intrinsic anisotropies depends on their relative amplitude. Thus, if the two velocity components $v_t$ and $v_r$, do not exhibit a well-defined dipolar modulation, while the rank correlation coefficient $r_{v_t v_r}$ does display such modulation, it suggests that the extrinsic anisotropies caused by the warp have a smaller amplitude compared to the intrinsic anisotropies resulting from genuine velocity perturbations. The rank correlation coefficient proves to be a more robust estimator of dipolar oscillations as it specifically measures the correlation between $v_t$ and $v_r$. { It is worth noting that other types of large-scale perturbations, such as tidal interactions with a satellite galaxy or the presence of a bar, are not expected to produce similar dipolar correlations between velocity components in the outskirts of a galaxy. This is because oscillations induced by a warp and detected by the VRM exhibit coherence in both velocity components, whereas other types of perturbations generally lack a physical mechanism to generate such dipolar correlations.
}

Estimating the amplitude of intrinsic velocity anisotropies can be challenging because, as previously mentioned, if the intrinsic perturbations correspond to rapid localized variations in the velocity field, the VRM may not be able to accurately capture them. It is, however, possible to place at least a lower limit on the intrinsic perturbations whose amplitude must be equal to or larger than that of the extrinsic ones, depending on whether oscillations in the radial and transversal component are detectable or not.

{  In addition, we will examine the correlation between each of the two velocity components and the LOS velocity dispersion map. It is important to emphasize that the LOS velocity dispersion map provides independent information obtained from observations, distinct from the LOS velocity field. By generating spatial maps of the other two correlation coefficients defined in Equation \ref{eq:rxy}, our objective is to gain insights into the relationships between the velocity components and the velocity dispersion. The underlying concept is straightforward: if there is a gradient in a velocity component field, it can be correlated with a local increase in the velocity dispersion. Consequently, the correlation coefficients will show higher values in these regions. However, in the case of a warp, such a correlation should not be present. Therefore, the simultaneous presence of an enhancement in all three correlation coefficients can be interpreted as a signature of the velocity gradient observed by the VRM, indicating its actual existence.
}

\subsection{Analysis of individual galaxies} 
The sample of galaxies considered in this work is the same as that used in the study by  \cite{SylosLabini_etal_2023}, where the interested reader can find a more detailed description of their properties. In particular, that work  provides extensive information on the TRM and VRM  analyses, including investigations conducted with different resolutions and tests conducted to assess the convergence of the quadrupole and octopole moments (see also \cite{Walter_etal_2008,deBlok_etal_2008} for further details on the THINGS sample and the TRM analysis). For M33 we refer to \cite{Corbelli_etal_2014}  for a discussion about the \HI\ observations.

In this section we present results for a galaxy with a warp visibile as a modulation in the correlation coefficient only (NGC 2903),  a galaxy (NGC 5457) for which there is a very strong correlation between the variance and the fluctuations in the velocity fields and finally a galaxy without any clear evidence for a deformed geometry (NGC 5194).  

We will discuss in details the results of the analysis for these three galaxies, while we will give the key features for all other results of the analysis of individual galaxies  (reported in the Appendix): in brief, the majority of the galaxies exhibit a very moderate warp and only in a few cases do we observe a strong warp. However, all galaxies are characterized by intrinsic fluctuations that are of the same order of, or more often larger than, extrinsic ones. 

{ Tab. \ref{tab1} summarizes our findings. It reports the Spearman correlation coefficient, $\cal{C}$, between the rank correlation velocity coefficient computed for the galactic data, $r_{v_t v_r}(R,\theta)$, and for the toy disc model, $r_{v_t v_r}^{tm}(R,\theta)$, constructed to have the same orientation angles measured by the  TRM of the real galactic data. As discussed in Sect.\ref{sect:warps} , only when $\cal{C}$ $> 0.2$ can we conclude that the warp is present.}  

{ We emphasize that an anisotropic gradient in the radial velocity cannot lead to a variation in the orientation angles and, therefore, cannot be mistaken for a warp. This distinction arises from the inherent nature of warps within the TRM framework, where a warp corresponds to a varying orientation of circular rings. As a result, warps cannot exhibit asymmetry in their orientation angles. Therefore, while large variations in the orientation angles and anisotropic gradients in the radial velocity can introduce complexities in the analysis, it is essential to differentiate between these effects and the distinct characteristics associated with warps. }

On the other hand we noted that in certain cases, the VRM may struggle to accurately reconstruct a complex input field, particularly when the perturbation field exhibits rapid  changes in localized patches. Consequently, it is not possible to straightforwardly estimate the amplitude of velocity perturbations, e.g. the radial and transversal dispersion in function of radius, as the reconstruction method can in some cases significantly impact the results.  By examining the convergence behavior of the moments, we can gain insights into the statistical properties and stability of the velocity field reconstruction. This analysis provides a valuable means to evaluate the reliability and validity of the VRM method and its results. \cite{SylosLabini_etal_2023} provide a comprehensive discussion on this topic, offering further guidance on assessing the robustness of VRM outcomes.


\subsubsection{NGC 2903} 
Fig. \ref{fig:NGC2903-1} depicts the behavior (panel (a)) of the inclination angle $i(R)$ and P.A. $\phi_0(R)$ as measured by the TRM for NGC 2903 (for more details, refer to \cite{deBlok_etal_2008,SylosLabini_etal_2023}). In the outer part of the disk, the P.A. $\phi_0(R)$ undergoes an approximate change of $10^\circ$, while the inclination angle $i(R)$ varies by about $5^\circ$.
The rotation curve $v_c(R)$ estimated by the TRM exhibits a gradual decline (panel (b)). It has a maximum value of $v_c(R) = 205$ km/s at $R = 125''$ and decreases to $v_c(R) = 179$ km/s at $R = 700''$.
The transverse velocity component $v_t(R,\theta)$ estimated by the VRM is, within the error bars, equal to $v_c(R)$. For $R > 600''$, a small correction has been computed due to the warp, as explained in Sect.\ref{sect:things}:   $v_t(R,\theta)$  is however more fluctuating toward the outermost regions of the disk, where indeed angular anisotropies can be observed in the two-dimensional map (see below).
{ The radial velocity component $v_r(R,\theta)$  exhibits a fluctuating behavior with an average value not exceeding 20 km s$^{-1}$ in agreement with 	\cite{DiTeodoro+Peek_2021,SylosLabini_etal_2023}. 
Note that error bars of the mean velocity profiles $v_t(R)$  and $v_r(R)$ in Fig. \ref{fig:NGC2903-1} are computed for each of the $N_r = 50$ rings, representing the standard deviation of the mean among the $N_a = 32$ angular cells into which each ring is divided.
}

\begin{figure*}
\gridline{\fig{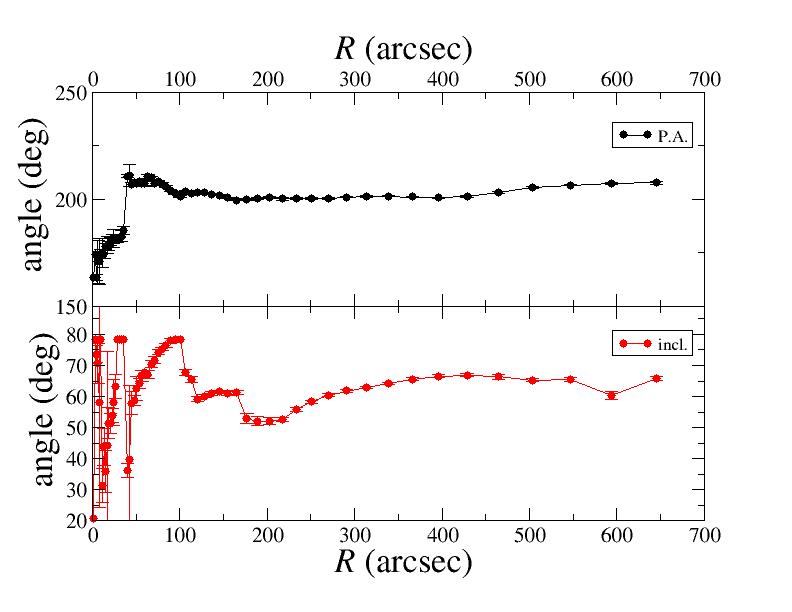}{0.45\textwidth}{(a)}
              \fig{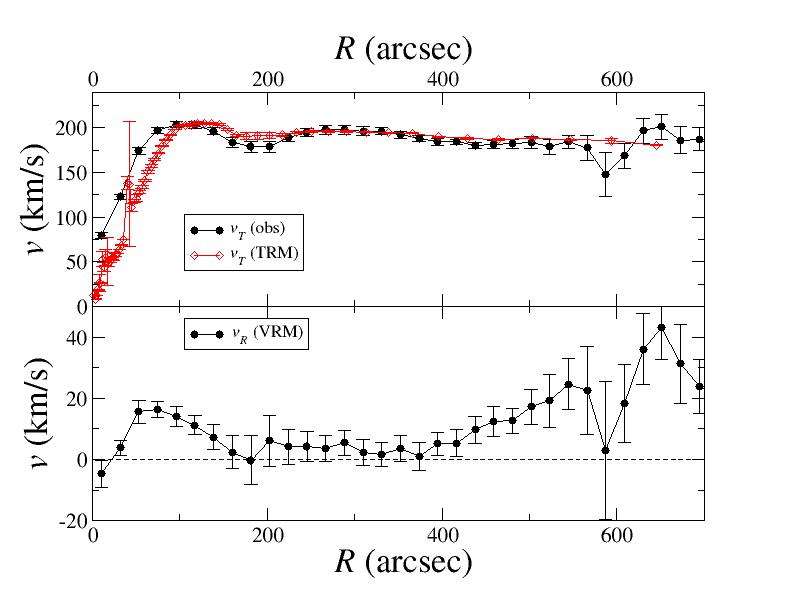}{0.45\textwidth}{(b)}
              }
  \gridline{
 \fig{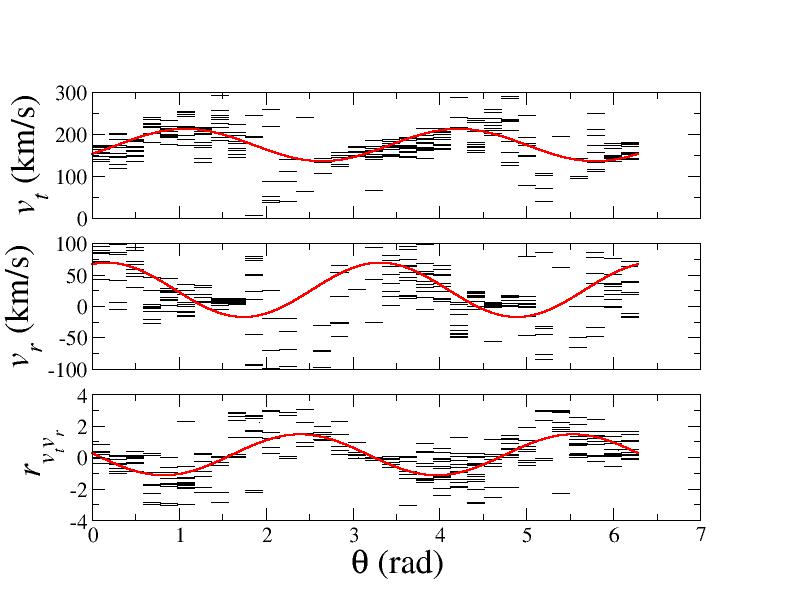}{0.45\textwidth}{(c)}
                \fig{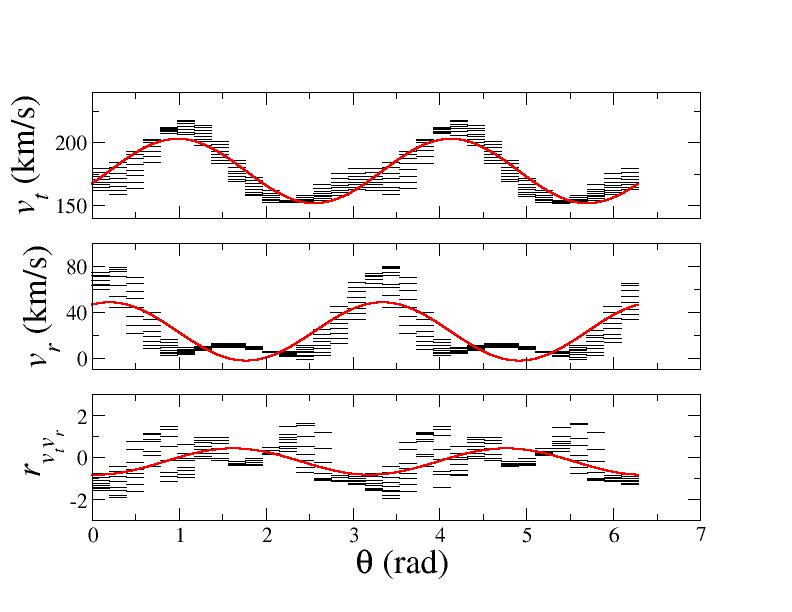}{0.45\textwidth}{(d)}}
\caption{ The panel respectively show for NGC 2903: 
 (a)  the inclination and P.A. (rescaled by $180^\circ$) as detected by the TRM;
 (b) the mean radial profiles $v_t(R)$ and $v_r(R)$ obtained by the VRM  and   the determination through the TRM of $v_c(R)$;
 (c) the transversal, radial and rank  velocity correlation coefficient  in the external region of the galaxy in function of the polar angle $\theta$ (the red line has frequency $\omega=2$);
 (d) the same but for the toy model.
 {  
Note that the external regions of the galaxy and of the toy model shown in panels (c) and (d) correspond to radial distances $r> 0.75 R_{\text{max}}$, where $R_{\text{max}}$ is the maximum radius reported in the respective velocity maps. This same limit is consistently applied throughout the analysis of all other galaxies considered in this work.
 }
 }  
\label{fig:NGC2903-1} 
\end{figure*}
%

\begin{figure*}
\gridline{\fig{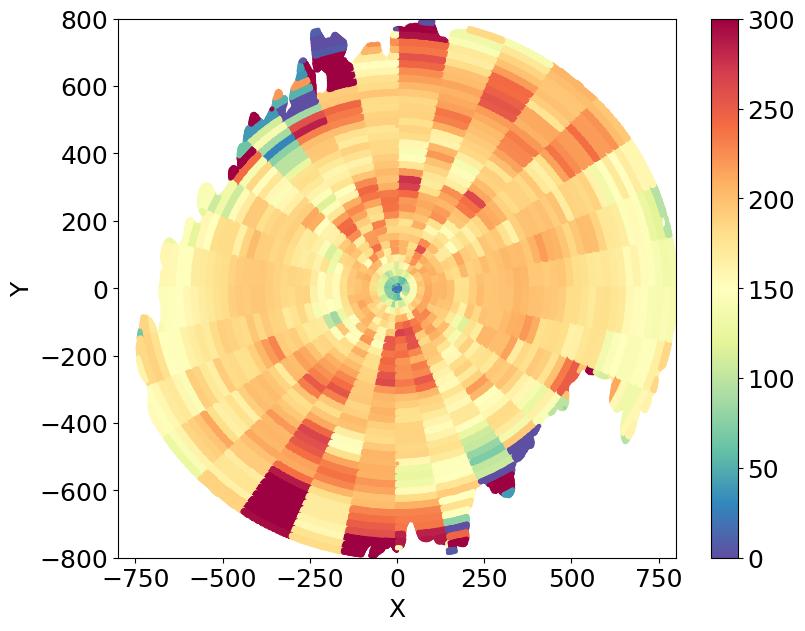}{0.3\textwidth}{(a)}
              \fig{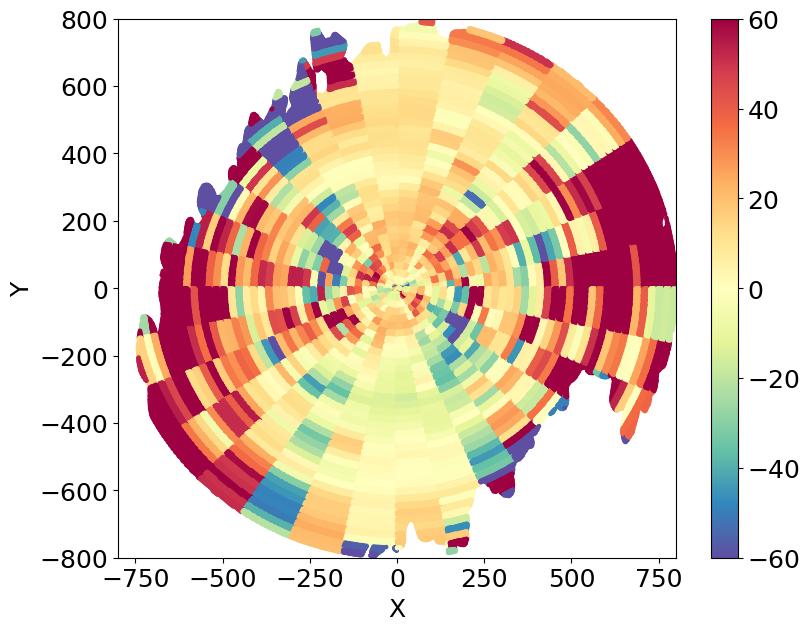}{0.3\textwidth}{(b)}
               \fig{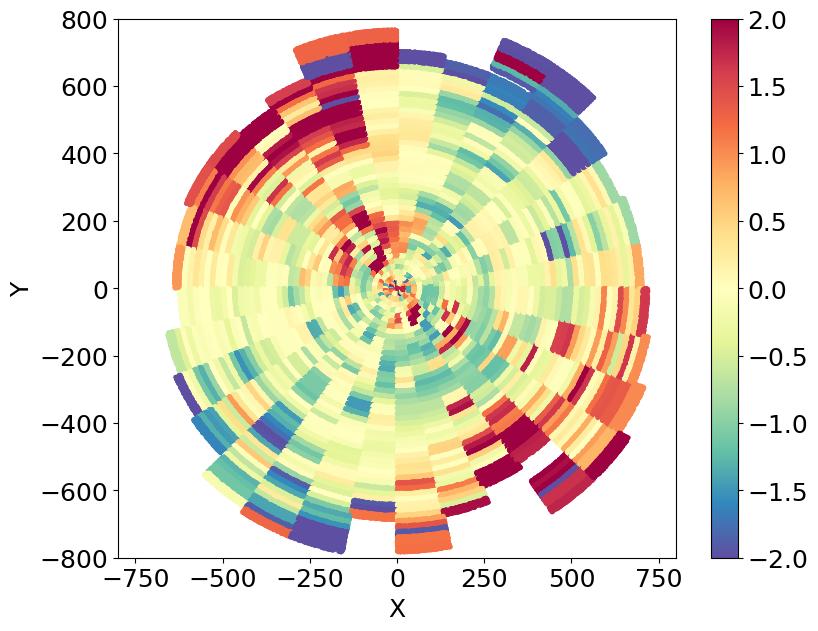}{0.3\textwidth}{(c)}}
\gridline{\fig{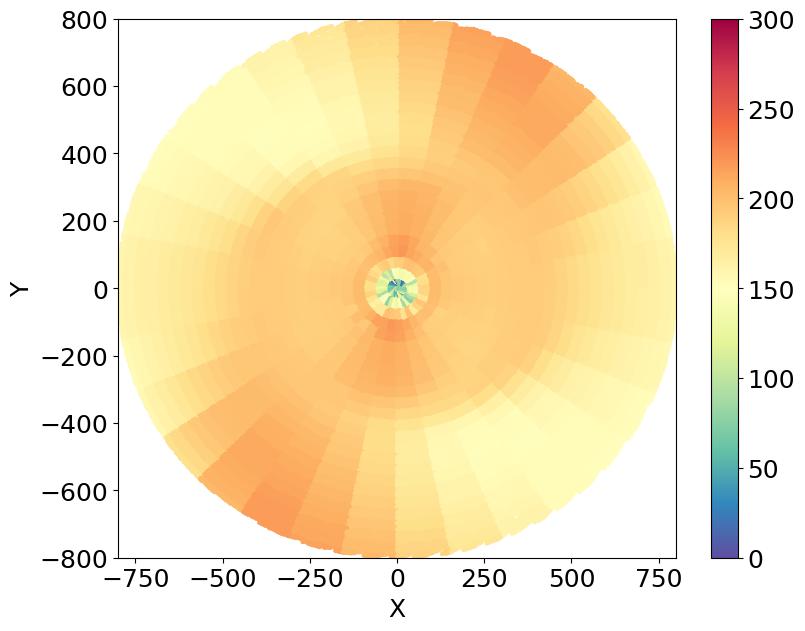}{0.3\textwidth}{(d)}
              \fig{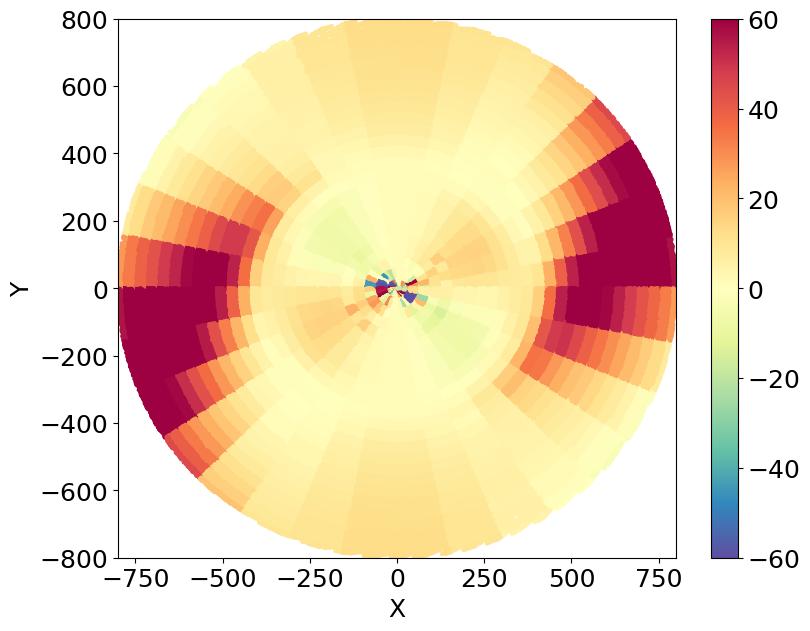}{0.3\textwidth}{(e)}
              \fig{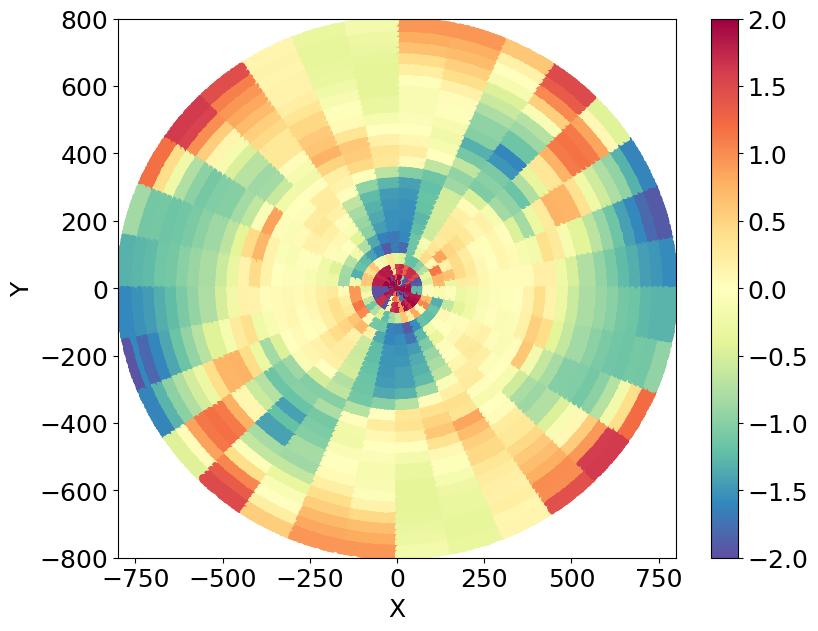}{0.3\textwidth}{(f)}}
\gridline{\fig{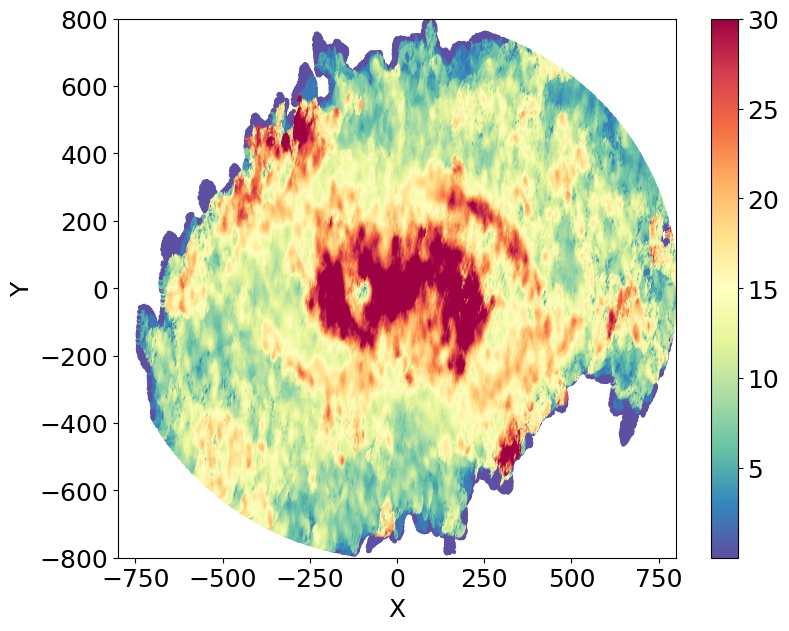}{0.3\textwidth}{(g)}
              \fig{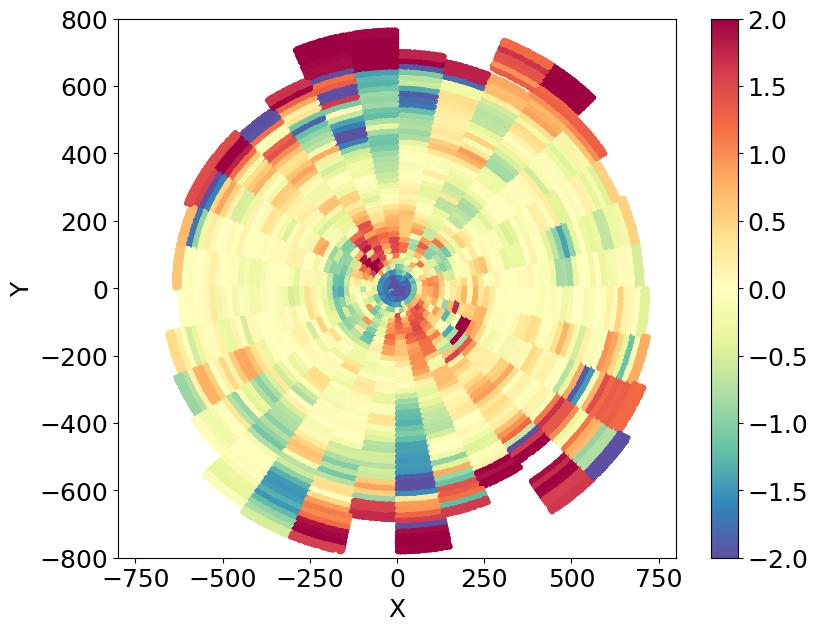}{0.3\textwidth}{(h)}
              \fig{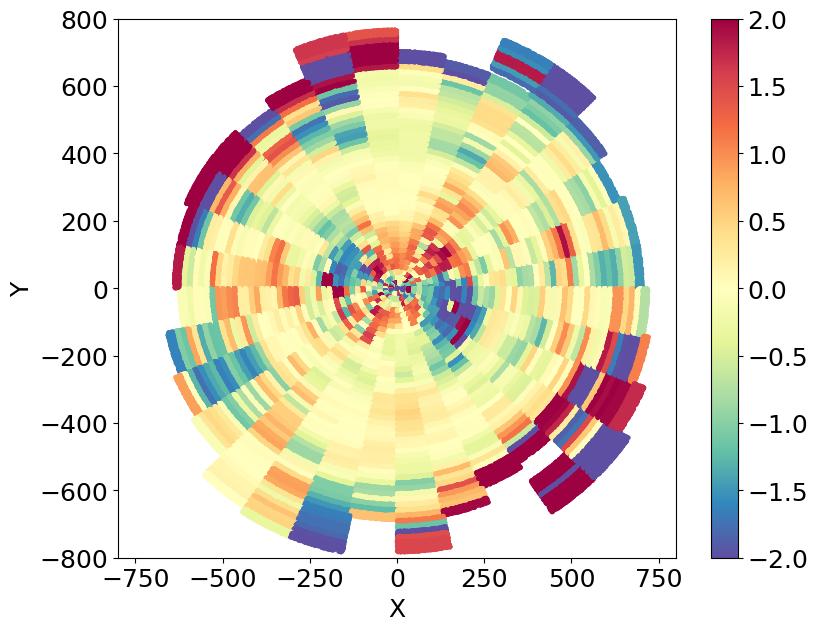}{0.3\textwidth}{(i)}}
\caption{The panels for the observed galaxy NGC 2903 respectively show: 
(a) the transversal velocity component  $v_t(R,\theta)$ map, 
(b) the radial velocity component $v_r(R,\theta)$ map, 
(c) the rank  velocity correlation coefficient  $r_{v_r v_t}(R, \theta)$ map. 
The same quantities but for the toy model 
(d) $v_t^{tm}(R,\theta)$, 
(e) $v_r^{tm}(R,\theta)$
(f)  $r_{v_r v_t}^{tm}(R, \theta)$.
For the observed galaxy 
(g) the velocity dispersion map $\sigma (R,\theta)$ map,  
(h) the rank   correlation coefficient $r_{\sigma v_t}(R, \theta)$ map,
(i)  the rank   correlation coefficient $r_{\sigma v_r}(R, \theta)$ map.
Velocities and standard deviations are in km s$^{-1}$.
} 
\label{fig:NGC2903-2} 
\end{figure*}

{ Finally panel (c) of Fig. \ref{fig:NGC2903-1}  presents the transversal, radial, and rank velocity correlation coefficients in the outer region of the galaxy as a function of the polar angle $\theta$, while panel (d) shows the same for the toy model. One notable observation is that the rank correlation coefficient, $r_{v_t v_r}(R, \theta)$, exhibits a clear dipolar modulation resembling $\sin(2\theta)$, which indicates the presence of a warp. Both the transverse velocity component, $v_t$, and the radial velocity component, $v_r$, display intrinsic perturbations beyond the extrinsic ones caused by the warp. In this case, we estimate ${\cal C} = 0.4$, suggesting the presence of a moderate warp. The analysis reveals that the intrinsic velocity perturbations significantly influence the velocity field, overshadowing the comparatively minor contribution from the warp.

 Fig.\ref{fig:NGC2903-2}  presents the velocity dispersion and velocity component maps for both the observed galaxy and the toy model, along with the corresponding rank correlation coefficients. By the visual inspection of the maps (a)-(f) one may conclude the presence of the dipolar modulation component  in the peripheries of the disk.   Finally, the positive correlations observed in $r_{\sigma v_t}(R, \theta)$ and $r_{\sigma v_r}(R, \theta)$ (see panels (h) and (i) of  Fig.\ref{fig:NGC2903-2}) indicate that the velocity field in the outer regions is perturbed. Additionally, the velocity dispersion field shows correlations with both the transverse and radial velocity fields in the inner disc, albeit in slightly different directions, suggesting the potential presence of a bar structure. }


\subsubsection{NGC 5457} 

This galaxy was not included in the sample considered by \cite{deBlok_etal_2008} because its global inclination angle is small, i.e. $i=30^\circ$.  According to the measurements obtained from the  TRM \citep{SylosLabini_etal_2023}, the P.A. $\phi_0(R)$ exhibit significant variations at small radii, specifically for $R<500$",   whereas  the inclination angle $i(R)$ at larger radii (refer to panel (a) Fig.\ref{fig:NGC5457-1}) displays a smooth decay in the outer disc's regions. In our toy model, we neglect the fluctuations in the inclination angle and assume it remains constant at small radii, while it shows a smooth variation of about $10^\circ$ at larger radii. The reason for this choice is that warps are common in the outer regions of galaxies and not in their inner disc, so that large variation of the inclination angle in the inner disc is more probably an artifact of the TRM analysis due to the incorrect assumption of neglecting radial motions. On the other hand, the P.A. shows a variation of approximately $20^\circ$ at small radii and then remains constant for $R>500$". 

The rotation curve $v_c(R)$ determined by the TRM displays substantial fluctuations around an approximately constant value of 120 km s$^{-1}$. Instead, the transverse velocity profile $v_t(R)$ from the VRM  reaches a maximum at $R\approx 600$" and then it presents a clear decrease.  The radial velocity profile $v_r(R)$ obtained from the  VRM also exhibit a behavior characterized by large fluctuations but with a small mean value smaller than 20 km s$^{-1}$.  { 
Large amplitude fluctuations characterize the velocity dispersion profile: as shown below they are strongly anisotropic. } 
%
\begin{figure*}
\gridline{\fig{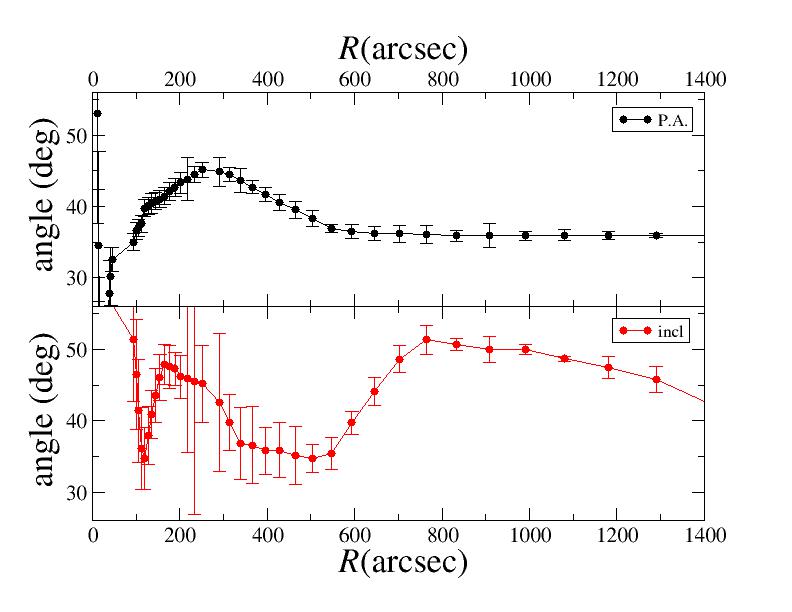}{0.45\textwidth}{(a)}
              \fig{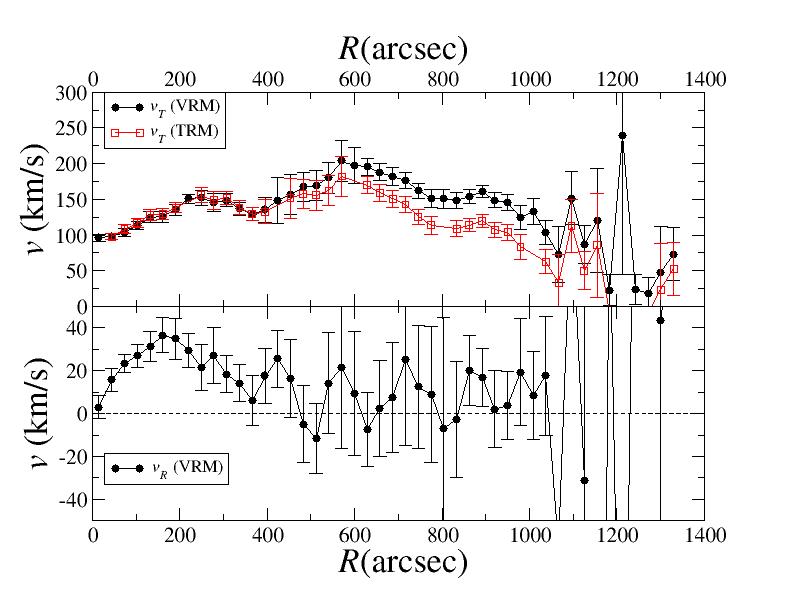}{0.45\textwidth}{(b)}
              }
  \gridline{
 \fig{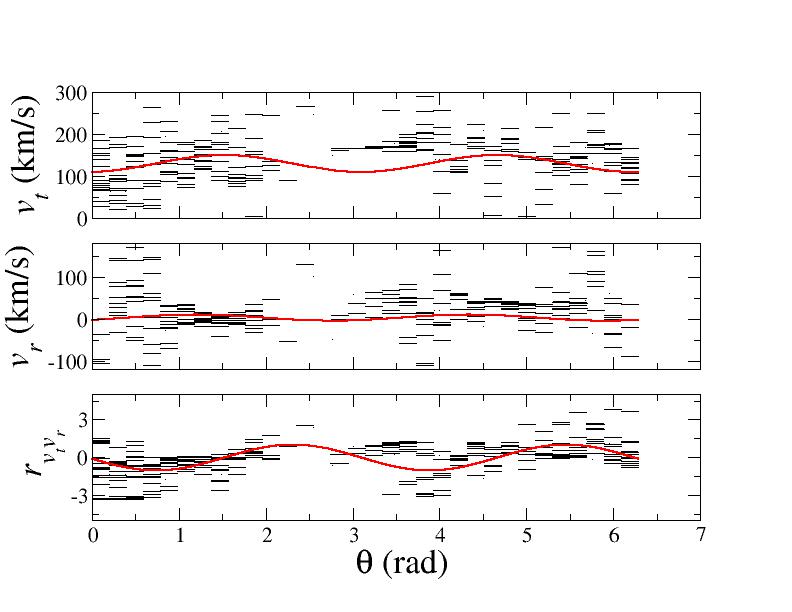}{0.45\textwidth}{(c)}
                \fig{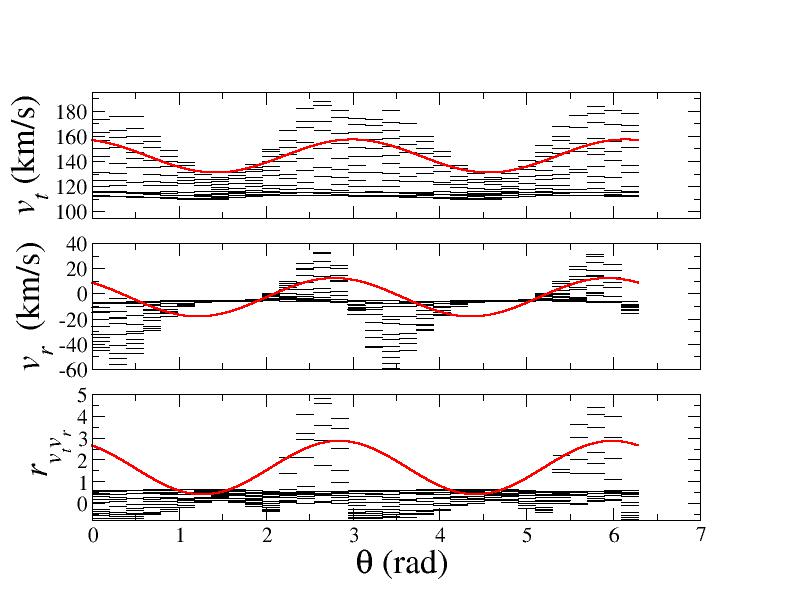}{0.45\textwidth}{(d)}}
     \caption{As  Fig.\ref{fig:NGC2903-1} but for NGC 5457.} 
\label{fig:NGC5457-1} 
\end{figure*}

Figure \ref{fig:NGC5457-2} reveal that the velocity field exhibits perturbations that cannot be solely attributed to a dipole modulation, providing instead evidence of a very weak signature of a warp. Neither $v_t$ nor $v_r$ display a dipolar oscillation, indicating that intrinsic velocity perturbations are larger than the contribution from the warp itself. { Indeed, in this case the correlation coefficient is  very small, i.e. ${\cal C} = 0.1$, indicating that the deformation of the warp is very small. } 
In contrast, the toy disc model exhibits dipole anisotropy in both velocity components, as well as in the rank correlation coefficient $r_{v_t v_r}^{tm}(R, \theta)$. Additionally, the radial velocity component shows a quadrupole pattern induced by variations in the inclination angle and P.A. Intrinsic anisotropies characterize the velocity field of the galaxy, with both velocity components displaying large correlated fluctuations in the direction corresponding to approximately $\theta \approx 30^\circ$. Notably, the velocity variance field also shows an enhancement in the same direction. Consequently, the correlation coefficient $r_{\sigma v_r}(R,\theta)$ exhibits a highly significant pattern in this direction, further emphasizing the presence of significant perturbations in the velocity field.

Considering that the amplitude of the extrinsic anisotropies is about 60-80 km/s and that they are not observed in the galaxy's case, we can deduce that the intrinsic anisotropies are of a similar magnitude. However, we conclude that the variations of the orientation angles measured by the TRM in this case are likely artifacts of the analysis, specifically the assumption that radial motions are negligible.

Indeed, by comparing the behaviors of $r_{v_t v_r}^{tm}$ and $r_{v_t v_r}$ in the outer regions, we find that they exhibit substantial differences. { From this, we conclude that a warp may be present in the galaxy, although it is has a negligible effect much smaller than that expected based on the variation of the orientation angles measured by the TRM.} Furthermore, we determine that the galaxy is primarily dominated by large intrinsic perturbations in both velocity components.
%
\begin{figure*}
\gridline{\fig{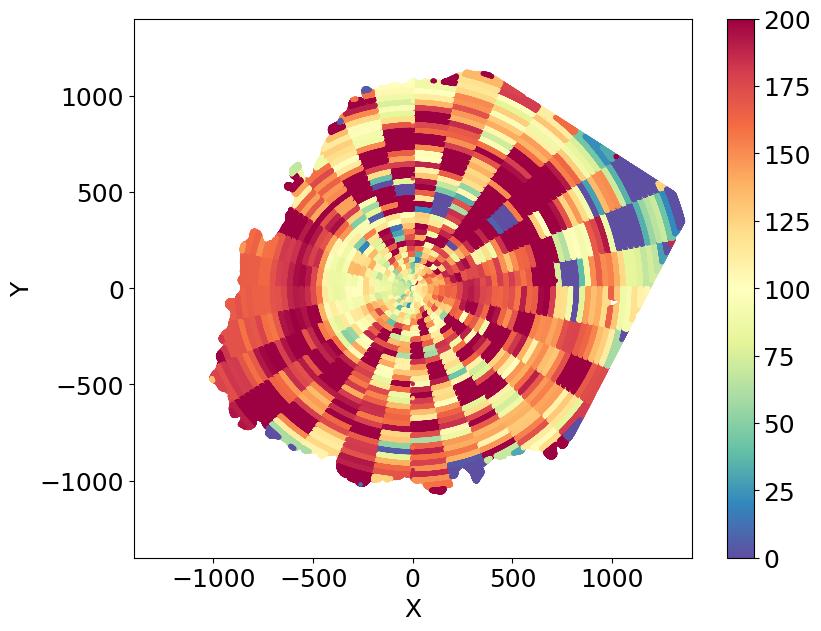}{0.3\textwidth}{(a)}
              \fig{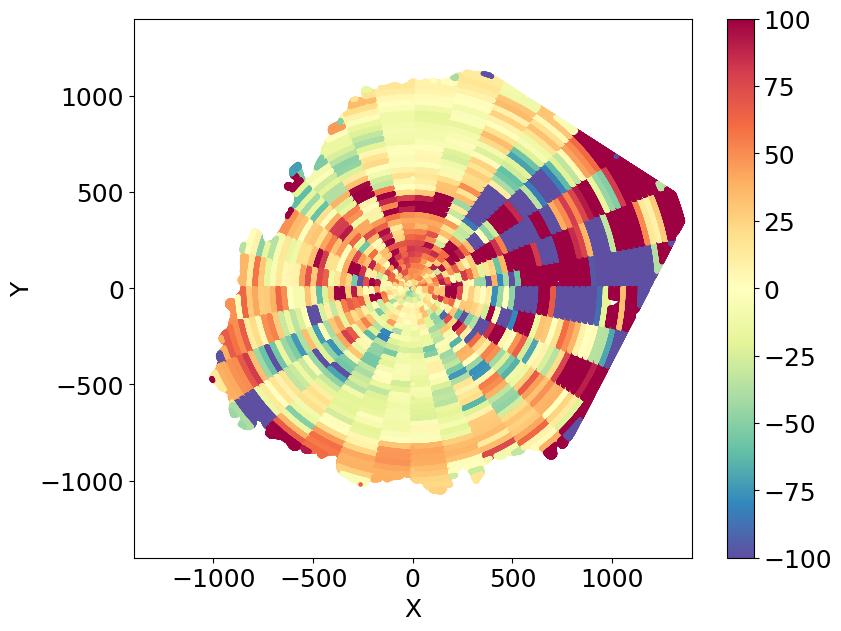}{0.3\textwidth}{(b)}
               \fig{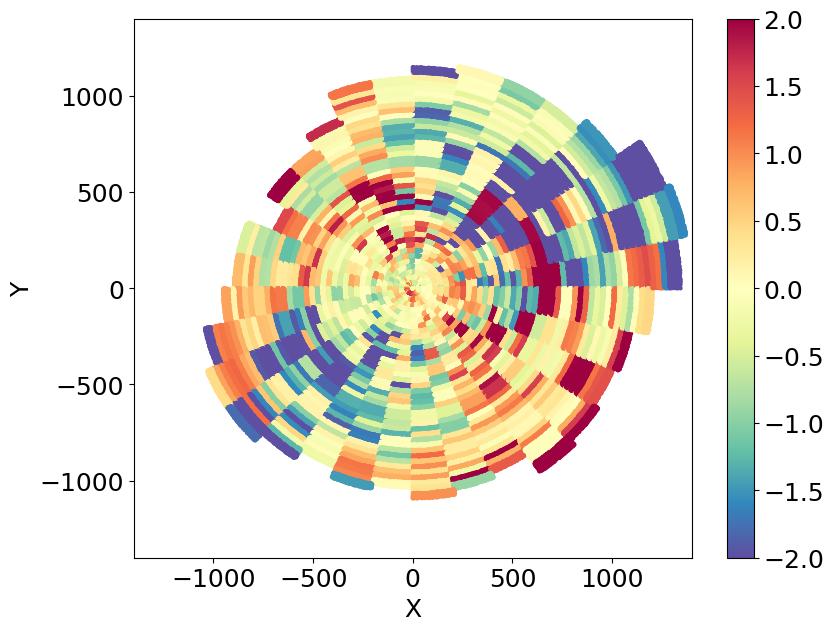}{0.3\textwidth}{(c)}}
\gridline{\fig{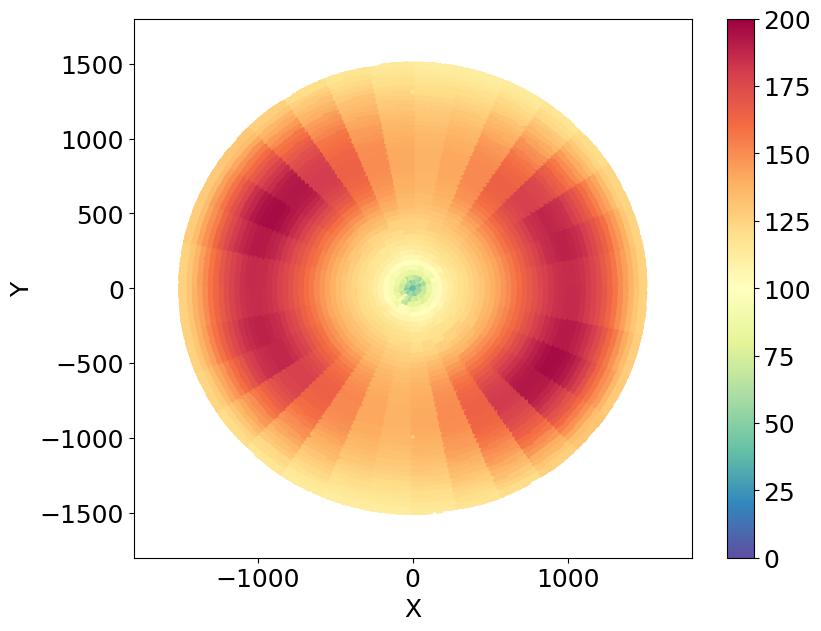}{0.3\textwidth}{(d)}
              \fig{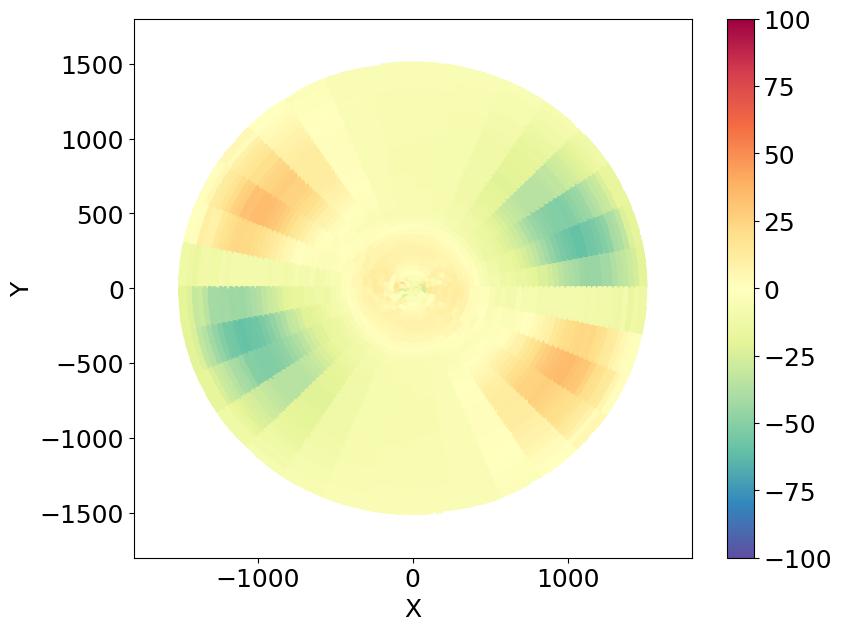}{0.3\textwidth}{(e)}
              \fig{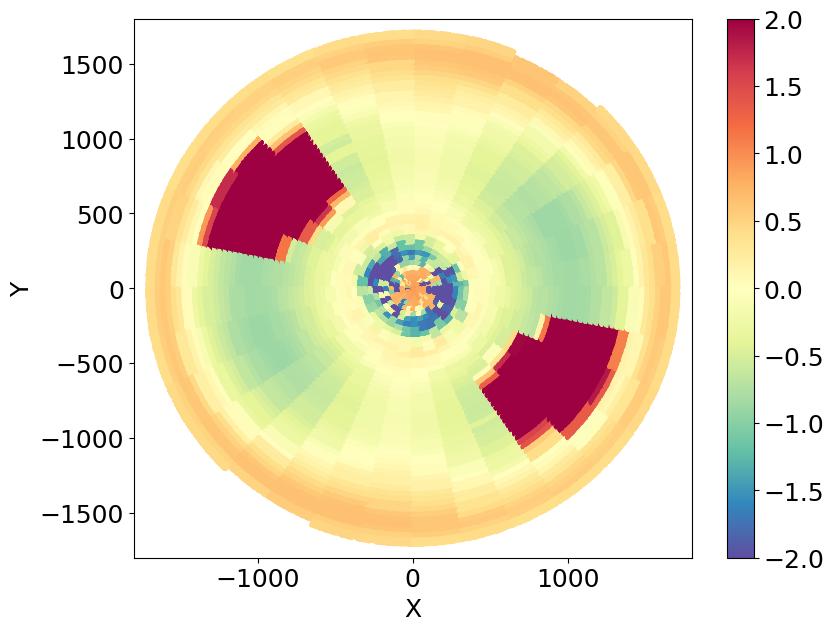}{0.3\textwidth}{(f)}}
\gridline{\fig{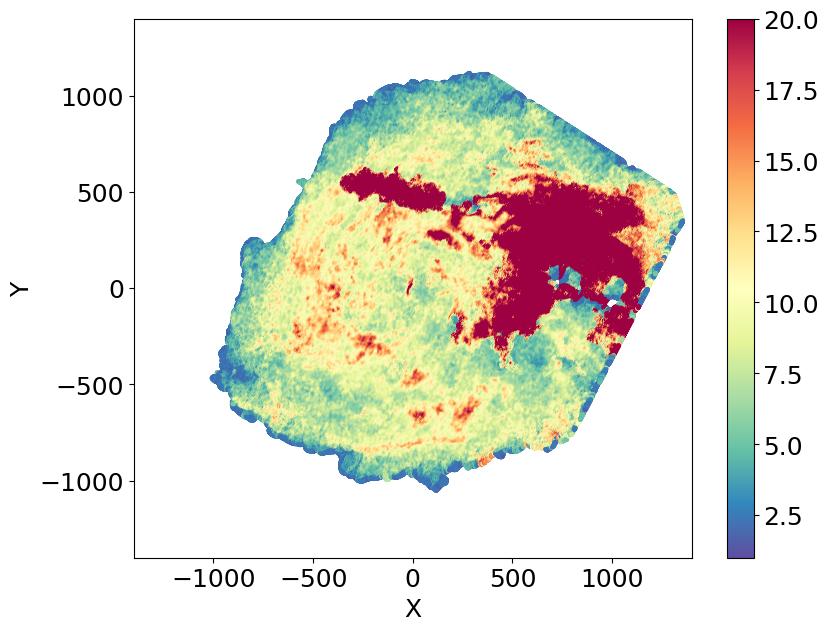}{0.3\textwidth}{(g)}
              \fig{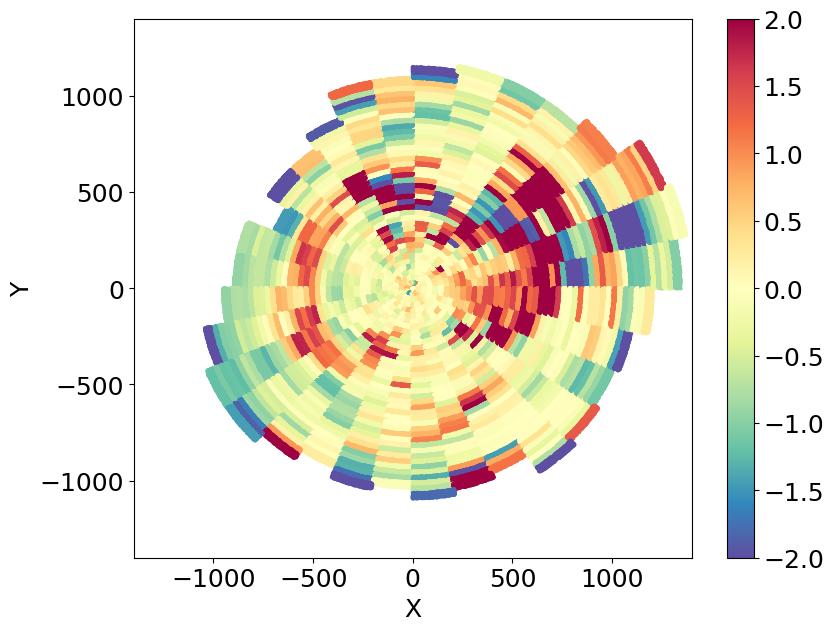}{0.3\textwidth}{(h)}
              \fig{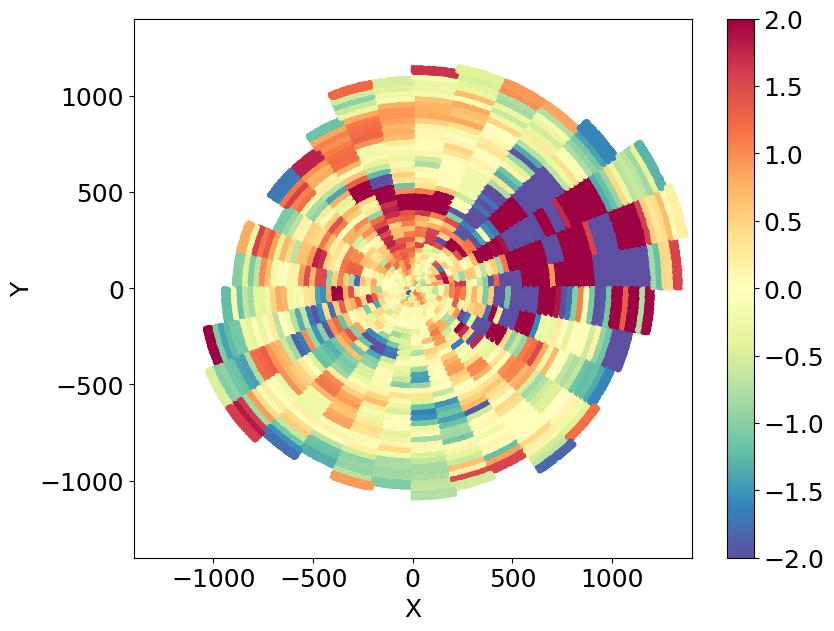}{0.3\textwidth}{(i)}}
\caption{As  Fig.\ref{fig:NGC2903-2} but for NGC 5457.} 
\label{fig:NGC5457-2} 
\end{figure*}
%


\subsubsection{NGC 5194} 

The proximity of a satellite to this galaxy has a significant influence on the M51's velocity field and the morphology of the neutral hydrogen distribution. The inclination angle  and the P.A. exhibits substantial variations of tens of degrees when transitioning from the inner to the outer regions of the disk, as illustrated in panel (a) of Fig. \ref{fig:NGC5194-1}. Assuming that the TRM accurately measured the variations in the orientation angles, the corrections to $v_r$ and $v_t$ are can larger than 100 km/s due to the large variations in the orientation angles (see panels (b) and (c) of Fig. \ref{fig:NGC5194-1}): such large variations suggest that the TRM analysis is problematic for this galaxy.  The velocity dispersion profile also shows significant fluctuations beyond $R \sim$ 200'', which can be attributed to the large perturbations induced by the presence of the satellite.
To investigate whether the observed variations in $i(R)$ and $\phi_0(R)$ correspond to a true geometric deformation resulting from a warp, the velocity field of 
the  corresponding toy disc model is analyzed. This model is constructed using the procedure outlined in Section \ref{sect:things}. 
{ We find that the correlation coefficient is  close to zero, i.e. ${\cal C} = 0.03$, corresponding to the absence of the warp.}

As depicted in Fig. \ref{fig:NGC5194-2} the velocity field of the toy disc model exhibits extensive and symmetric anisotropy patterns different from those observed in the actual galaxy. 
Particularly, in the direction of approximately $200^\circ$, there is a clear indication of the satellite's presence in both velocity components: $v_t$ demonstrates significant perturbations, while $v_r$ shows an expanding trend. Conversely, in the opposite direction, an extended tail characterizes the system, accompanied by a positive $v_r$. The velocity variance field further supports the reality of the signal by displaying significant perturbations aligned with the satellite's direction. Based on the features observed in the velocity rank correlation field, we can conclude that the presence of a warp can be excluded in this case and that the galaxy velocity field is perturbed by intrinsic velocity fluctuations, the largest ones due the the satellite. The perturbations affect the external regions of the disk, while its inner part is less perturbed. However, in the inner regions velocity fluctuations of order of 50 km/s characterize both velocity components maps. 

\begin{figure*}
\gridline{\fig{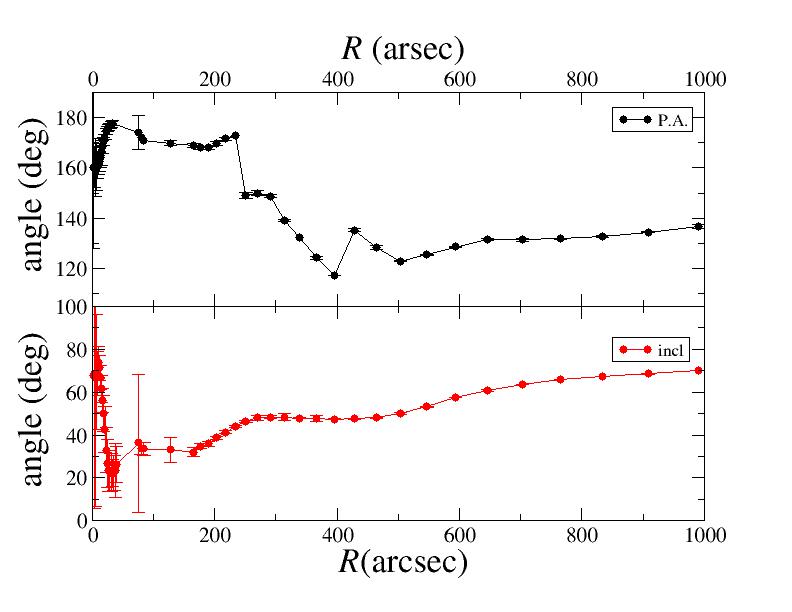}{0.45\textwidth}{(a)}
              \fig{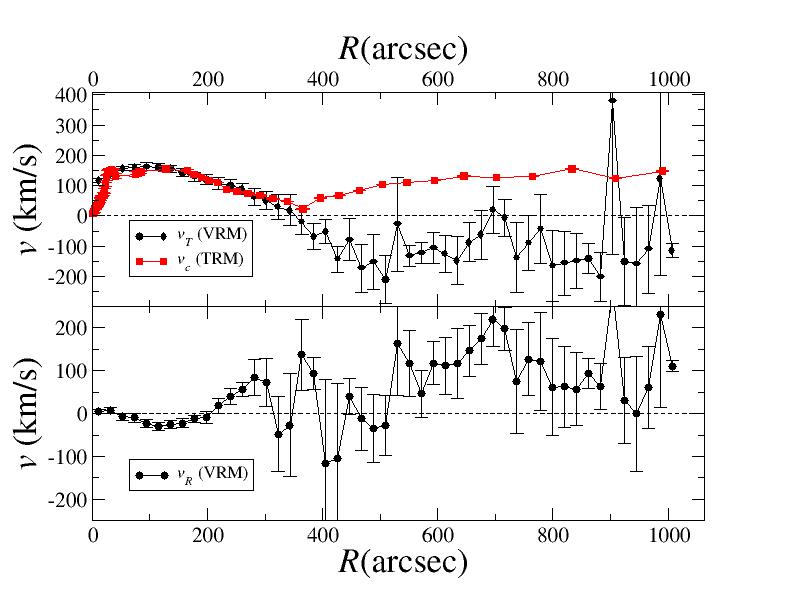}{0.45\textwidth}{(b)}
              }
  \gridline{
 \fig{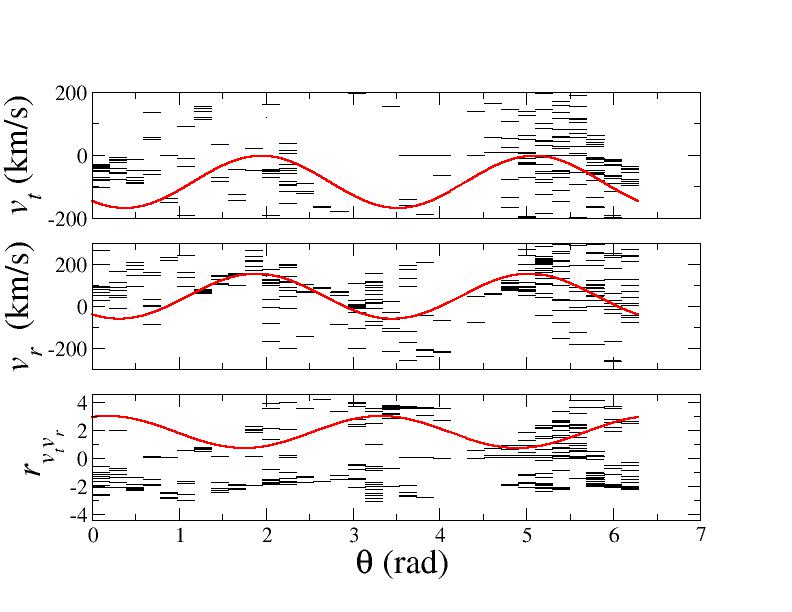}{0.45\textwidth}{(c)}
                \fig{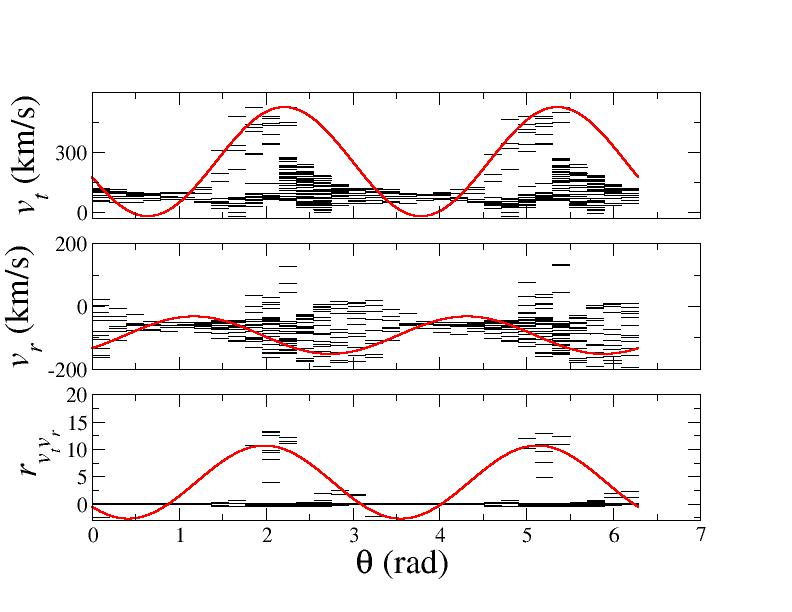}{0.45\textwidth}{(d)}}
     \caption{As  Fig.\ref{fig:NGC2903-1} but for NGC 5194.} 
\label{fig:NGC5194-1} 
\end{figure*}
%

\begin{figure*}
\gridline{\fig{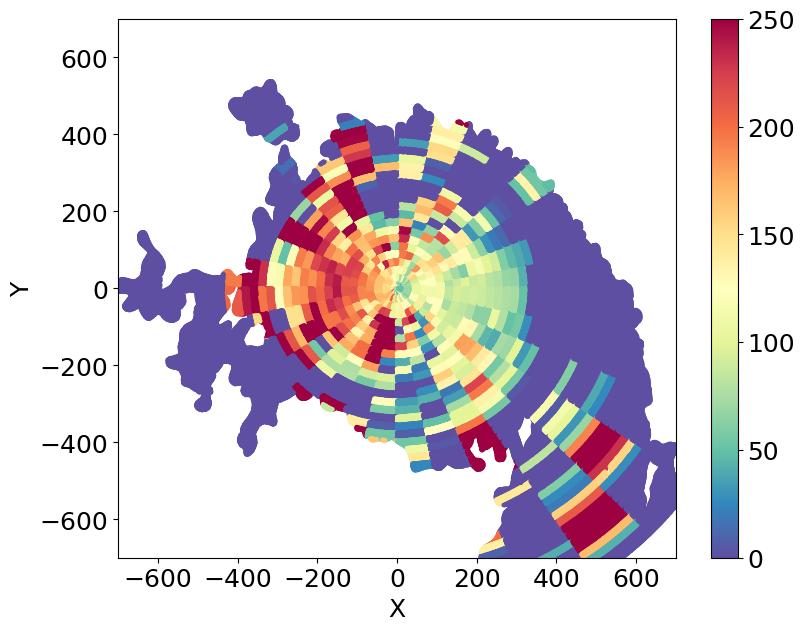}{0.3\textwidth}{(a)}
              \fig{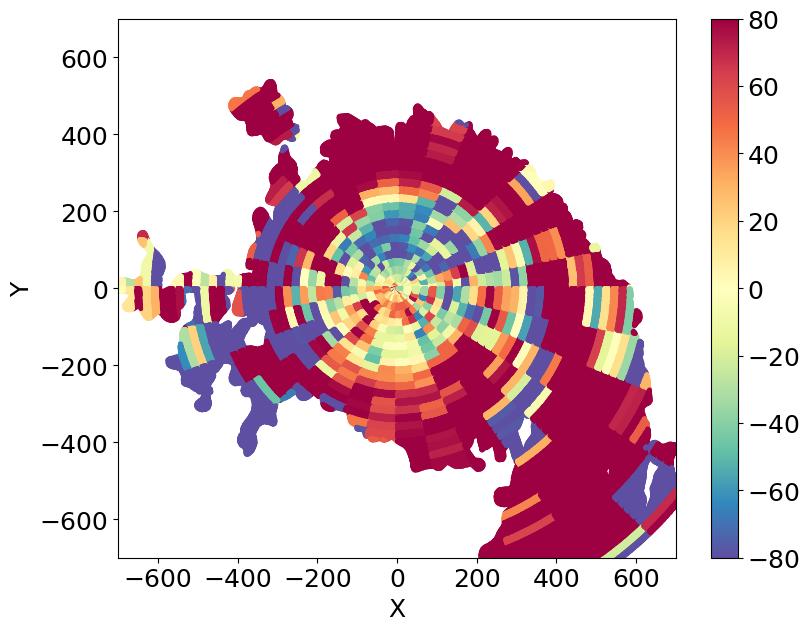}{0.3\textwidth}{(b)}
               \fig{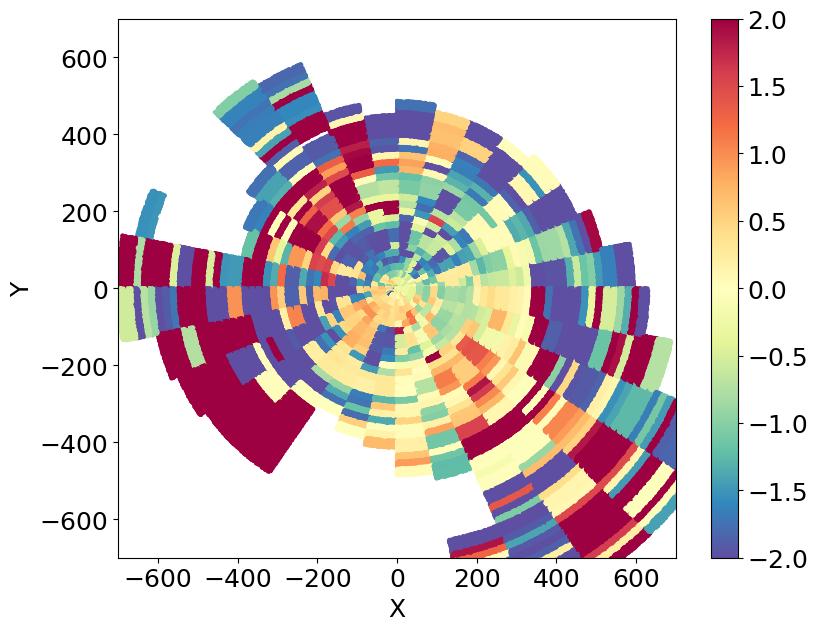}{0.3\textwidth}{(c)}}
\gridline{\fig{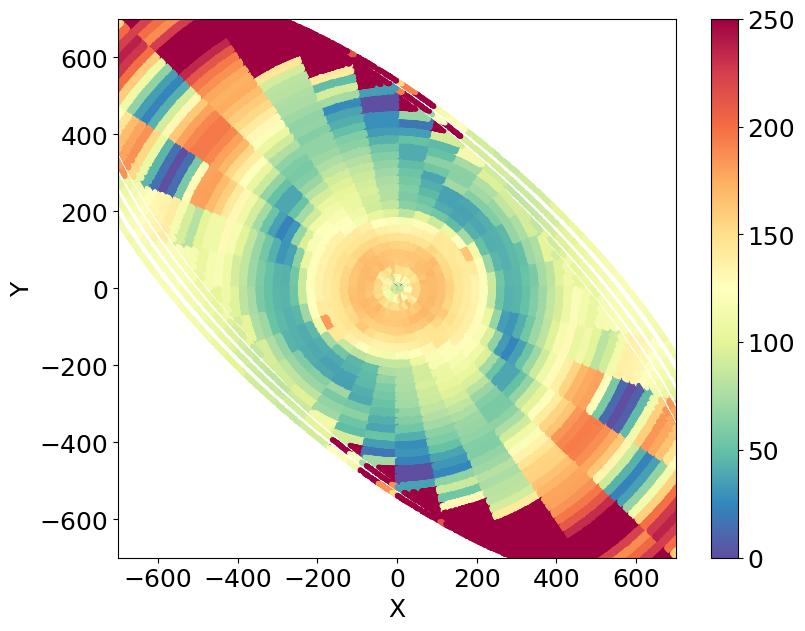}{0.3\textwidth}{(d)}
              \fig{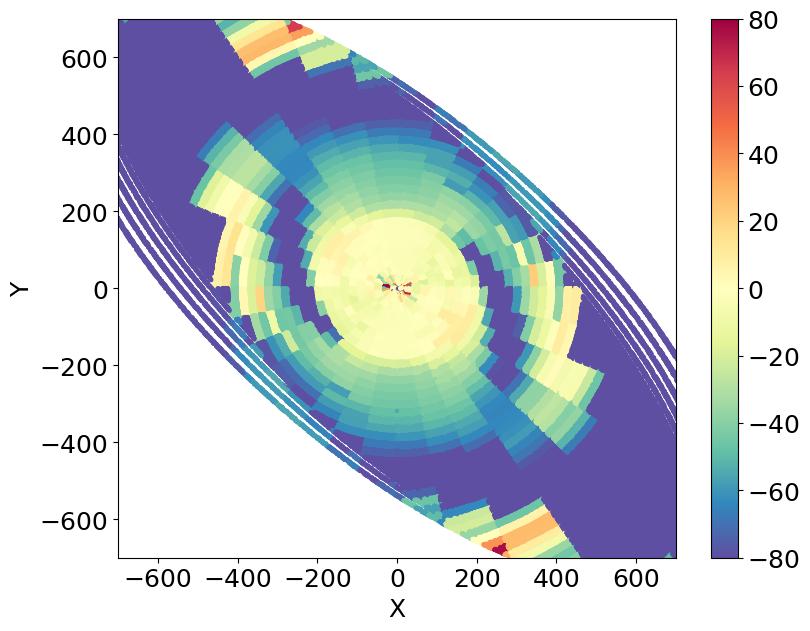}{0.3\textwidth}{(e)}
              \fig{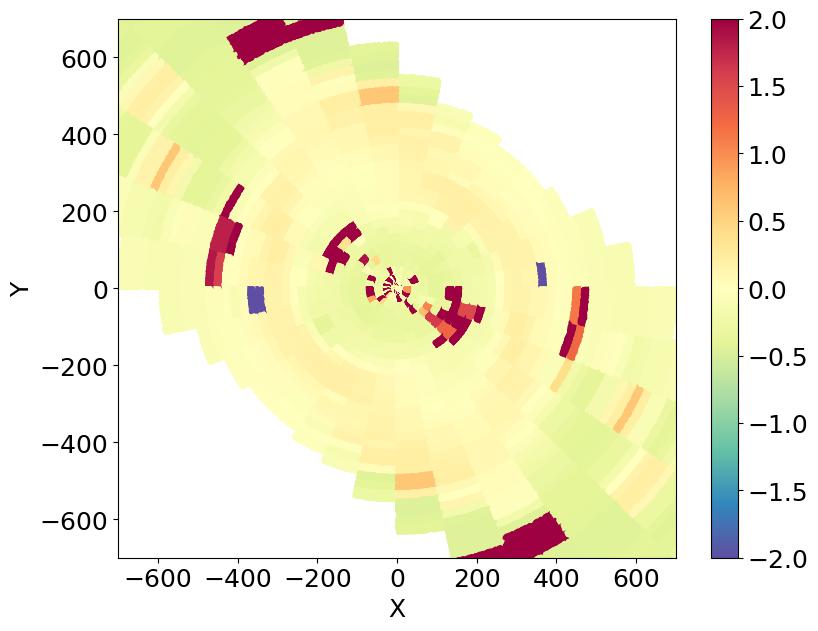}{0.3\textwidth}{(f)}}
\gridline{\fig{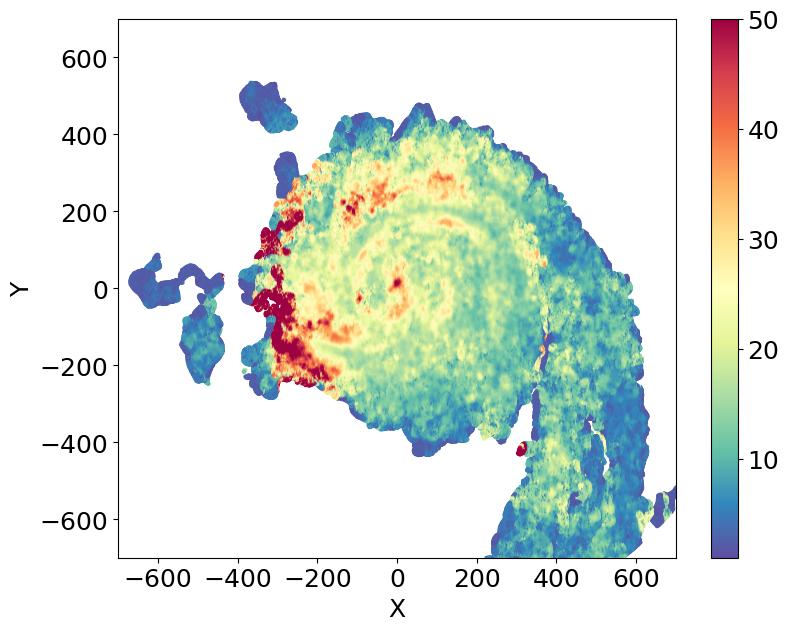}{0.3\textwidth}{(g)}
              \fig{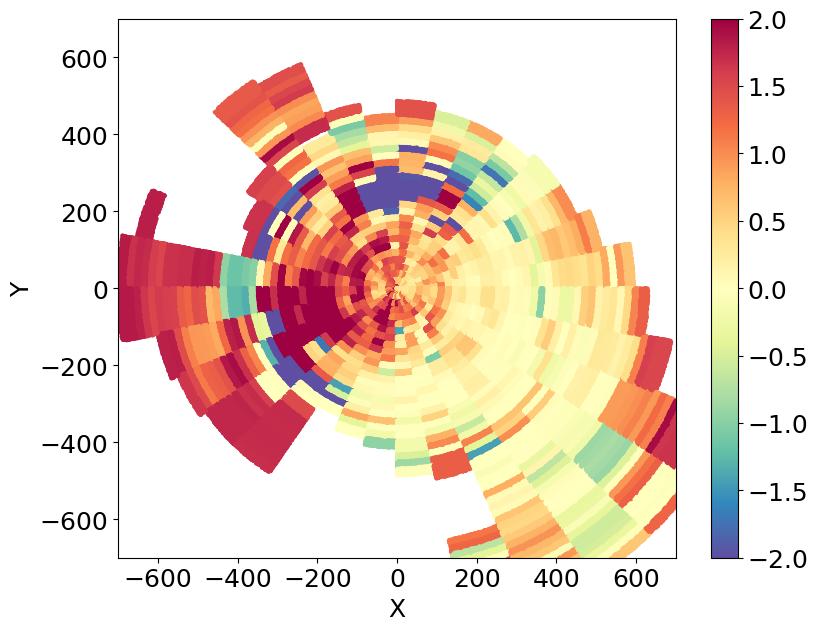}{0.3\textwidth}{(h)}
              \fig{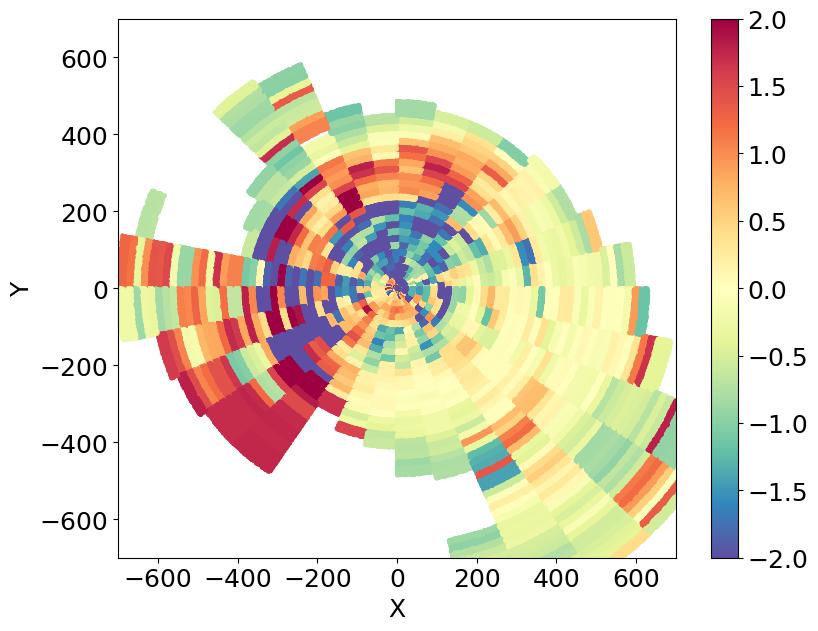}{0.3\textwidth}{(i)}}
\caption{As  Fig.\ref{fig:NGC2903-2} but for NGC 5194.} 
\label{fig:NGC5194-2} 
\end{figure*}
%

\begin{table}
\begin{center}
\begin{tabular}{ |c|c| } 
 \hline
 Galaxy  & $\cal{C}$  \\ 
 \hline 
 NGC 628 & 0.50  \\ 
 NGC 925 & 0.42  \\ 
 NGC 2366 & 0.04  \\ 
 NGC 2403 & 0.44  \\ 
 NGC 2841 & 0.25  \\ 
 NGC 2903 & 0.45  \\ 
 NGC 2976 & 0.14  \\ 
 NGC 3031 & 0.18  \\ 
 NGC 3184 & 0.33  \\ 
 NGC 3198 & 0.40  \\ 
 NGC 3351 & 0.42  \\ 
 NGC 3521 & 0.40  \\ 
 NGC 3261 & 0.08  \\ 
 NGC 3627 & 0.22  \\ 
 NGC 4736 & 0.34  \\ 
 NGC 4826 & 0.08  \\ 
 NGC 5055 & 0.36  \\ 
 NGC 5194 & 0.03  \\ 
 NGC 5236 & 0.19  \\ 
 NGC 5457 & 0.10  \\ 
 NGC 6946 & 0.52  \\ 
 NGC 7331 & 0.16  \\ 
 NGC 7793 & 0.51  \\ 
 DDO 154 & 0.09  \\ 
 M 33 & 0.18 \\ 
 \hline
\end{tabular}
\end{center}
\caption{ The Spearman correlation coefficient, $\cal{C}$, computed between the rank correlation velocity coefficient for the galactic data, $r_{v_t v_r}(R,\theta)$, and the toy disc model, $r_{v_t v_r}^{tm}(R,\theta)$, which is constructed to have the same orientation angles measured by the  TRM of the real galactic data. A value ${\cal C} <0.2$ is interpreted as signaling the absence of warp, otherwise a warp is confirmed to be present in the outer regions of the galactic discs.}
\label{tab1}
\end{table}


\section{Discussion and conclusions} 
\label{sect:discussion}

{  The reconstruction of the velocity field of an external galaxy from the observation of the line-of-sight velocity (LOS)  component is generally done by applying methods that must be necessarily based on some assumptions.} The most common hypothesis is that radial velocities are null, and the geometry of the galactic disc may be deformed by a warp. This method, generally named the titled-ring model (TRM), allows the reconstruction of the transversal velocity profile in a series of rings that may have different orientations  \citep{Warner_etal_1973,Rogstad_etal_1974}: the variation of the inclination angle and/or the position angle of each ring with radius corresponds to the geometric properties of the warp. Attempts based on the same technique, where the radial velocity is not a priori set to zero, enable in principle the determination of radial flows, i.e. the radial velocity profile. However, it remains unclear whether this method can accurately measure all radial velocities due to their degeneracy with warps. Indeed, geometric deformation and radial flows are degenerate in the projected image of a galaxy if we know only the LOS velocity component as it occur for the observations of external galaxies.

Alternatively, there are methods that assume the disc to be flat and reconstruct the transversal velocity field  placing limits on non-circular motions \citep{Barnes+Sellwood_2003,Spekkens+Sellwood_2007,Sellwood+Sanchez_2010,deNaray_etal_2012,Sellwood+Spekkens_2015,Sellwood_etal_2021}. Recently, the velocity ring model (VRM) was introduced, which, under the assumption of a flat disc, can reconstruct coarse-grained two-dimensional maps of both the transversal and radial velocity components. That is, this methods allows for the reconstruction not only of the  transversal and radial velocity profiles averaged over concentric rings, but also it can be measure both velocity components in angular sectors so to trace the distribution of angular anisotropies and to relate that to spatial structures. The VRM has recently  been applied to the galaxies of the THINGS sample: a detailed discussion can be found in \cite{SylosLabini_etal_2023}. 

In this work we have further developed these previous results studying in more details the effects of a warp on the velocity fields reconstructed by the VRM. Specifically, through tests conducted with toy disc galaxy models, we have demonstrated that the VRM, when applied to cases where the disc is warped, provides a clear signal indicating the existence of a warp. Although the application of the VRM to a warped disc contradicts its basic assumption of a flat disc, it aligns with the reality that, for an external galaxy, one does not a priori know whether the disc is flat or not. We have thus found that in toy disc models a warp, when analyzed with the VRM, corresponds to a dipolar angular  correlation between the radial and transversal velocity components. In addition, such an angular modulation corresponds to a smooth and slow change of both velocity components with the polar angle and can be very accurately detected by the VRM analysis. 
{  We have illustrated, through the analysis of toy galactic disc models, that the presence of warps becomes evident in the study of velocity fields reconstructed by the VRM.  In particular,  we have shown that  the spatial properties of velocity anisotropies, obtained in the VRM  by assuming a flat disc, offer a distinct and definitive signature of the potential presence of a warp.  Thus if the disc is intrinsically warped, applying VRM allows us to break the degeneracy between geometric deformations of the disc and radial motions and to measure the amplitude of velocity perturbations.}

We applied these findings to the galaxies in the THINGS sample plus M33, to characterize the nature of their velocity fields. Specifically, we compared the velocity field of each galaxy in our sample obtained with the VRM to that of a toy disc model, measured as well by using the VRM. Such a toy disc model is not flat but have the same geometric deformations (i.e., behavior of the orientation angles with radius) as well as the same circular velocity found by the TRM in the specific galaxy under consideration. The study of the  toy model constructed with the properties detected by the TRM can be though  as a null-hypothesis test. By comparing the  velocity maps of the real and of the synthetic galaxy we can identify the features of the velocity maps that cannot be solely explained as a warp or exclude that a warp is present in the velocity distribution. From this analysis we conclude that most galaxies in the sample exhibit a moderate warp signature in their outer regions, albeit with varying intensities. However, some galaxies do not conform to a simple warp deformation and instead display perturbed velocity fields. In these cases, the results from the TRM analysis may suggest the presence of a warp, whereas it may not be evident when analyzing the correlation between the velocity components.

In both scenarios, whether a warp is observed or not, the VRM can be used to determine the  velocity profiles of the transverse and radial velocity components averaged over concentric rings. When a warp is detected, corrections to the velocity profiles can be computed to account for the deviations from these mean behaviors that are attributable to the warp itself. Furthermore, our findings consistently indicate that intrinsic variations in the velocity fields are always present in the galaxies of the sample we have considered. If a warp is also present, such variations add to the extrinsic velocity anisotropies associated with warps. These intrinsic variations can be connected to local perturbations such as bars, spiral arms, or interactions with satellites, etc. 

{ It is worth noting that the VRM reconstruction method, as well as the test developed in this work to disentangle warps from non-circular motions, can be applied, for instance, to the analysis of gas near a supermassive black hole \citep{Zhang_etal_2025} to determine the presence of inflows and their possible amplitude, as well as to the study of gas kinematics in protoplanetary disks \citep{Zuleta_etal_2024}.
}

In general, in order to develop a dynamical galactic model from the measurements of velocity fields in galaxies, it  is crucial to quantify the intensity of the intrinsic kinematic perturbations. We have found that these  typically  have an amplitude which exceed that of the extrinsic velocity perturbations which are  due to the geometrical deformation in the VRM analysis, and thus are artifacts of an inconsistent assumption. It is interesting to observe that in our own Galaxy both a warp and intrinsic velocity perturbations are present. Indeed, there is clear evidence that the Milky Way's disc is warped, as indicated by studies such as \cite{Levine_etal_2006, Kalberla_etal_2007, Reyle_etal_2009}. Moreover, the data of the Gaia mission \citep{Gaia_2016}, thanks to the six-dimensional mapping of stars positions and velocities,  have recently unveiled large-scale gradients in all velocity components and deviations from axisymmetry (see,  e.g., \cite{Antoja_etal_2018, Lopez-Corredoira_Sylos-Labini_2019, Antoja_etal_2021, Recio-Blanco_etal_2022, Katz_etal_2022, Wang_etal_2023}). The Milky way has thus both a complex geometrical structure and non-trivial velocity fields, a situation that is quite common for the galaxies in our sample.

In addition to the previously mentioned findings, our analysis has revealed that a very useful information is provided by  spatial correlation between the radial velocity component derived from the VRM analysis and the velocity dispersion, i.e. the second moment map of the galaxy which is generally also provided by \HI\ observations.  This correlation offers further evidence for the existence of intrinsic perturbations in a galaxy's velocity field.
The correlation between the asymmetric features observed in the radial velocity component and the velocity dispersion arises from the fact that both are generated by local perturbations. This correlation helps to break the degeneracy between non-circular motions and warps, allowing us to distinguish between these two different features. For this reason it provides valuable insights into the underlying dynamics and structures within a galaxy.

In summary, the results obtained enable both to determine whether a galactic disc is warped and to the quantify of the amplitude and directions of non-circular flows within a galaxy. This information is crucial for characterizing galactic dynamics and understanding the processes involved in galactic evolution. By studying and quantifying galaxy geometrical deformations and  non-circular flows, we gain a deeper understanding of the intricate interplay between various physical mechanisms and their impact on galactic structures and kinematics.

\section*{Acknowledgments} 
We thank S\'ebastien Comor\'on  for useful comments on a preliminary version of the paper and for a number of analysis of toy dic models with the code {\tt kinemestry}. 
We also thank Roberto Capuzzo-Dolcetta, Edvige Corbelli and Michael Joyce for useful discussions and David Thilker for the use of the data of M33.


\begin{thebibliography}{}
\expandafter\ifx\csname natexlab\endcsname\relax\def\natexlab#1{#1}\fi
\providecommand{\url}[1]{\href{#1}{#1}}
\providecommand{\dodoi}[1]{doi:~\href{http://doi.org/#1}{\nolinkurl{#1}}}
\providecommand{\doeprint}[1]{\href{http://ascl.net/#1}{\nolinkurl{http://ascl.net/#1}}}
\providecommand{\doarXiv}[1]{\href{https://arxiv.org/abs/#1}{\nolinkurl{https://arxiv.org/abs/#1}}}

\bibitem[{{Antoja} {et~al.}(2018){Antoja}, {Helmi}, {Romero-G{\'o}mez}, {Katz},
  {Babusiaux}, {Drimmel}, {Evans}, {Figueras}, {Poggio}, {Reyl{\'e}}, {Robin},
  {Seabroke}, \& {Soubiran}}]{Antoja_etal_2018}
{Antoja}, T., {Helmi}, A., {Romero-G{\'o}mez}, M., {et~al.} 2018, Nature, 561,
  360, \dodoi{10.1038/s41586-018-0510-7}

\bibitem[{{Barnes} \& {Sellwood}(2003)}]{Barnes+Sellwood_2003}
{Barnes}, E.~I., \& {Sellwood}, J.~A. 2003, Astronom.J., 125, 1164,
  \dodoi{10.1086/346142}

\bibitem[{{Chemin} {et~al.}(2006){Chemin}, {Carignan}, {Drouin}, \&
  {Freeman}}]{Chemin_etal_2006}
{Chemin}, L., {Carignan}, C., {Drouin}, N., \& {Freeman}, K.~C. 2006,
  Astron.J., 132, 2527, \dodoi{10.1086/508859}

\bibitem[{{Corbelli} \& {Schneider}(1997)}]{Corbelli_Schneider_1997}
{Corbelli}, E., \& {Schneider}, S.~E. 1997, Astrophys.J., 479, 244,
  \dodoi{10.1086/303849}

\bibitem[{{Corbelli} {et~al.}(2014){Corbelli}, {Thilker}, {Zibetti},
  {Giovanardi}, \& {Salucci}}]{Corbelli_etal_2014}
{Corbelli}, E., {Thilker}, D., {Zibetti}, S., {Giovanardi}, C., \& {Salucci},
  P. 2014, Astron.Astrophys., 572, A23, \dodoi{10.1051/0004-6361/201424033}

\bibitem[{{De Blok} {et~al.}(2008){De Blok}, {Walter}, {Brinks},
  {Trachternach}, {Oh}, \& {Kennicutt}}]{deBlok_etal_2008}
{De Blok}, W.~J.~G., {Walter}, F., {Brinks}, E., {et~al.} 2008, Astronom.J.,
  136, 2648, \dodoi{10.1088/0004-6256/136/6/2648}

\bibitem[{{Di Teodoro} \& {Peek}(2021)}]{DiTeodoro+Peek_2021}
{Di Teodoro}, E.~M., \& {Peek}, J.~E.~G. 2021, Astrophys.J., 923, 220,
  \dodoi{10.3847/1538-4357/ac2cbd}

\bibitem[{{Erroz-Ferrer} {et~al.}(2015){Erroz-Ferrer}, {Knapen}, {Leaman},
  {Cisternas}, {Font}, {Beckman}, {Sheth}, {Mu{\~n}oz-Mateos},
  {D{\'\i}az-Garc{\'\i}a}, {Bosma}, {Athanassoula}, {Elmegreen}, {Ho}, {Kim},
  {Laurikainen}, {Martinez-Valpuesta}, {Meidt}, \&
  {Salo}}]{Erroz-Ferrer_etal_2015a}
{Erroz-Ferrer}, S., {Knapen}, J.~H., {Leaman}, R., {et~al.} 2015,
  Mon.Not.R.Astr.Soc., 451, 1004, \dodoi{10.1093/mnras/stv924}

\bibitem[{{Franx} {et~al.}(1994){Franx}, {van Gorkom}, \& {de
  Zeeuw}}]{Franx_etal_1994}
{Franx}, M., {van Gorkom}, J.~H., \& {de Zeeuw}, T. 1994, Astrophys.J., 436,
  642, \dodoi{10.1086/174939}

\bibitem[{{Fraternali} {et~al.}(2001){Fraternali}, {Oosterloo}, {Sancisi}, \&
  {van Moorsel}}]{Fraternali_etal_2001}
{Fraternali}, F., {Oosterloo}, T., {Sancisi}, R., \& {van Moorsel}, G. 2001,
  Astrophys.J.Lett., 562, L47, \dodoi{10.1086/338102}

\bibitem[{{Gaia Collaboration} {et~al.}(2016){Gaia Collaboration}, {Prusti},
  {de Bruijne}, {Brown}, {Vallenari}, {Babusiaux}, {Bailer-Jones}, {Bastian},
  {Biermann}, {Evans}, \& et~al.}]{Gaia_2016}
{Gaia Collaboration}, {Prusti}, T., {de Bruijne}, J.~H.~J., {et~al.} 2016,
  Astron.Astrophys., 595, A1, \dodoi{10.1051/0004-6361/201629272}

\bibitem[{{Gaia Collaboration} {et~al.}(2021){Gaia Collaboration}, {Antoja},
  {McMillan}, {Kordopatis}, {Ramos}, {Helmi}, {Balbinot}, {Cantat-Gaudin},
  {Chemin}, {Figueras}, {Jordi}, {Khanna}, {Romero-G{\'o}mez}, {Seabroke},
  {Brown}, {Vallenari}, {Prusti}, {de Bruijne}, {Babusiaux}, {Biermann},
  {Creevey}, {Evans}, {Eyer}, {Hutton}, {Jansen}, {Klioner}, {Lammers},
  {Lindegren}, {Luri}, {Mignard}, {Panem}, {Pourbaix}, {Randich}, {Sartoretti},
  {Soubiran}, {Walton}, {Arenou}, {Bailer-Jones}, {Bastian}, {Cropper},
  {Drimmel}, {Katz}, {Lattanzi}, {van Leeuwen}, {Bakker}, {Casta{\~n}eda}, {De
  Angeli}, {Ducourant}, {Fabricius}, {Fouesneau}, {Fr{\'e}mat}, {Guerra},
  {Guerrier}, {Guiraud}, {Jean-Antoine Piccolo}, {Masana}, {Messineo},
  {Mowlavi}, {Nicolas}, {Nienartowicz}, {Pailler}, {Panuzzo}, {Riclet}, {Roux},
  {Sordo}, {Tanga}, {Th{\'e}venin}, {Gracia-Abril}, {Portell}, {Teyssier},
  {Altmann}, {Andrae}, {Bellas-Velidis}, {Benson}, {Berthier}, {Blomme},
  {Brugaletta}, {Burgess}, {Busso}, {Carry}, {Cellino}, {Cheek}, {Clementini},
  {Damerdji}, {Davidson}, {Delchambre}, {Dell'Oro},
  {Fern{\'a}ndez-Hern{\'a}ndez}, {Galluccio}, {Garc{\'\i}a-Lario},
  {Garcia-Reinaldos}, {Gonz{\'a}lez-N{\'u}{\~n}ez}, {Gosset}, {Haigron},
  {Halbwachs}, {Hambly}, {Harrison}, {Hatzidimitriou}, {Heiter},
  {Hern{\'a}ndez}, {Hestroffer}, {Hodgkin}, {Holl}, {Jan{\ss}en}, {Jevardat de
  Fombelle}, {Jordan}, {Krone-Martins}, {Lanzafame}, {L{\"o}ffler}, {Lorca},
  {Manteiga}, {Marchal}, {Marrese}, {Moitinho}, {Mora}, {Muinonen}, {Osborne},
  {Pancino}, {Pauwels}, {Recio-Blanco}, {Richards}, {Riello}, {Rimoldini},
  {Robin}, {Roegiers}, {Rybizki}, {Sarro}, {Siopis}, {Smith}, {Sozzetti},
  {Ulla}, {Utrilla}, {van Leeuwen}, {van Reeven}, {Abbas}, {Abreu Aramburu},
  {Accart}, {Aerts}, {Aguado}, {Ajaj}, {Altavilla}, {{\'A}lvarez}, {{\'A}lvarez
  Cid-Fuentes}, {Alves}, {Anderson}, {Varela}, {Audard}, {Baines}, {Baker},
  {Balaguer-N{\'u}{\~n}ez}, {Balog}, {Barache}, {Barbato}, {Barros}, {Barstow},
  {Bartolom{\'e}}, {Bassilana}, {Bauchet}, {Baudesson-Stella}, {Becciani},
  {Bellazzini}, {Bernet}, {Bertone}, {Bianchi}, {Blanco-Cuaresma}, {Boch},
  {Bombrun}, {Bossini}, {Bouquillon}, {Bragaglia}, {Bramante}, {Breedt},
  {Bressan}, {Brouillet}, {Bucciarelli}, {Burlacu}, {Busonero}, {Butkevich},
  {Buzzi}, {Caffau}, {Cancelliere}, {C{\'a}novas}, {Carballo}, {Carlucci},
  {Carnerero}, {Carrasco}, {Casamiquela}, {Castellani}, {Castro-Ginard},
  {Castro Sampol}, {Chaoul}, {Charlot}, {Chiavassa}, {Cioni}, {Comoretto},
  {Cooper}, {Cornez}, {Cowell}, {Crifo}, {Crosta}, {Crowley}, {Dafonte},
  {Dapergolas}, {David}, {David}, {de Laverny}, {De Luise}, {De March}, {De
  Ridder}, {de Souza}, {de Teodoro}, {de Torres}, {del Peloso}, {del Pozo},
  {Delgado}, {Delgado}, {Delisle}, {Di Matteo}, {Diakite}, {Diener},
  {Distefano}, {Dolding}, {Eappachen}, {Enke}, {Esquej}, {Fabre}, {Fabrizio},
  {Faigler}, {Fedorets}, {Fernique}, {Fienga}, {Fouron}, {Fragkoudi}, {Fraile},
  {Franke}, {Gai}, {Garabato}, {Garcia-Gutierrez}, {Garc{\'\i}a-Torres},
  {Garofalo}, {Gavras}, {Gerlach}, {Geyer}, {Giacobbe}, {Gilmore}, {Girona},
  {Giuffrida}, {Gomez}, {Gonzalez-Santamaria}, {Gonz{\'a}lez-Vidal}, {Granvik},
  {Guti{\'e}rrez-S{\'a}nchez}, {Guy}, {Hauser}, {Haywood}, {Hidalgo}, {Hilger},
  {H{\l}adczuk}, {Hobbs}, {Holland}, {Huckle}, {Jasniewicz}, {Jonker},
  {Juaristi Campillo}, {Julbe}, {Karbevska}, {Kervella}, {Kochoska},
  {Kontizas}, {Korn}, {Kostrzewa-Rutkowska}, {Kruszy{\'n}ska}, {Lambert},
  {Lanza}, {Lasne}, {Le Campion}, {Le Fustec}, {Lebreton}, {Lebzelter},
  {Leccia}, {Leclerc}, {Lecoeur-Taibi}, {Liao}, {Licata}, {Lindstr{\o}m},
  {Lister}, {Livanou}, {Lobel}, {Madrero Pardo}, {Managau}, {Mann}, {Marchant},
  {Marconi}, {Marcos Santos}, {Marinoni}, {Marocco}, {Marshall}, {Martin Polo},
  {Mart{\'\i}n-Fleitas}, {Masip}, {Massari}, {Mastrobuono-Battisti}, {Mazeh},
  {Messina}, {Michalik}, {Millar}, {Mints}, {Molina}, {Molinaro}, {Moln{\'a}r},
  {Montegriffo}, {Mor}, {Morbidelli}, {Morel}, {Morris}, {Mulone}, {Munoz},
  {Muraveva}, {Murphy}, {Musella}, {Noval}, {Ord{\'e}novic}, {Orr{\`u}},
  {Osinde}, {Pagani}, {Pagano}, {Palaversa}, {Palicio}, {Panahi}, {Pawlak},
  {Pe{\~n}alosa Esteller}, {Penttil{\"a}}, {Piersimoni}, {Pineau}, {Plachy},
  {Plum}, {Poggio}, {Poretti}, {Poujoulet}, {Pr{\v{s}}a}, {Pulone}, {Racero},
  {Ragaini}, {Rainer}, {Raiteri}, {Rambaux}, {Ramos-Lerate}, {Re Fiorentin},
  {Regibo}, {Reyl{\'e}}, {Ripepi}, {Riva}, {Rixon}, {Robichon}, {Robin},
  {Roelens}, {Rohrbasser}, {Rowell}, {Royer}, {Rybicki}, {Sadowski},
  {Sagrist{\`a} Sell{\'e}s}, {Sahlmann}, {Salgado}, {Salguero}, {Samaras},
  {Sanchez Gimenez}, {Sanna}, {Santove{\~n}a}, {Sarasso}, {Schultheis},
  {Sciacca}, {Segol}, {Segovia}, {S{\'e}gransan}, {Semeux}, {Siddiqui},
  {Siebert}, {Siltala}, {Slezak}, {Smart}, {Solano}, {Solitro}, {Souami},
  {Souchay}, {Spagna}, {Spoto}, {Steele}, {Steidelm{\"u}ller}, {Stephenson},
  {S{\"u}veges}, {Szabados}, {Szegedi-Elek}, {Taris}, {Tauran}, {Taylor},
  {Teixeira}, {Thuillot}, {Tonello}, {Torra}, {Torra}, {Turon}, {Unger},
  {Vaillant}, {van Dillen}, {Vanel}, {Vecchiato}, {Viala}, {Vicente},
  {Voutsinas}, {Weiler}, {Wevers}, {Wyrzykowski}, {Yoldas}, {Yvard}, {Zhao},
  {Zorec}, {Zucker}, {Zurbach}, \& {Zwitter}}]{Antoja_etal_2021}
{Gaia Collaboration}, {Antoja}, T., {McMillan}, P.~J., {et~al.} 2021,
  Astron.Astrophys., 649, A8, \dodoi{10.1051/0004-6361/202039714}

\bibitem[{{Garc{\'\i}a-Ruiz} {et~al.}(2002){Garc{\'\i}a-Ruiz}, {Sancisi}, \&
  {Kuijken}}]{Garcia-Ruiz_etal_2002}
{Garc{\'\i}a-Ruiz}, I., {Sancisi}, R., \& {Kuijken}, K. 2002,
  Astron.Astrophys., 394, 769, \dodoi{10.1051/0004-6361:20020976}

\bibitem[{{Gentile} {et~al.}(2007){Gentile}, {Salucci}, {Klein}, \&
  {Granato}}]{Gentile_etal_2007}
{Gentile}, G., {Salucci}, P., {Klein}, U., \& {Granato}, G.~L. 2007,
  Mon.Not.R.Astr.Soc., 375, 199, \dodoi{10.1111/j.1365-2966.2006.11283.x}

\bibitem[{{Hunter} {et~al.}(2012){Hunter}, {Ficut-Vicas}, {Ashley}, {Brinks},
  {Cigan}, {Elmegreen}, {Heesen}, {Herrmann}, {Johnson}, {Oh}, {Rupen},
  {Schruba}, {Simpson}, {Walter}, {Westpfahl}, {Young}, \&
  {Zhang}}]{Hunter_etal_2012}
{Hunter}, D.~A., {Ficut-Vicas}, D., {Ashley}, T., {et~al.} 2012, Astronom.J.,
  144, 134, \dodoi{10.1088/0004-6256/144/5/134}

\bibitem[{{Kalberla} {et~al.}(2007){Kalberla}, {Dedes}, {Kerp}, \&
  {Haud}}]{Kalberla_etal_2007}
{Kalberla}, P.~M.~W., {Dedes}, L., {Kerp}, J., \& {Haud}, U. 2007,
  Astron.Astrophys., 469, 511, \dodoi{10.1051/0004-6361:20066362}

\bibitem[{{Katz} {et~al.}(2022){Katz}, {Sartoretti}, {Guerrier}, {Panuzzo},
  {Seabroke}, {Th{\'e}venin}, {Cropper}, {Benson}, {Blomme}, {Haigron},
  {Marchal}, {Smith}, {Baker}, {Chemin}, {Damerdji}, {David}, {Dolding},
  {Fr{\'e}mat}, {Gosset}, {Jan{\ss}en}, {Jasniewicz}, {Lobel}, {Plum},
  {Samaras}, {Snaith}, {Soubiran}, {Vanel}, {Zwitter}, {Antoja}, {Arenou},
  {Babusiaux}, {Brouillet}, {Caffau}, {Di Matteo}, {Fabre}, {Fabricius},
  {Frakgoudi}, {Haywood}, {Huckle}, {Hottier}, {Lasne}, {Leclerc},
  {Mastrobuono-Battisti}, {Royer}, {Teyssier}, {Zorec}, {Crifo}, {Jean-Antoine
  Piccolo}, {Turon}, \& {Viala}}]{Katz_etal_2022}
{Katz}, D., {Sartoretti}, P., {Guerrier}, A., {et~al.} 2022, arXiv e-prints,
  arXiv:2206.05902.
\newblock \doarXiv{2206.05902}

\bibitem[{{Kuzio de Naray} {et~al.}(2012){Kuzio de Naray}, {Arsenault},
  {Spekkens}, {Sellwood}, {McDonald}, {Simon}, \& {Teuben}}]{deNaray_etal_2012}
{Kuzio de Naray}, R., {Arsenault}, C.~A., {Spekkens}, K., {et~al.} 2012,
  Mon.Not.R.Astr.Soc., 427, 2523, \dodoi{10.1111/j.1365-2966.2012.22126.x}

\bibitem[{{Levine} {et~al.}(2006){Levine}, {Blitz}, \&
  {Heiles}}]{Levine_etal_2006}
{Levine}, E.~S., {Blitz}, L., \& {Heiles}, C. 2006, Astrophys.J., 643, 881,
  \dodoi{10.1086/503091}

\bibitem[{{L{\'o}pez-Corredoira} \& {Sylos
  Labini}(2019)}]{Lopez-Corredoira_Sylos-Labini_2019}
{L{\'o}pez-Corredoira}, M., \& {Sylos Labini}, F. 2019, Astron.Astrophys., 621,
  A48, \dodoi{10.1051/0004-6361/201833849}

\bibitem[{{Peters} {et~al.}(2017){Peters}, {van der Kruit}, {Allen}, \&
  {Freeman}}]{Peters_etal_2017}
{Peters}, S.~P.~C., {van der Kruit}, P.~C., {Allen}, R.~J., \& {Freeman}, K.~C.
  2017, Mon.Not.R.Astr.Soc., 464, 2, \dodoi{10.1093/mnras/stw1774}

\bibitem[{Press {et~al.}(2007)Press, Teukolsky, Vetterling, \&
  Flannery}]{Numerical_Recepies}
Press, W.~H., Teukolsky, S.~A., Vetterling, W.~T., \& Flannery, B.~P. 2007,
  Numerical Recipes 3rd Edition: The Art of Scientific Computing, 3rd edn.
  (USA: Cambridge University Press)

\bibitem[{{Recio-Blanco} {et~al.}(2022){Recio-Blanco}, {de Laverny}, {Palicio},
  {Kordopatis}, {{\'A}lvarez}, {Schultheis}, {Contursi}, {Zhao}, {Torralba
  Elipe}, {Ordenovic}, {Manteiga}, {Dafonte}, {Oreshina-Slezak}, {Bijaoui},
  {Fremat}, {Seabroke}, {Pailler}, {Spitoni}, {Poggio}, {Creevey}, {Abreu
  Aramburu}, {Accart}, {Andrae}, {Bailer-Jones}, {Bellas-Velidis}, {Brouillet},
  {Brugaletta}, {Burlacu}, {Carballo}, {Casamiquela}, {Chiavassa}, {Cooper},
  {Dapergolas}, {Delchambre}, {Dharmawardena}, {Drimmel}, {Edvardsson},
  {Fouesneau}, {Garabato}, {Garcia-Lario}, {Garcia-Torres}, {Gavel}, {Gomez},
  {Gonzalez-Santamaria}, {Hatzidimitriou}, {Heiter}, {Jean-Antoine Piccolo},
  {Kontizas}, {Korn}, {Lanzafame}, {Lebreton}, {Le Fustec}, {Licata},
  {Lindstrom}, {Livanou}, {Lobel}, {Lorca}, {Magdaleno Romeo}, {Marocco},
  {Marshall}, {Mary}, {Nicolas}, {Pallas-Quintela}, {Panem}, {Pichon},
  {Riclet}, {Robin}, {Rybizki}, {Santovena}, {Silvelo}, {Smart}, {Sarro},
  {Sordo}, {Soubiran}, {Suvege}, {Ulla}, {Vallenari}, {Zorec}, {Utrilla}, \&
  {Bakker}}]{Recio-Blanco_etal_2022}
{Recio-Blanco}, A., {de Laverny}, P., {Palicio}, P.~A., {et~al.} 2022, arXiv
  e-prints, arXiv:2206.05541.
\newblock \doarXiv{2206.05541}

\bibitem[{{Reshetnikov} \& {Combes}(1998)}]{Reshetnikov+Combes_1998}
{Reshetnikov}, V., \& {Combes}, F. 1998, Astron.Astrophys., 337, 9

\bibitem[{{Reshetnikov} {et~al.}(2016){Reshetnikov}, {Mosenkov}, {Moiseev},
  {Kotov}, \& {Savchenko}}]{Reshetnikov_etal_2016}
{Reshetnikov}, V.~P., {Mosenkov}, A.~V., {Moiseev}, A.~V., {Kotov}, S.~S., \&
  {Savchenko}, S.~S. 2016, Mon.Not.R.Astr.Soc., 461, 4233,
  \dodoi{10.1093/mnras/stw1623}

\bibitem[{{Reyl{\'e}} {et~al.}(2009){Reyl{\'e}}, {Marshall}, {Robin}, \&
  {Schultheis}}]{Reyle_etal_2009}
{Reyl{\'e}}, C., {Marshall}, D.~J., {Robin}, A.~C., \& {Schultheis}, M. 2009,
  Astron.Astrophys., 495, 819, \dodoi{10.1051/0004-6361/200811341}

\bibitem[{{Rogstad} {et~al.}(1974){Rogstad}, {Lockhart}, \&
  {Wright}}]{Rogstad_etal_1974}
{Rogstad}, D.~H., {Lockhart}, I.~A., \& {Wright}, M.~C.~H. 1974, Astrophys.J.,
  193, 309, \dodoi{10.1086/153164}

\bibitem[{{S{\'a}nchez-Saavedra} {et~al.}(2003){S{\'a}nchez-Saavedra},
  {Battaner}, {Guijarro}, {L{\'o}pez-Corredoira}, \&
  {Castro-Rodr{\'{\i}}guez}}]{Sanchez-Saavedra_etal_2003}
{S{\'a}nchez-Saavedra}, M.~L., {Battaner}, E., {Guijarro}, A.,
  {L{\'o}pez-Corredoira}, M., \& {Castro-Rodr{\'{\i}}guez}, N. 2003,
  Astron.Astrophys., 399, 457, \dodoi{10.1051/0004-6361:20021751}

\bibitem[{{Sancisi}(1976)}]{Sancisi_1976}
{Sancisi}, R. 1976, Astron.Astrohys., 53, 159

\bibitem[{{Schmidt} {et~al.}(2016){Schmidt}, {Bigiel}, {Klessen}, \& {de
  Blok}}]{Schmidt_etal_2016}
{Schmidt}, T.~M., {Bigiel}, F., {Klessen}, R.~S., \& {de Blok}, W.~J.~G. 2016,
  Mon.Not.R.Astr.Soc., 457, 2642, \dodoi{10.1093/mnras/stw011}

\bibitem[{{Schoenmakers} {et~al.}(1997){Schoenmakers}, {Franx}, \& {de
  Zeeuw}}]{Schoenmakers_etal_1997}
{Schoenmakers}, R.~H.~M., {Franx}, M., \& {de Zeeuw}, P.~T. 1997,
  Mon.Not.R.Astr.Soc., 292, 349, \dodoi{10.1093/mnras/292.2.349}

\bibitem[{{Schwarzkopf} \& {Dettmar}(2001)}]{Schwarzkopf+Dettmar_2001}
{Schwarzkopf}, U., \& {Dettmar}, R.-J. 2001, Astron.Astrophys., 373, 402,
  \dodoi{10.1051/0004-6361:20010548}

\bibitem[{{Sellwood} \& {S{\'a}nchez}(2010)}]{Sellwood+Sanchez_2010}
{Sellwood}, J.~A., \& {S{\'a}nchez}, R.~Z. 2010, Mon.Not.R.Astr.Soc., 404,
  1733, \dodoi{10.1111/j.1365-2966.2010.16430.x}

\bibitem[{{Sellwood} \& {Spekkens}(2015)}]{Sellwood+Spekkens_2015}
{Sellwood}, J.~A., \& {Spekkens}, K. 2015, arXiv e-prints, arXiv:1509.07120.
\newblock \doarXiv{1509.07120}

\bibitem[{Sellwood {et~al.}(2021)Sellwood, Spekkens, \&
  Eckel}]{Sellwood_etal_2021}
Sellwood, J.~A., Spekkens, K., \& Eckel, C.~S. 2021, Monthly Notices of the
  Royal Astronomical Society, 502, 3843, \dodoi{10.1093/mnras/stab009}

\bibitem[{{Simon} {et~al.}(2005){Simon}, {Bolatto}, {Leroy}, {Blitz}, \&
  {Gates}}]{Simon_etal_2005}
{Simon}, J.~D., {Bolatto}, A.~D., {Leroy}, A., {Blitz}, L., \& {Gates}, E.~L.
  2005, Astrophys.J., 621, 757, \dodoi{10.1086/427684}

\bibitem[{{Spekkens} \& {Sellwood}(2007)}]{Spekkens+Sellwood_2007}
{Spekkens}, K., \& {Sellwood}, J.~A. 2007, Astrophys.J., 664, 204,
  \dodoi{10.1086/518471}

\bibitem[{{Sylos Labini} {et~al.}(2025){Sylos Labini}, {Capuzzo-Dolcetta}, {De
  Marzo}, \& {Straccamore}}]{SylosLabini_etal_2025}
{Sylos Labini}, F., {Capuzzo-Dolcetta}, R., {De Marzo}, G., \& {Straccamore},
  M. 2025, Astron.Astrophys., 693, A248, \dodoi{10.1051/0004-6361/202452556}

\bibitem[{{Sylos Labini} {et~al.}(2023){Sylos Labini}, {Straccamore}, {De
  Marzo}, \& {Comer{\'o}n}}]{SylosLabini_etal_2023}
{Sylos Labini}, F., {Straccamore}, M., {De Marzo}, G., \& {Comer{\'o}n}, S.
  2023, Mon.Not.R.Astr.Soc., 524, 1560, \dodoi{10.1093/mnras/stad1916}

\bibitem[{{Trachternach} {et~al.}(2008){Trachternach}, {de Blok}, {Walter},
  {Brinks}, \& {Kennicutt}}]{Trachternach_etal_2008}
{Trachternach}, C., {de Blok}, W.~J.~G., {Walter}, F., {Brinks}, E., \&
  {Kennicutt}, Jr., R.~C. 2008, Astronom.J., 136, 2720,
  \dodoi{10.1088/0004-6256/136/6/2720}

\bibitem[{{Walter} {et~al.}(2008){Walter}, {Brinks}, {de Blok}, {Bigiel},
  {Kennicutt}, {Thornley}, \& {Leroy}}]{Walter_etal_2008}
{Walter}, F., {Brinks}, E., {de Blok}, W.~J.~G., {et~al.} 2008, Astronom.J.,
  136, 2563, \dodoi{10.1088/0004-6256/136/6/2563}

\bibitem[{{Wang} \& {Lilly}(2023)}]{Wang+Lilly_2023}
{Wang}, E., \& {Lilly}, S.~J. 2023, Astrophys.J., 944, 143,
  \dodoi{10.3847/1538-4357/acaf31}

\bibitem[{{Wang} {et~al.}(2023){Wang}, {Chrob{\'a}kov{\'a}},
  {L{\'o}pez-Corredoira}, \& {Sylos Labini}}]{Wang_etal_2023}
{Wang}, H.-F., {Chrob{\'a}kov{\'a}}, {\v{Z}}., {L{\'o}pez-Corredoira}, M., \&
  {Sylos Labini}, F. 2023, Astrophys.J., 942, 12,
  \dodoi{10.3847/1538-4357/aca27c}

\bibitem[{{Warner} {et~al.}(1973){Warner}, {Wright}, \&
  {Baldwin}}]{Warner_etal_1973}
{Warner}, P.~J., {Wright}, M.~C.~H., \& {Baldwin}, J.~E. 1973,
  Mon.Not.R.Astr.Soc., 163, 163, \dodoi{10.1093/mnras/163.2.163}

\bibitem[{Wong {et~al.}(2004)Wong, Blitz, \& Bosma}]{Wong_etal_2004}
Wong, T., Blitz, L., \& Bosma, A. 2004, Astrophys.J., 605, 183,
  \dodoi{10.1086/382215}

\bibitem[{{Zhang} {et~al.}(2025){Zhang}, {Bureau}, {Ruffa}, {Cappellari},
  {Davis}, {Dominiak}, {Elford}, {Iguchi}, {Lelli}, {Sarzi}, \&
  {Williams}}]{Zhang_etal_2025}
{Zhang}, H., {Bureau}, M., {Ruffa}, I., {et~al.} 2025, Mon.Not.R.Astr.Soc.,
  \dodoi{10.1093/mnras/staf055}

\bibitem[{{Zuleta} {et~al.}(2024){Zuleta}, {Birnstiel}, \&
  {Teague}}]{Zuleta_etal_2024}
{Zuleta}, A., {Birnstiel}, T., \& {Teague}, R. 2024, Astron.Astrophys., 692,
  A56, \dodoi{10.1051/0004-6361/202451145}

\end{thebibliography}
\bibliographystyle{aasjournal}

\newpage
\clearpage 
\section*{Appendix}



\subsection*{NGC 628} 

The inclination angle $i(R)$ and P.A. $\phi_0(R)$ exhibit substantial variations of tens of degrees when transitioning from the inner to the outer regions of the disk, as illustrated in panel (a) of Fig. \ref{fig:NGC628-1}.
Panel (b) of the figure presents both the observed and corrected velocity profiles, assuming that the TRM accurately measured the variations in the orientation angles. In this case, due to the large variations in the orientation angles, the corrections can be on the order of 100 km/s. Panel (c) displays the corrections themselves.
The velocity dispersion profile also demonstrates significant fluctuations in the same direction, $\theta \approx \pi/2$ where the radial velocity shows a rapid increase toward the disc peripheries. 
To investigate whether the observed variations in $i(R)$ and $\phi_0(R)$ correspond to a geometric deformation resulting from a warp, the velocity field of a corresponding toy disc model is analyzed.

As depicted in Figs \ref{fig:NGC628-1}-\ref{fig:NGC628-2}, the velocity field of the toy disc model exhibits extensive and symmetrical anisotropies that differ from those observed in the actual galaxy, NGC 628. The velocity components of NGC 628 do not display the dipolar oscillation observed in the toy model with the same orientation angles. However, the velocity rank correlation coefficient does show a dipolar oscillation, although it may not be as clear as in the toy model. Note that the different shape of the disc of NGC 628 and of the toy model can be attributed to the fact that in the former case a constant inclination angle has been employed to reconstruct  the image in the galaxy coordinates. 
{   We find that the correlation coefficient is  ${\cal C} = 0.5$}. 

Based on these comparisons { and on the value of the correlation coefficient ${\cal C}$}, we can conclude that a warp is most likely present in NGC 628. However, its amplitude in terms of the variation of the orientation angles is not as strong as to produce a change of $\Delta P.A. \approx 70^\circ$. In this situation, the corrected velocity profiles are also affected by the larger variations of the orientation angle detected by the TRM and the actual profiles of NGC 628 are likely to lie between the uncorrected and corrected ones. In addition, we may conclude that large velocity gradients of the oder of $\sim 50-70$ km s$^{-1}$ are present in the outer regions of the dic of NGC 628.
\begin{figure*}
\gridline{\fig{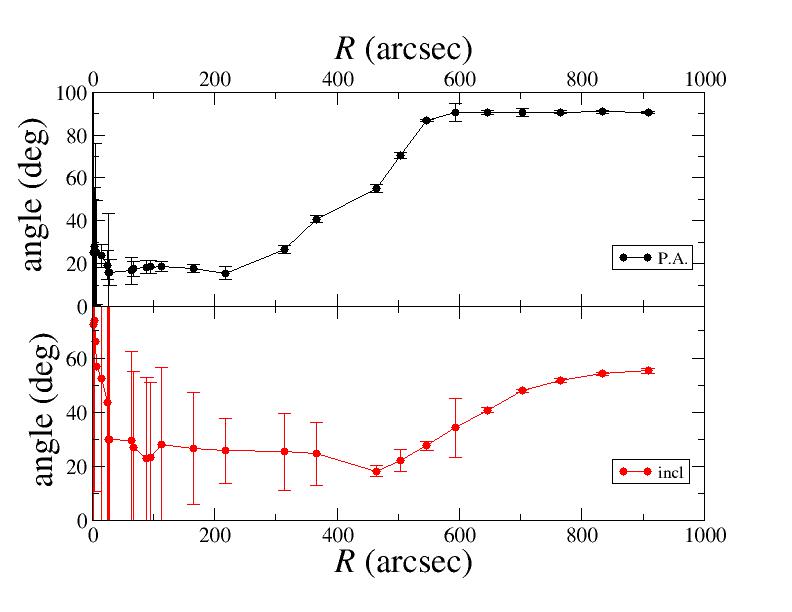}{0.45\textwidth}{(a)}
              \fig{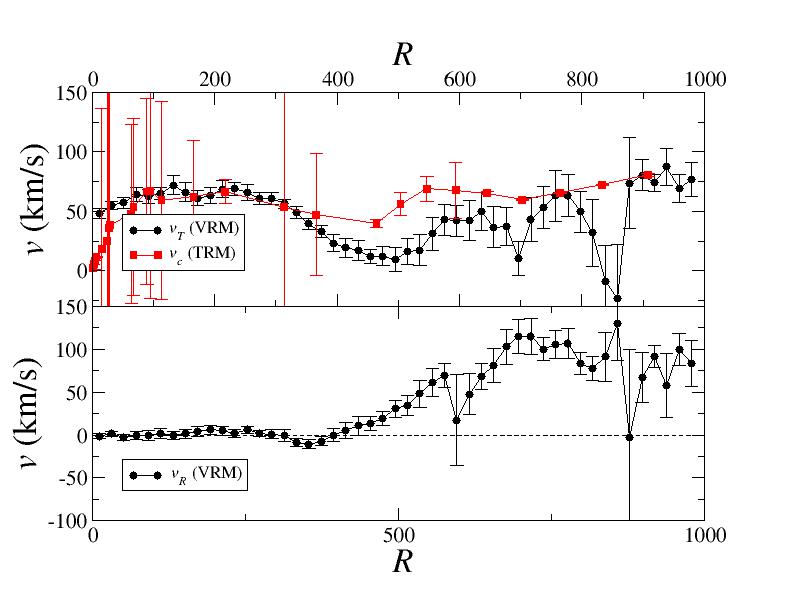}{0.45\textwidth}{(b)}
              }
  \gridline{
 \fig{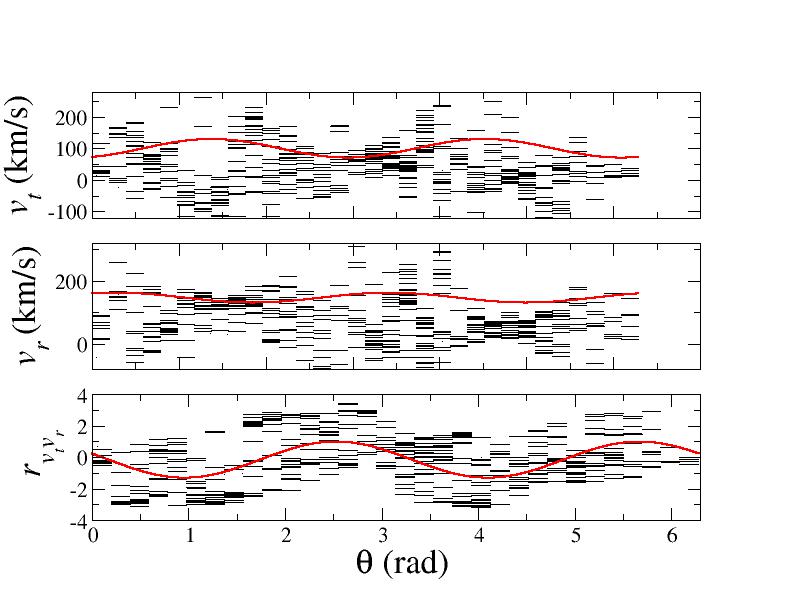}{0.45\textwidth}{(c)}
                \fig{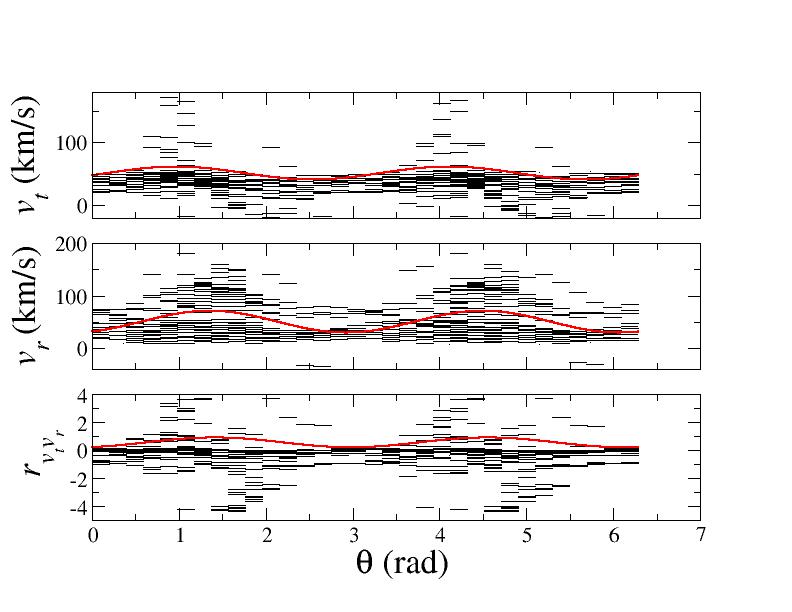}{0.45\textwidth}{(d)}}
     \caption{As  Fig.\ref{fig:NGC2903-1} but for NGC 628.} 
\label{fig:NGC628-1} 
\end{figure*}
%

\begin{figure*}
\gridline{\fig{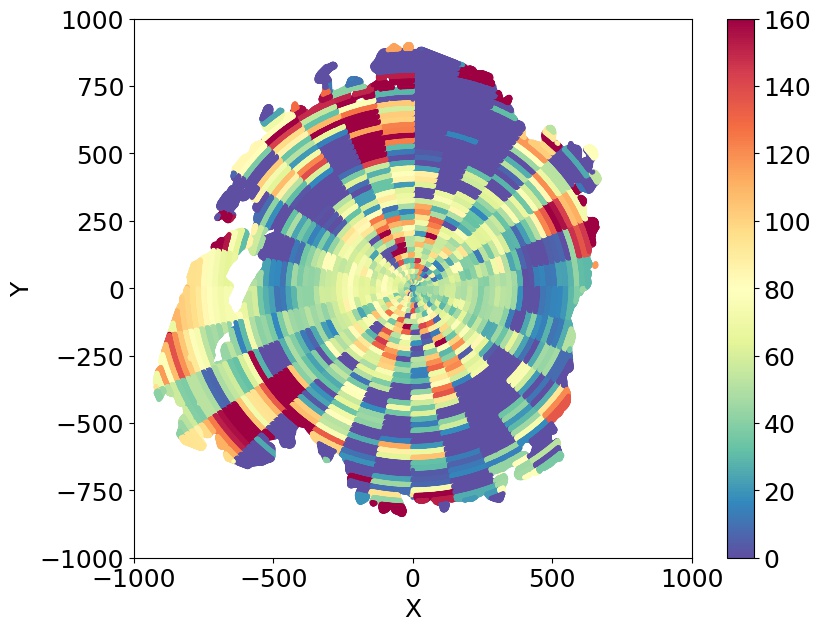}{0.3\textwidth}{(a)}
              \fig{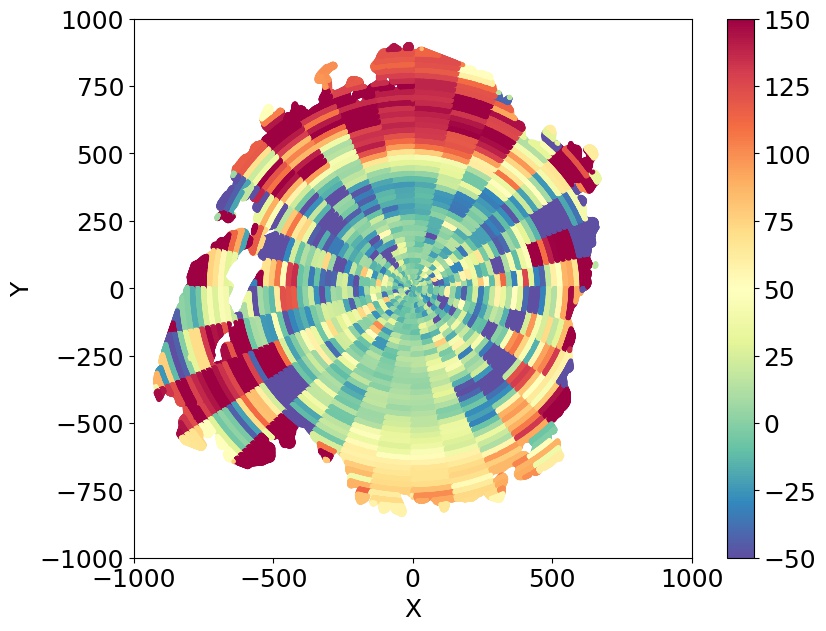}{0.3\textwidth}{(b)}
               \fig{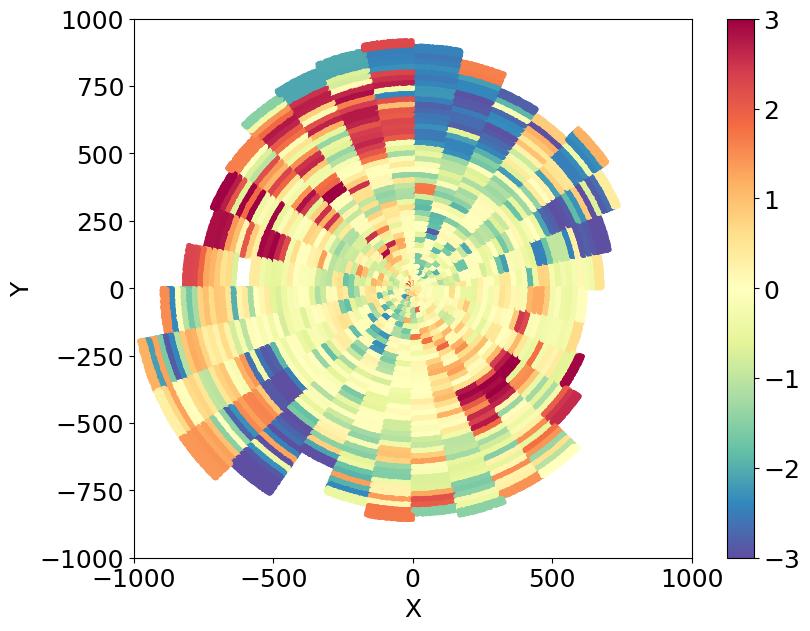}{0.3\textwidth}{(c)}}
\gridline{\fig{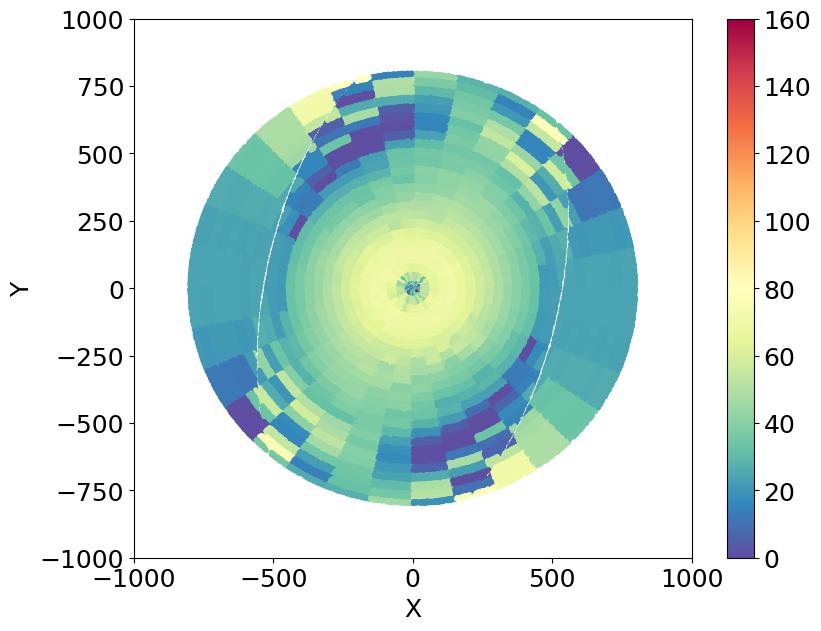}{0.3\textwidth}{(d)}
              \fig{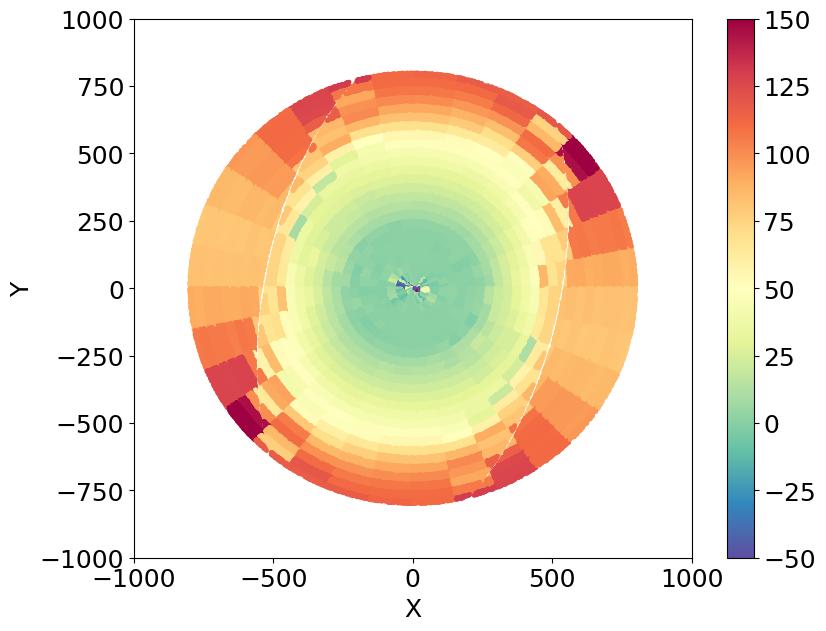}{0.3\textwidth}{(e)}
              \fig{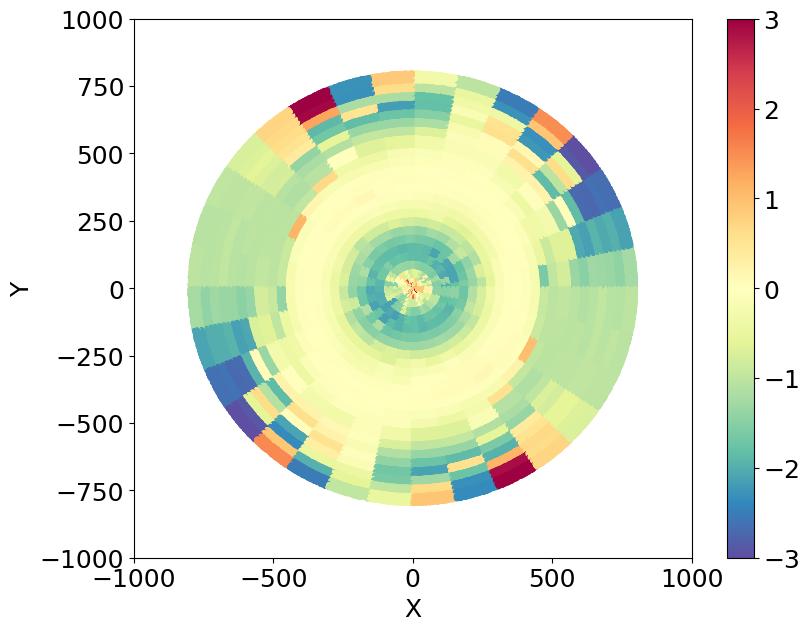}{0.3\textwidth}{(f)}}
\gridline{\fig{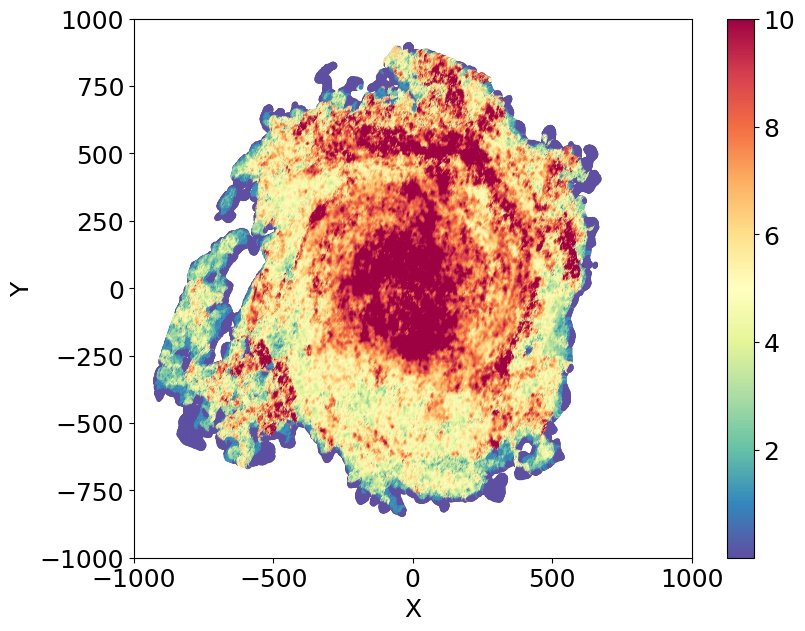}{0.3\textwidth}{(g)}
              \fig{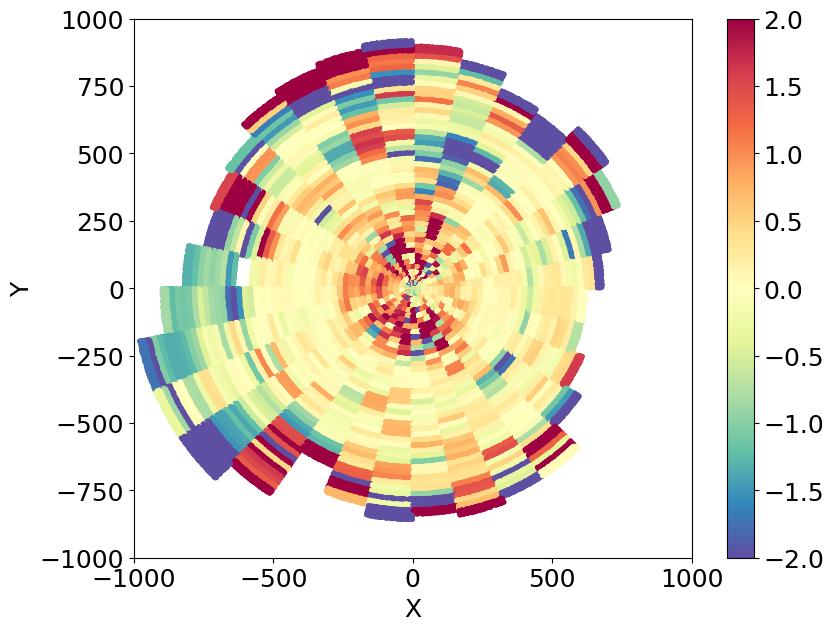}{0.3\textwidth}{(h)}
              \fig{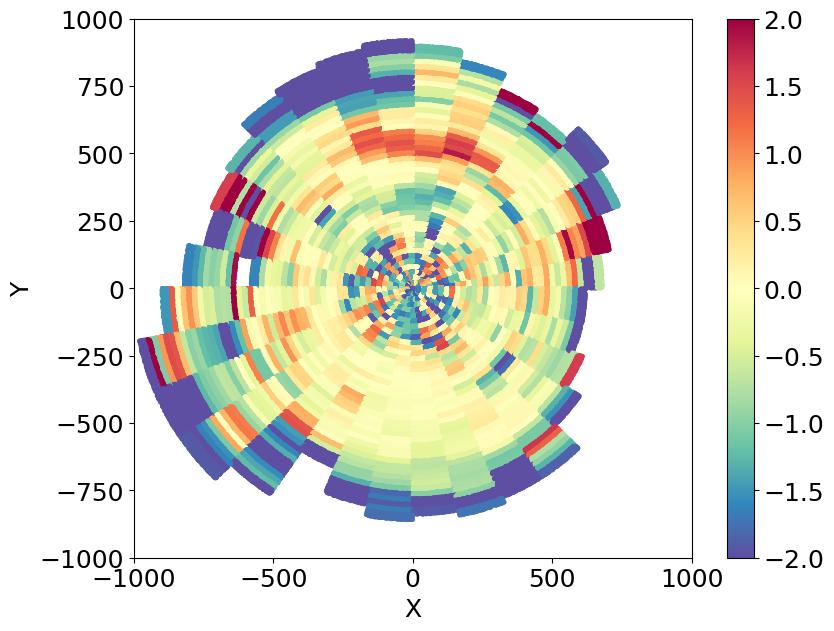}{0.3\textwidth}{(i)}}
\caption{As  Fig.\ref{fig:NGC2903-2} but for NGC 628.} 
\label{fig:NGC628-2} 
\end{figure*}
%

%

\subsection*{NGC 925} 

When transitioning from the inner to the outer regions of the disc in NGC 925, significant changes are observed in the inclination angle $i(R)$ and the position angle $\phi_0(R)$. The inclination angle undergoes a large change of approximately 30 degrees, while the position angle changes by around 20 degrees (see panel (a) of Fig. \ref{fig:NGC925-1}). 
{ The transverse velocity component $v_t(R)$ approaches the circular velocity $v_c(R)$ at small radii, while exhibiting a fluctuating behavior at larger radii. The  radial velocity component $v_r(R)$ oscillates around zero at small radii but shows significant fluctuations towards larger radii. }These fluctuations correspond to velocity anisotropies that can be identified in the two-dimensional maps depicted in Fig. \ref{fig:NGC925-2}. The velocity dispersion, on average, exhibits a smooth decay, but it is also affected by considerable fluctuations.

In the toy disc model, the dipolar modulation is evident in the middle region of the disc, rather than the outer region. This is because it is in this middle region where the orientation angles detected by the TRM show the largest variation. It is worth noting that the toy model has a limiting radius of 500'', while NGC 925 extends asymmetrically up to approximately 750''. The velocity rank correlation coefficient of the observed galaxy, beyond the modulation similar to that of the toy model in its middle region, exhibits a more complex structure that is not revealed by the toy disc model. {  This suggests that intrinsic velocity perturbations dominate the velocity field, even though the correlation coefficient is ${\cal C} = 0.4$, which is compatible with the presence of a moderate warp.}
Furthermore, positive correlations between the radial velocity and the velocity dispersion can be detected in both the inner and outer regions of the disc. These correlations indicate that variations in the radial velocity are associated with changes in the velocity dispersion.
\begin{figure*}
\gridline{\fig{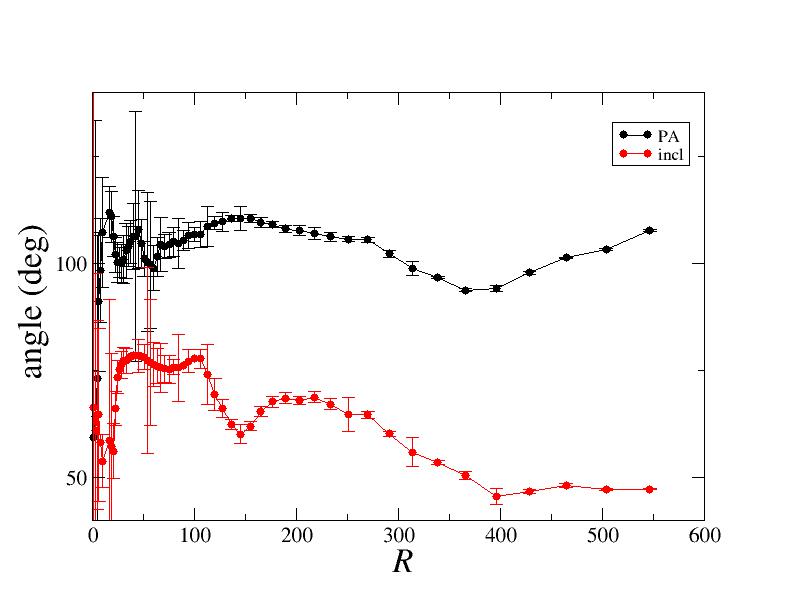}{0.45\textwidth}{(a)}
              \fig{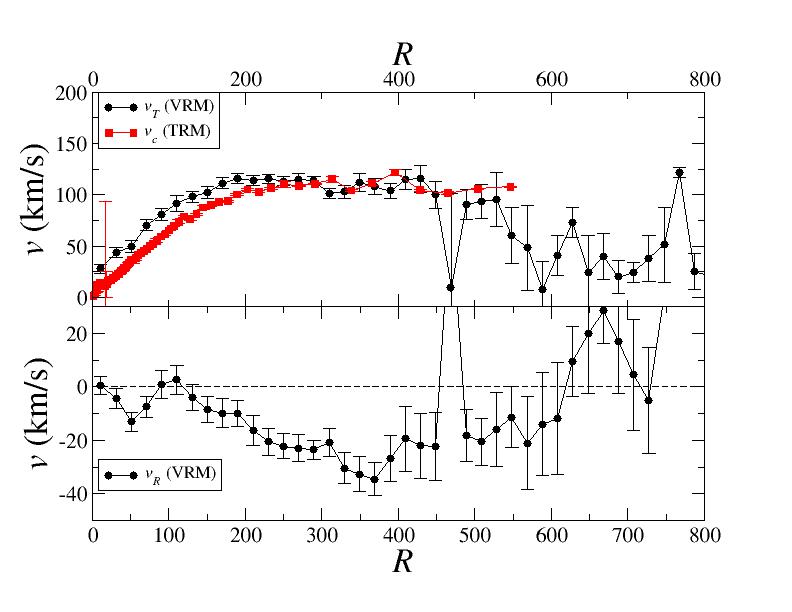}{0.45\textwidth}{(b)}
              }
  \gridline{
 \fig{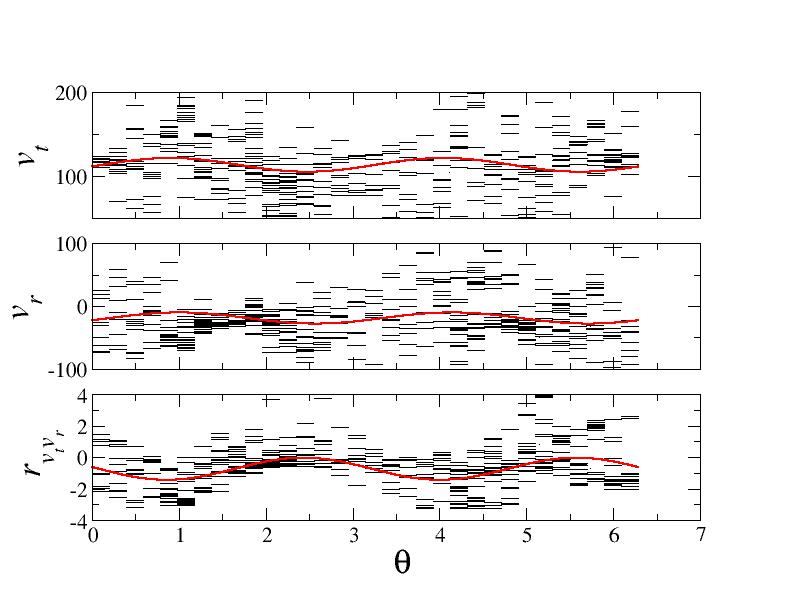}{0.45\textwidth}{(c)}
                \fig{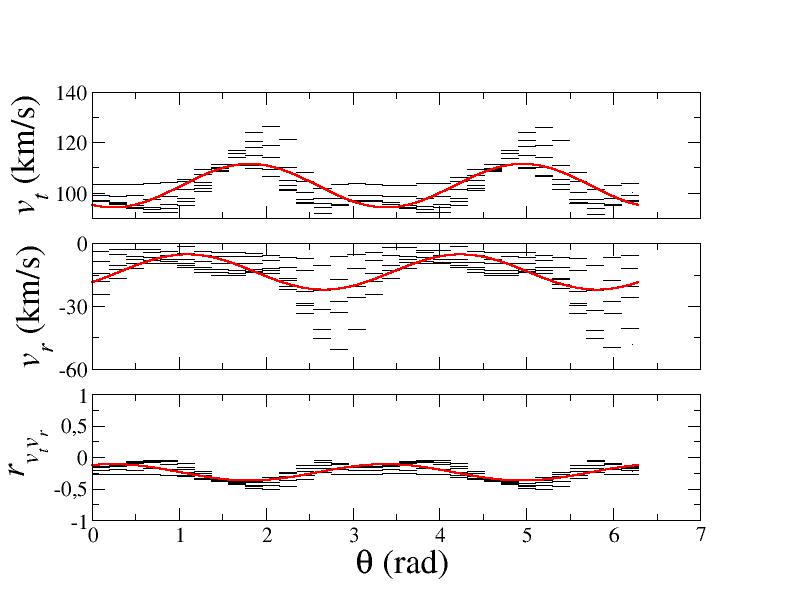}{0.45\textwidth}{(d)}}
     \caption{As  Fig.\ref{fig:NGC2903-1} but for NGC 925.} 
\label{fig:NGC925-1} 
\end{figure*}
%

\begin{figure*}
\gridline{\fig{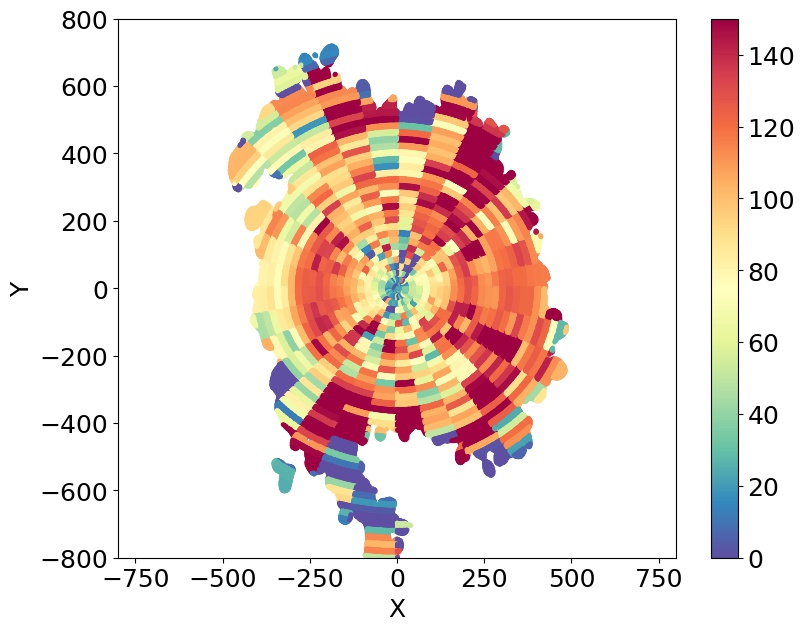}{0.3\textwidth}{(a)}
              \fig{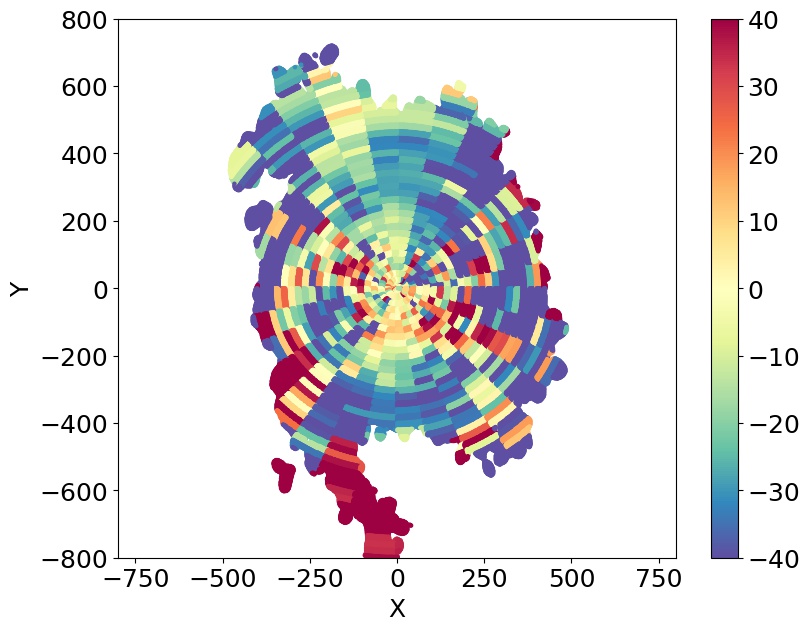}{0.3\textwidth}{(b)}
               \fig{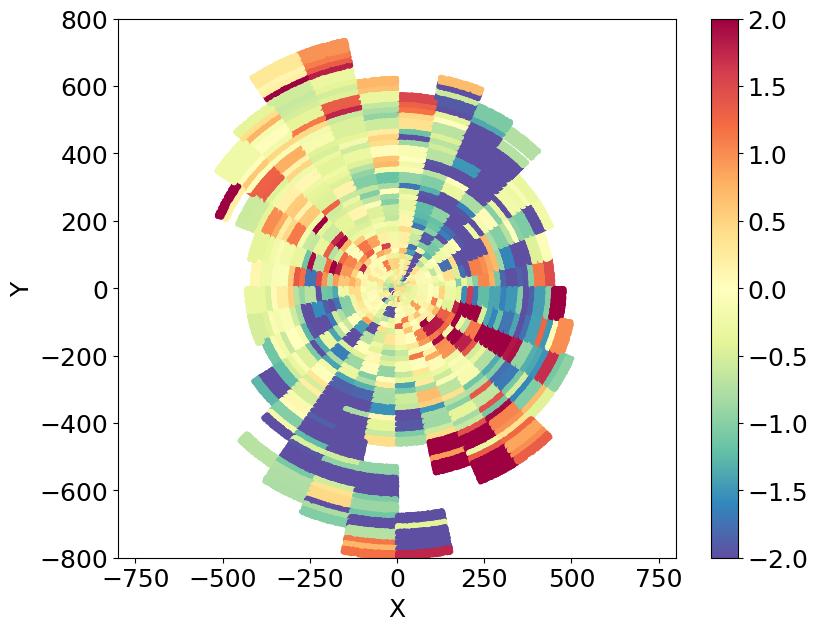}{0.3\textwidth}{(c)}}
\gridline{\fig{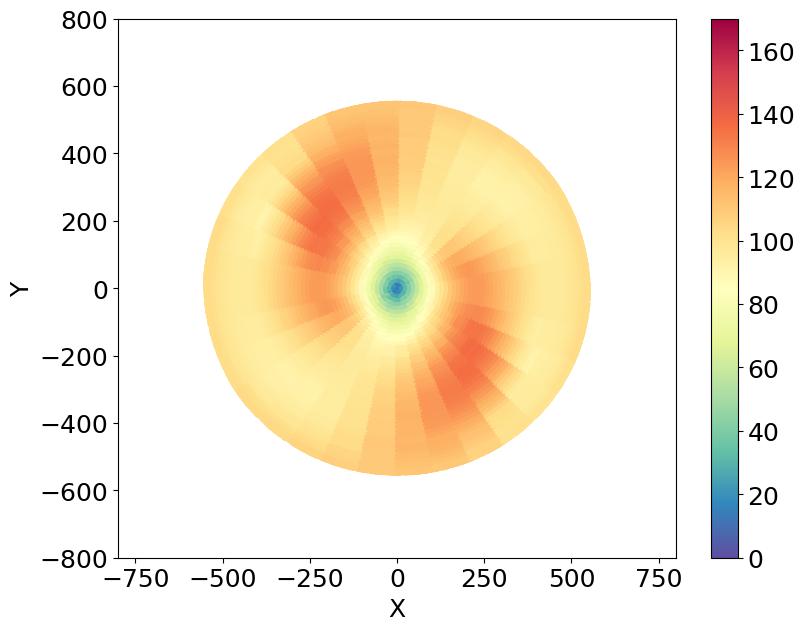}{0.3\textwidth}{(d)}
              \fig{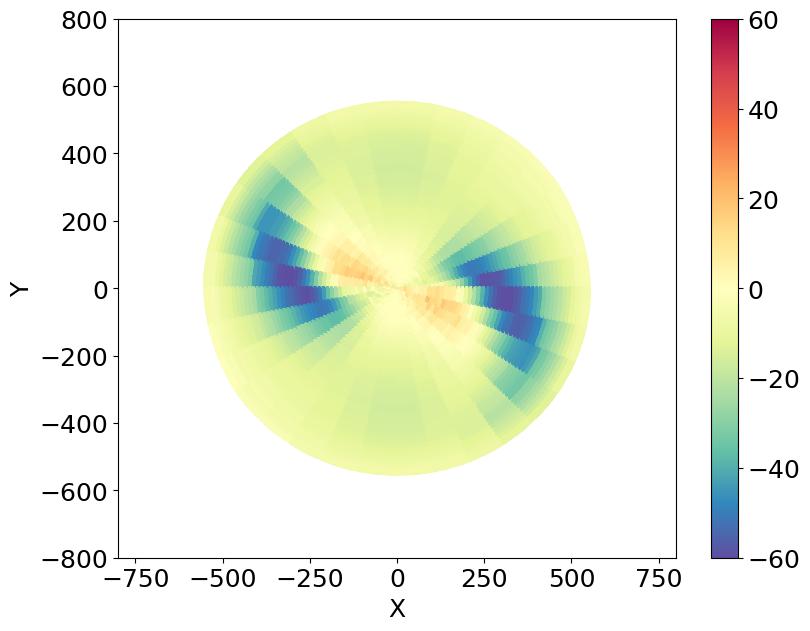}{0.3\textwidth}{(e)}
              \fig{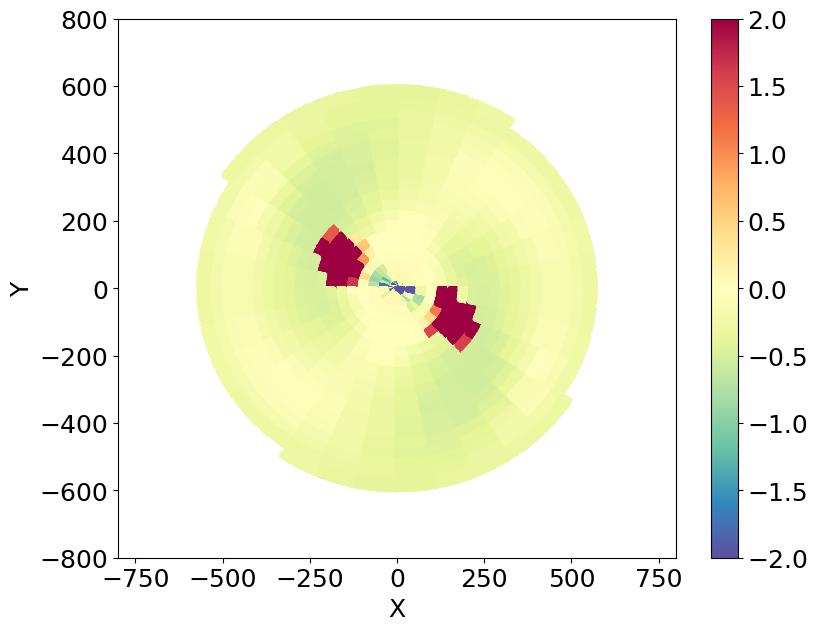}{0.3\textwidth}{(f)}}
\gridline{\fig{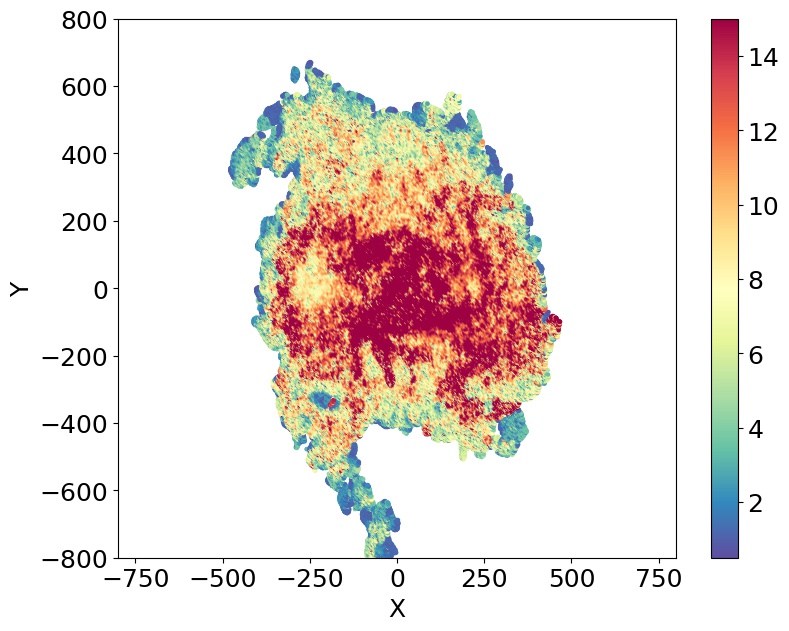}{0.3\textwidth}{(g)}
              \fig{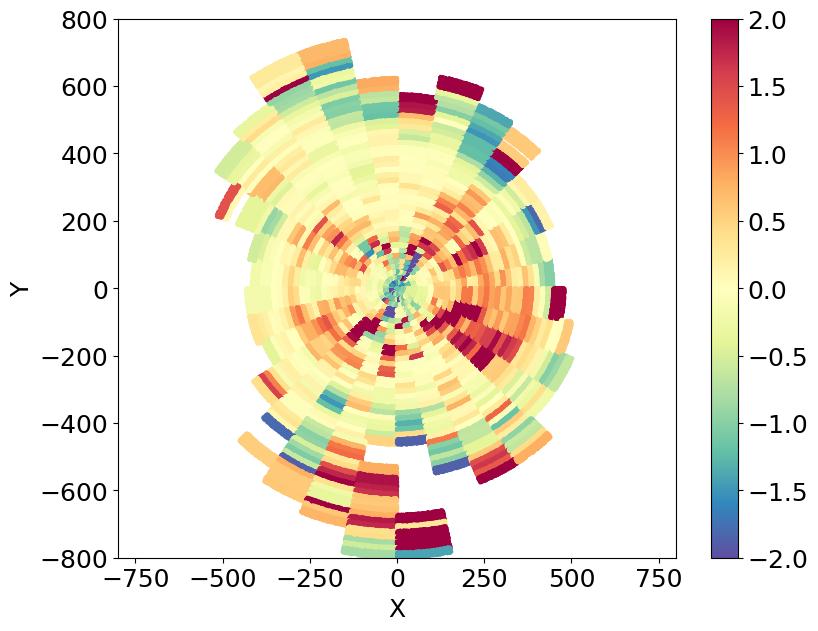}{0.3\textwidth}{(h)}
              \fig{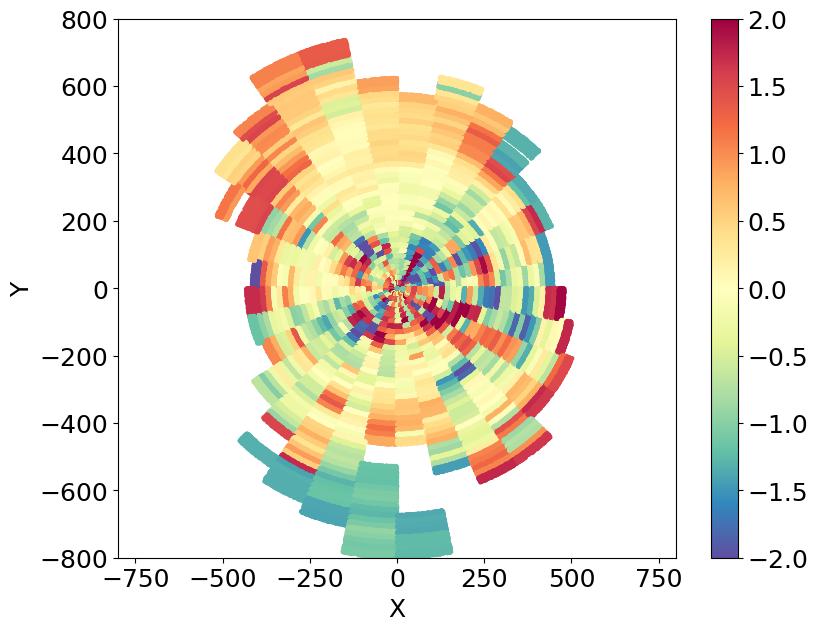}{0.3\textwidth}{(i)}}
\caption{As  Fig.\ref{fig:NGC2903-2} but for NGC 925.} 
\label{fig:NGC925-2} 
\end{figure*}


\subsection*{NGC 2366} 
The inclination angle of NGC 2366 remains nearly constant across the disc,  while the P.A. shows a variation of about 10$^\circ$ in the middle of the disc, i.e., for 100''$<R<$200''. This variation is indicative of a moderate warp, as shown in Fig. \ref{fig:NGC2366-1}. Upon close inspection of the velocity component profiles and their rank correlation coefficient, there is no clear signature of dipolar oscillations, and thus the reality of the warp is uncertain. { Indeed, the correlation coefficient is ${\cal C} = 0.04$, a value compatible with no warp}. The same conclusion is reached by analyzing the maps in Fig. \ref{fig:NGC2366-2}. Instead, there is unambiguous evidence for the presence of relatively large intrinsic perturbations in the velocity field. Finally, the velocity dispersion field is characterized by diffuse fluctuations, but there is a clear correlation with the perturbations in the velocity components maps.
\begin{figure*}
\gridline{\fig{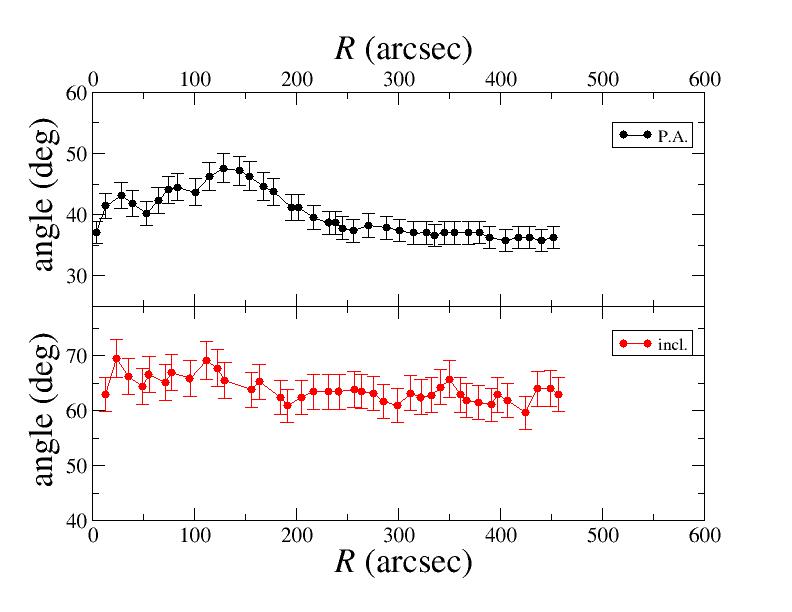}{0.45\textwidth}{(a)}
              \fig{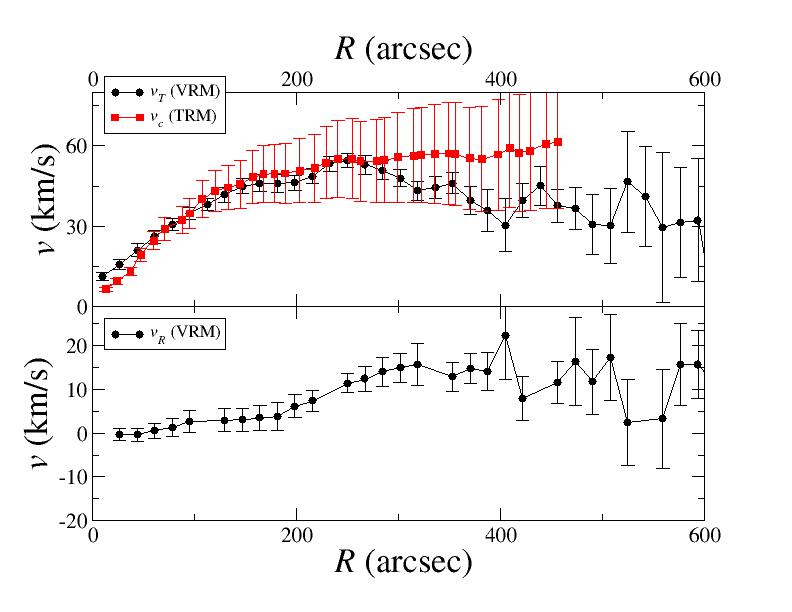}{0.45\textwidth}{(b)}
              }
  \gridline{
 \fig{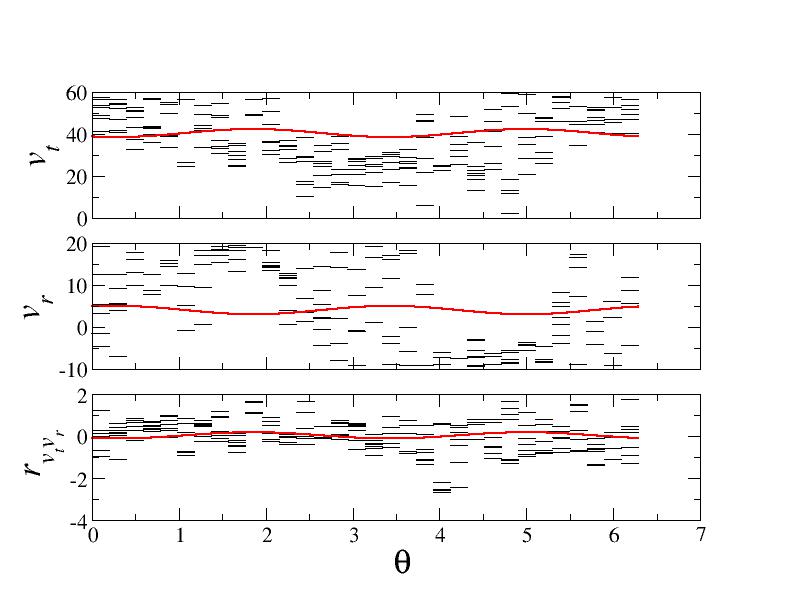}{0.45\textwidth}{(c)}
                \fig{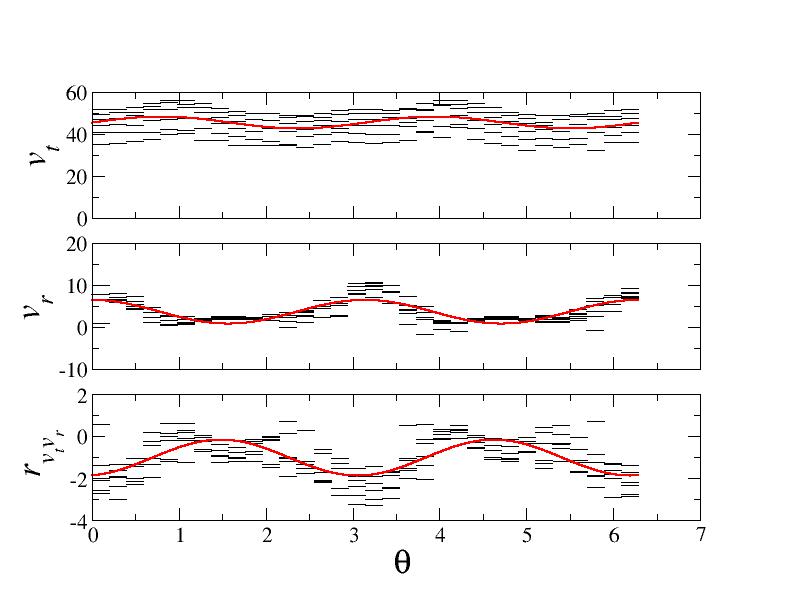}{0.45\textwidth}{(d)}}
     \caption{As  Fig.\ref{fig:NGC2903-1} but for NGC 2366.} 
\label{fig:NGC2366-1} 
\end{figure*}
%
\begin{figure*}
\gridline{\fig{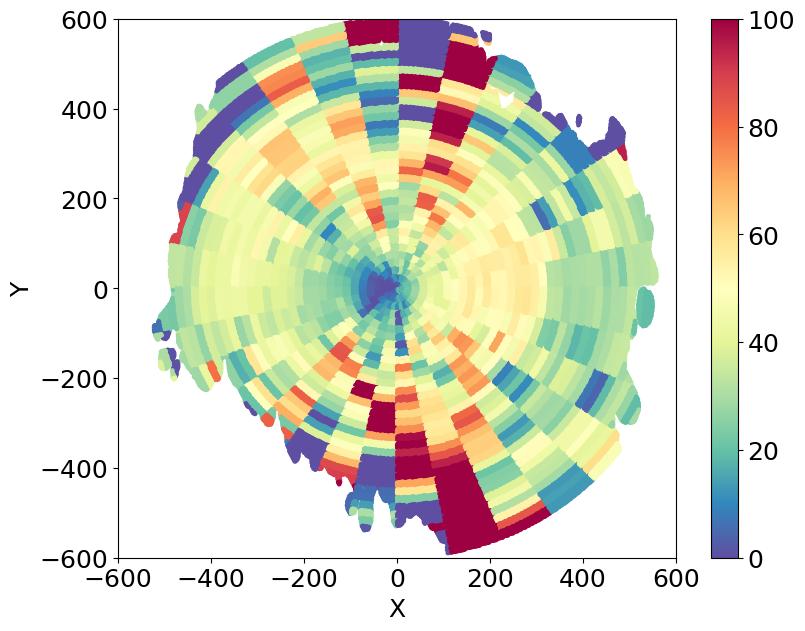}{0.3\textwidth}{(a)}
              \fig{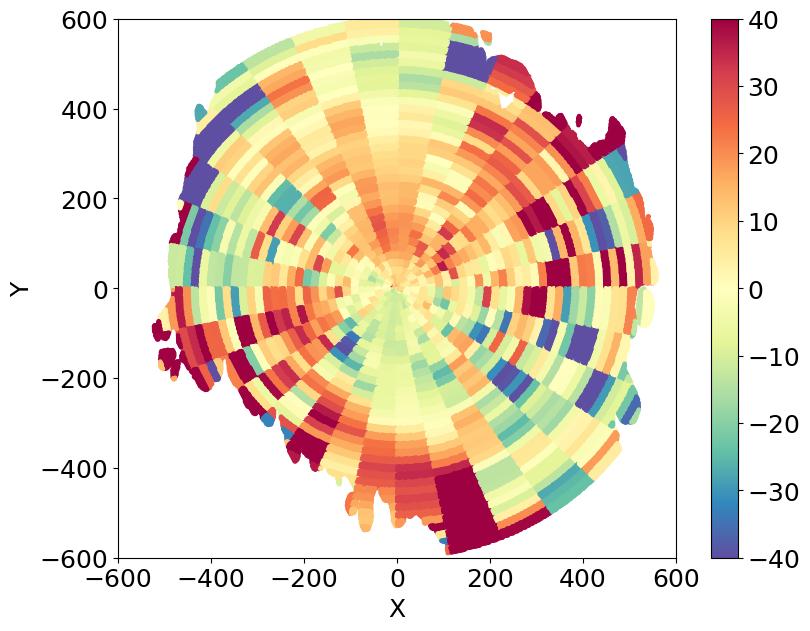}{0.3\textwidth}{(b)}
               \fig{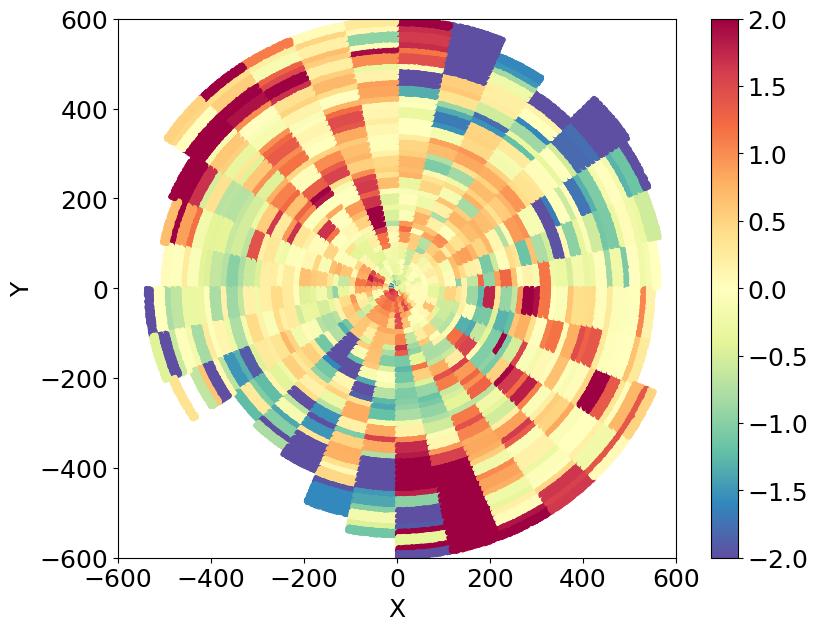}{0.3\textwidth}{(c)}}
\gridline{\fig{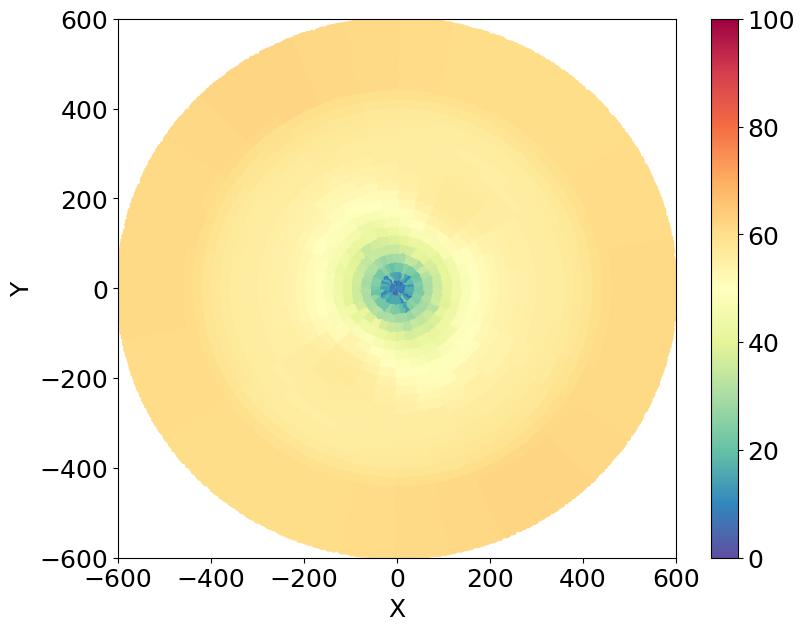}{0.3\textwidth}{(d)}
              \fig{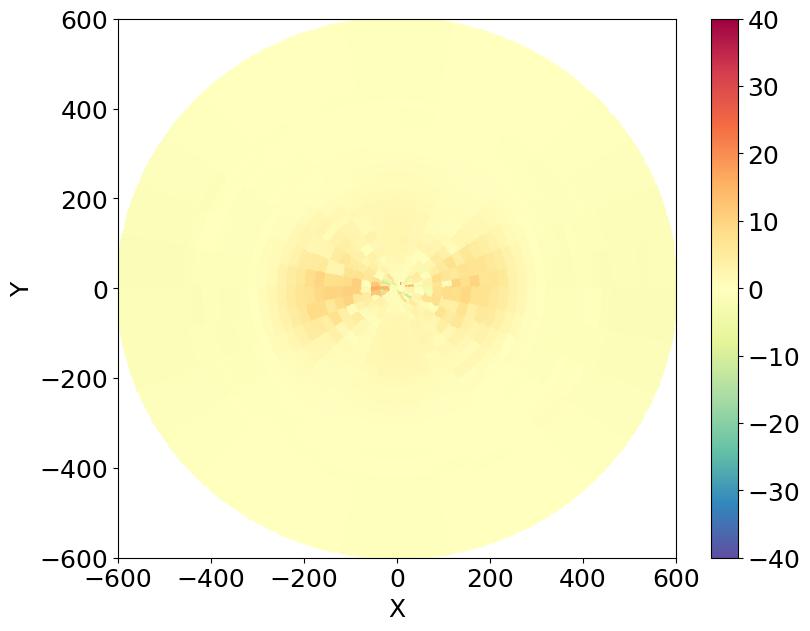}{0.3\textwidth}{(e)}
              \fig{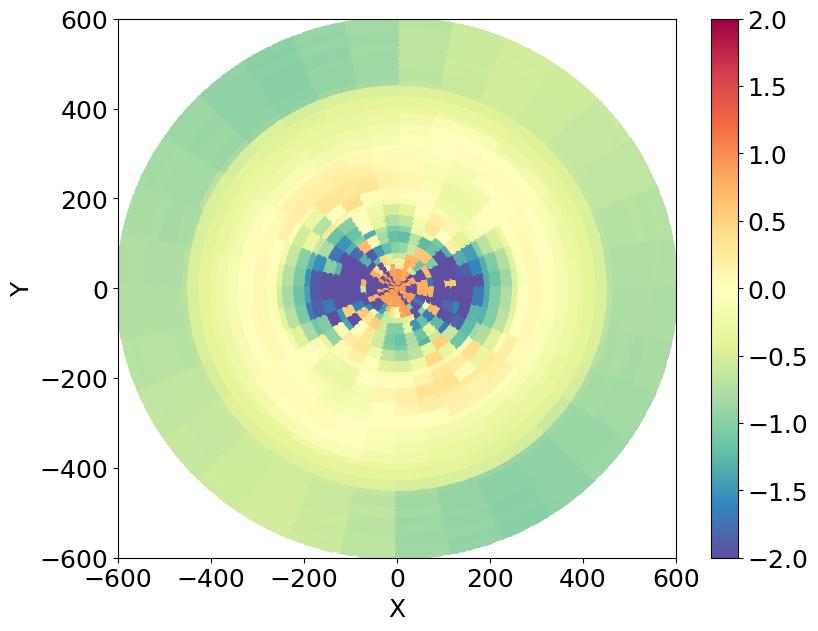}{0.3\textwidth}{(f)}}
\gridline{\fig{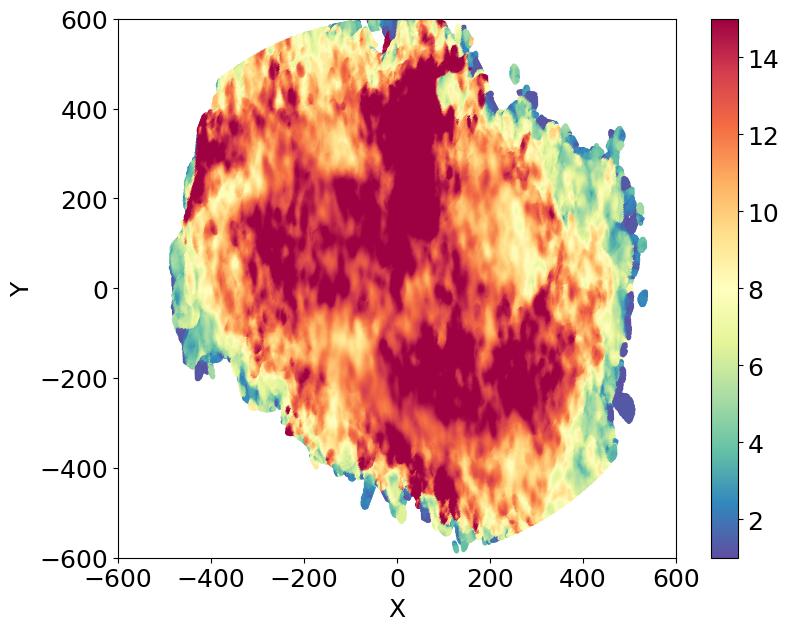}{0.3\textwidth}{(g)}
              \fig{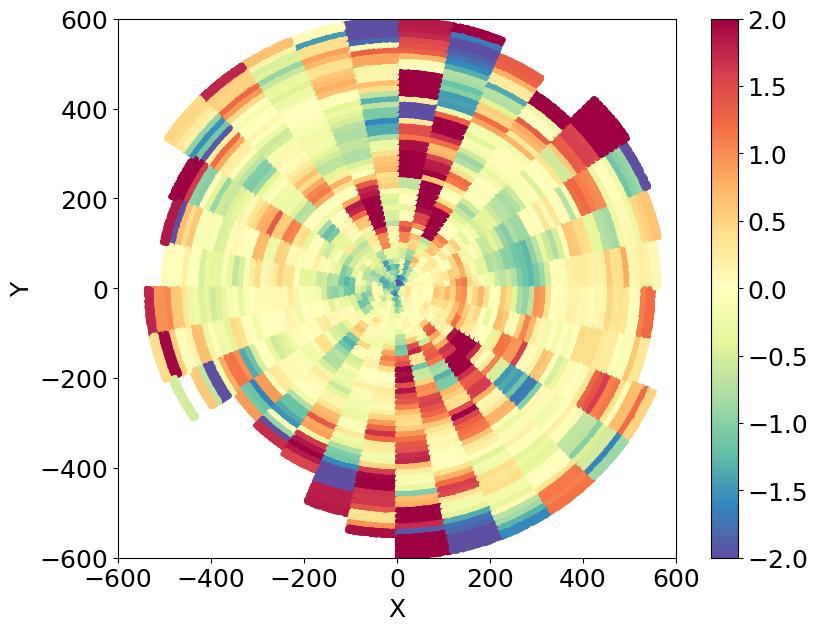}{0.3\textwidth}{(h)}
              \fig{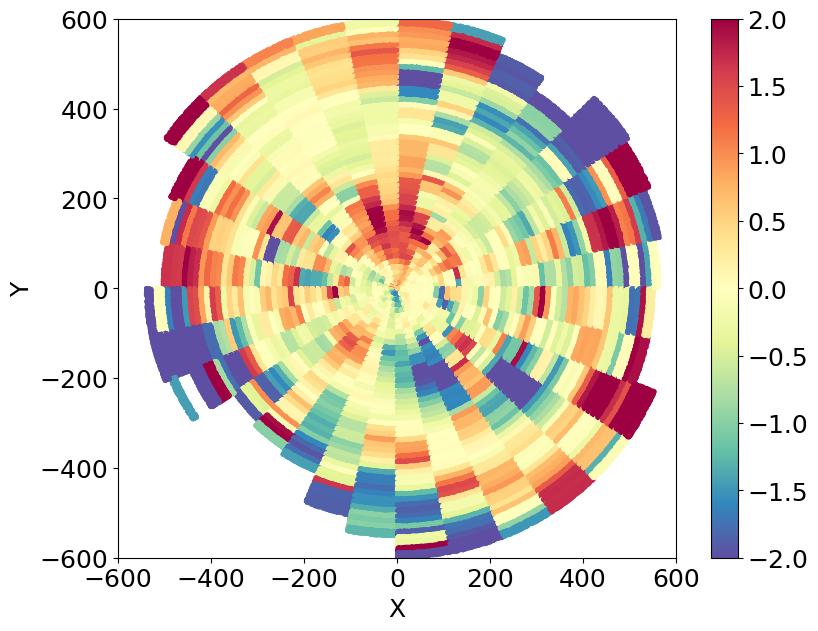}{0.3\textwidth}{(i)}}
\caption{As  Fig.\ref{fig:NGC2903-2} but for NGC 2366.} 
\label{fig:NGC2366-2} 
\end{figure*}


\subsection*{NGC 2403} 

In this case both the  inclination angle $i(R)$ and the P.A.  $\phi_0(R)$ show  variations smaller than $10^\circ$ (see panel (a) of Fig.\ref{fig:NGC2403-1}).
{ The  the transversal  velocity $v_t(R)$ coincides, within the error bars, with $v_c(R)$. The radial velocity fluctuates and its amplitude is smaller than $\sim 20$ km s$^{-1}$.} 
The velocity dispersion profile displays a smooth decays with small fluctuations. 
The small variations in the orientation angles give rise to  symmetrical anisotropies of small amplitude; in this case the velocity rank correlation coefficients of the  toy model and of the actual galaxy have similar patterns of fluctuations (see Fig.\ref{fig:NGC2403-1}-\ref{fig:NGC2403-2}) that are well represented by a dipolar oscillation. Even the velocity components show a dipolar oscillation (see Fig.\ref{fig:NGC2403-2}). 
{ The correlation coefficient is ${\cal C} = 0.4$, indicating the presence of a moderate warp}.
It is interesting to note that  the rank   correlation coefficients $r_{\sigma v_t}(R, \theta)$ and $r_{\sigma v_r}(R, \theta)$ show an positive correlation along the same direction, an evidence that favors the presence of a bar-like structure.
\begin{figure*}
\gridline{\fig{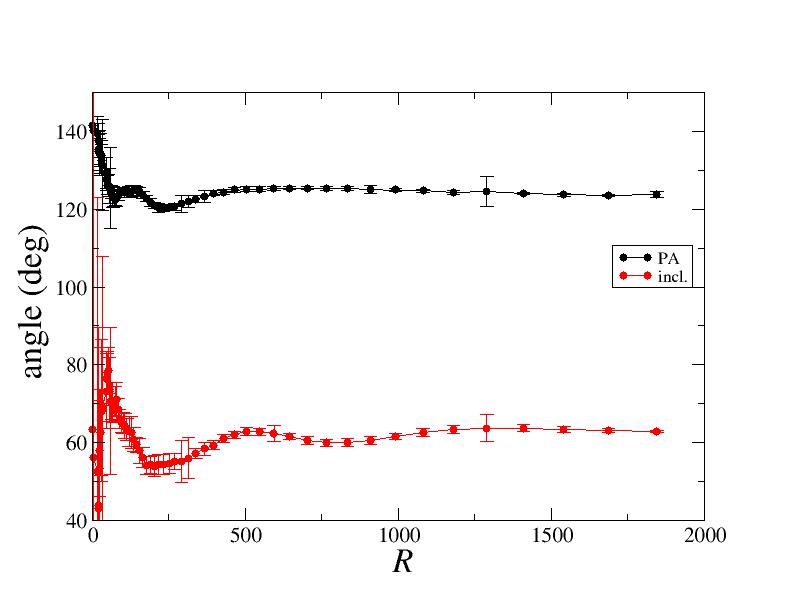}{0.45\textwidth}{(a)}
              \fig{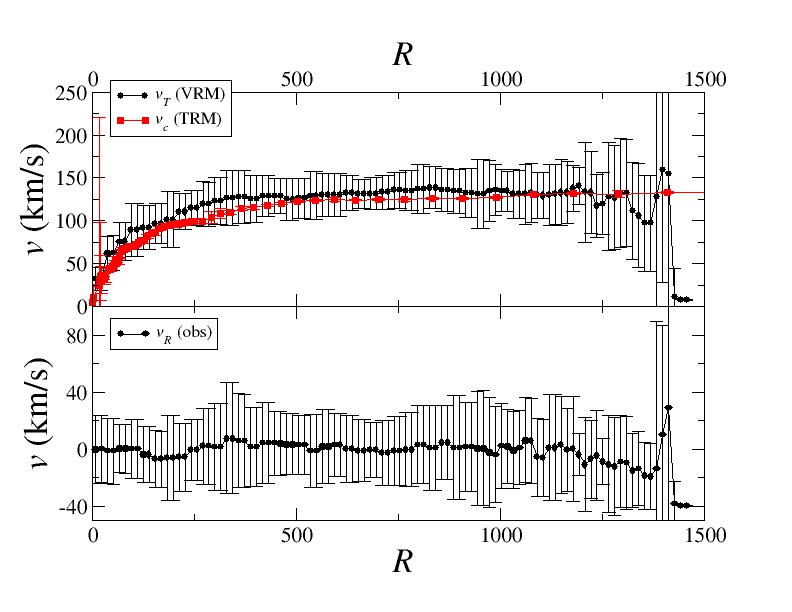}{0.45\textwidth}{(b)}
              }
  \gridline{
 \fig{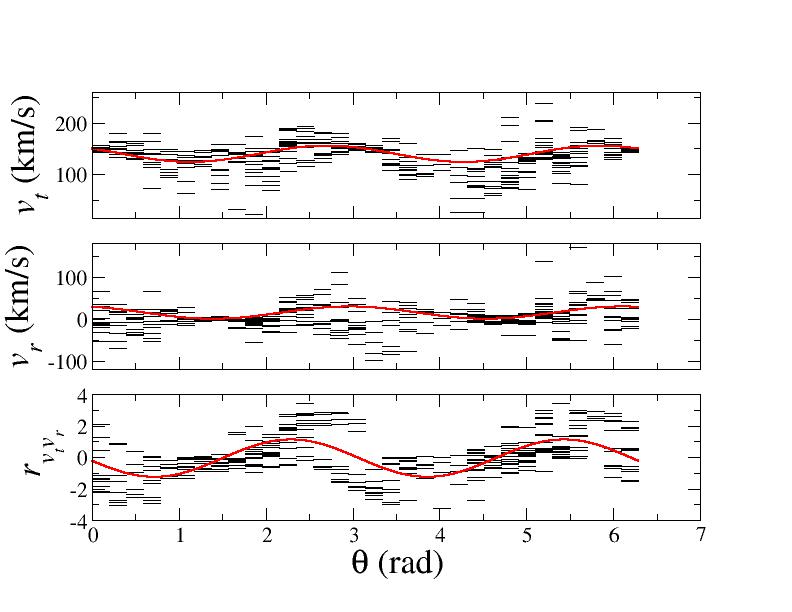}{0.45\textwidth}{(c)}
                \fig{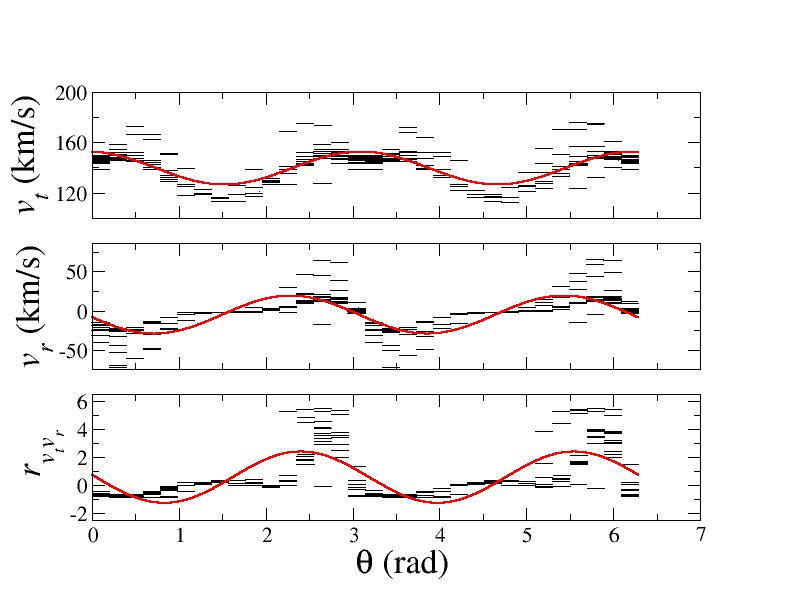}{0.45\textwidth}{(d)}}
     \caption{As  Fig.\ref{fig:NGC2903-1} but for NGC 2403.} 
\label{fig:NGC2403-1} 
\end{figure*}
%

\begin{figure*}
\gridline{\fig{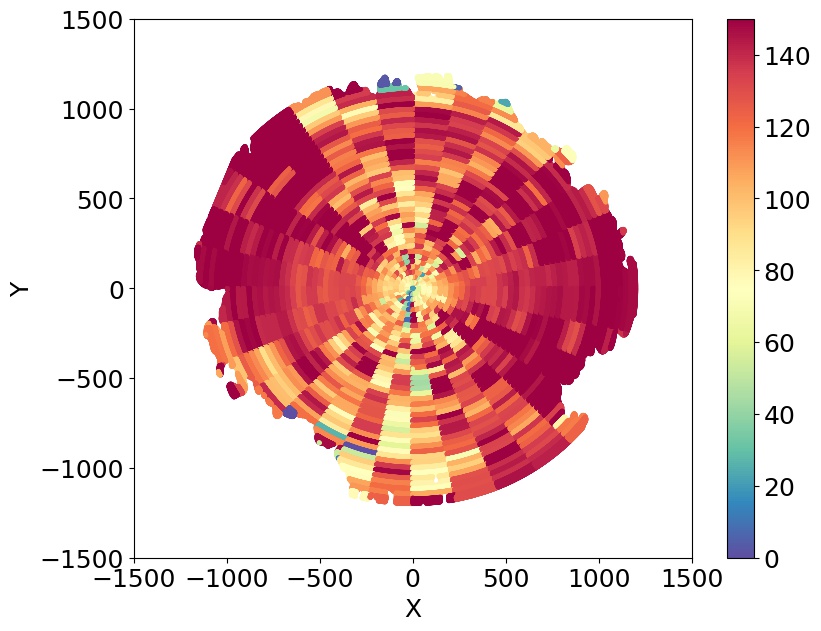}{0.3\textwidth}{(a)}
              \fig{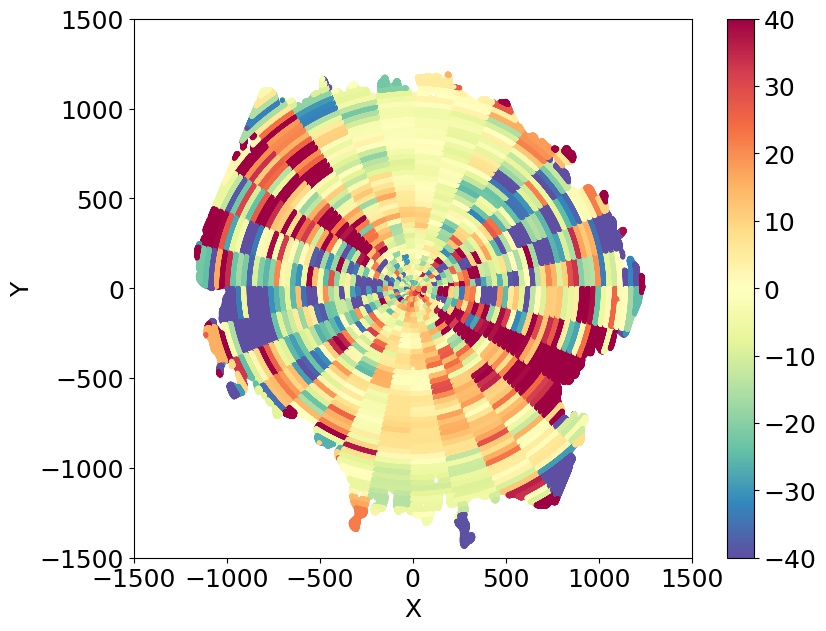}{0.3\textwidth}{(b)}
               \fig{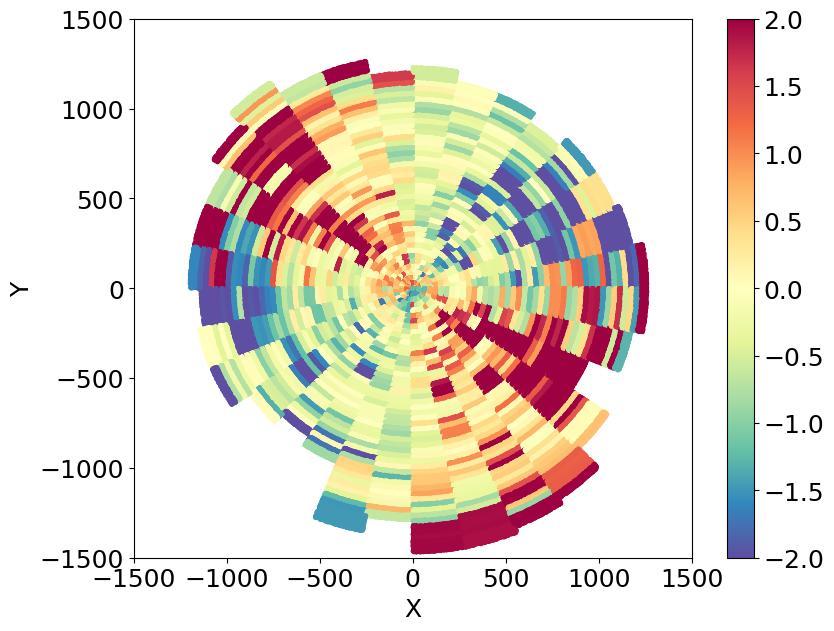}{0.3\textwidth}{(c)}}
\gridline{\fig{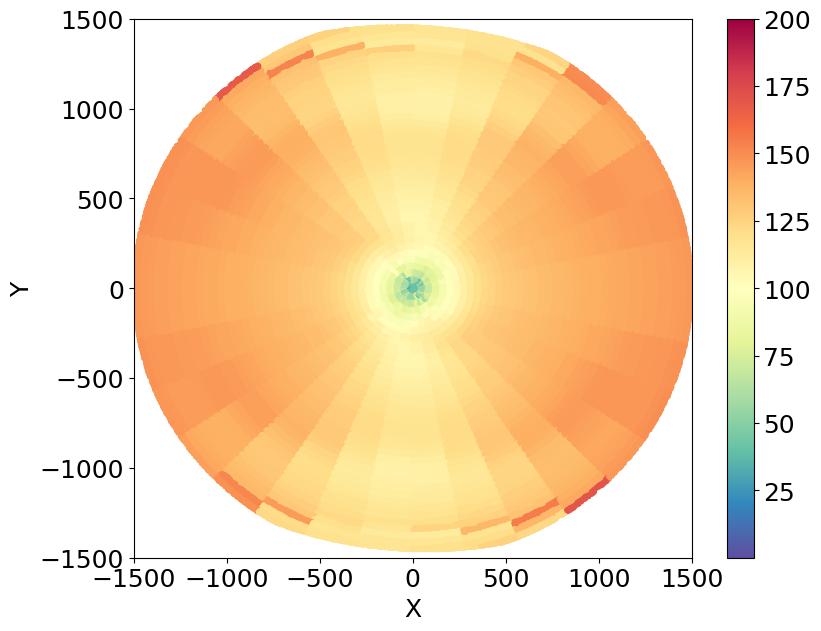}{0.3\textwidth}{(d)}
              \fig{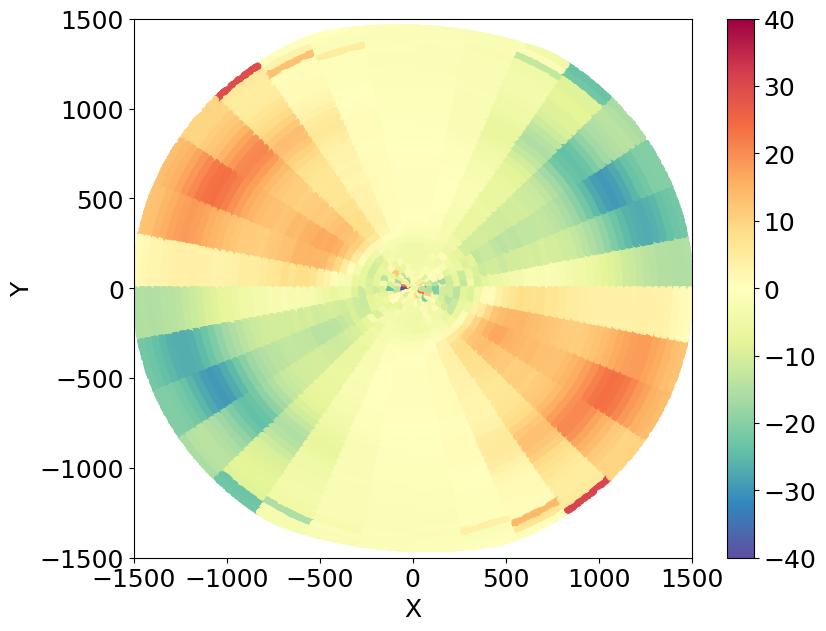}{0.3\textwidth}{(e)}
              \fig{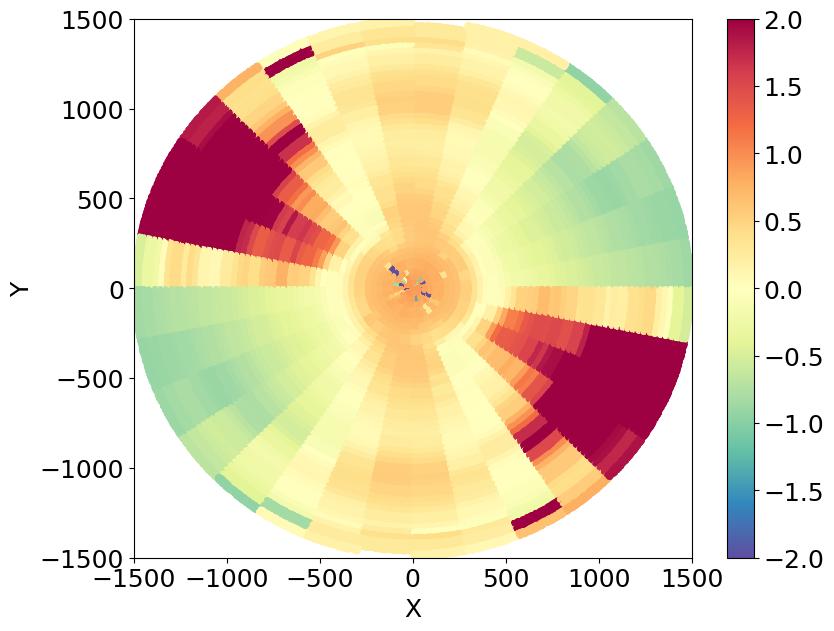}{0.3\textwidth}{(f)}}
\gridline{\fig{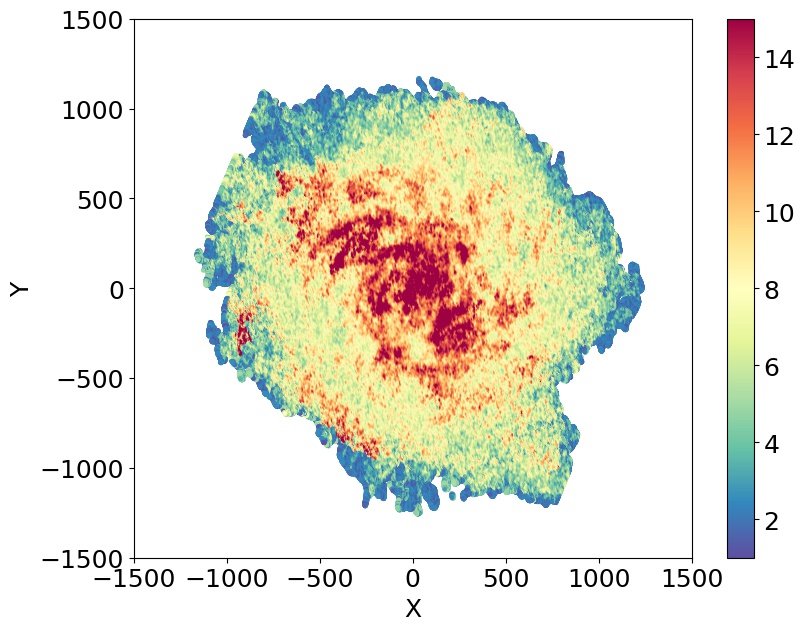}{0.3\textwidth}{(g)}
              \fig{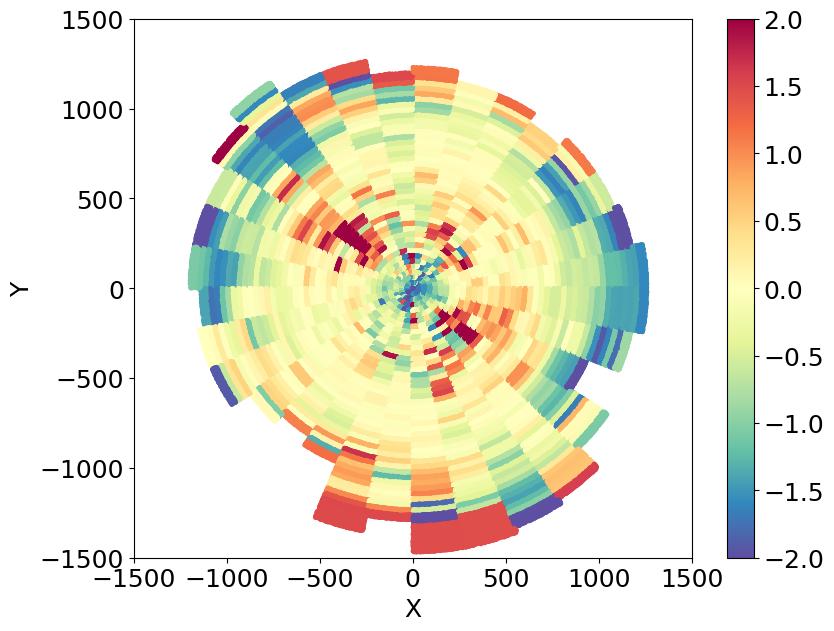}{0.3\textwidth}{(h)}
              \fig{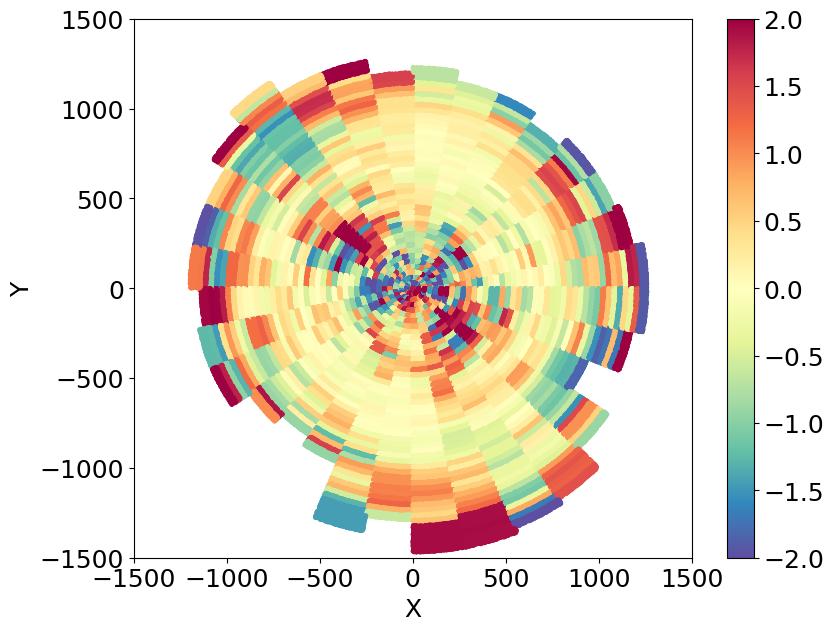}{0.3\textwidth}{(i)}}
\caption{As  Fig.\ref{fig:NGC2903-2} but for NGC 2403.} 
\label{fig:NGC2403-2} 
\end{figure*}


\subsection*{NGC 2841}

This galaxy exhibits the presence of a moderate warp, which is consistent with the variations observed in the orientation angles from the inner to the outer part of the disc (see panel (a) of Fig. \ref{fig:NGC2841-1}).
{ The correlation coefficient is ${\cal C} = 0.25$}. 
Beyond the dipolar modulation induced by the warp (mostly visible in the velocity rank correlation coefficient), we find evidence for intrinsic velocity perturbations of a few tens of km s$^{-1}$. The velocity dispersion field shows anisotropic patterns in the direction $\theta \approx 135^\circ$, correlating with both transversal and radial velocity fluctuations and indicating the existence of a bar-like structure (Fig.\ref{fig:NGC2841-2}).
%
\begin{figure*}
\gridline{\fig{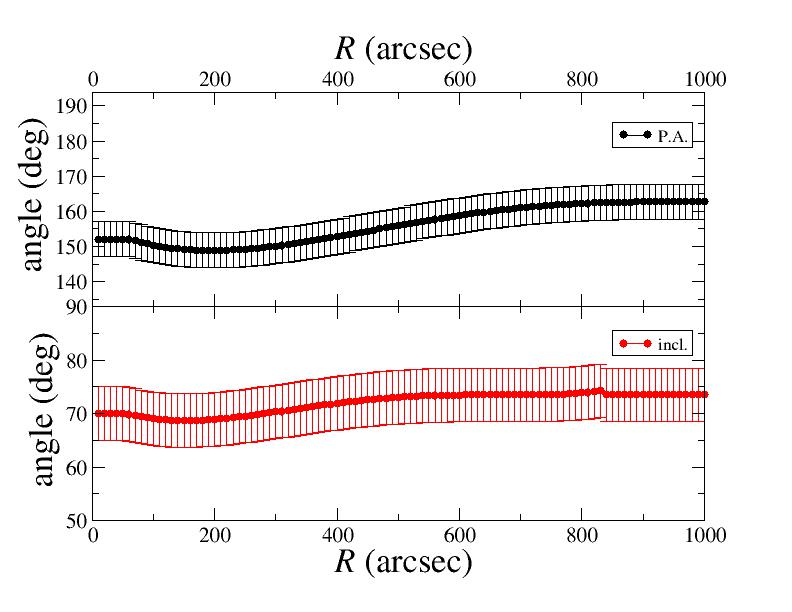}{0.45\textwidth}{(a)}
              \fig{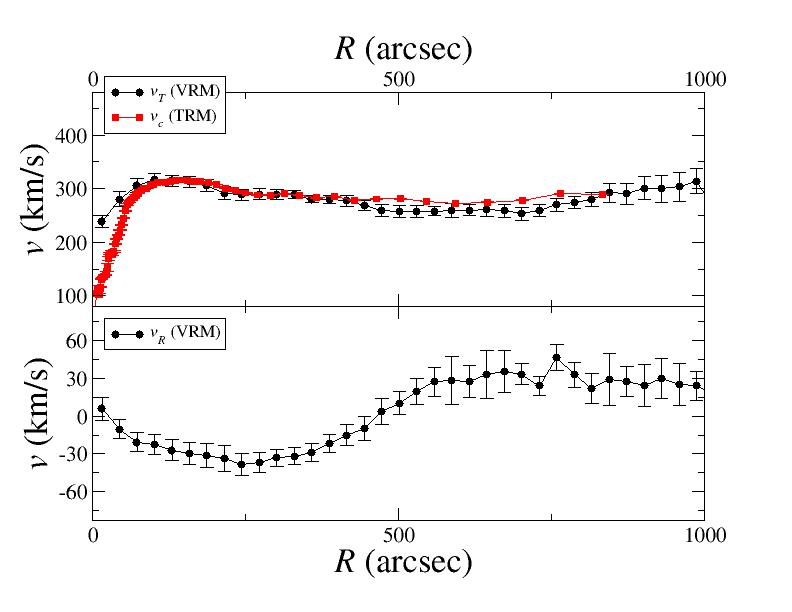}{0.45\textwidth}{(b)}
              }
  \gridline{
 \fig{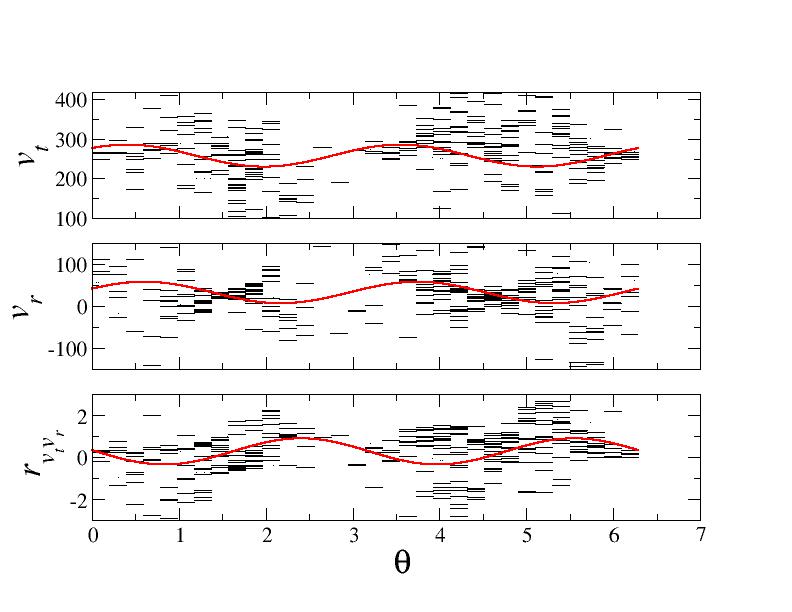}{0.45\textwidth}{(c)}
                \fig{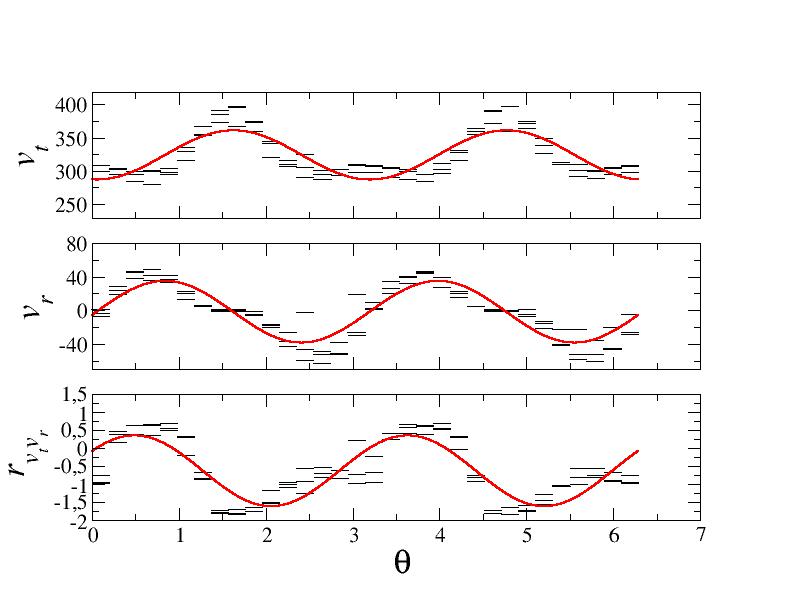}{0.45\textwidth}{(d)}}
     \caption{As  Fig.\ref{fig:NGC2903-1} but for NGC 2841.} 
\label{fig:NGC2841-1} 
\end{figure*}
%

\begin{figure*}
\gridline{\fig{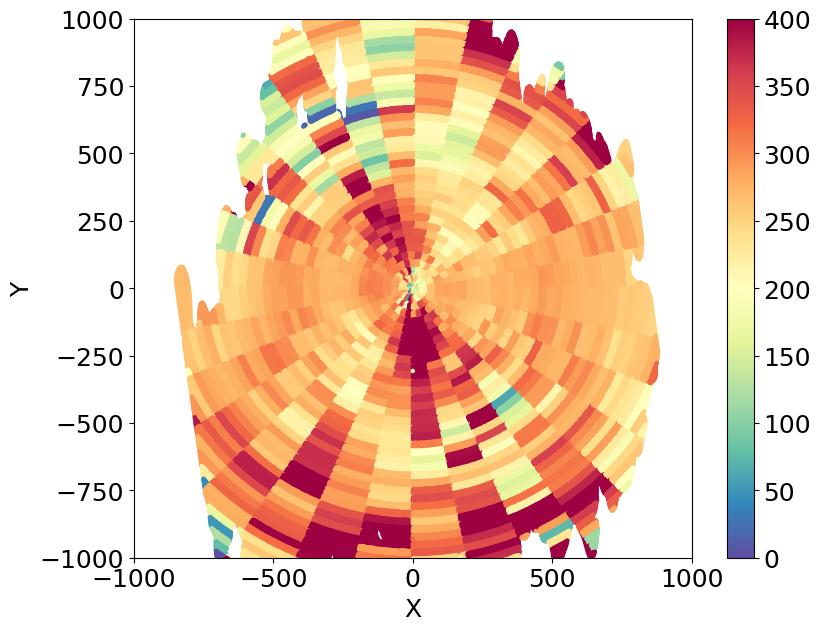}{0.3\textwidth}{(a)}
              \fig{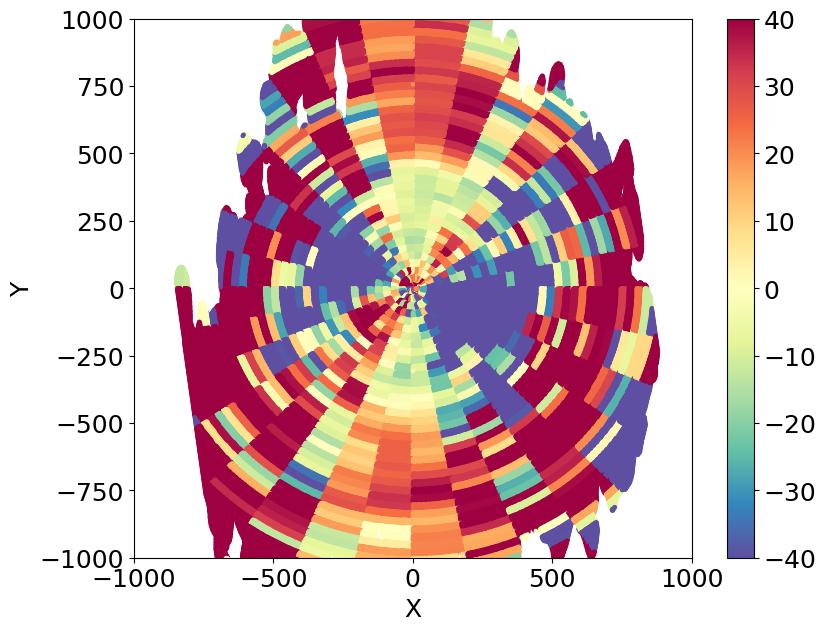}{0.3\textwidth}{(b)}
               \fig{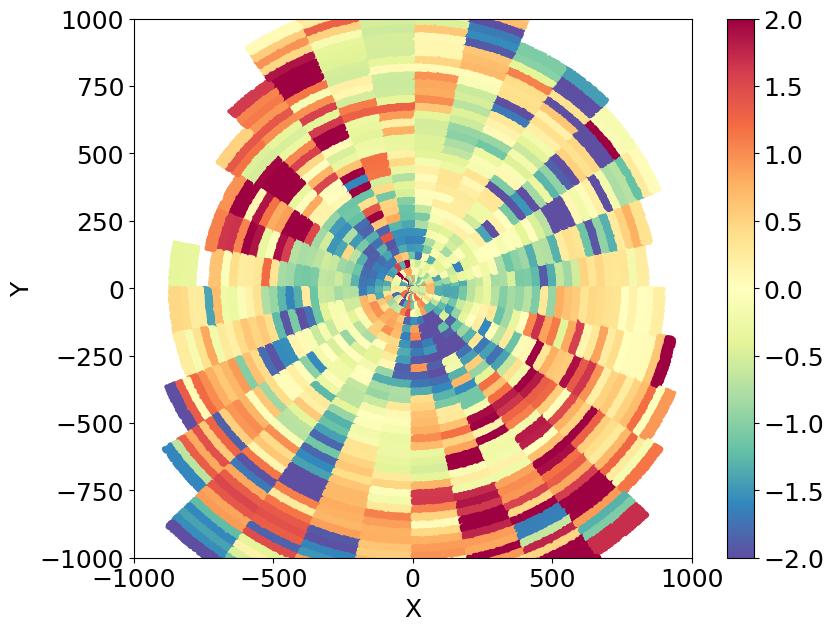}{0.3\textwidth}{(c)}}
\gridline{\fig{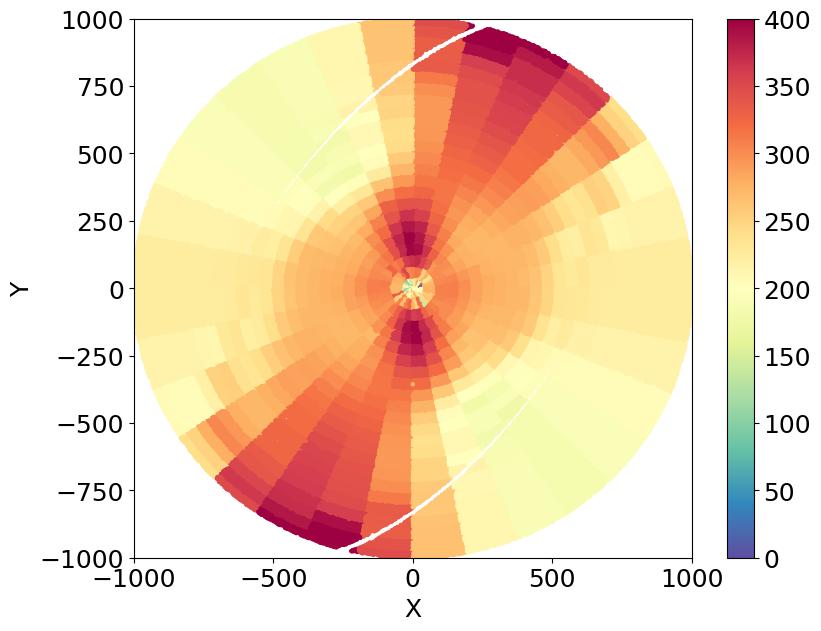}{0.3\textwidth}{(d)}
              \fig{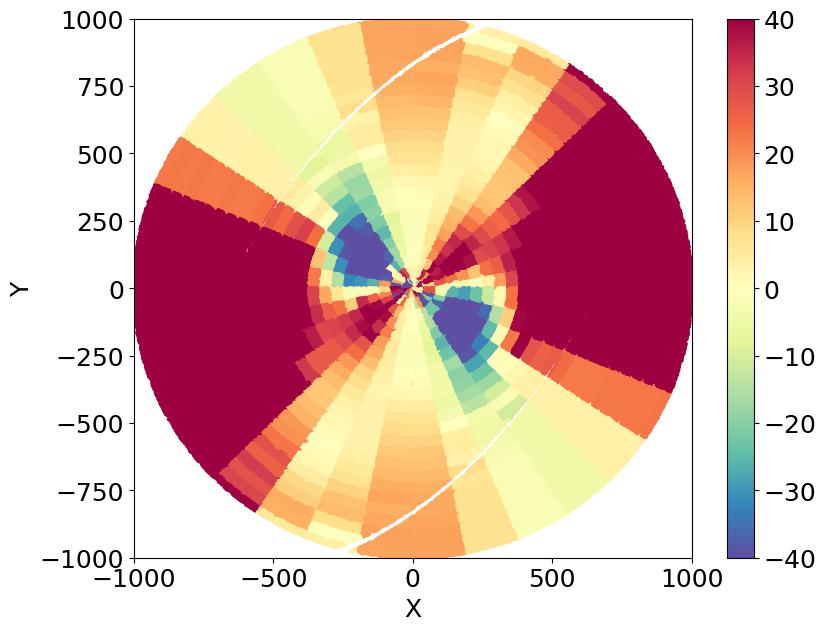}{0.3\textwidth}{(e)}
              \fig{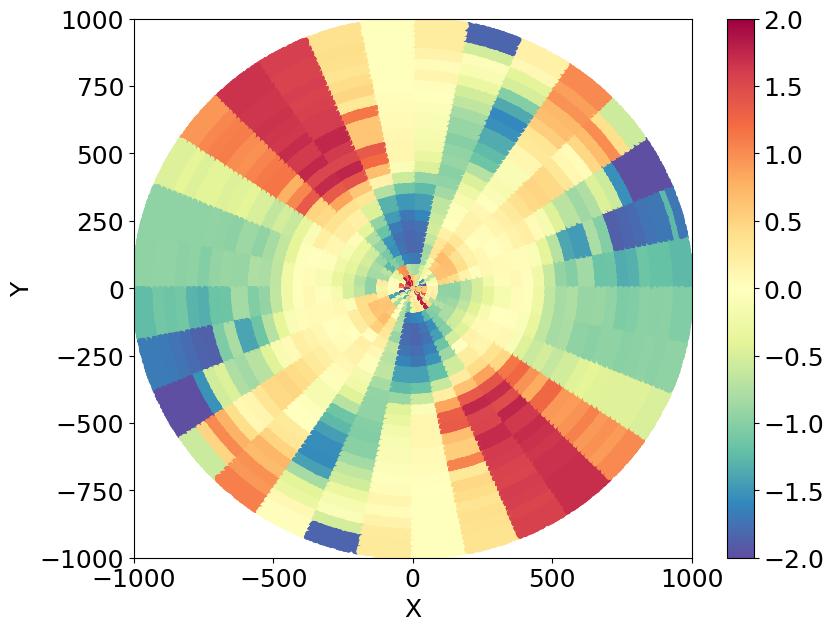}{0.3\textwidth}{(f)}}
\gridline{\fig{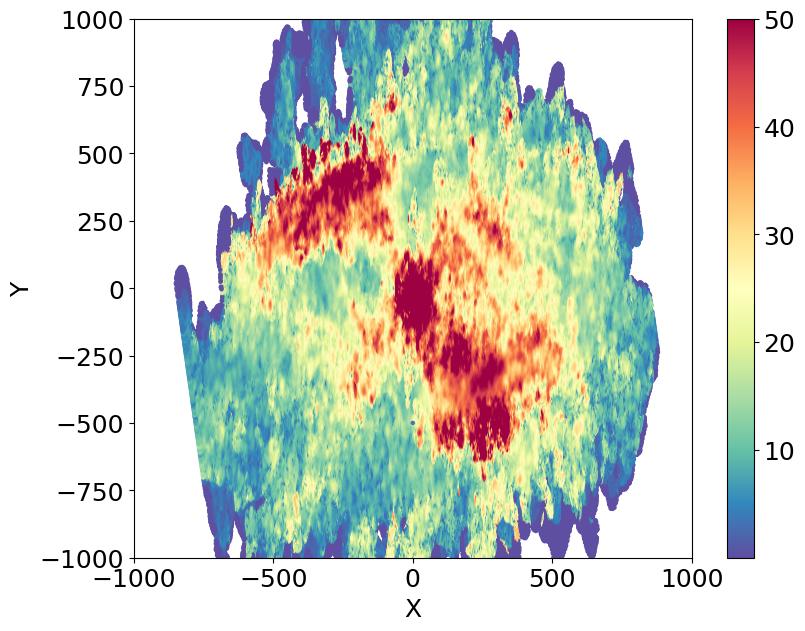}{0.3\textwidth}{(g)}
              \fig{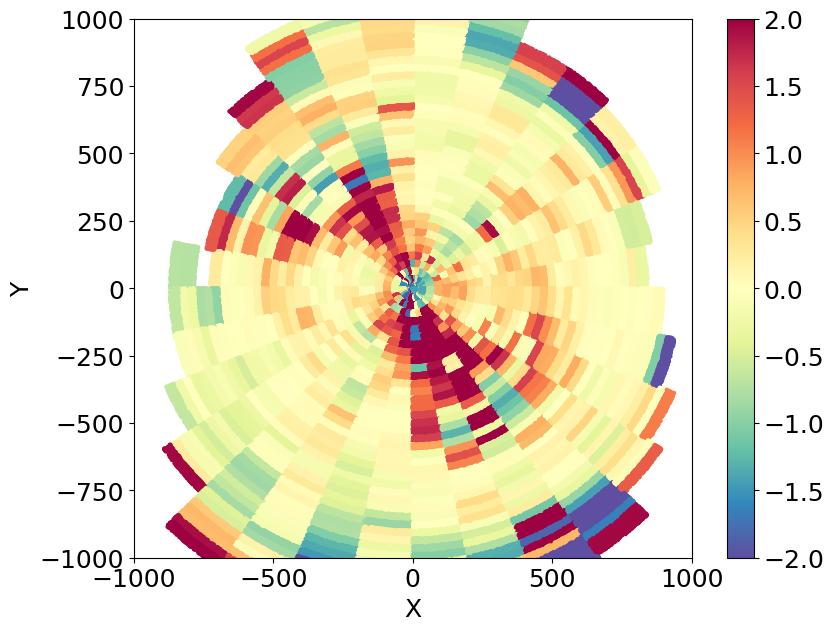}{0.3\textwidth}{(h)}
              \fig{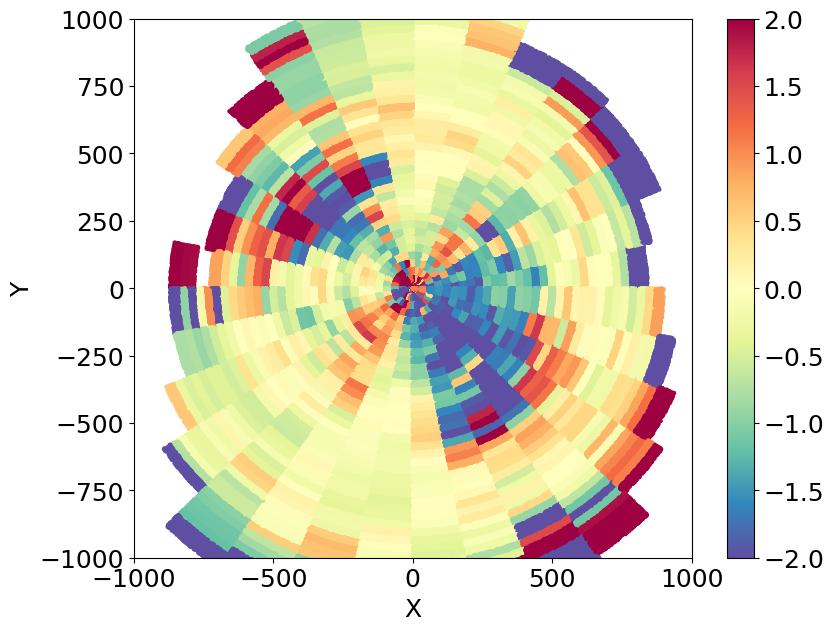}{0.3\textwidth}{(i)}}
\caption{As  Fig.\ref{fig:NGC2903-2} but for NGC 2841.} 
\label{fig:NGC2841-2} 
\end{figure*}
\clearpage


\subsection*{NGC 2976} 

Both orientation angles exhibit significant variations; however, these variations do not correspond to a well-defined dipolar modulation, neither in the rank correlation coefficient nor in the velocity components (see Figs. \ref{fig:NGC2976-1} - \ref{fig:NGC2976-2}). Therefore, we conclude that the velocity of NGC 2976 is predominantly influenced by intrinsic velocity perturbations rather than slow, warp-induced variations. { Indeed, the correlation coefficient is ${\cal C} = 0.1$ compatibile with no warp}.  This conclusion is supported by the velocity dispersion field, which does not show a smooth decay as a function of the distance from the center.
\begin{figure*}
\gridline{\fig{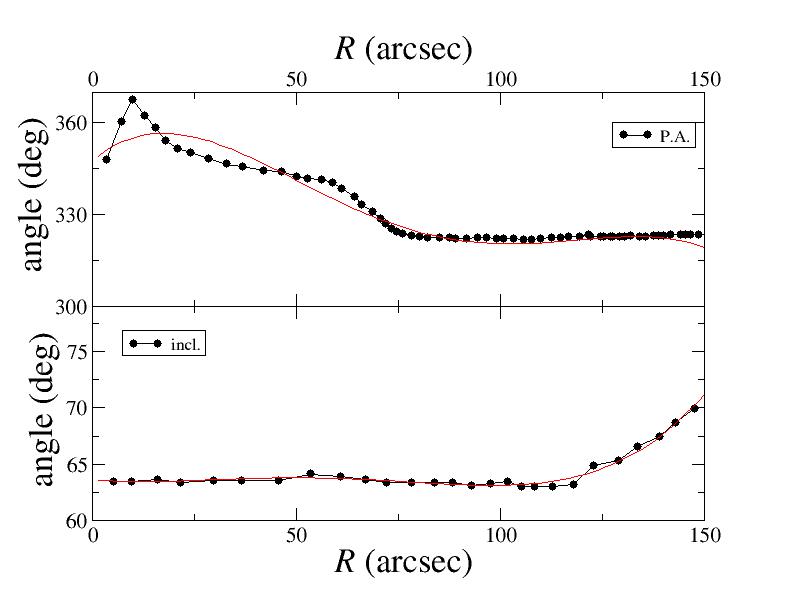}{0.45\textwidth}{(a)}
              \fig{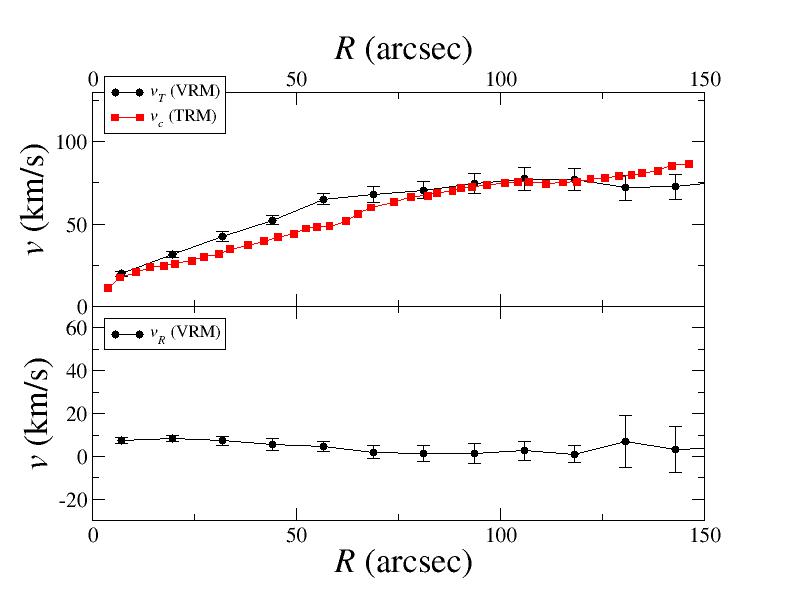}{0.45\textwidth}{(b)}
              }
  \gridline{
 \fig{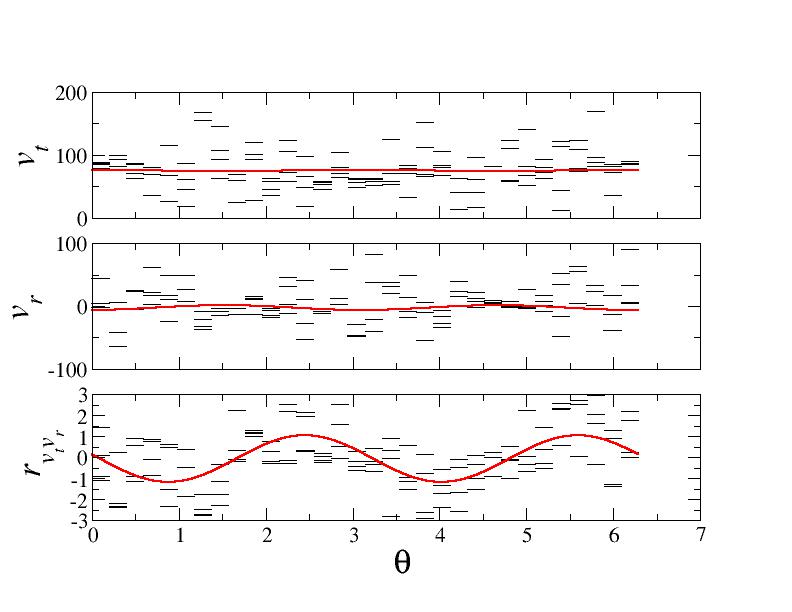}{0.45\textwidth}{(c)}
                \fig{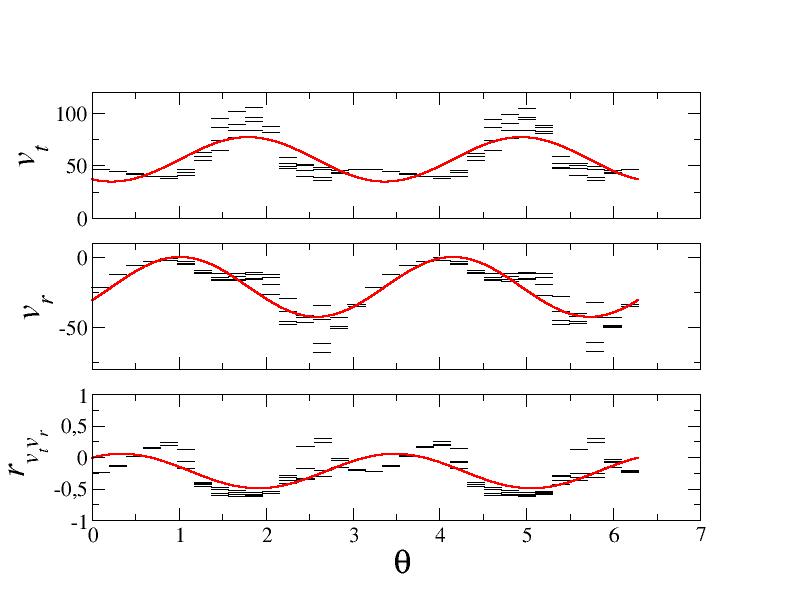}{0.45\textwidth}{(d)}}
     \caption{As  Fig.\ref{fig:NGC2903-1} but for NGC 2976.} 
\label{fig:NGC2976-1} 
\end{figure*}

\begin{figure*}
\gridline{\fig{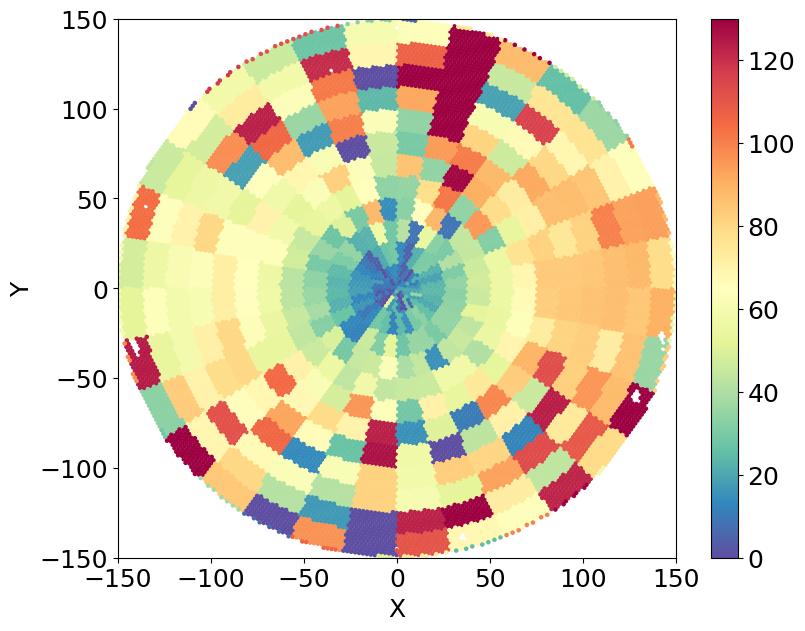}{0.3\textwidth}{(a)}
              \fig{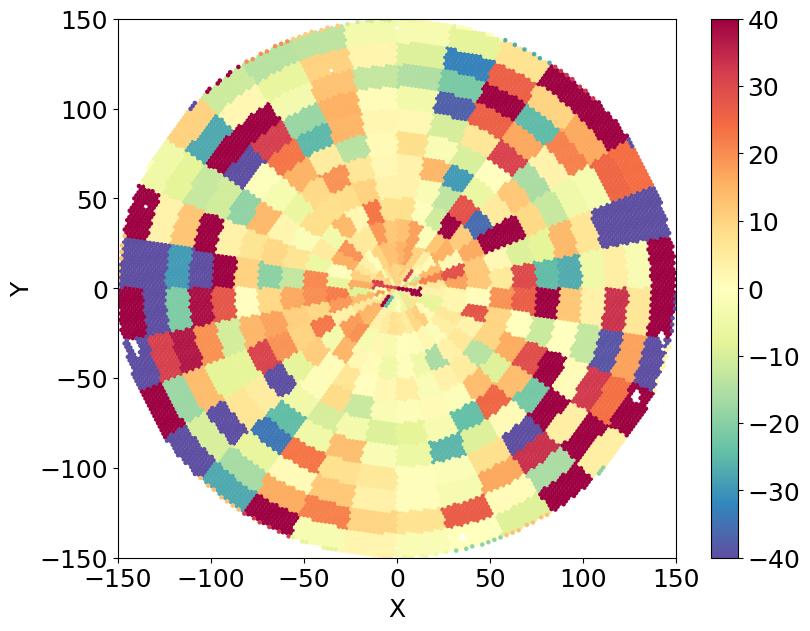}{0.3\textwidth}{(b)}
               \fig{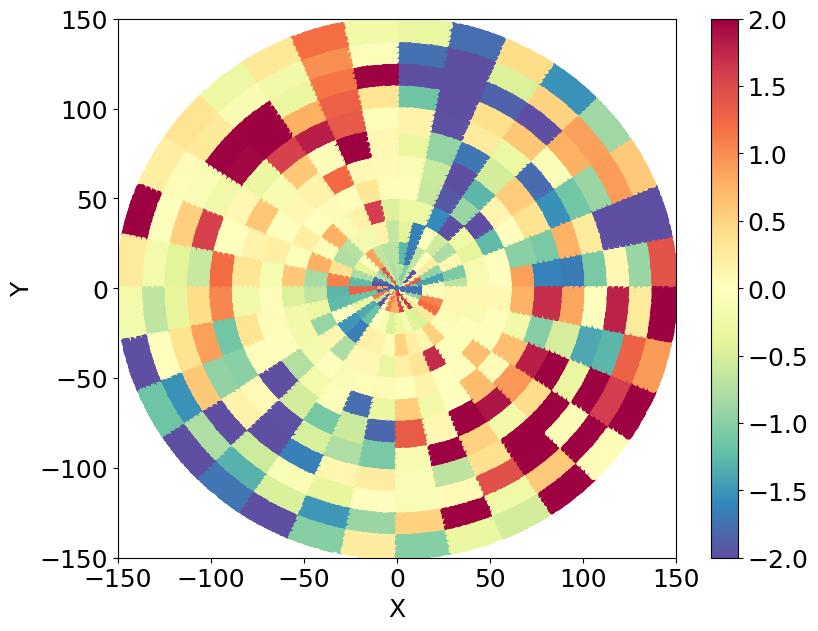}{0.3\textwidth}{(c)}}
\gridline{\fig{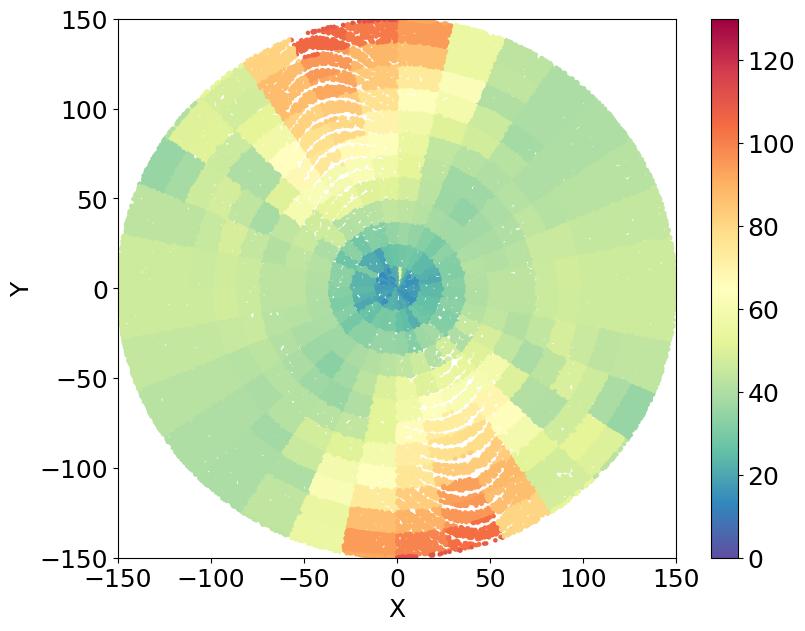}{0.3\textwidth}{(d)}
              \fig{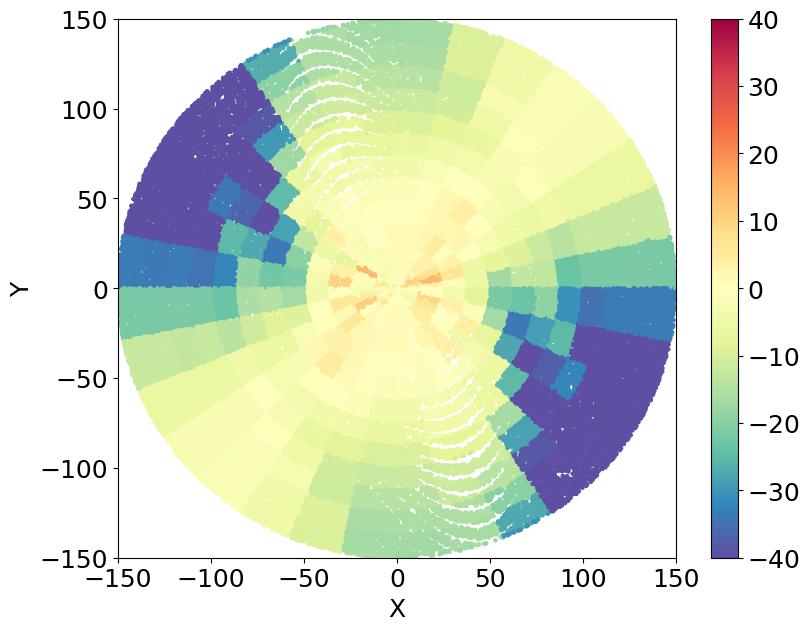}{0.3\textwidth}{(e)}
              \fig{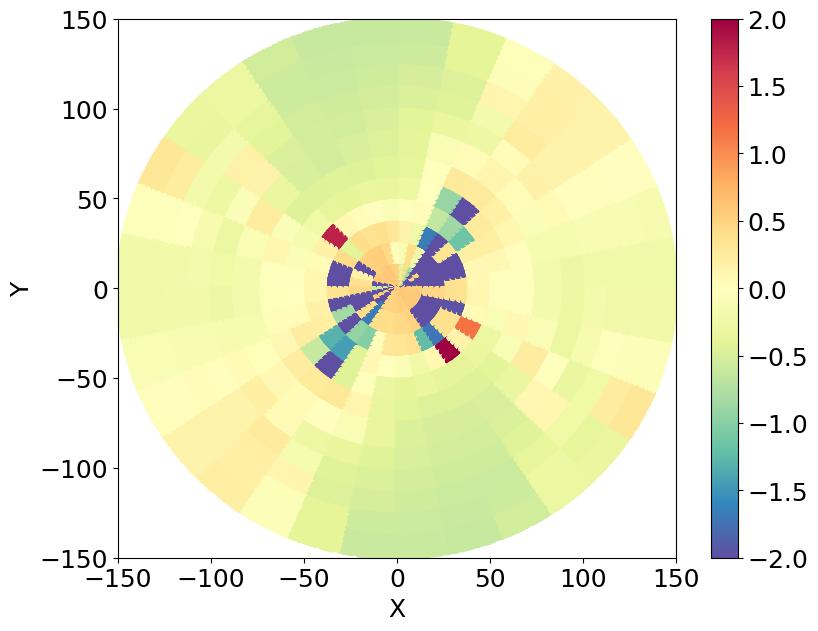}{0.3\textwidth}{(f)}}
\gridline{\fig{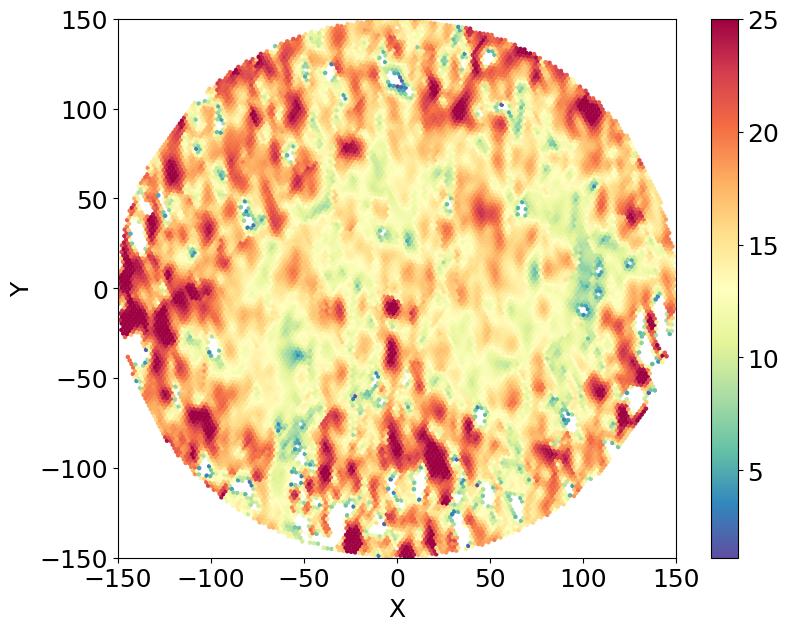}{0.3\textwidth}{(g)}
              \fig{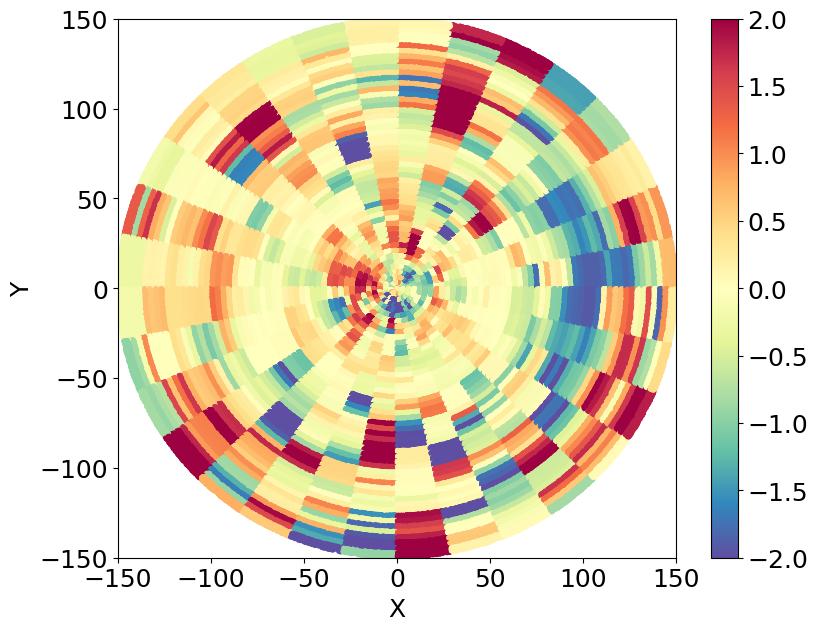}{0.3\textwidth}{(h)}
              \fig{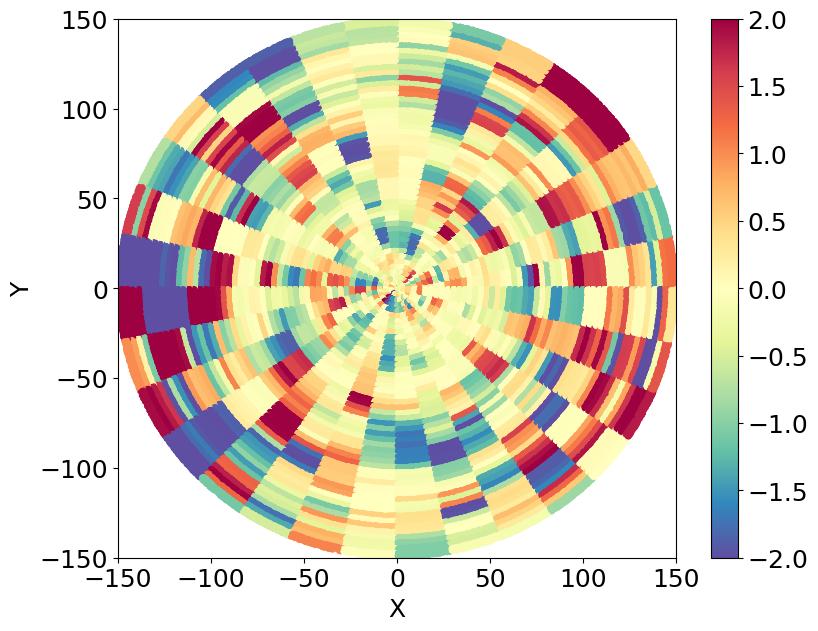}{0.3\textwidth}{(i)}}
\caption{As  Fig.\ref{fig:NGC2903-2} but for NGC 2976.} 
\label{fig:NGC2841-2} 
\end{figure*}


\subsection*{NGC 3031} 
Both orientations angle display a variation of a few degrees in the inner disk (Figure \ref{fig:NGC3031-1}). The velocity components maps (Figure \ref{fig:NGC3031-2}) do not reveal 
the presence of a warp in the inner disc and neither does it the rank correlation coefficient. 
{ The correlation coefficient is ${\cal C} = 0.18$ marginally compatibile with a warp}.
Instead, asymmetric patters of fluctuations are shown by the velocity component maps. The velocity dispersion is higher in the center. 
\begin{figure*}
\gridline{\fig{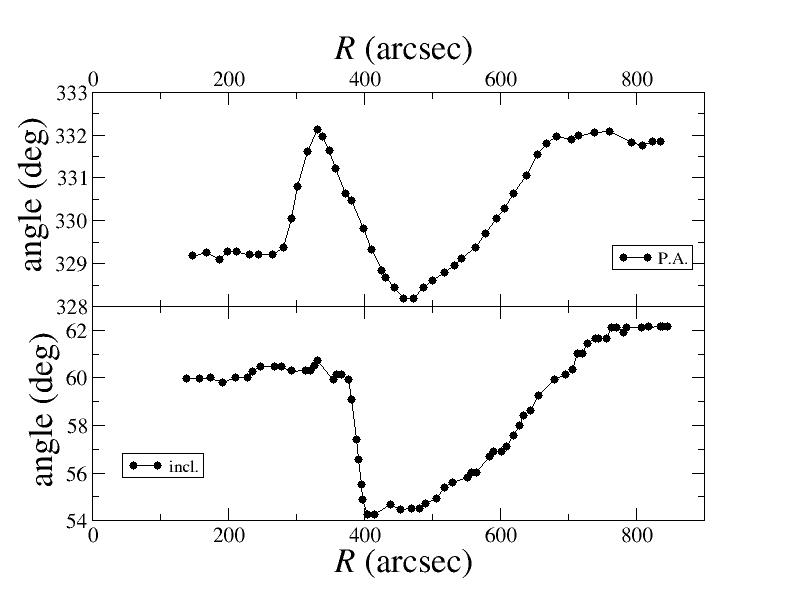}{0.45\textwidth}{(a)}
              \fig{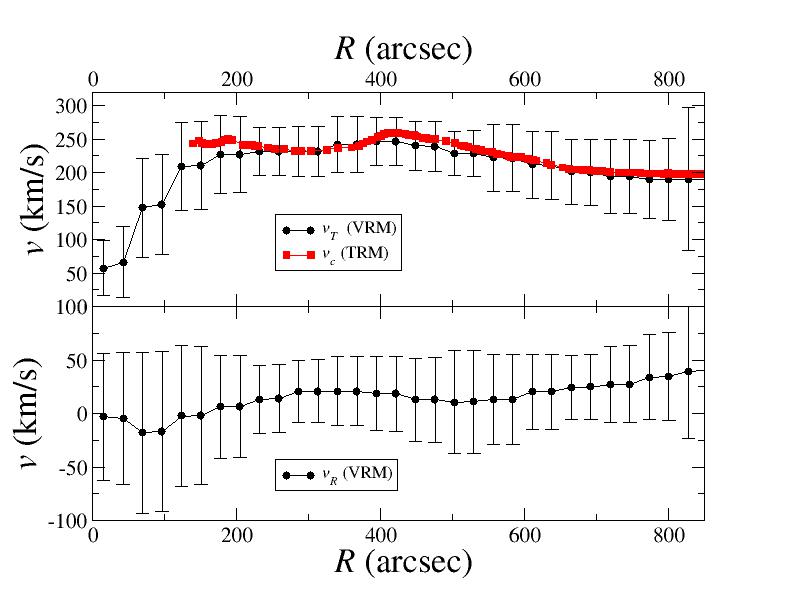}{0.45\textwidth}{(b)}
              }
  \gridline{
 \fig{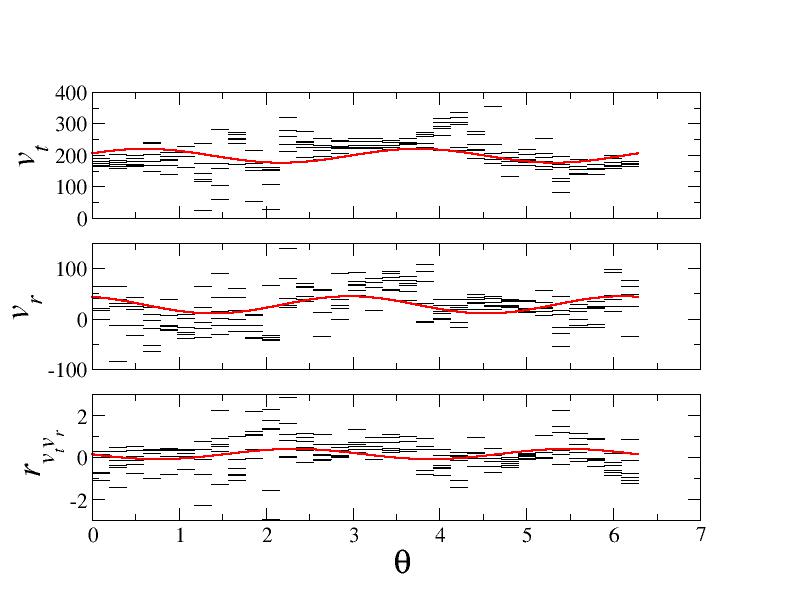}{0.45\textwidth}{(c)}
                \fig{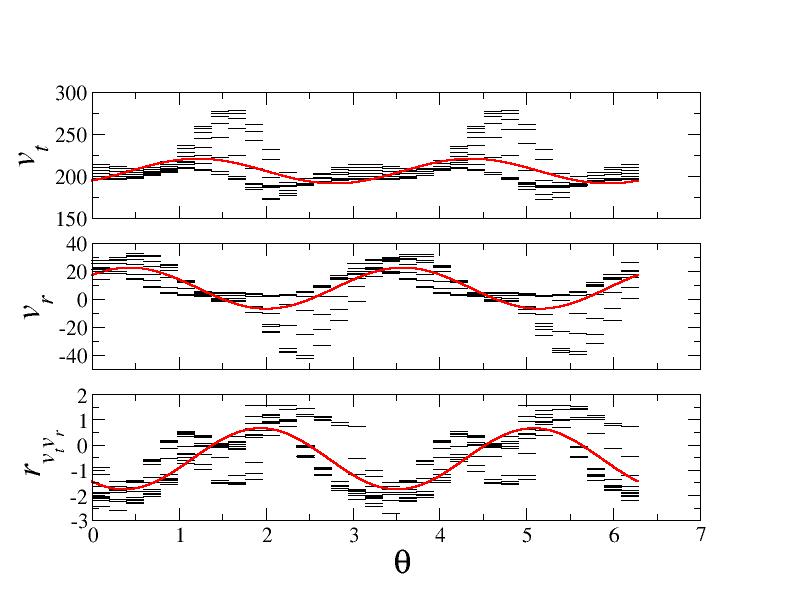}{0.45\textwidth}{(d)}}
     \caption{As  Fig.\ref{fig:NGC2903-1} but for NGC 3031.} 
\label{fig:NGC3031-1} 
\end{figure*}
%

\begin{figure*}
\gridline{\fig{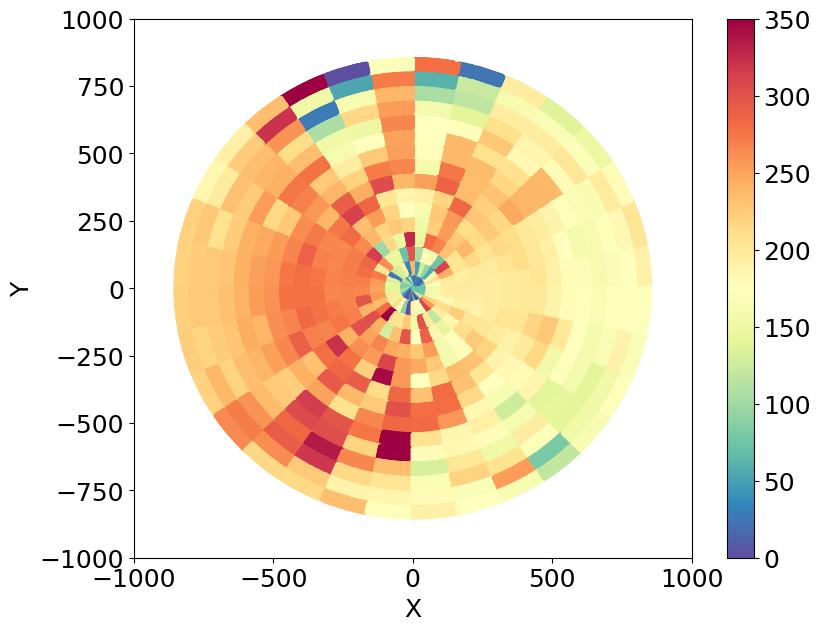}{0.3\textwidth}{(a)}
              \fig{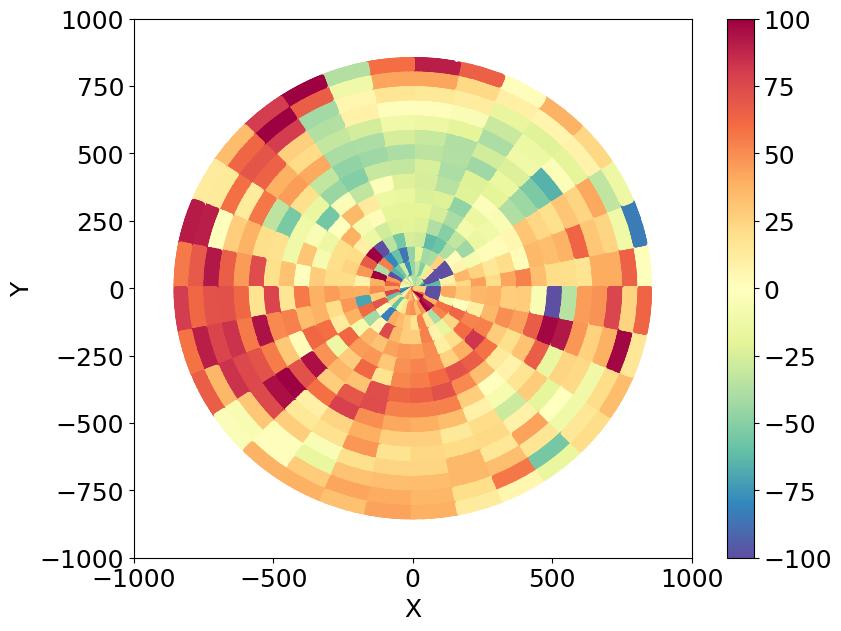}{0.3\textwidth}{(b)}
               \fig{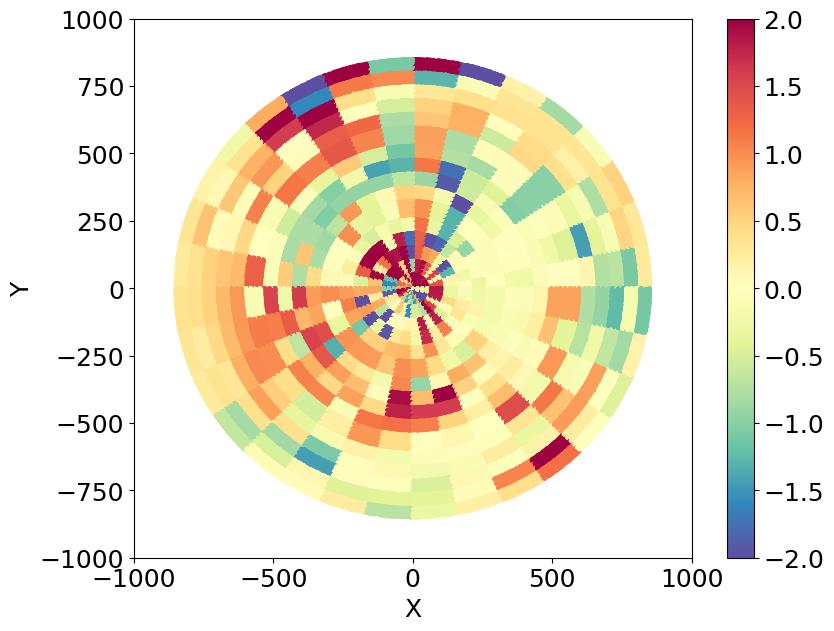}{0.3\textwidth}{(c)}}
\gridline{\fig{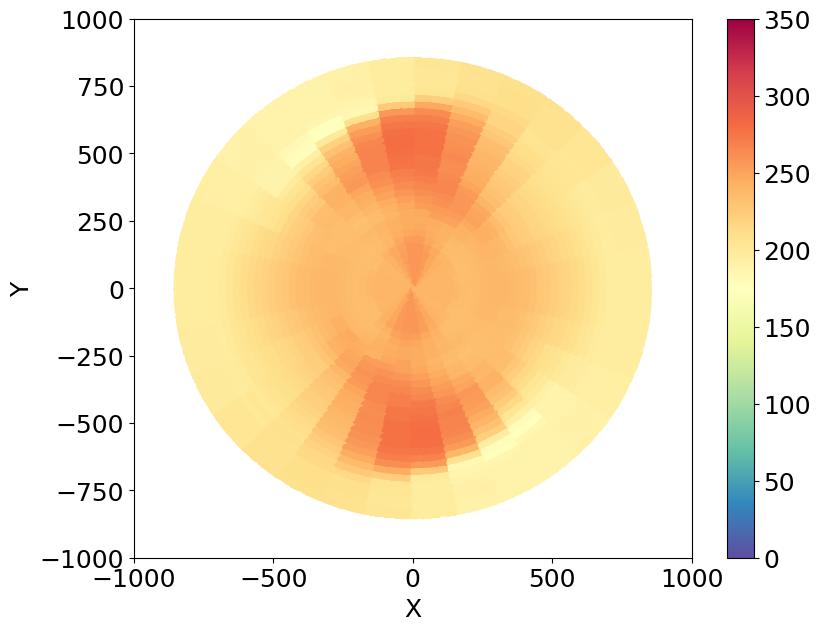}{0.3\textwidth}{(d)}
              \fig{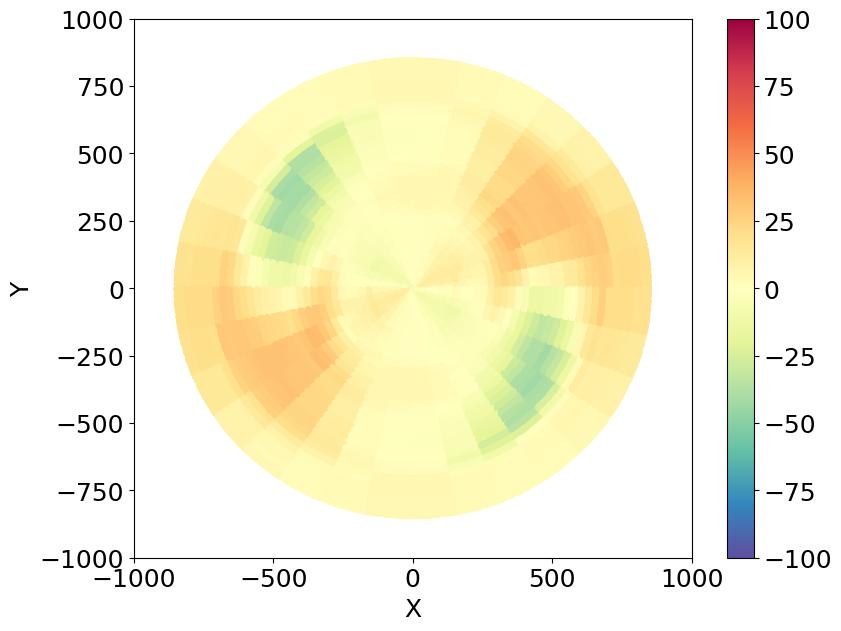}{0.3\textwidth}{(e)}
              \fig{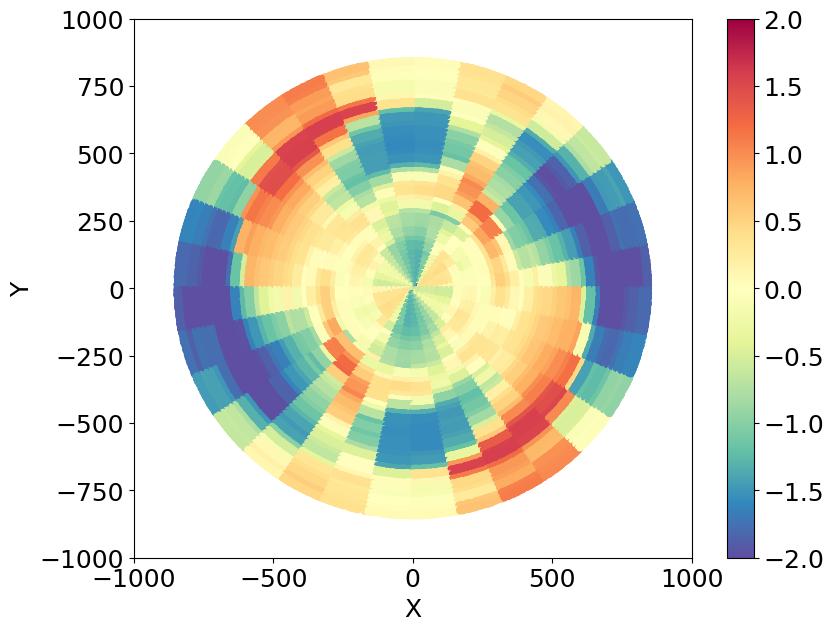}{0.3\textwidth}{(f)}}
\gridline{\fig{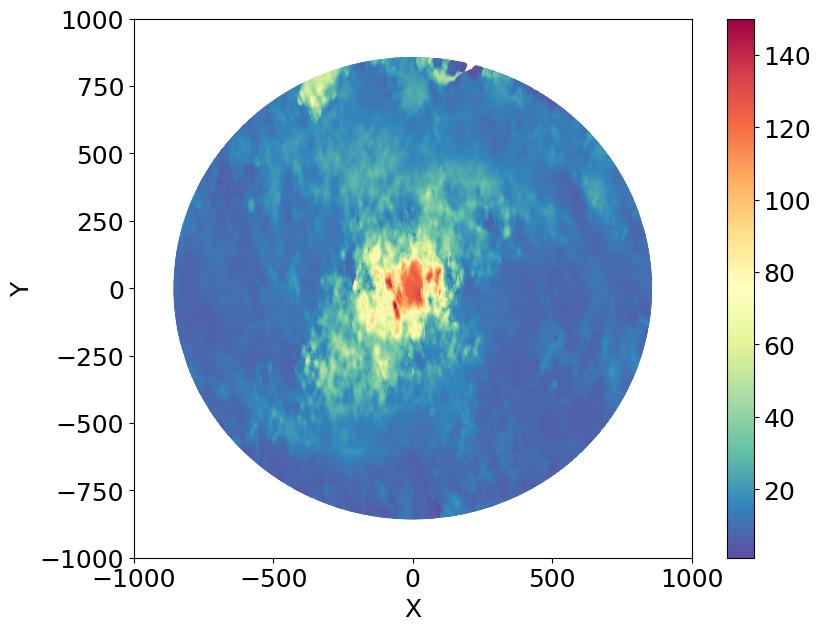}{0.3\textwidth}{(g)}
              \fig{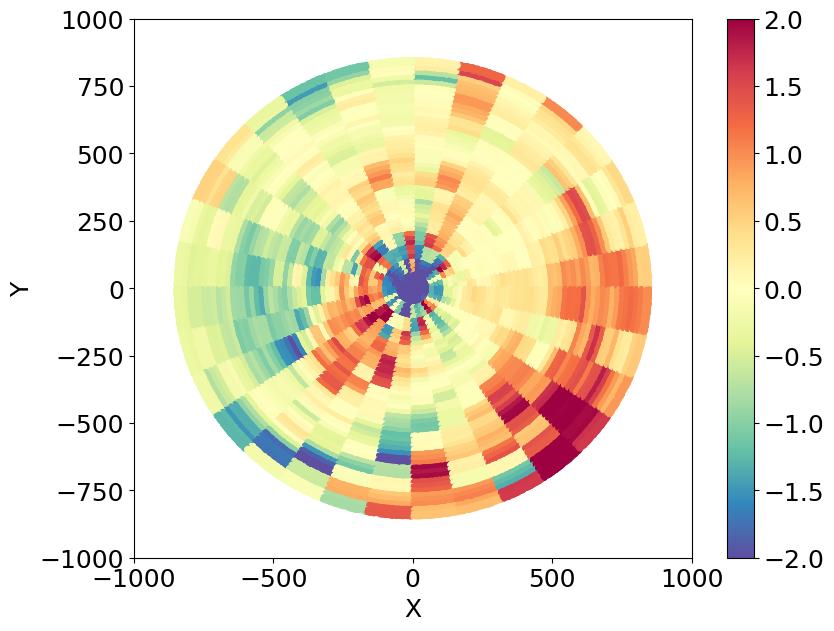}{0.3\textwidth}{(h)}
              \fig{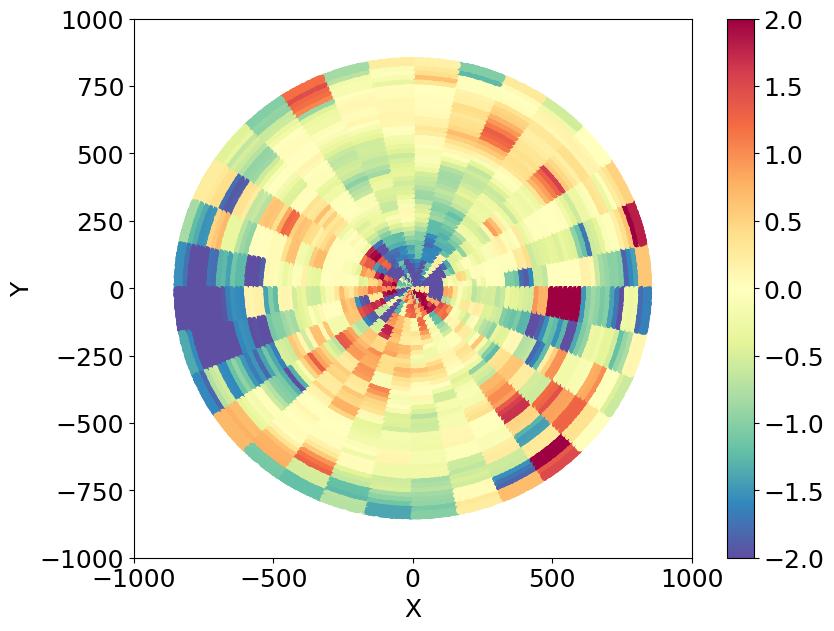}{0.3\textwidth}{(i)}}
\caption{As  Fig.\ref{fig:NGC2903-2} but for NGC 3031.} 
\label{fig:NGC3031-2} 
\end{figure*}


\subsection*{NGC 3184} 
The inclination angle $i(R)$ is less than 40$^\circ$, and its determination with the TRM is problematic; the same holds for the P.A. (see Fig.\ref{fig:NGC3184-1}). The orientation angles show large variations in the inner disc, which are probably artifacts of the TRM method while they show a monotonic decay for $R>150$".  The rank correlation coefficient shows a dipolar modulation, although the signal is much noisier than that of the corresponding toy model (see Fig.\ref{fig:NGC3184-2}). The velocity components, on the other hand, are dominated by intrinsic perturbations and do not reveal the presence of a dipolar modulation. Thus, if a warp is present, it has a small amplitude, and the velocity field is still dominated by intrinsic perturbations: { the correlation coefficient is ${\cal C} = 0.18$ marginally compatibile with a warp}.

The velocity dispersion field shows a monotonic decay and is rather regular and smooth.
%
\begin{figure*}
\gridline{\fig{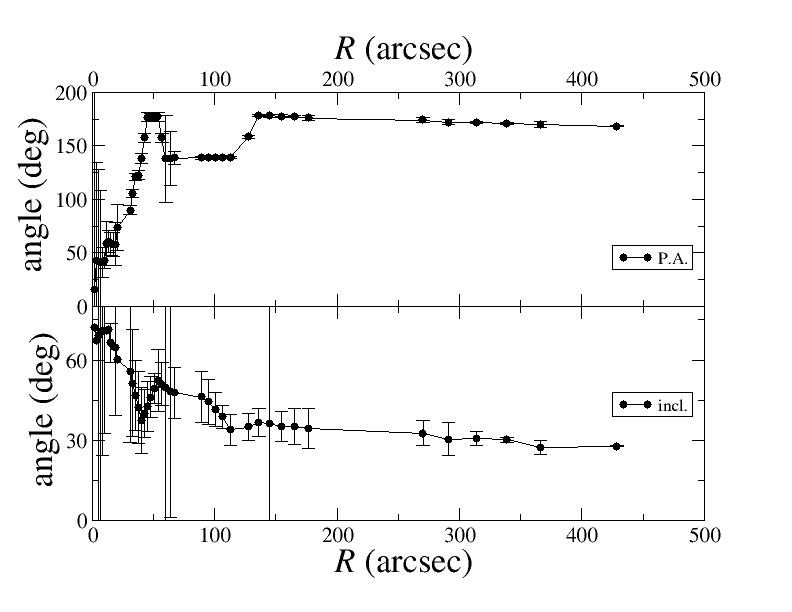}{0.45\textwidth}{(a)}
              \fig{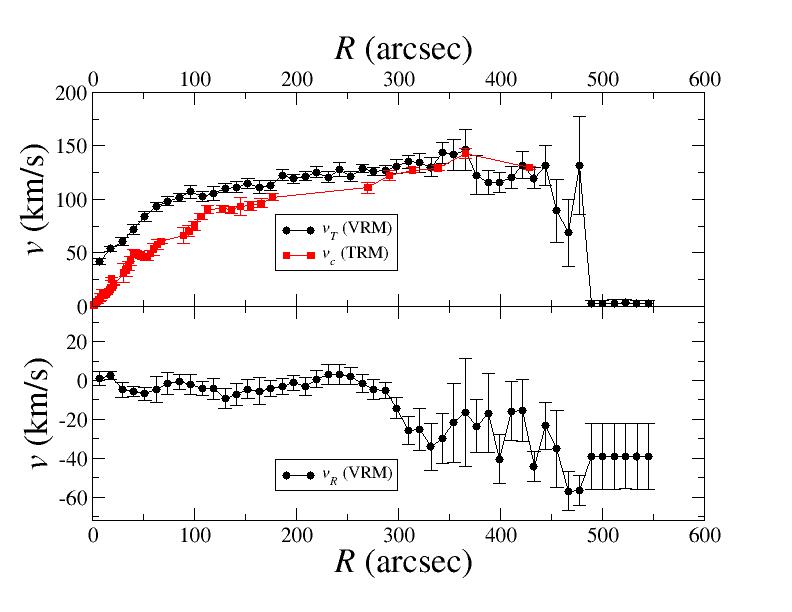}{0.45\textwidth}{(b)}
              }
  \gridline{
 \fig{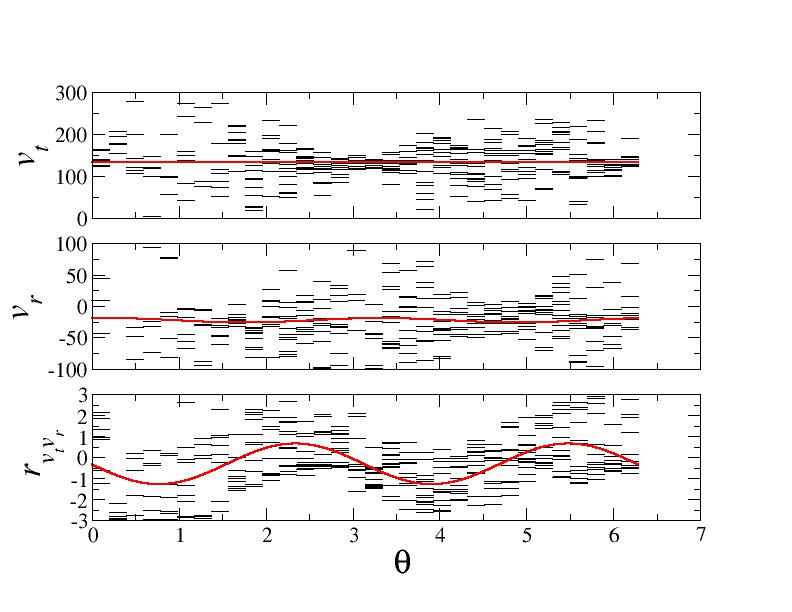}{0.45\textwidth}{(c)}
                \fig{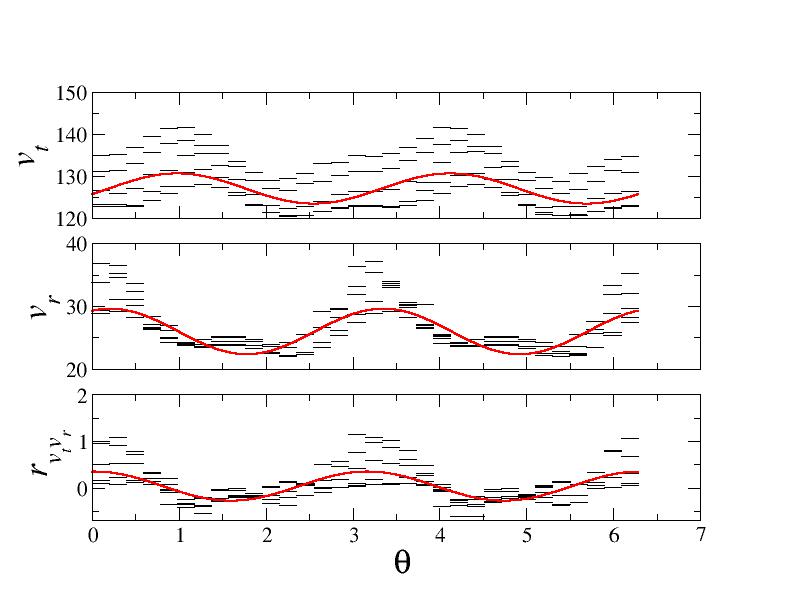}{0.45\textwidth}{(d)}}
     \caption{As  Fig.\ref{fig:NGC2903-1} but for NGC 3184.} 
\label{fig:NGC3184-1} 
\end{figure*}
%
\begin{figure*}
\gridline{\fig{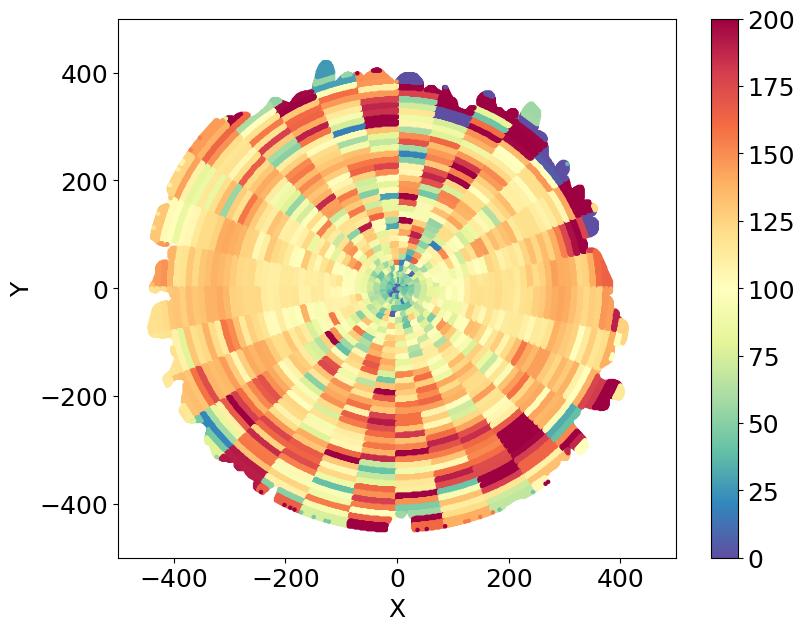}{0.3\textwidth}{(a)}
              \fig{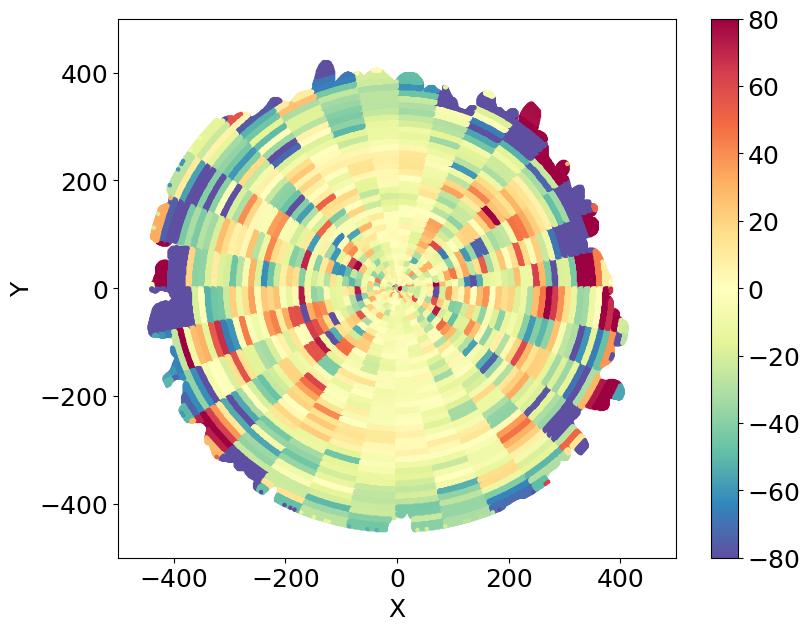}{0.3\textwidth}{(b)}
               \fig{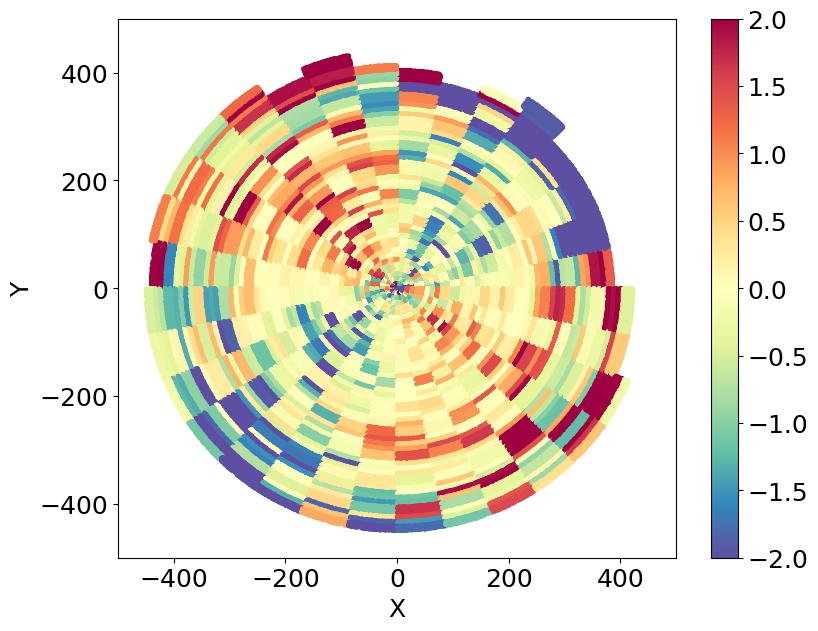}{0.3\textwidth}{(c)}}
\gridline{\fig{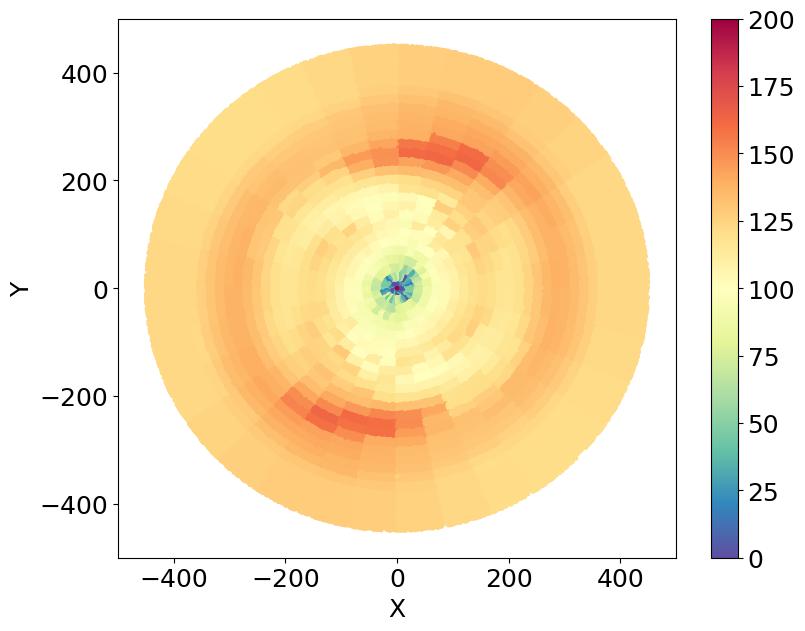}{0.3\textwidth}{(d)}
              \fig{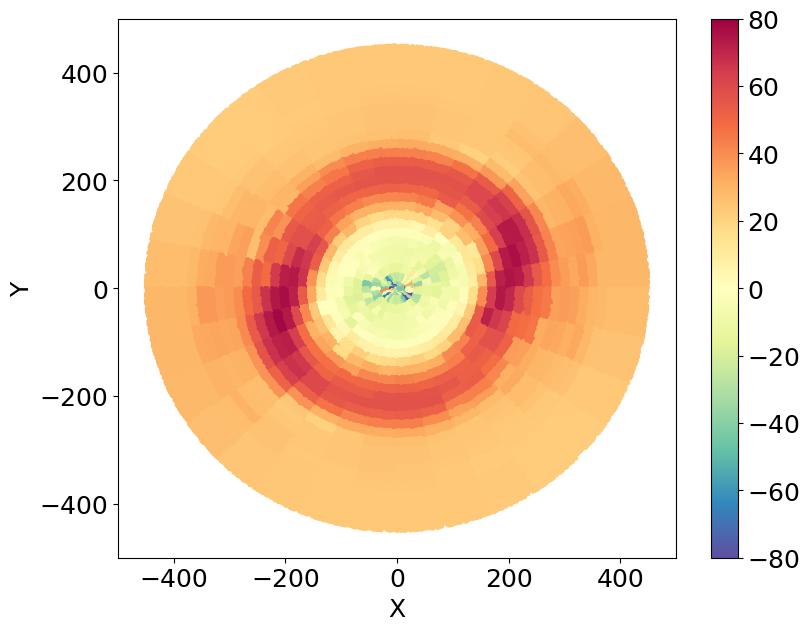}{0.3\textwidth}{(e)}
              \fig{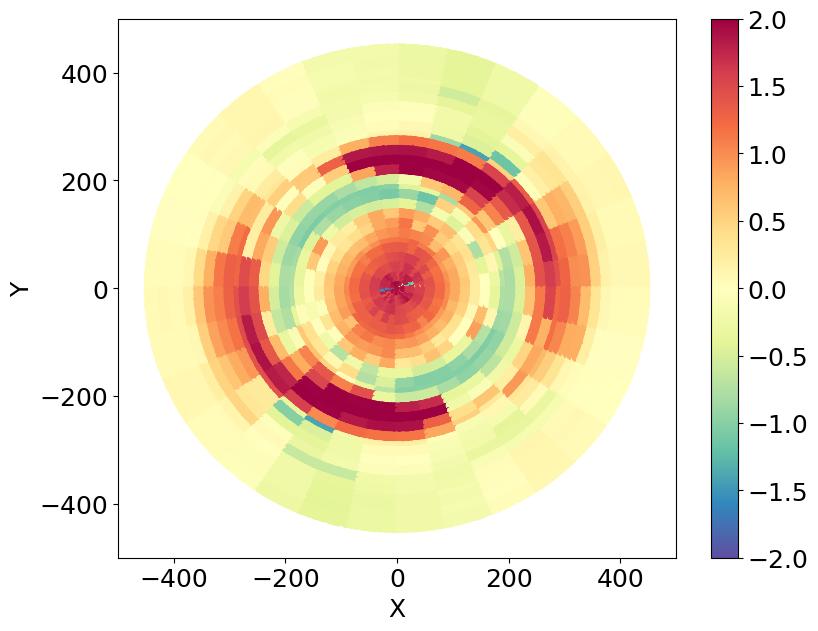}{0.3\textwidth}{(f)}}
\gridline{\fig{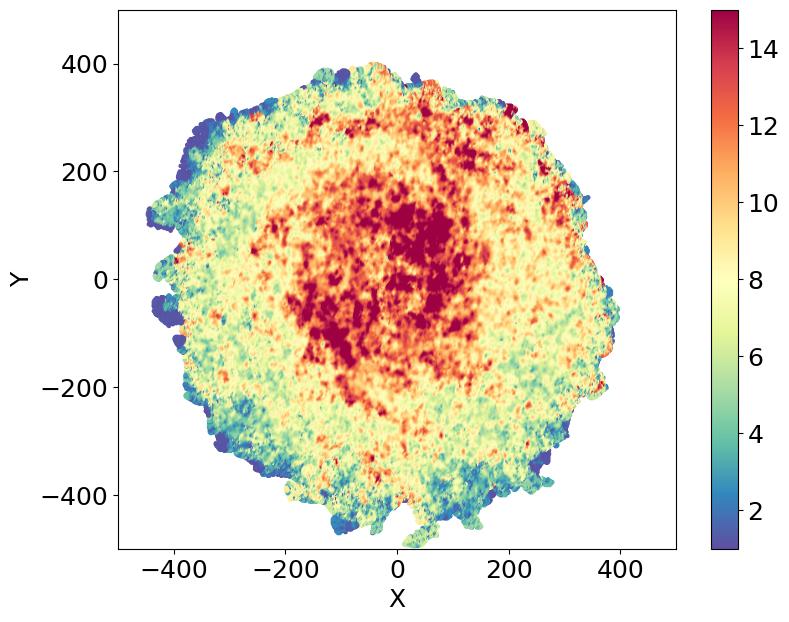}{0.3\textwidth}{(g)}
              \fig{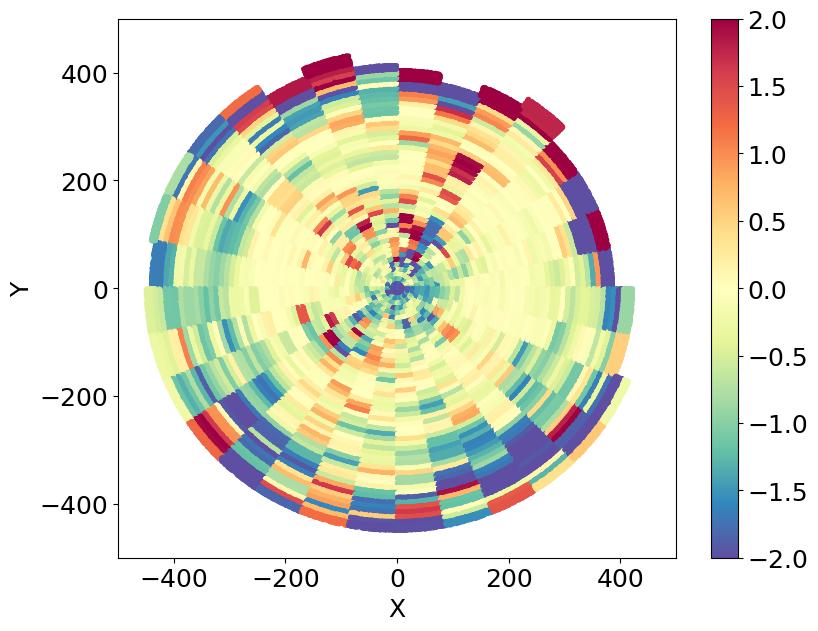}{0.3\textwidth}{(h)}
              \fig{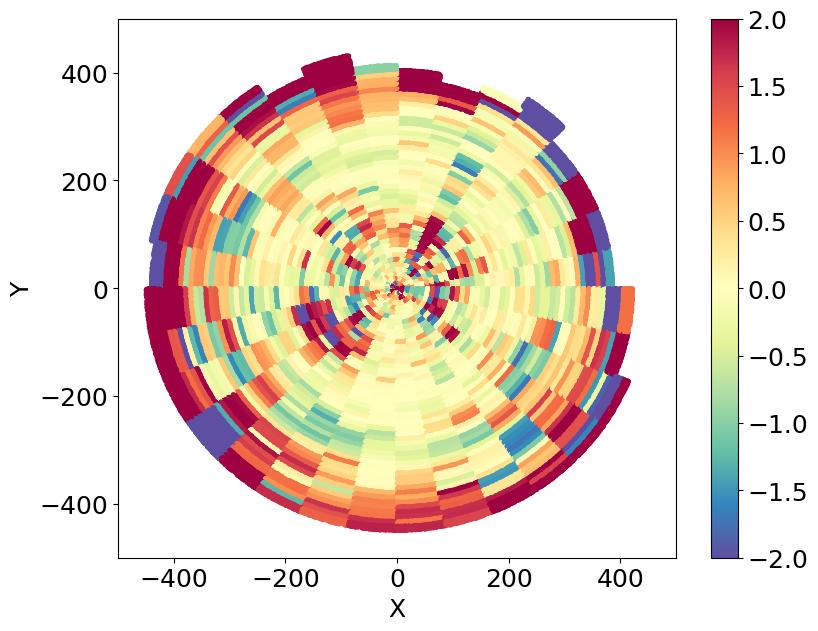}{0.3\textwidth}{(i)}}
\caption{As  Fig.\ref{fig:NGC2903-2} but for NGC 3184.} 
\label{fig:NGC3184-2} 
\end{figure*}

\clearpage


\subsection*{NGC 3198}

In this case, both the inclination angle $i(R)$ and the P.A.  show smooth variations smaller than $10^\circ$ (see Figure \ref{fig:NGC3198-1}).
Even for this galaxy, the dipolar modulation can be detected only in the behavior of the rank correlation coefficient, while the velocity components are dominated by intrinsic fluctuations. 
{ The correlation coefficient is ${\cal C}$ = 0.40 suggesting the presence of a  warp}.
The maps of the velocity components show moderate anisotropies  (see Figure \ref{fig:NGC3198-2}).
The velocity dispersion field is smooth, and its profile displays a gradual decay.

\begin{figure*}
\gridline{\fig{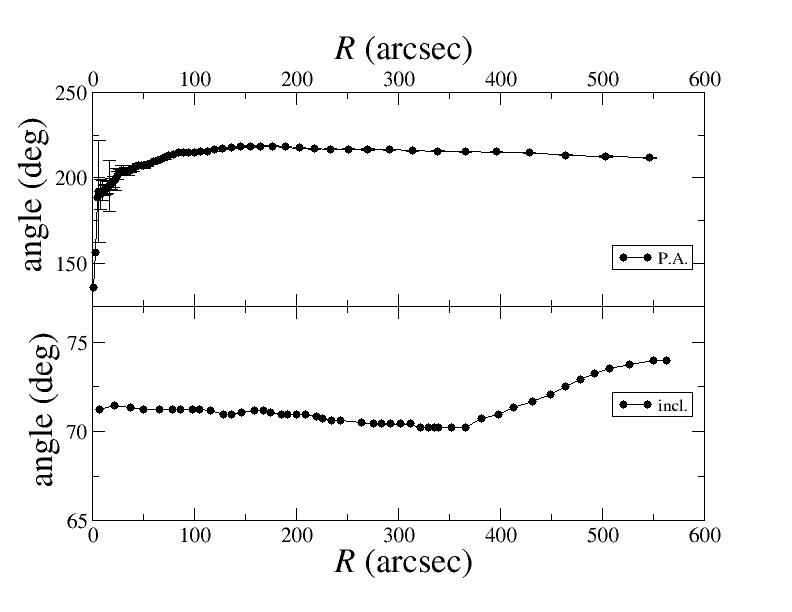}{0.45\textwidth}{(a)}
              \fig{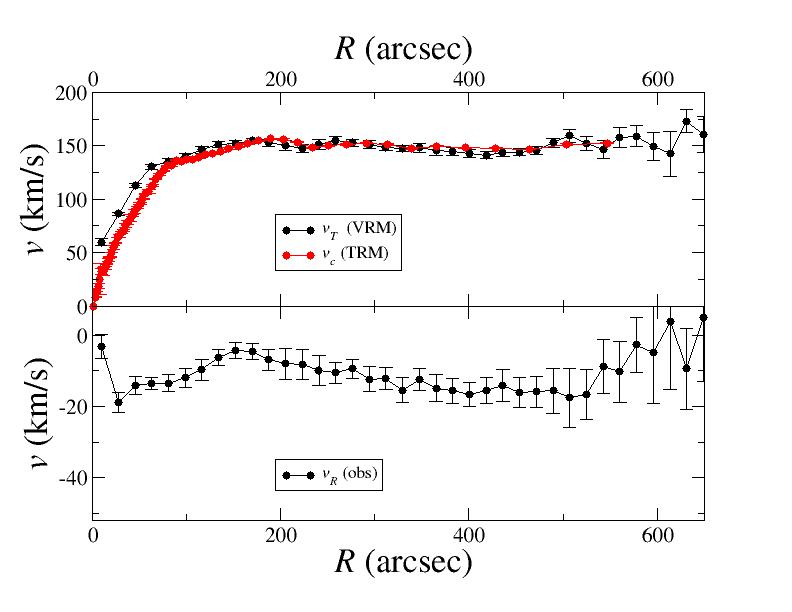}{0.45\textwidth}{(b)}
              }
  \gridline{
 \fig{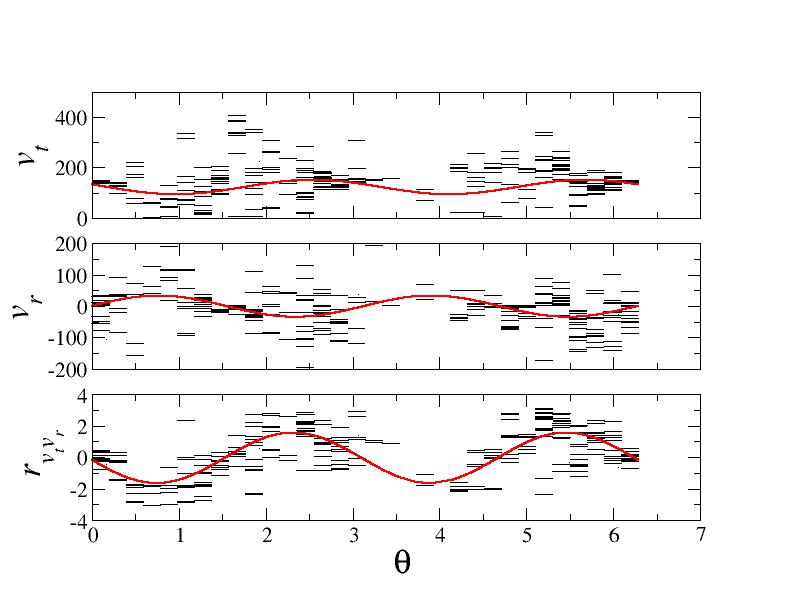}{0.45\textwidth}{(c)}
                \fig{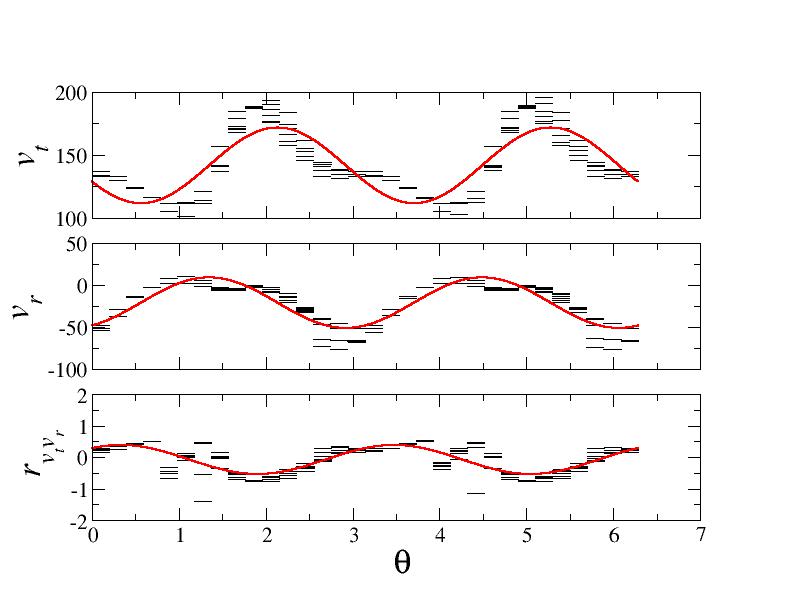}{0.45\textwidth}{(d)}}
     \caption{As  Fig.\ref{fig:NGC2903-1} but for NGC 3198.} 
\label{fig:NGC3198-1} 
\end{figure*}
%
\begin{figure*}
\gridline{\fig{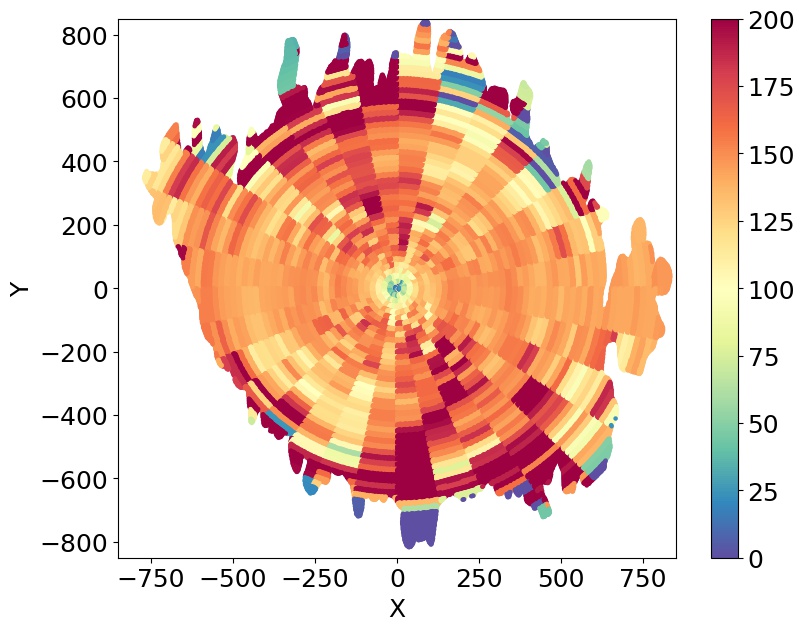}{0.3\textwidth}{(a)}
              \fig{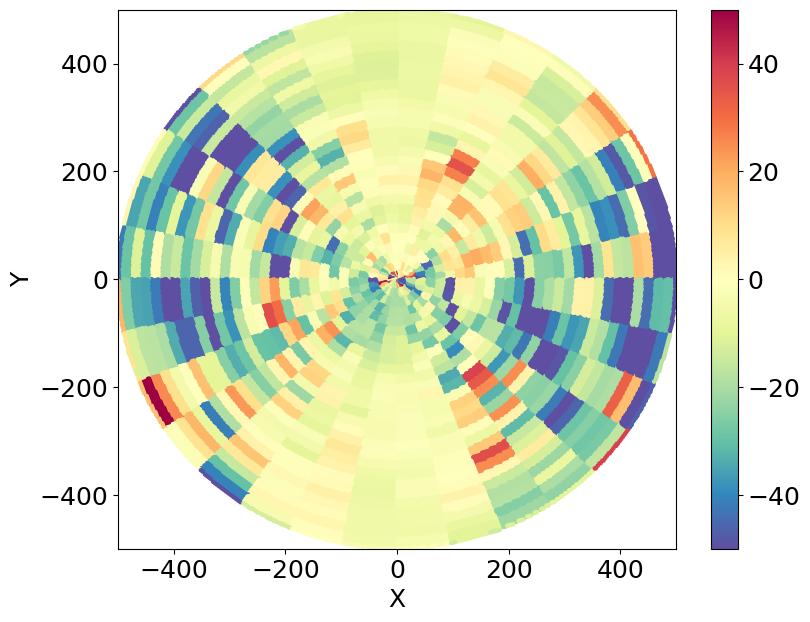}{0.3\textwidth}{(b)}
               \fig{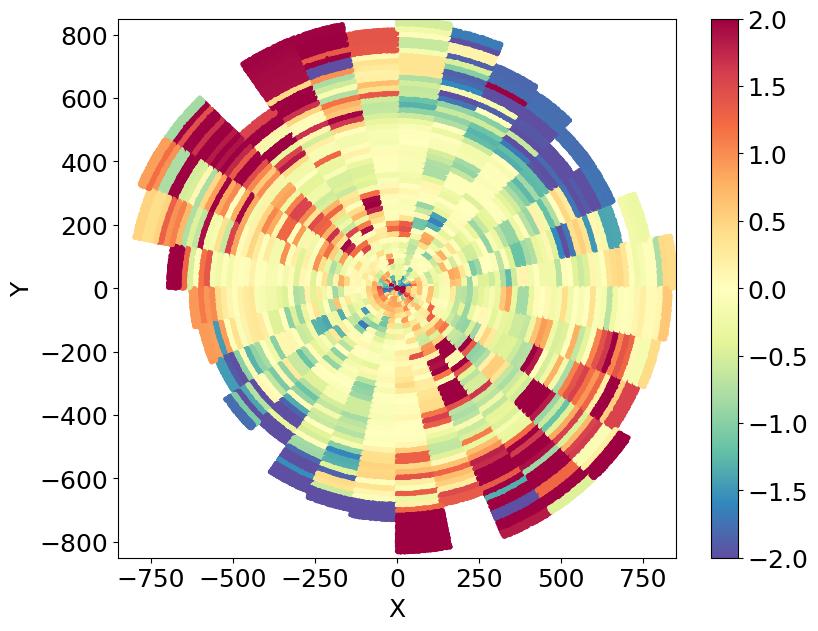}{0.3\textwidth}{(c)}}
\gridline{\fig{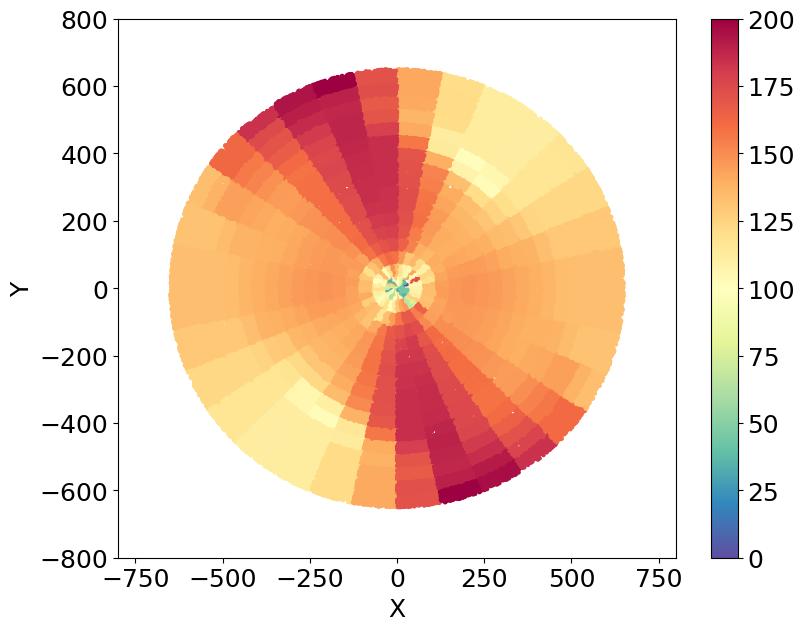}{0.3\textwidth}{(d)}
              \fig{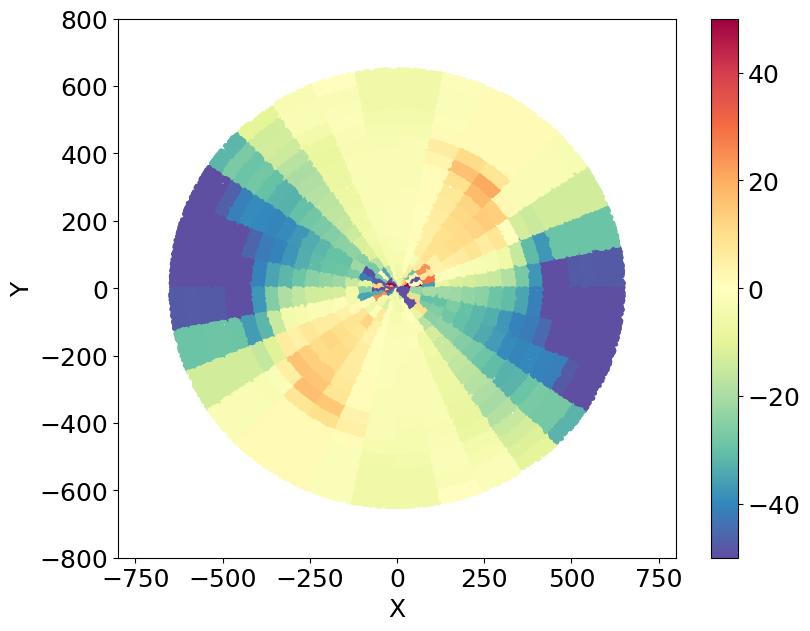}{0.3\textwidth}{(e)}
              \fig{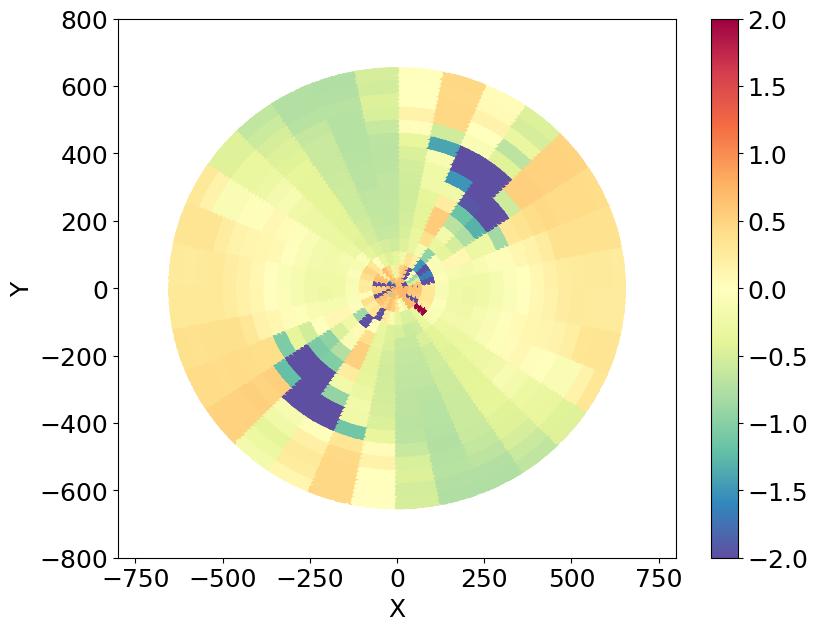}{0.3\textwidth}{(f)}}
\gridline{\fig{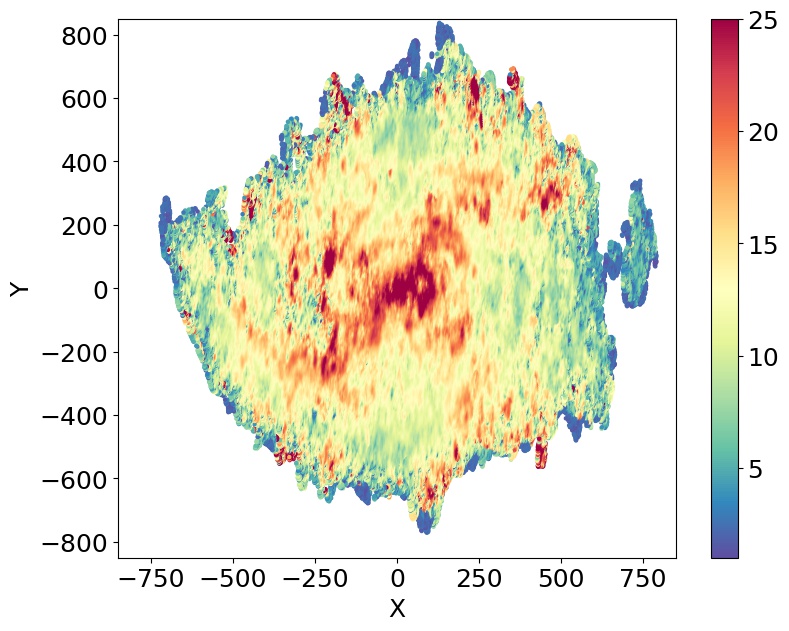}{0.3\textwidth}{(g)}
              \fig{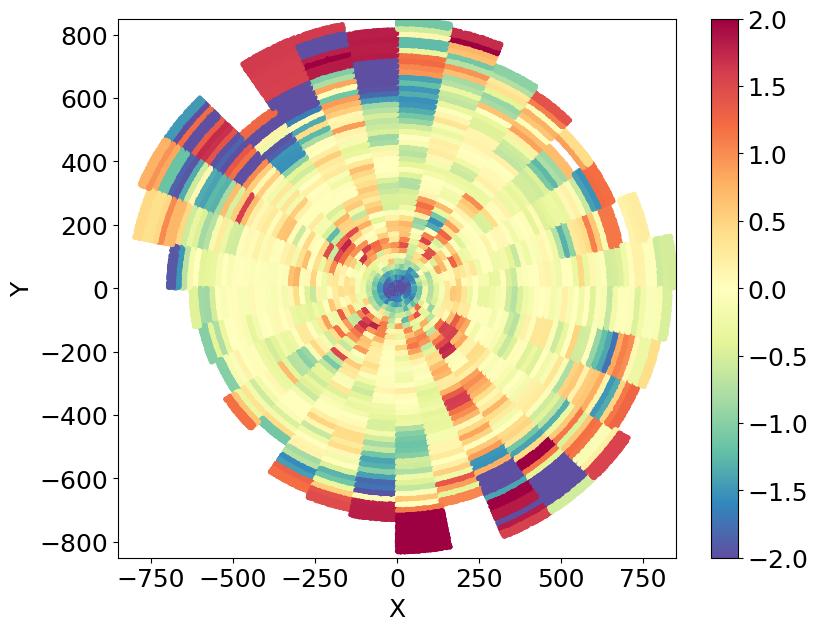}{0.3\textwidth}{(h)}
              \fig{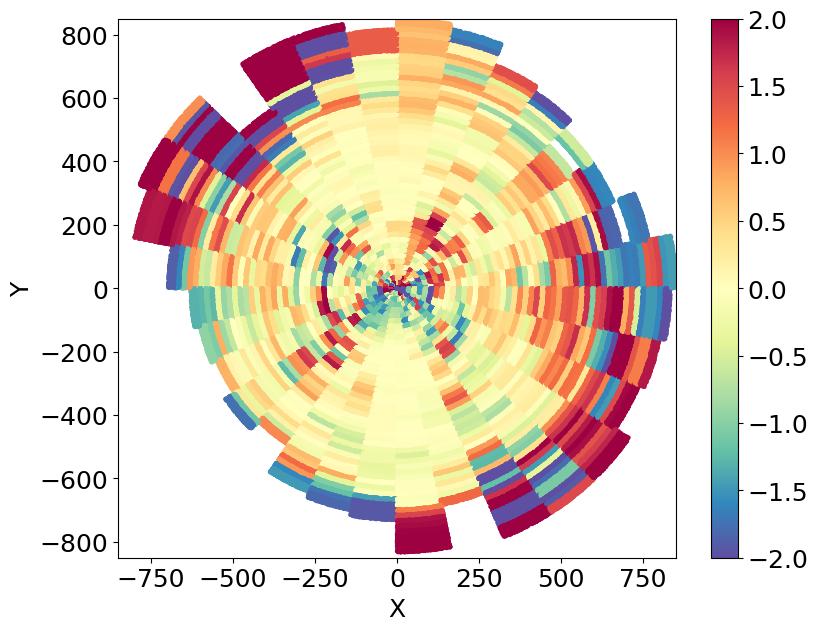}{0.3\textwidth}{(i)}}
\caption{As  Fig.\ref{fig:NGC2903-2} but for NGC 3198.} 
\label{fig:NGC3198-2} 
\end{figure*}
%


\subsection*{NGC 3351}

Even in this case, the inclination angle is around $40^\circ$, so the results of the TRM can be affected by the low inclination of the galaxy. Both orientation angles show a moderate change going from the inner to the outer disc. The velocity rank correlation coefficient shows a very noisy dipolar modulation and { the correlation coefficient is ${\cal C}$ = 0.42 suggesting the presence of a  warp}.
Both components are dominated by intrinsic fluctuations (see Figure \ref{fig:NGC3351-1}). The velocity component 2D maps reveal that the velocity field is quite smooth. This is coherent with the behavior of the velocity dispersion field that does not show large anisotropies (see Fig\ref{fig:NGC3351-2}).
\begin{figure*}
\gridline{\fig{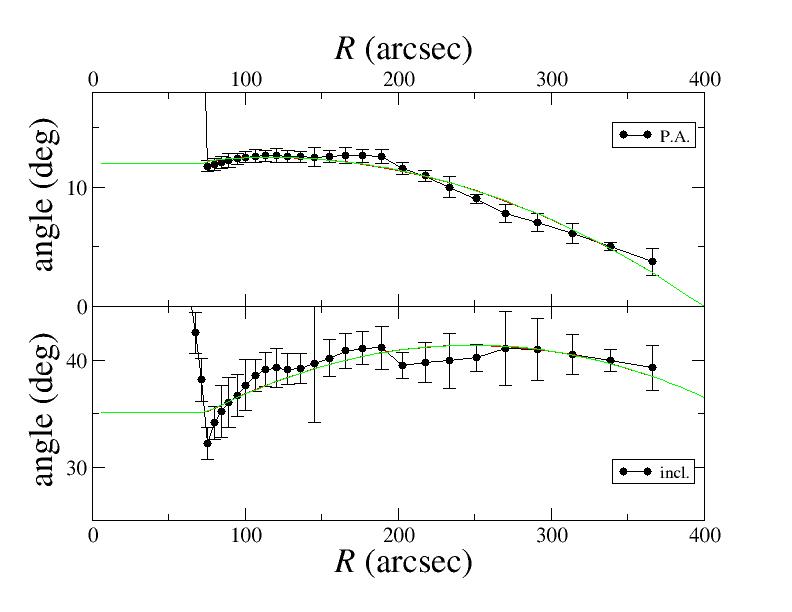}{0.45\textwidth}{(a)}
              \fig{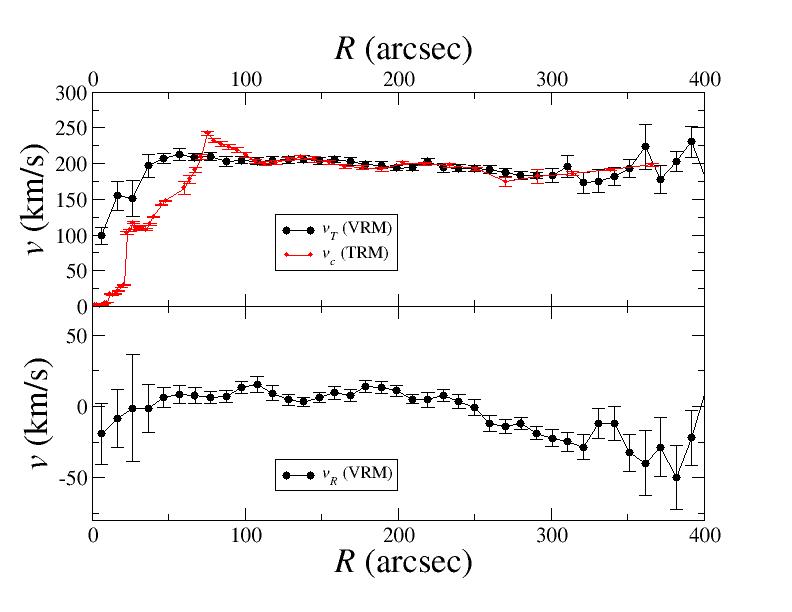}{0.45\textwidth}{(b)}
              }
  \gridline{
 \fig{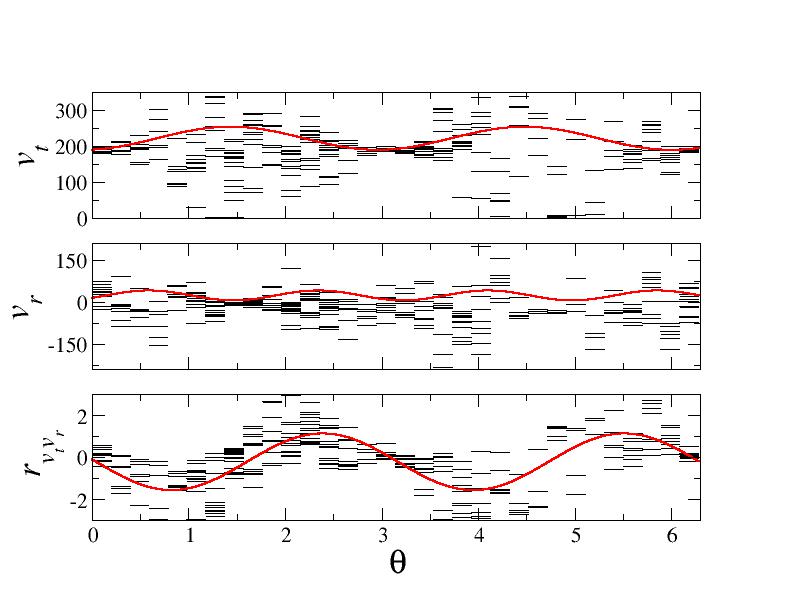}{0.45\textwidth}{(c)}
                \fig{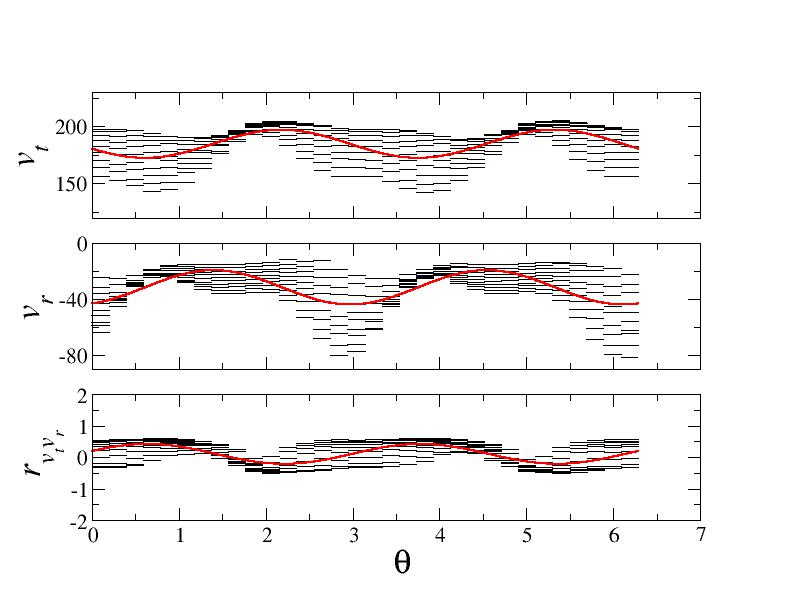}{0.45\textwidth}{(d)}}
     \caption{As  Fig.\ref{fig:NGC2903-1} but for NGC 3351.} 
\label{fig:NGC3351-1} 
\end{figure*}
%
\begin{figure*}
\gridline{\fig{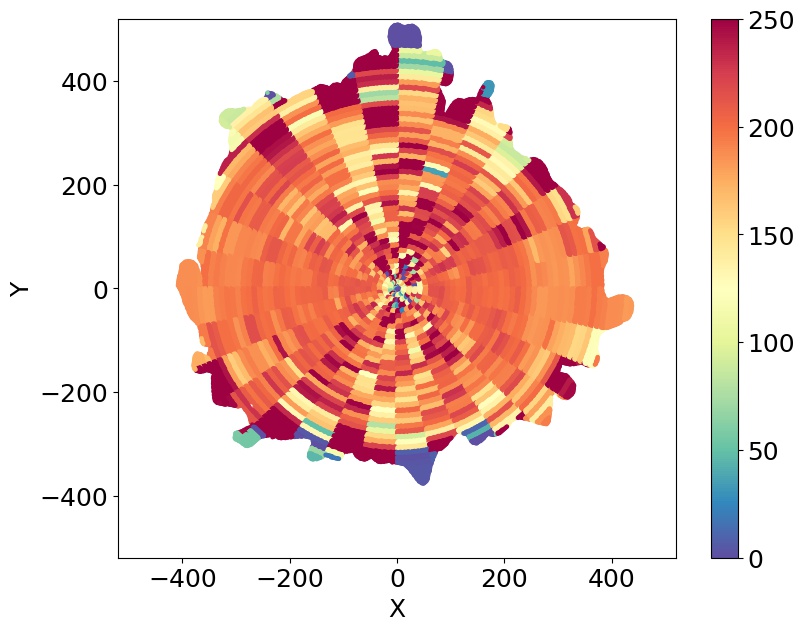}{0.3\textwidth}{(a)}
              \fig{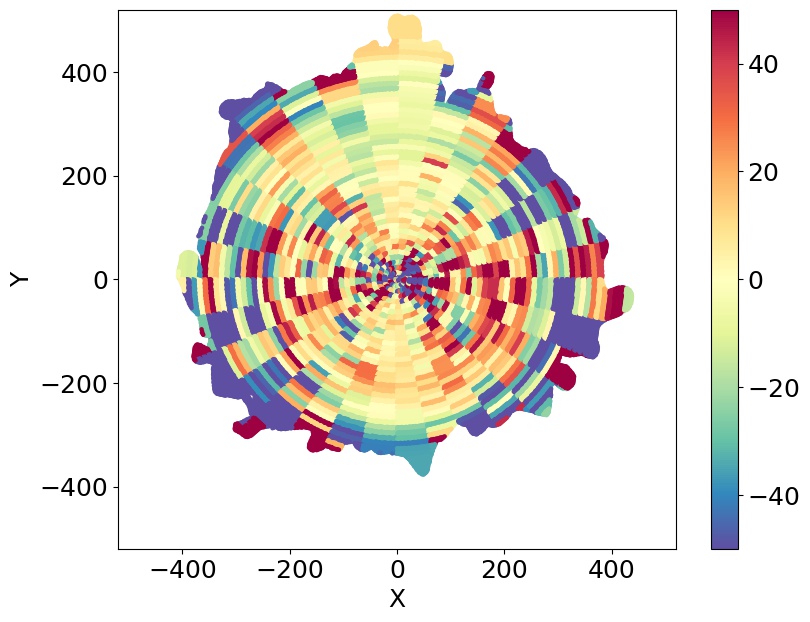}{0.3\textwidth}{(b)}
               \fig{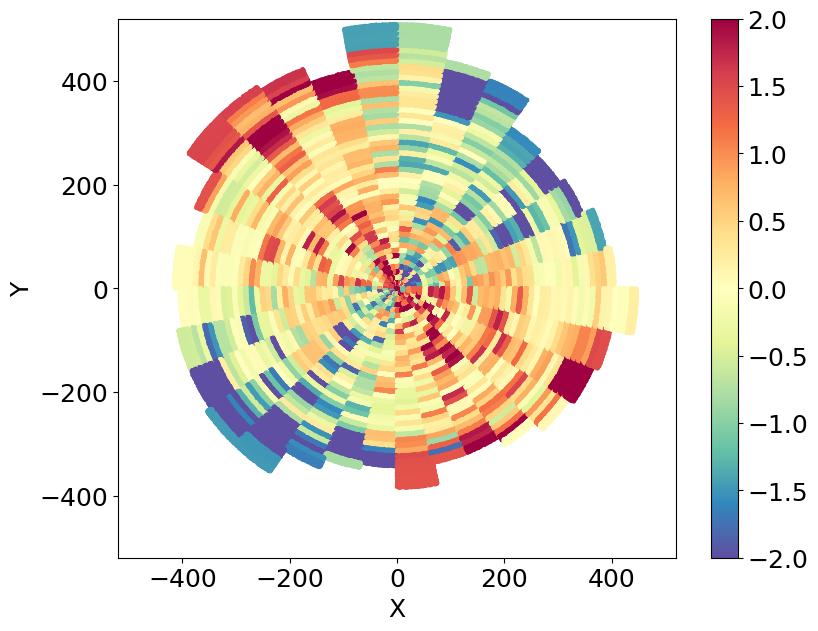}{0.3\textwidth}{(c)}}
\gridline{\fig{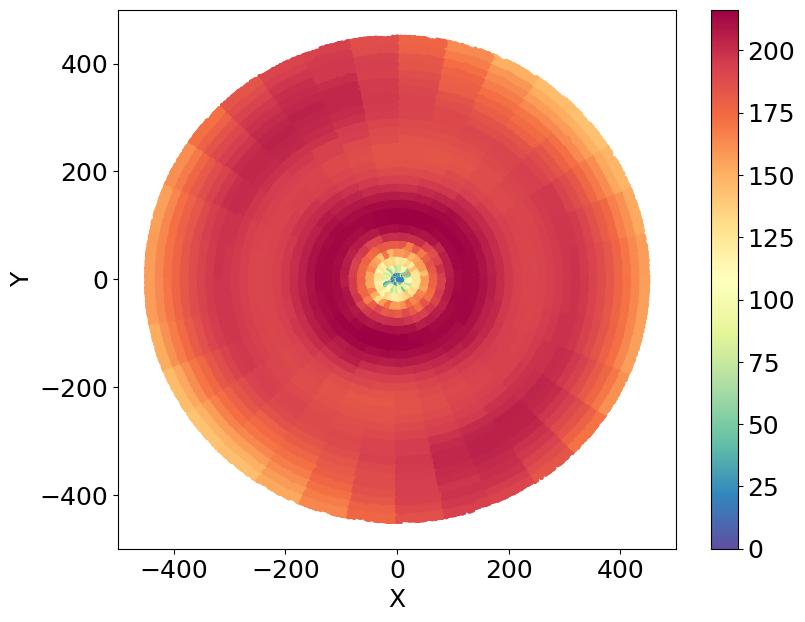}{0.3\textwidth}{(d)}
              \fig{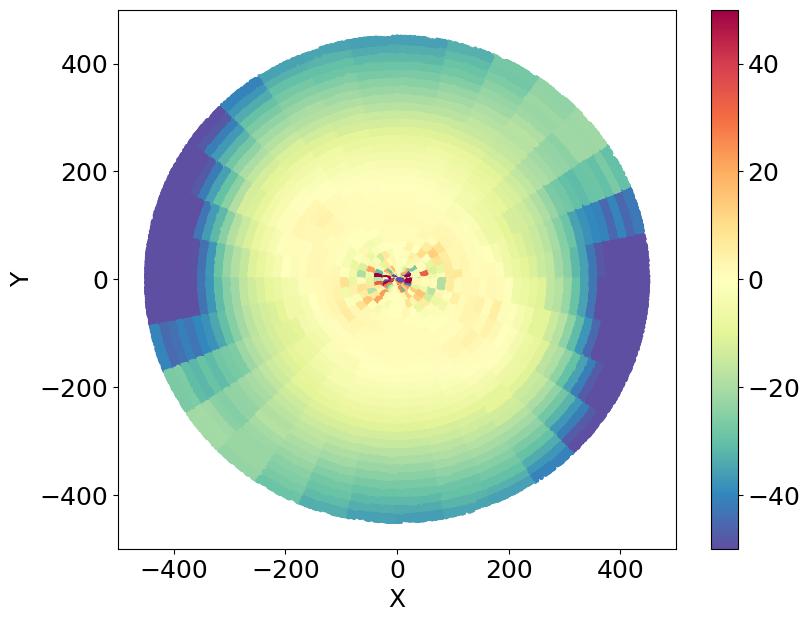}{0.3\textwidth}{(e)}
              \fig{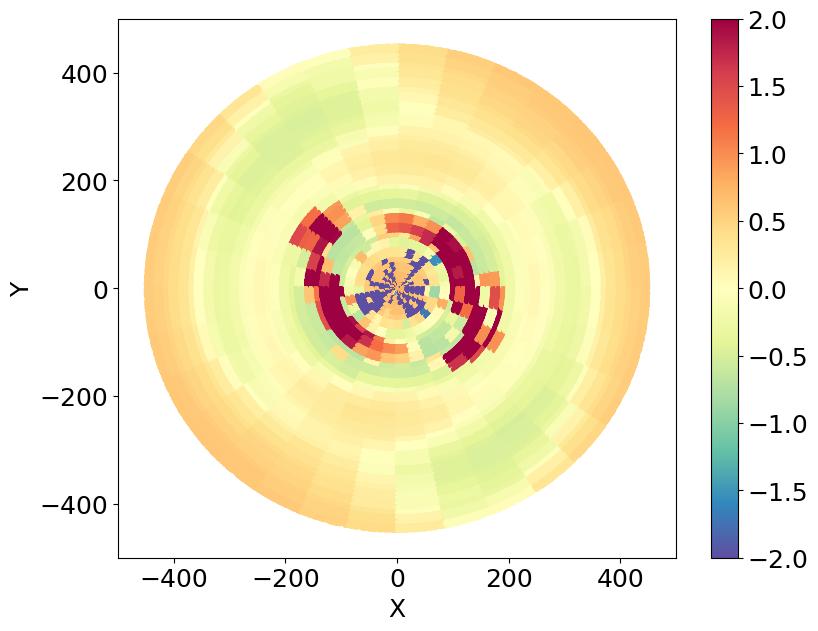}{0.3\textwidth}{(f)}}
\gridline{\fig{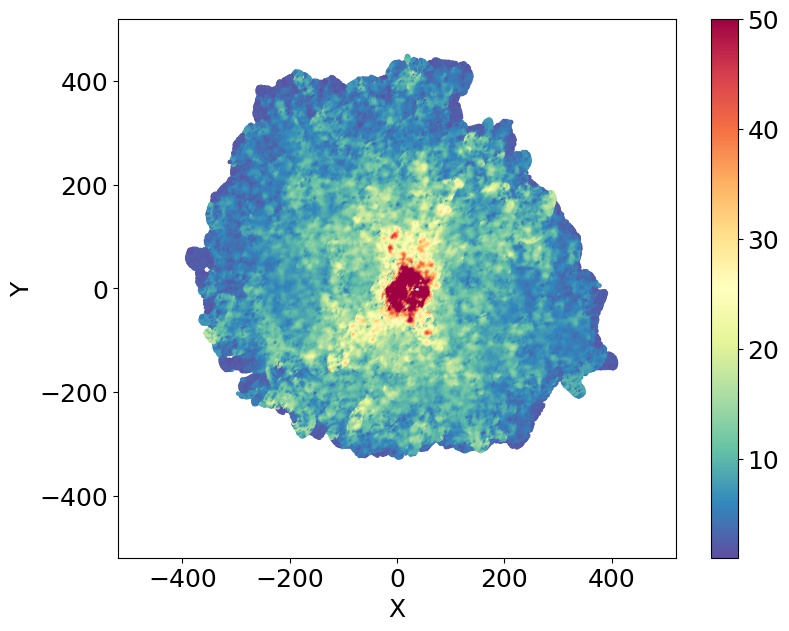}{0.3\textwidth}{(g)}
              \fig{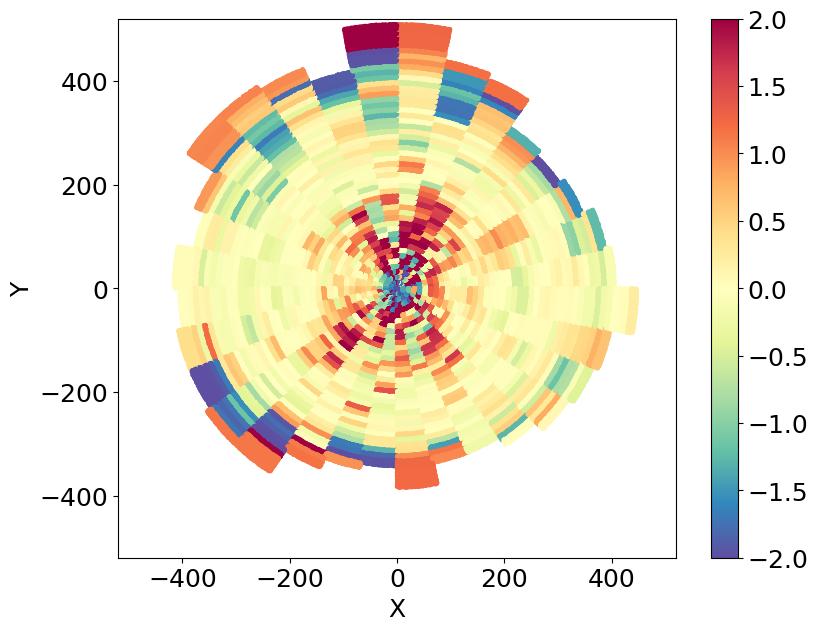}{0.3\textwidth}{(h)}
              \fig{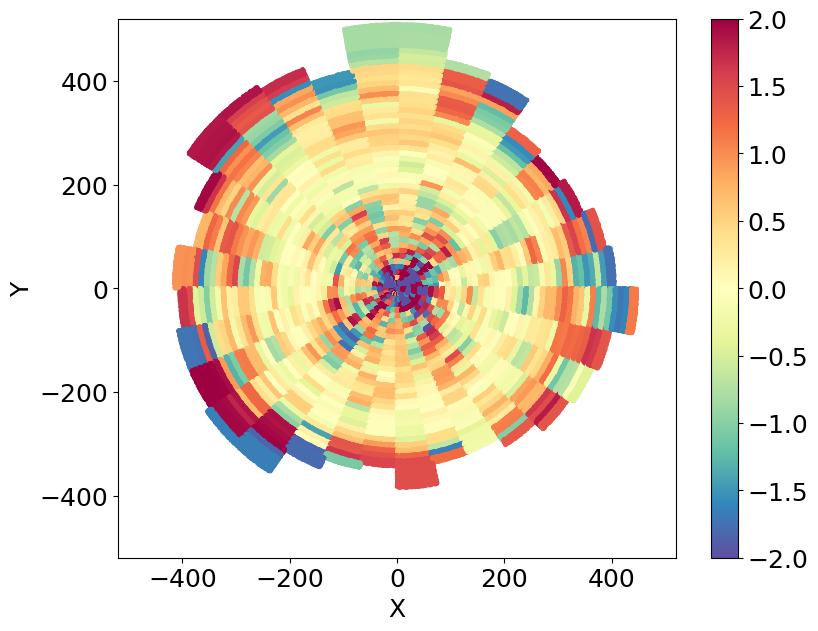}{0.3\textwidth}{(i)}}
\caption{As  Fig.\ref{fig:NGC2903-2} but for NGC 3351.} 
\label{fig:NGC3351-2} 
\end{figure*}


\subsection*{NGC 3521} 

Even for NGC 3521, the variation of the orientation angle is moderate, i.e., $< 10^\circ$. A small amplitude warp may thus be present and corresponds to the (noisy) dipolar modulation of the velocity rank correlation coefficient (Fig. \ref{fig:NGC3521-1}). 
{ This is confirmed by   the correlation coefficient that is ${\cal C}$ = 0.40}. However, in this case, intrinsic fluctuations are much larger than extrinsic ones and perturbations in the velocity dispersion field are clearly anisotropic. A correlation between $\sigma$ and the $v_r$ can be noticed in the $r_{\sigma v_r}(R, \theta)$ map (Figure \ref{fig:NGC3521-2}).
\begin{figure*}
\gridline{\fig{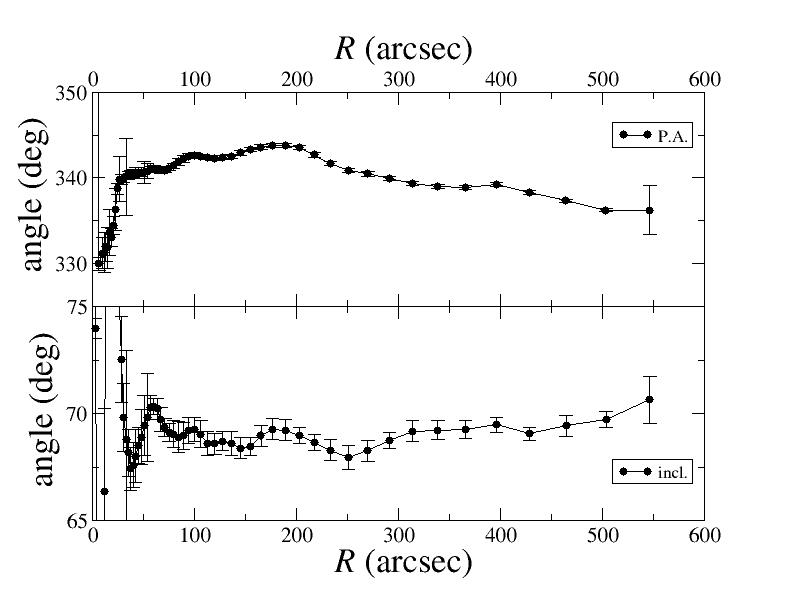}{0.45\textwidth}{(a)}
              \fig{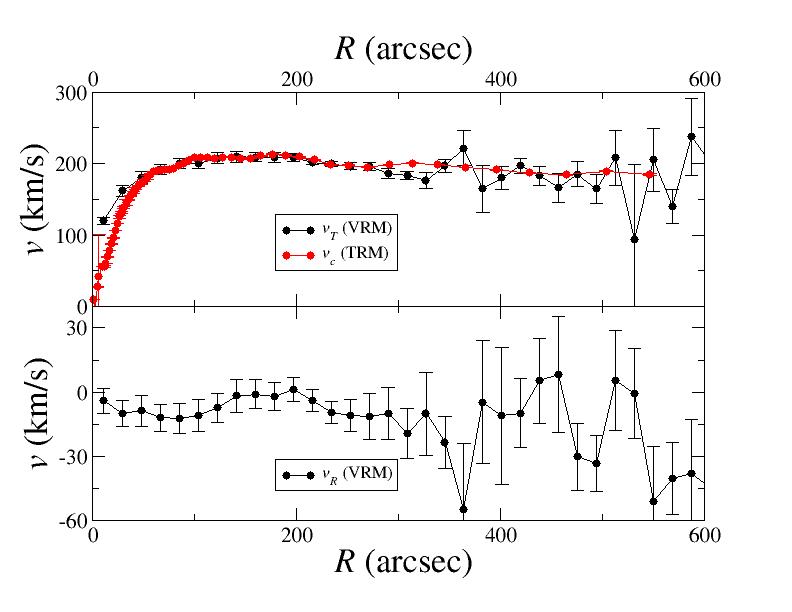}{0.45\textwidth}{(b)}
              }
  \gridline{
 \fig{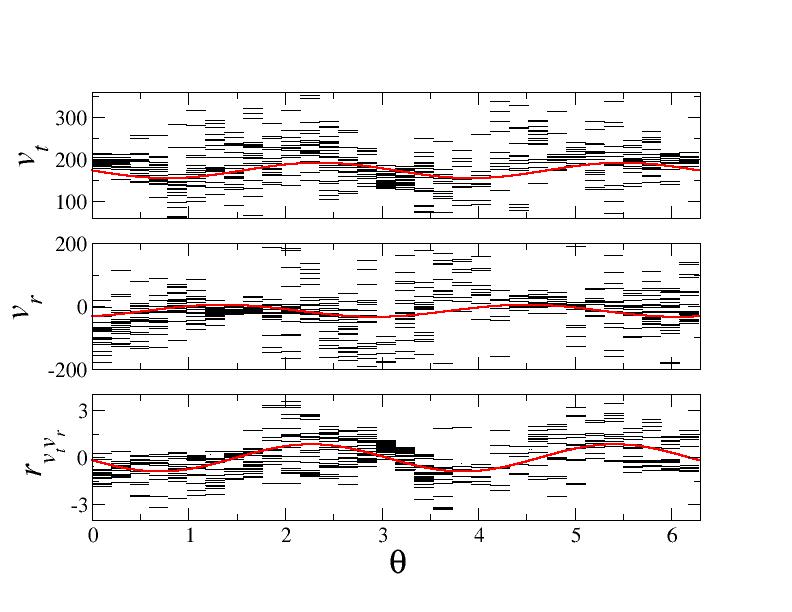}{0.45\textwidth}{(c)}
                \fig{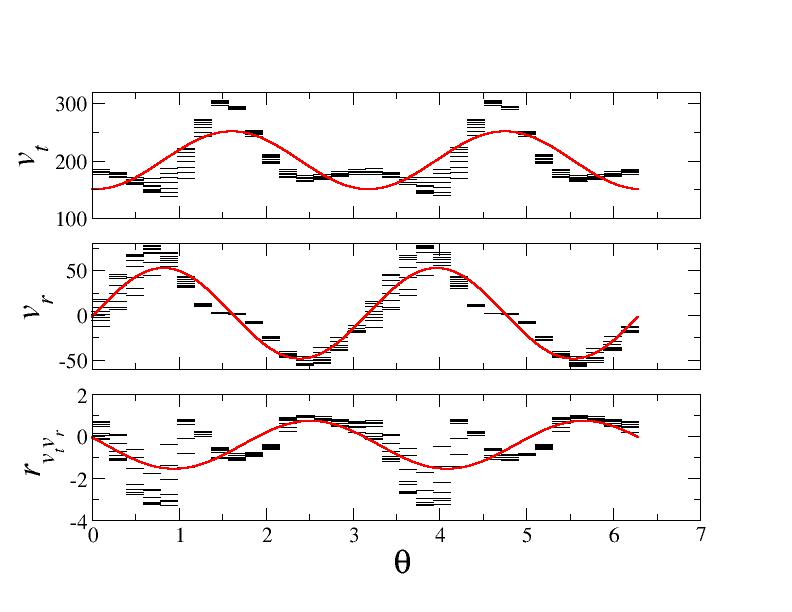}{0.45\textwidth}{(d)}}
     \caption{As  Fig.\ref{fig:NGC2903-1} but for NGC 3521.} 
\label{fig:NGC3521-1} 
\end{figure*}
%

\begin{figure*}
\gridline{\fig{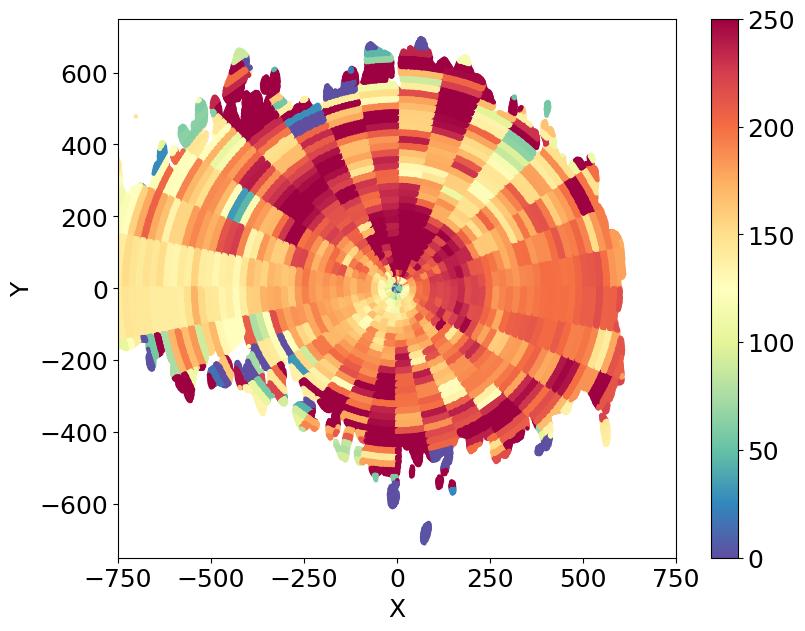}{0.3\textwidth}{(a)}
              \fig{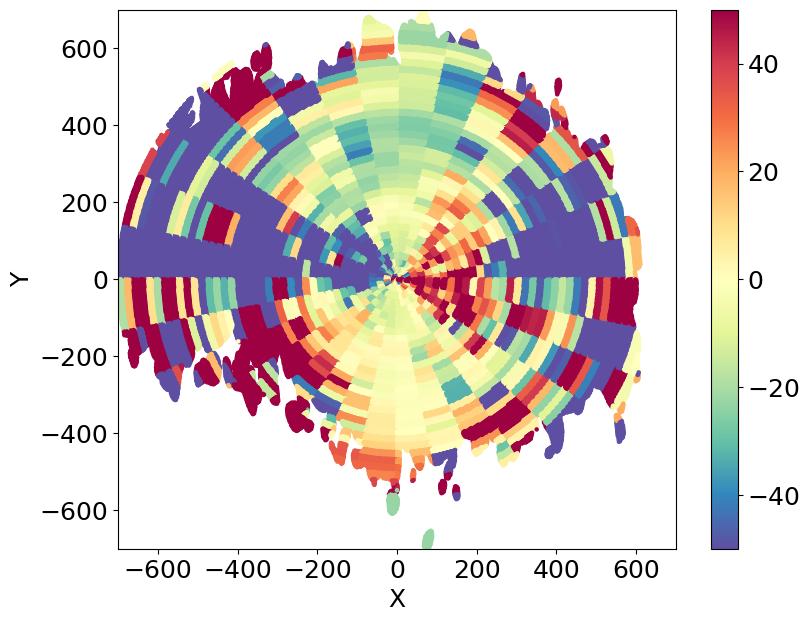}{0.3\textwidth}{(b)}
               \fig{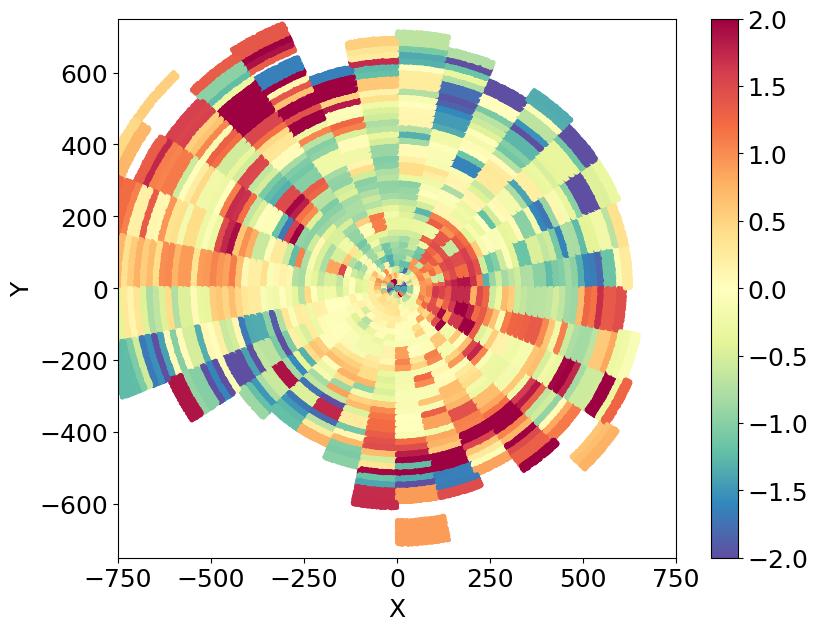}{0.3\textwidth}{(c)}}
\gridline{\fig{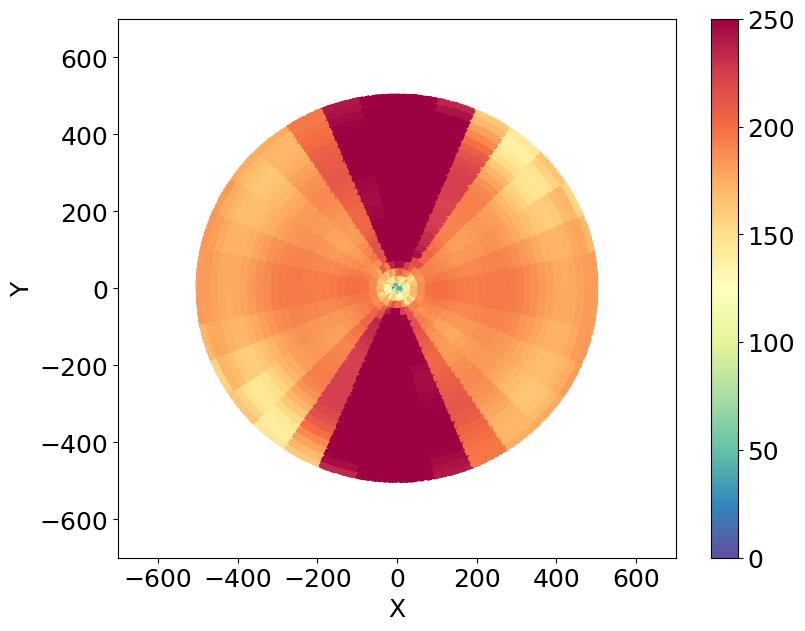}{0.3\textwidth}{(d)}
              \fig{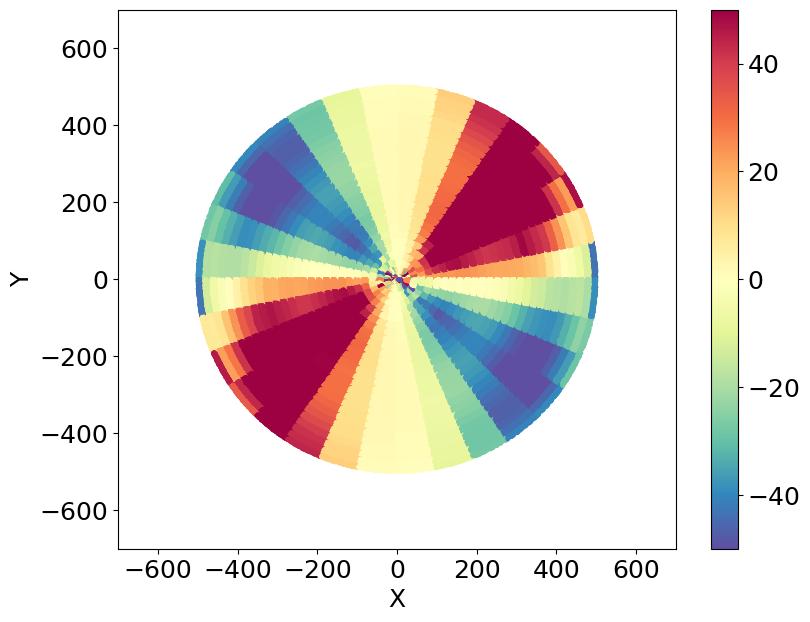}{0.3\textwidth}{(e)}
              \fig{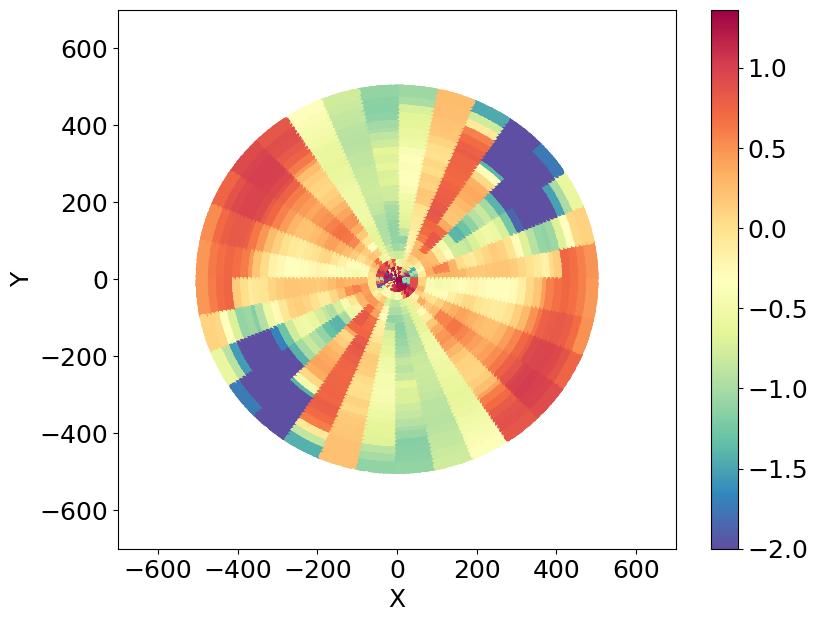}{0.3\textwidth}{(f)}}
\gridline{\fig{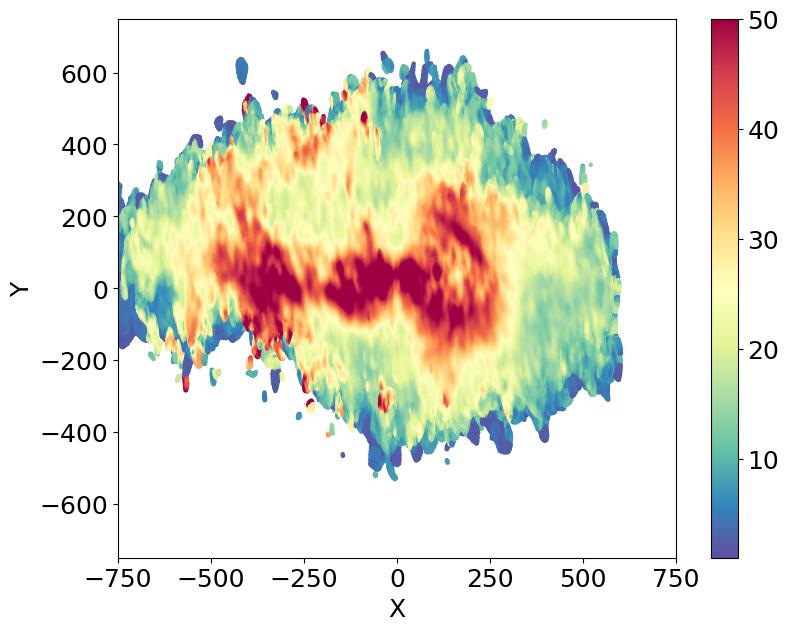}{0.3\textwidth}{(g)}
              \fig{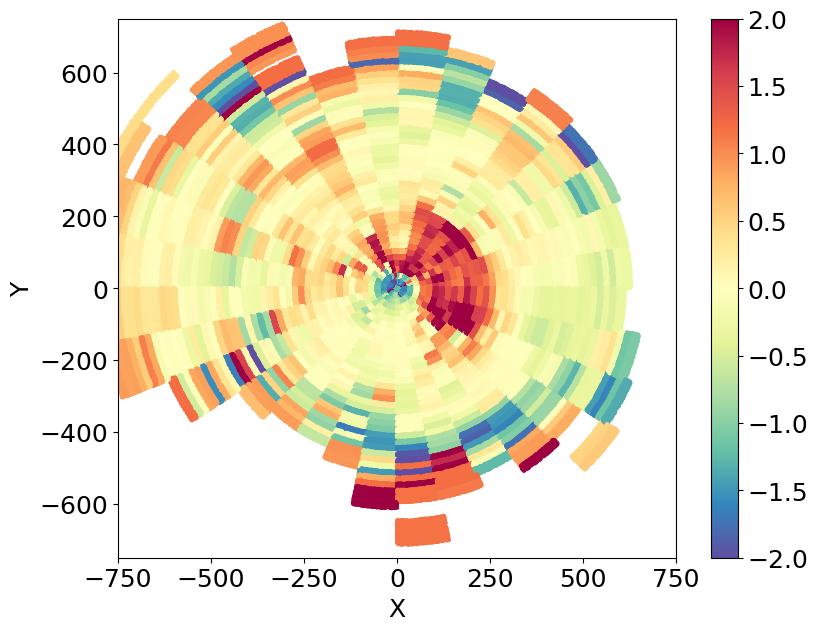}{0.3\textwidth}{(h)}
              \fig{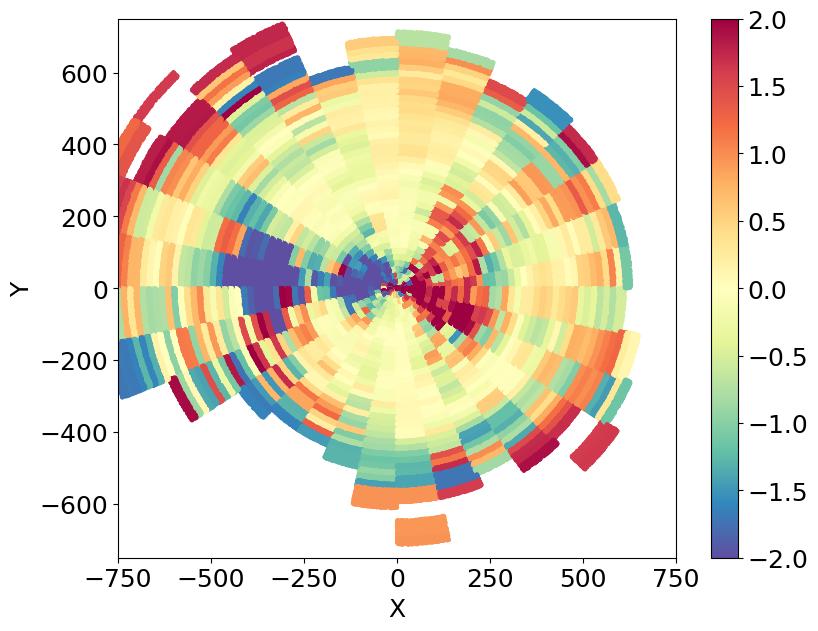}{0.3\textwidth}{(i)}}
\caption{As  Fig.\ref{fig:NGC2903-2} but for NGC 3521.} 
\label{fig:NGC3521-2} 
\end{figure*}
%


\subsection*{NGC 3621} 

The main feature of NGC3621 is a highly anisotropic velocity dispersion field (see Figs. \ref{fig:NGC3621-1}-\ref{fig:NGC3621-2}) that correlates with the radial velocity map. Thus, the velocity field is highly perturbed in an anisotropic way, and a warp, if present, is not clearly detectable because extrinsic fluctuations are much smaller than intrinsic ones. {  Indeed,  the correlation coefficient is ${\cal C}$ = 0.08, suggesting that no warp is present.}. 
\begin{figure*}
\gridline{\fig{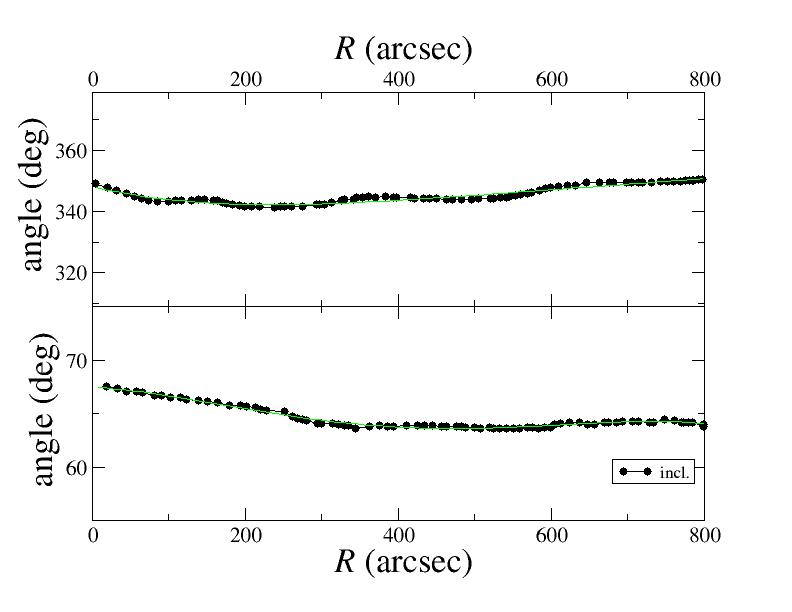}{0.45\textwidth}{(a)}
              \fig{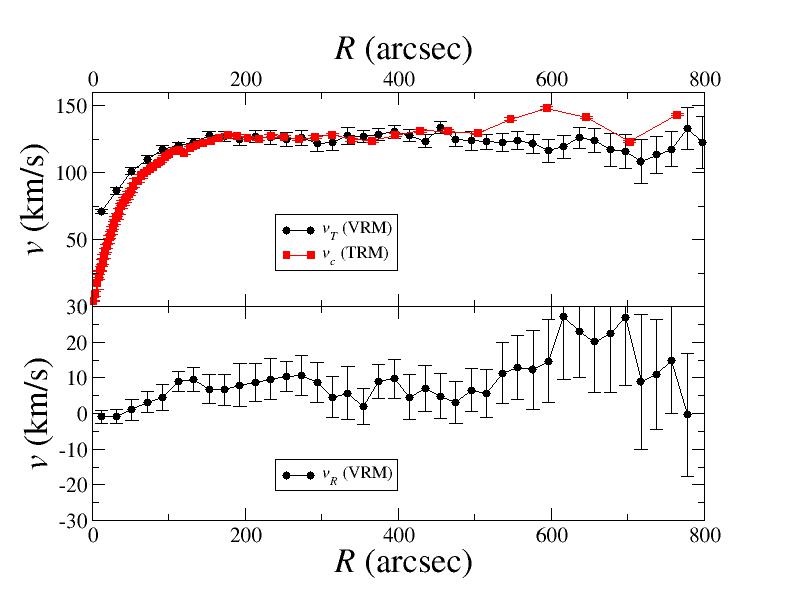}{0.45\textwidth}{(b)}
              }
  \gridline{
 \fig{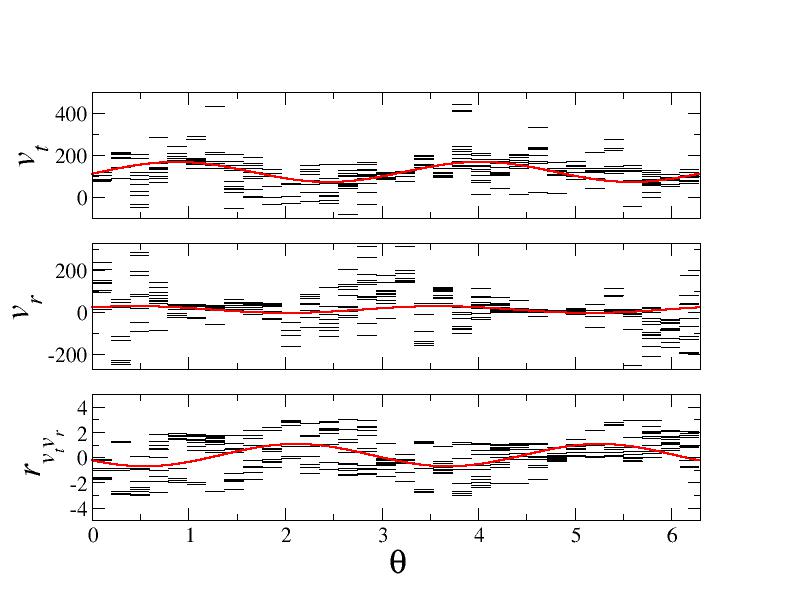}{0.45\textwidth}{(c)}
                \fig{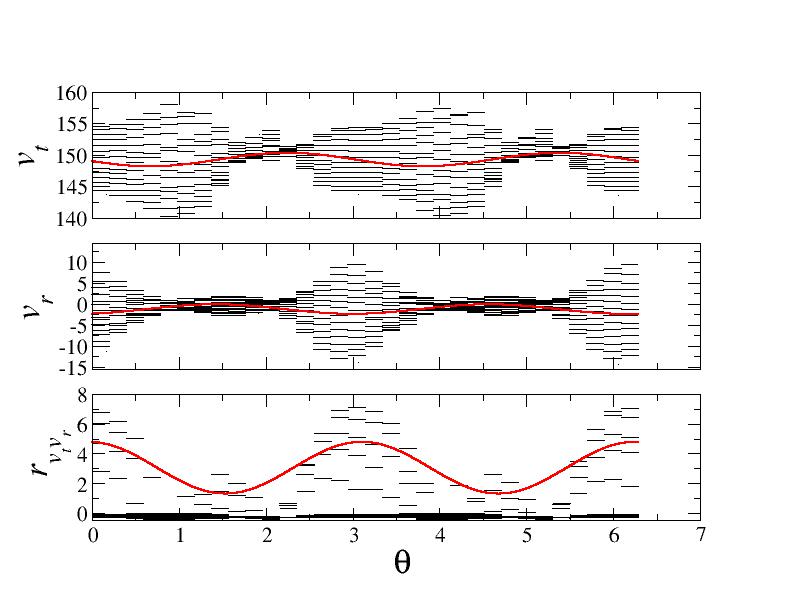}{0.45\textwidth}{(d)}}
     \caption{As  Fig.\ref{fig:NGC2903-1} but for NGC 3621.} 
\label{fig:NGC3621-1} 
\end{figure*}
%
\begin{figure*}
\gridline{\fig{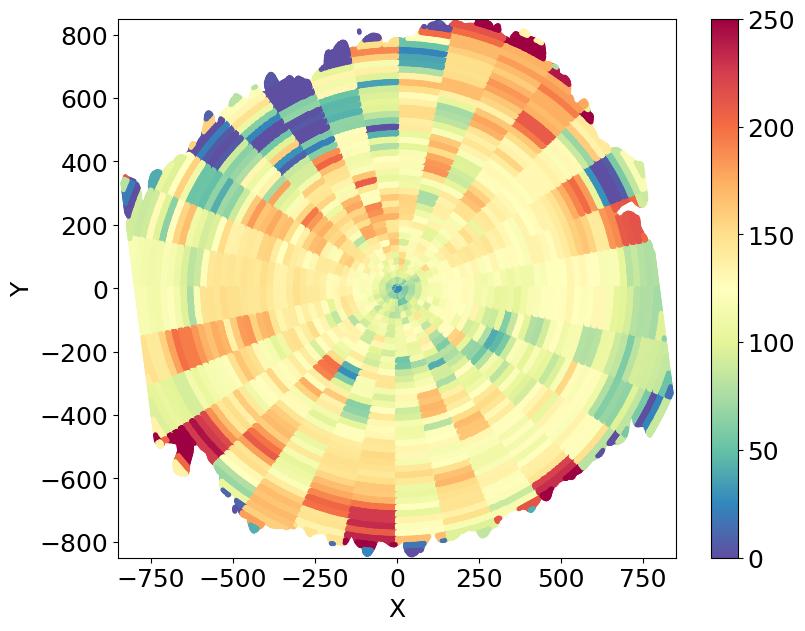}{0.3\textwidth}{(a)}
              \fig{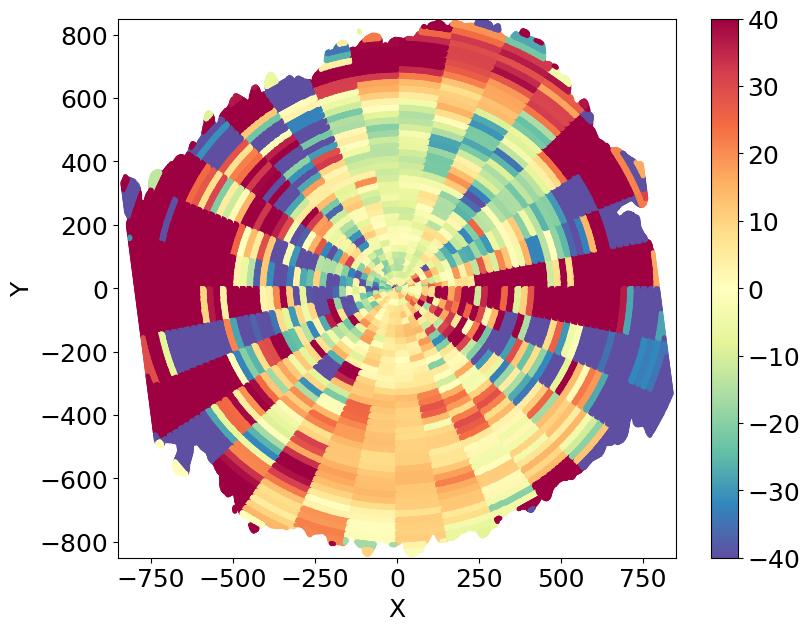}{0.3\textwidth}{(b)}
               \fig{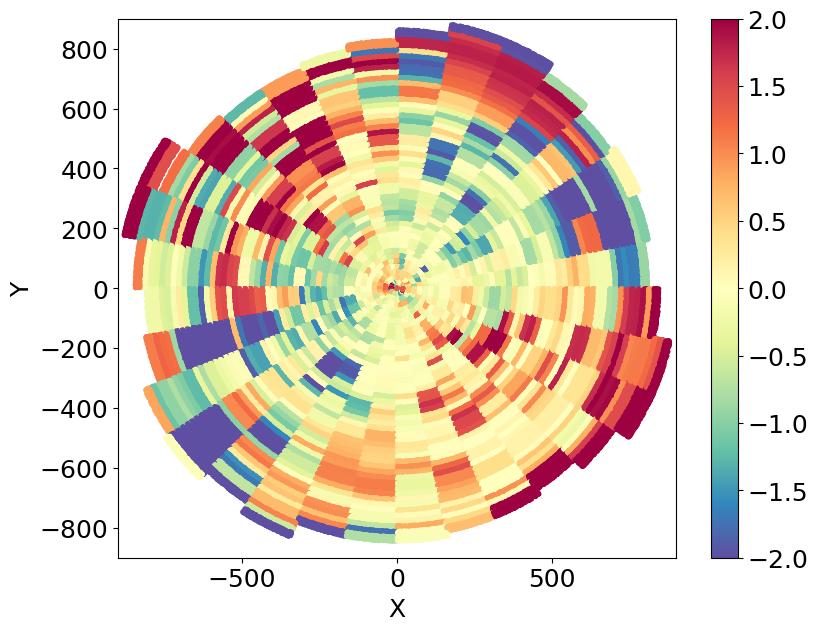}{0.3\textwidth}{(c)}}
\gridline{\fig{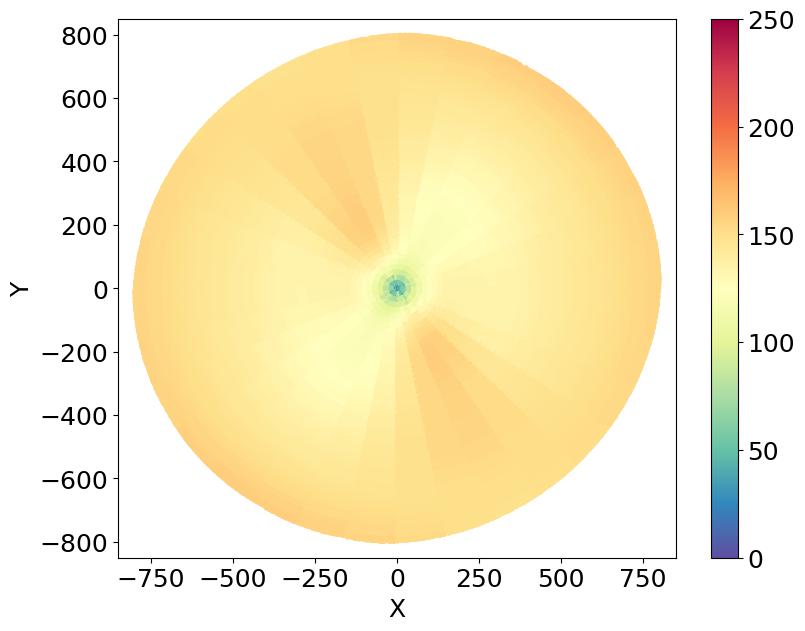}{0.3\textwidth}{(d)}
              \fig{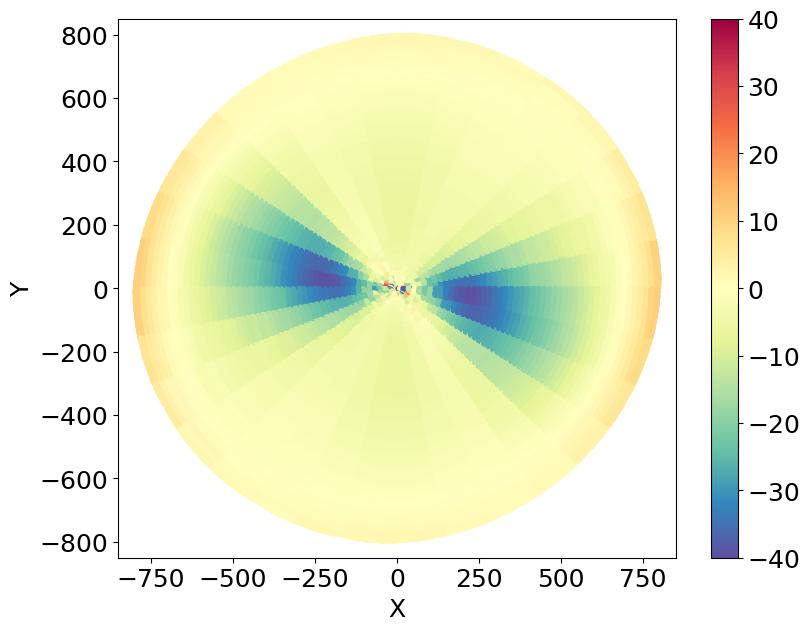}{0.3\textwidth}{(e)}
              \fig{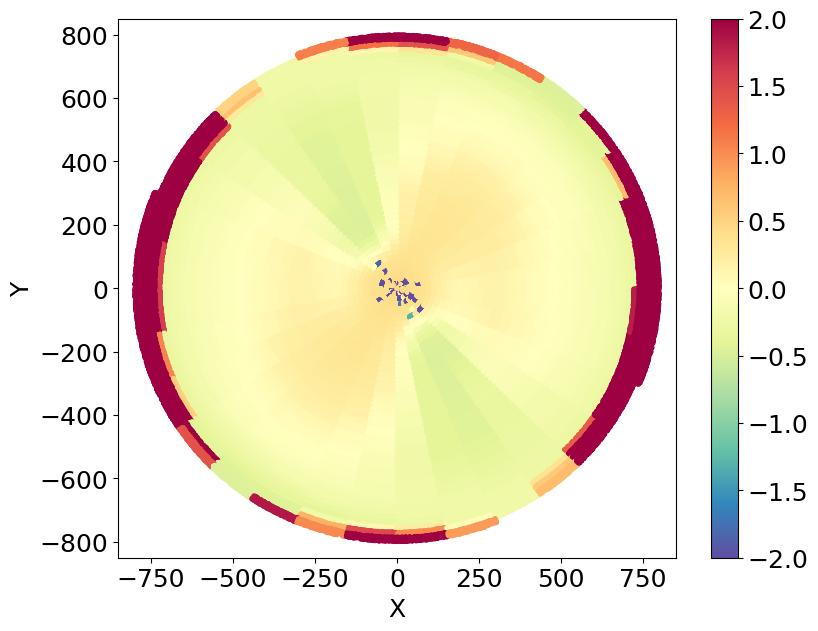}{0.3\textwidth}{(f)}}
\gridline{\fig{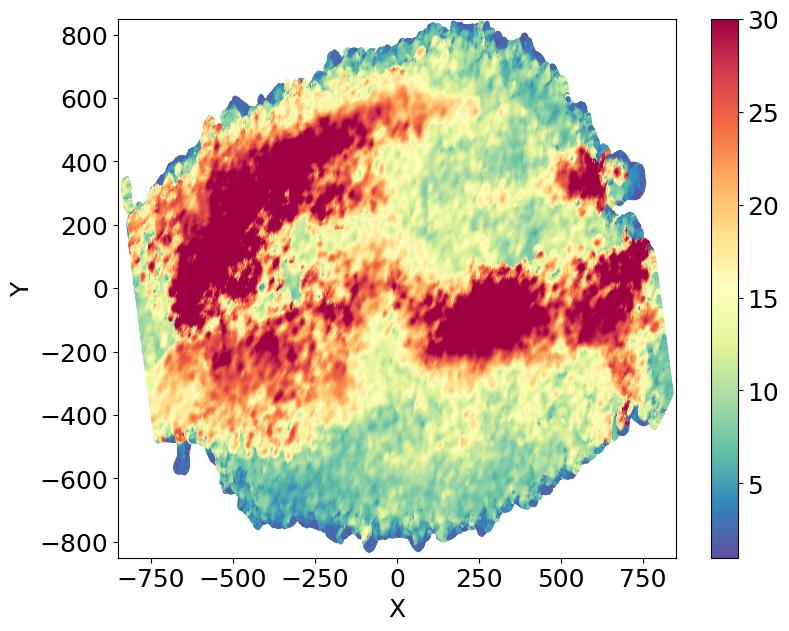}{0.3\textwidth}{(g)}
              \fig{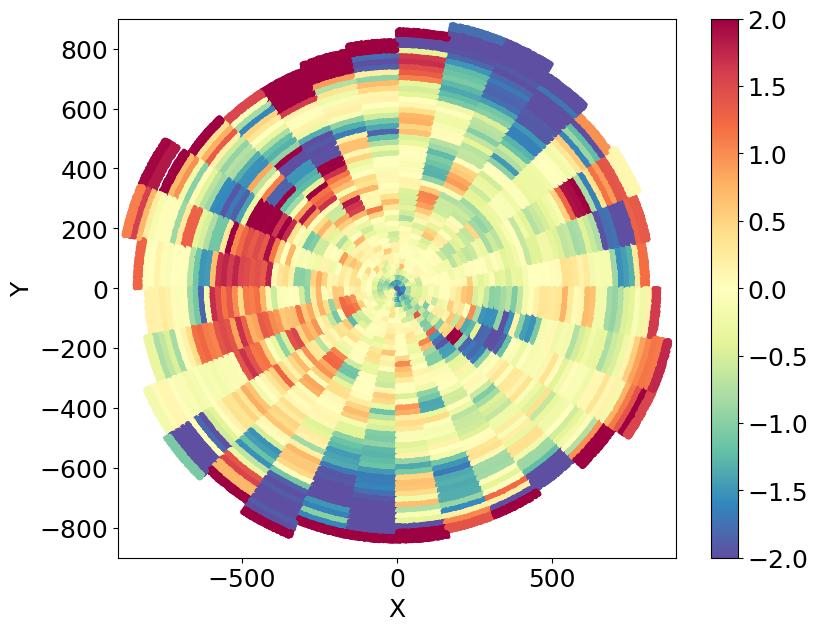}{0.3\textwidth}{(h)}
              \fig{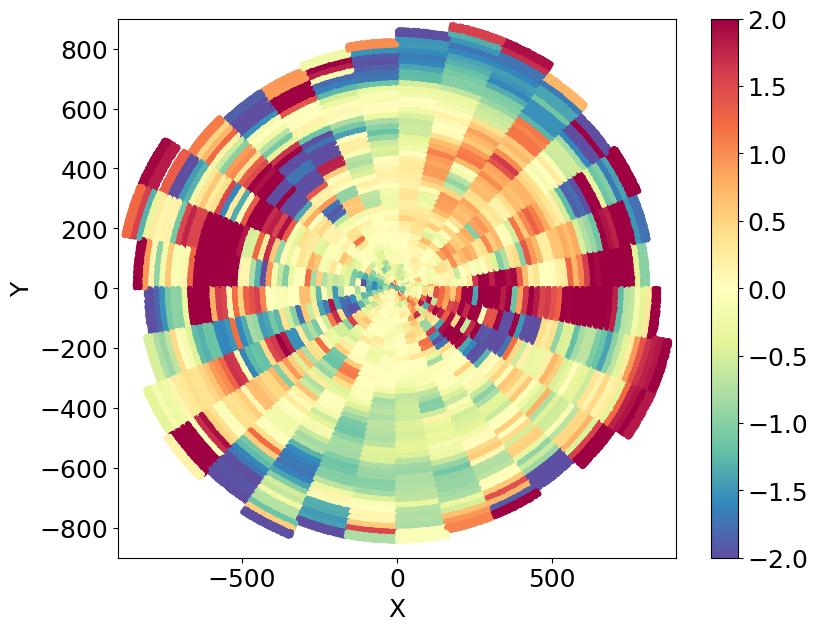}{0.3\textwidth}{(i)}}
\caption{As  Fig.\ref{fig:NGC2903-2} but for NGC 3621.} 
\label{fig:NGC3621-2} 
\end{figure*}
%


\subsection*{NGC 3627} 

Even NGC 3627 shows very weak evidence of being warped. Intrinsic velocity perturbations are larger than extrinsic ones, if present at all. This galaxy is interacting with other galaxies in the Leo Triplet, and this is the source of the anisotropic velocity field. The velocity dispersion field also shows a highly anisotropic pattern that correlates with the transversal and radial velocity maps (Figs.\ref{fig:NGC3627-1}-\ref{fig:NGC3627-2}).
{  The correlation coefficient is  low, i.e., ${\cal C}$ = 0.22, and thus if  warp is present it is a marginal feature}. 

\begin{figure*}
\gridline{\fig{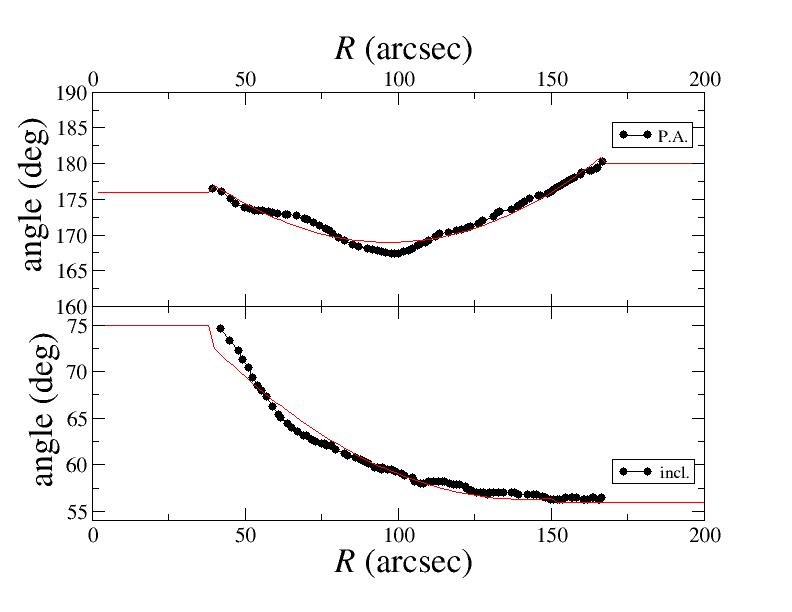}{0.45\textwidth}{(a)}
              \fig{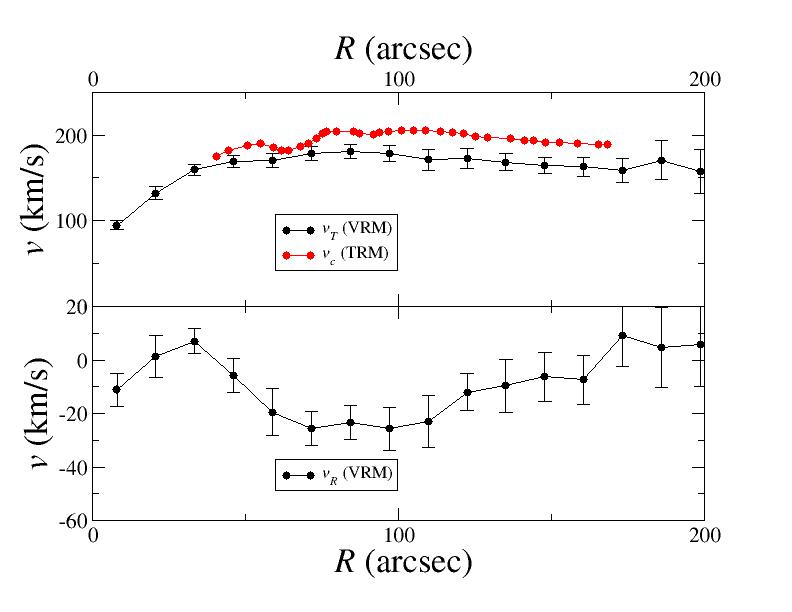}{0.45\textwidth}{(b)}
              }
  \gridline{
 \fig{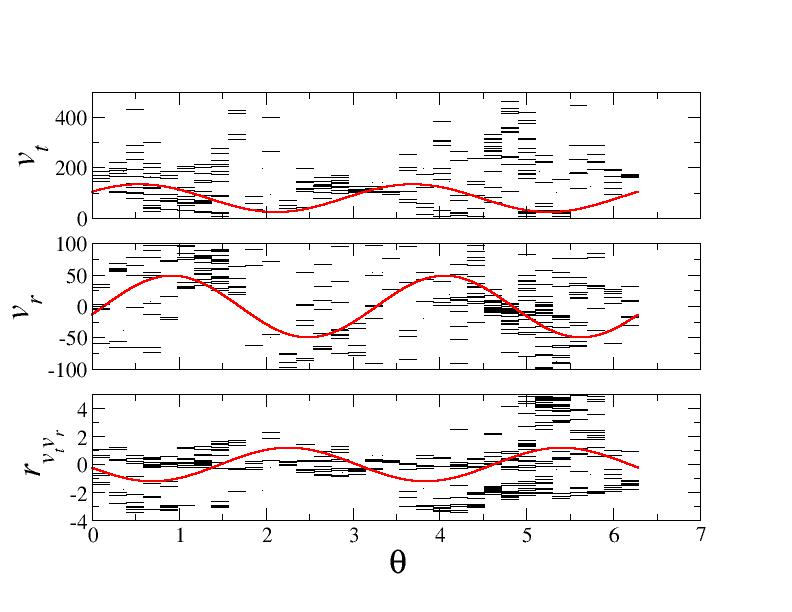}{0.45\textwidth}{(c)}
                \fig{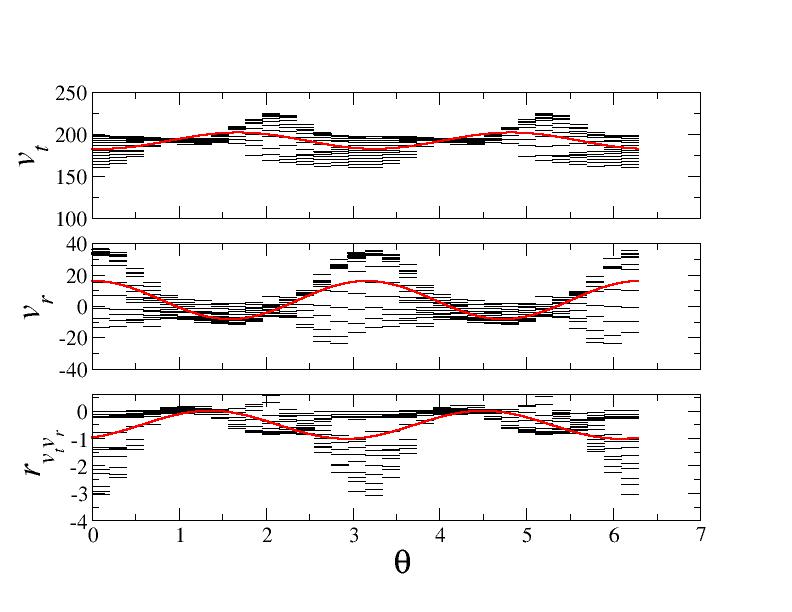}{0.45\textwidth}{(d)}}
     \caption{As  Fig.\ref{fig:NGC2903-1} but for NGC 3627.} 
\label{fig:NGC3627-1} 
\end{figure*}
%
\begin{figure*}
\gridline{\fig{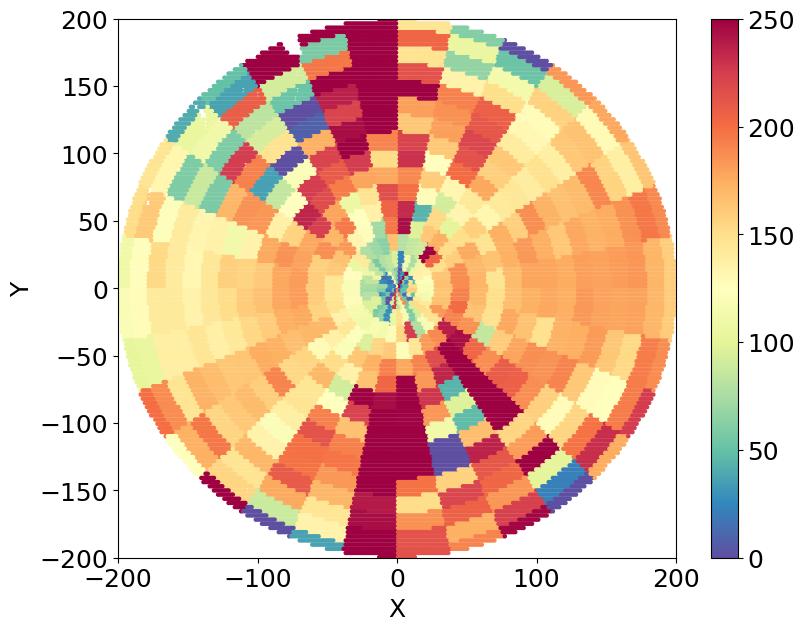}{0.3\textwidth}{(a)}
              \fig{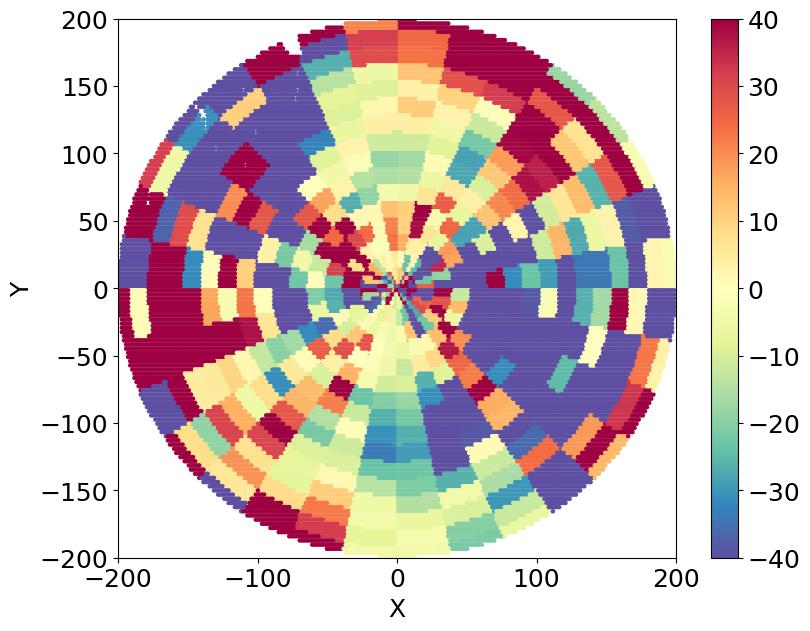}{0.3\textwidth}{(b)}
               \fig{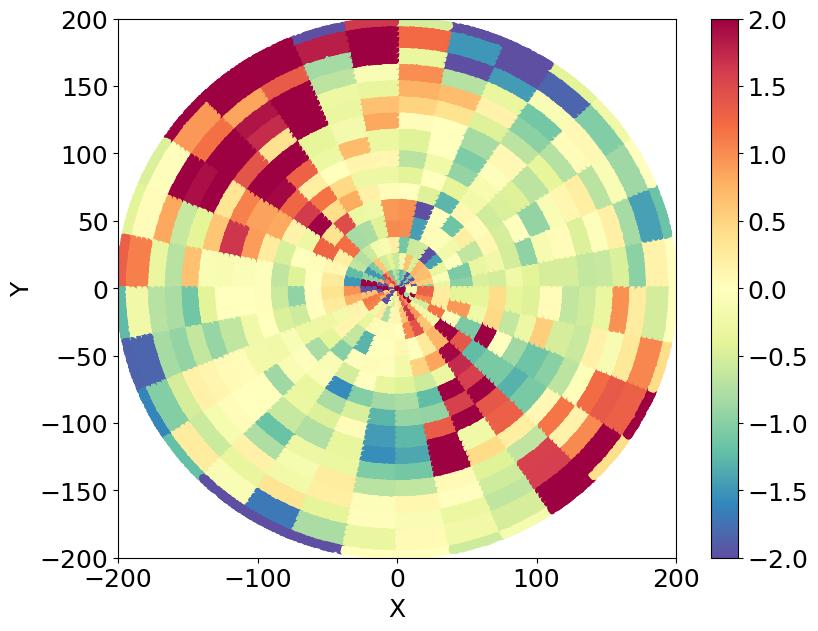}{0.3\textwidth}{(c)}}
\gridline{\fig{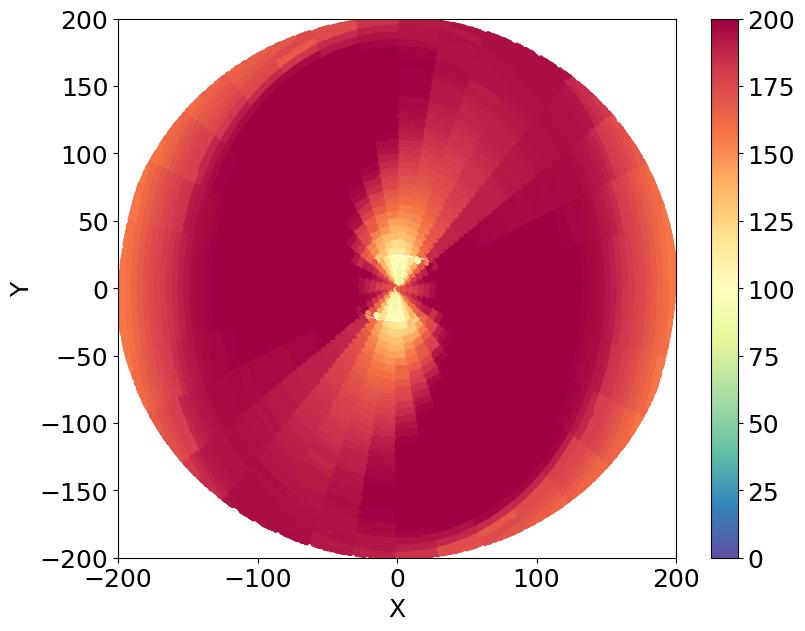}{0.3\textwidth}{(d)}
              \fig{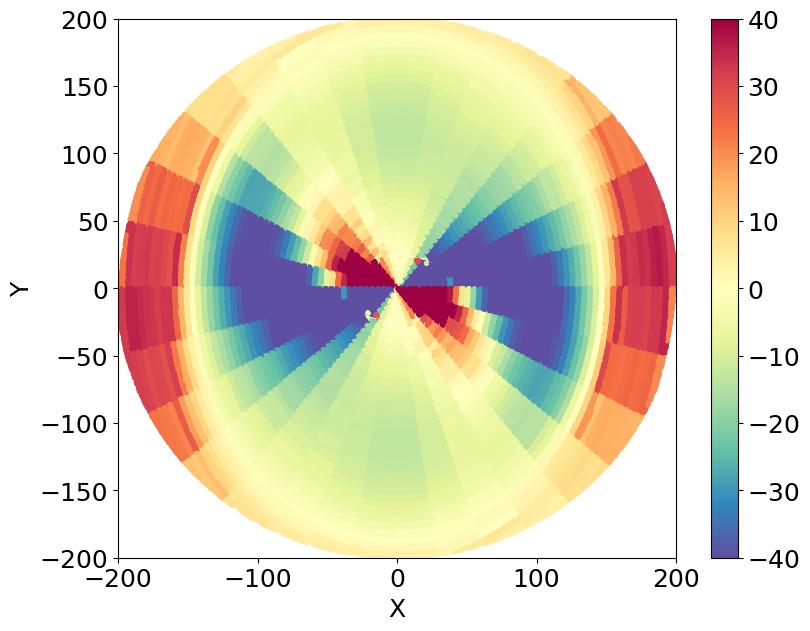}{0.3\textwidth}{(e)}
              \fig{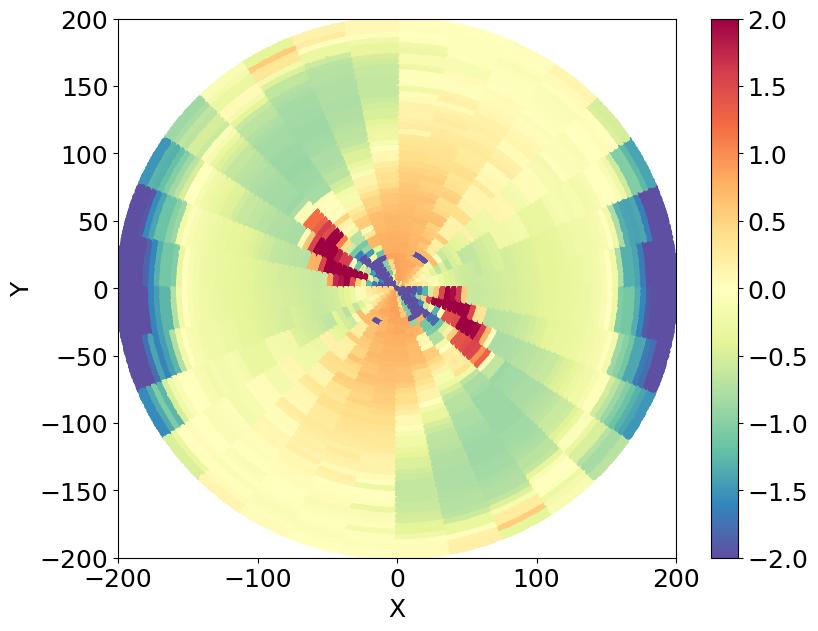}{0.3\textwidth}{(f)}}
\gridline{\fig{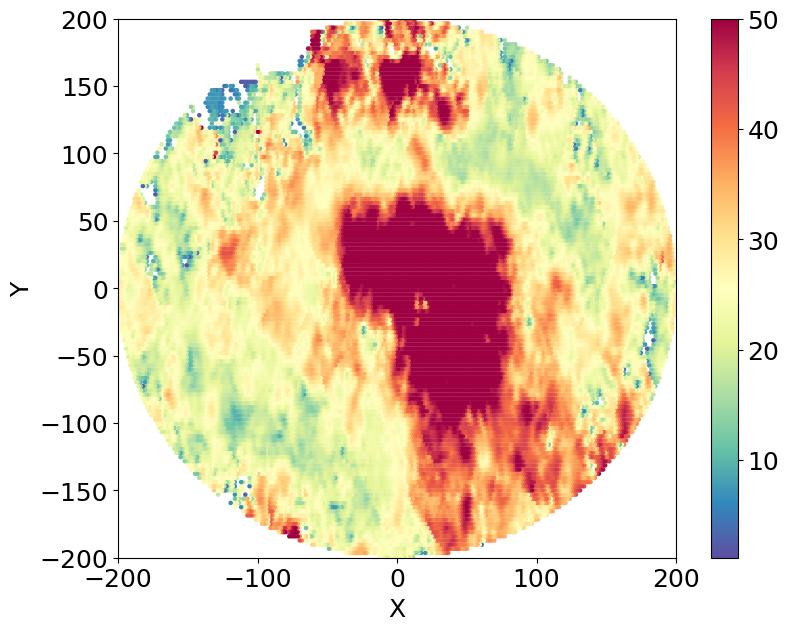}{0.3\textwidth}{(g)}
              \fig{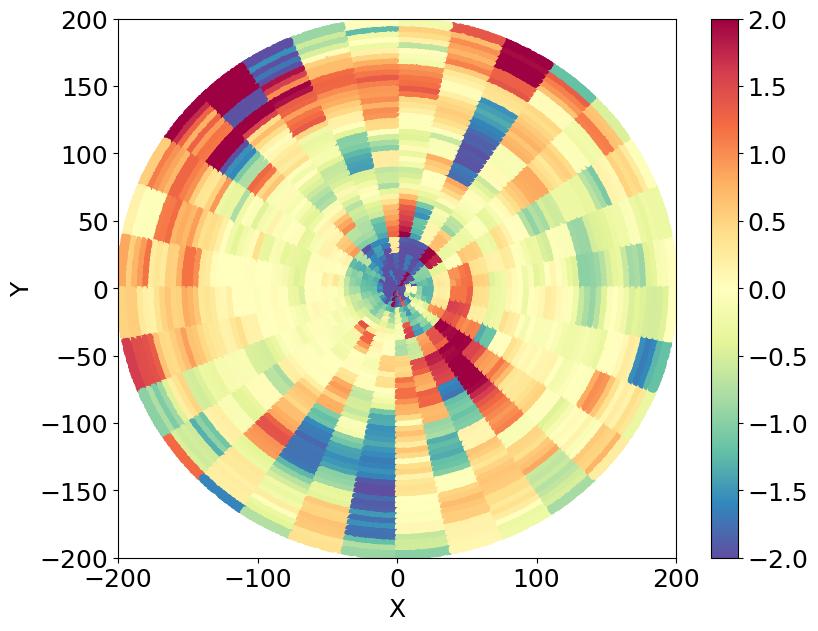}{0.3\textwidth}{(h)}
              \fig{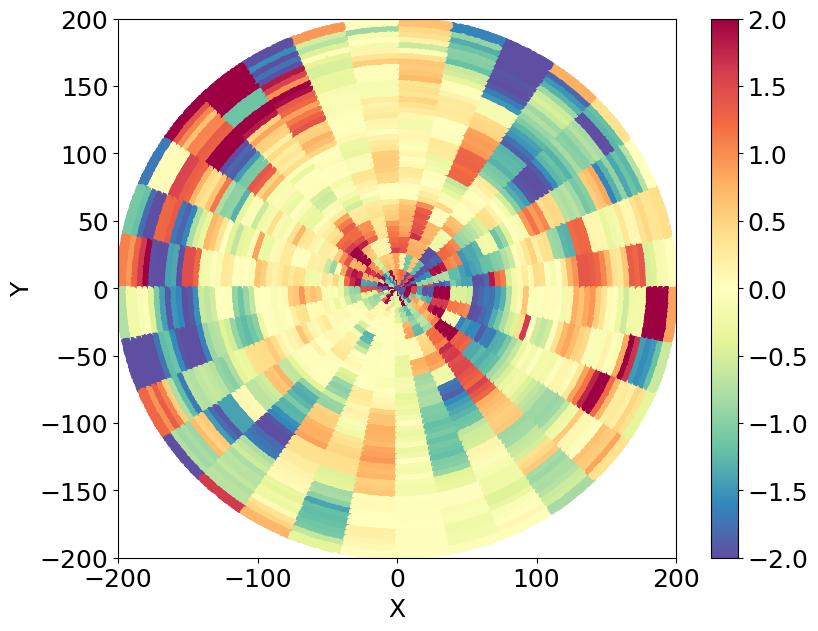}{0.3\textwidth}{(i)}}
\caption{As  Fig.\ref{fig:NGC2903-2} but for NGC 3627.} 
\label{fig:NGC3627-2} 
\end{figure*}
%


\subsection*{NGC 4736} 

In this case both the  inclination angle $i(R)$ and the P.A. show  variations smaller than $10^\circ$ (see panel (a) of Fig. \ref{fig:NGC4736-1}).
The modulation of the velocity rank correlation coefficient is  visibile so that we can conclude that the galaxy is  warped: { the correlation coefficient is ${\cal C}$ = 0.37}. Intrinsic velocity perturbations dominate both velocity component maps. The 
rank   correlation coefficient  $r_{\sigma v_r}(R, \theta)$ shows a positive value in correspondence to the anisotropic pattern in the velocity dispersion map for $\theta \approx 180^\circ$
(see Fig.\ref{fig:NGC4736-2}). 
%
\begin{figure*}
\gridline{\fig{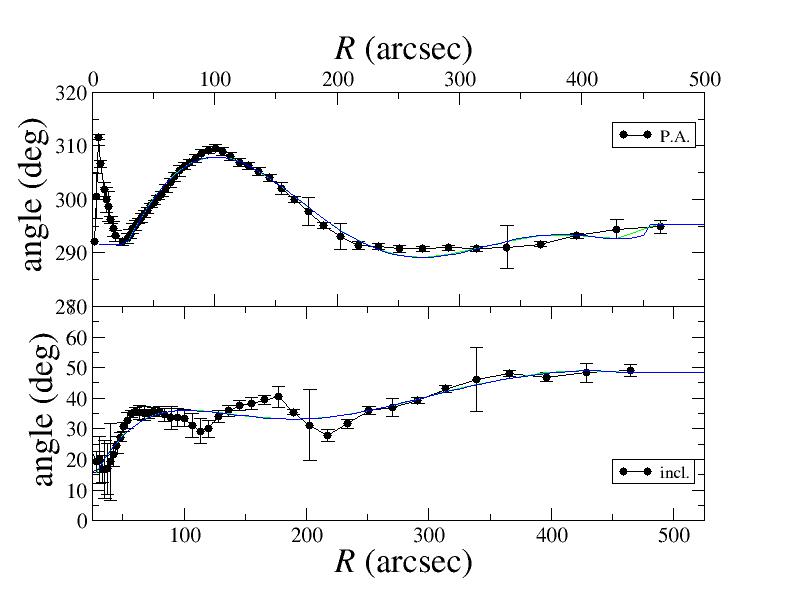}{0.45\textwidth}{(a)}
              \fig{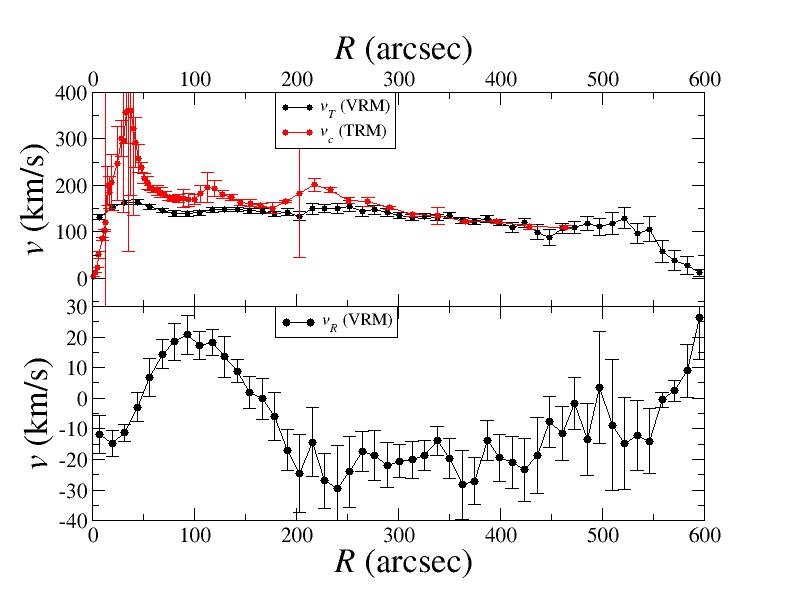}{0.45\textwidth}{(b)}
              }
  \gridline{
 \fig{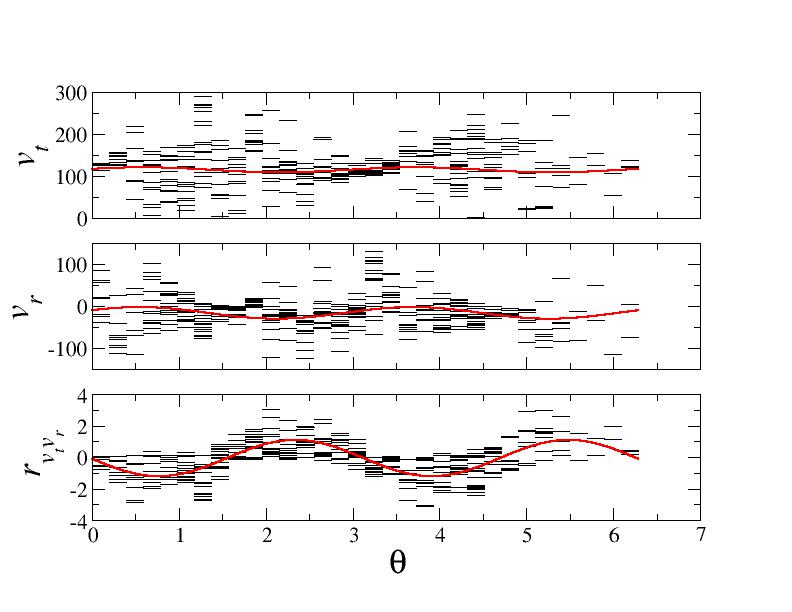}{0.45\textwidth}{(c)}
                \fig{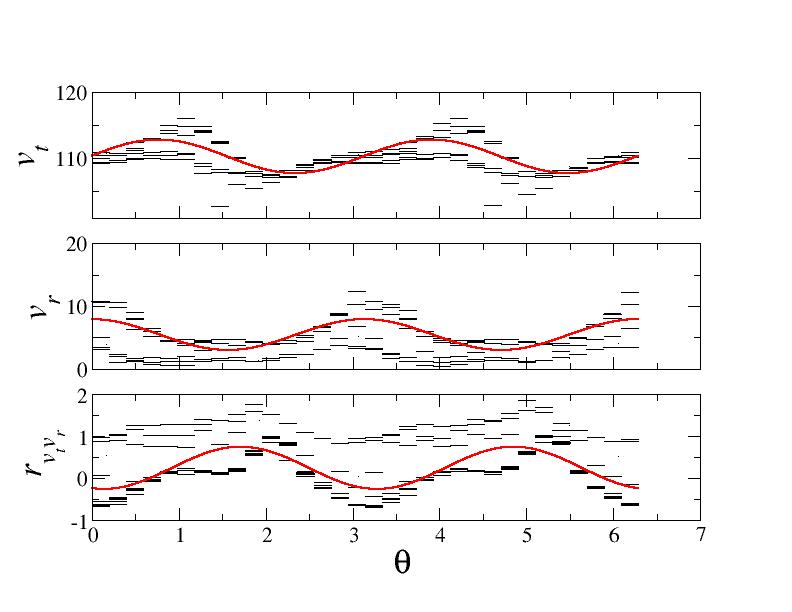}{0.45\textwidth}{(d)}}
     \caption{As  Fig.\ref{fig:NGC2903-1} but for NGC 4736.} 
\label{fig:NGC4736-1} 
\end{figure*}
%
\begin{figure*}
\gridline{\fig{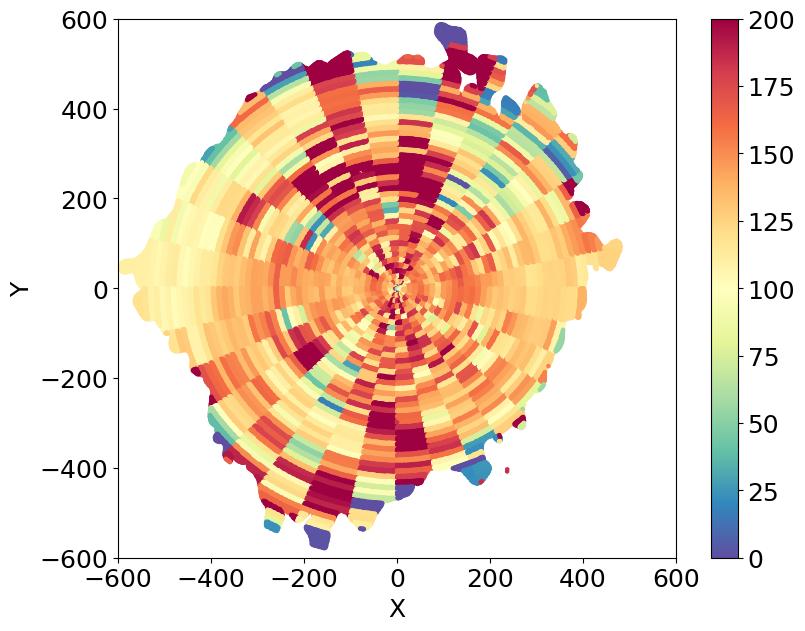}{0.3\textwidth}{(a)}
              \fig{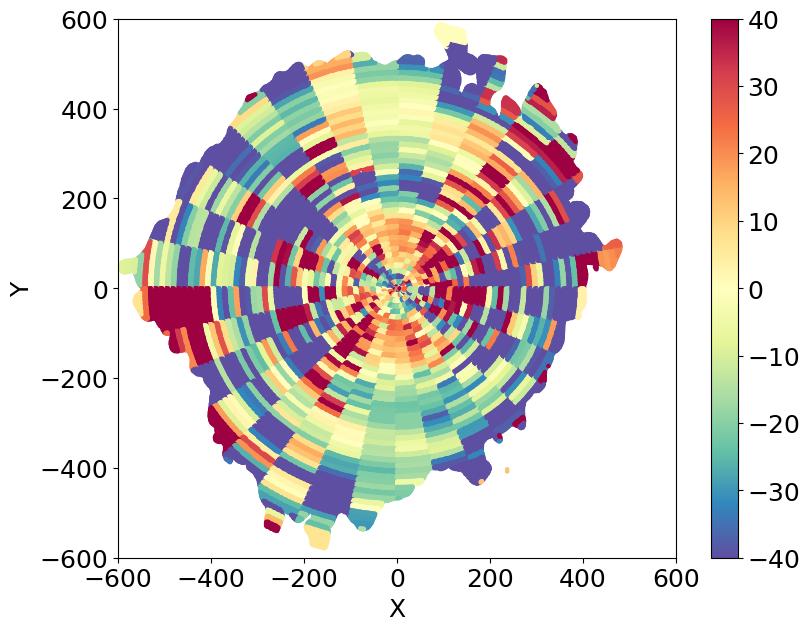}{0.3\textwidth}{(b)}
               \fig{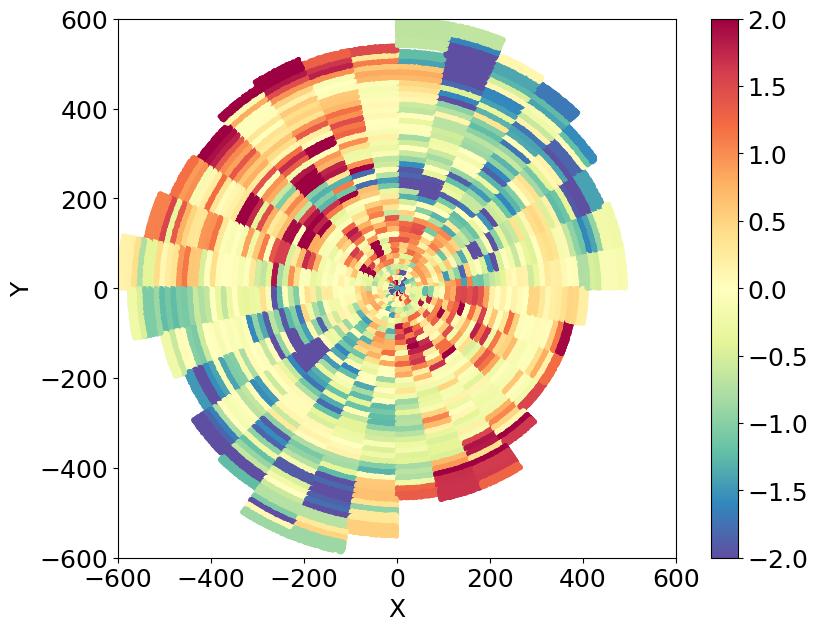}{0.3\textwidth}{(c)}}
\gridline{\fig{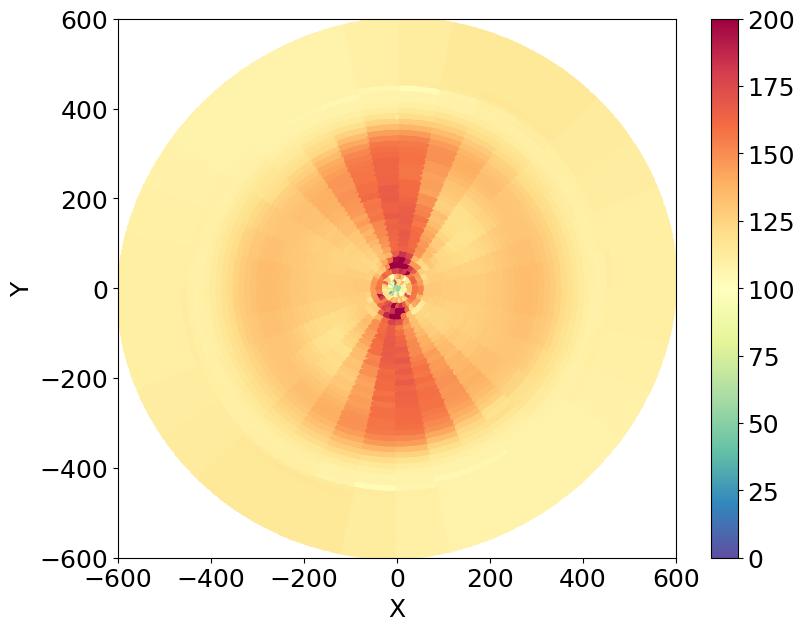}{0.3\textwidth}{(d)}
              \fig{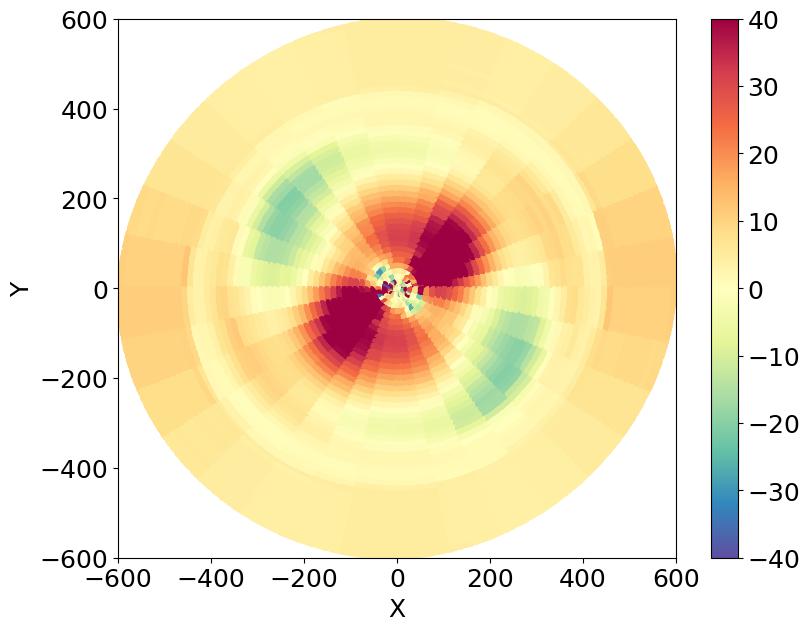}{0.3\textwidth}{(e)}
              \fig{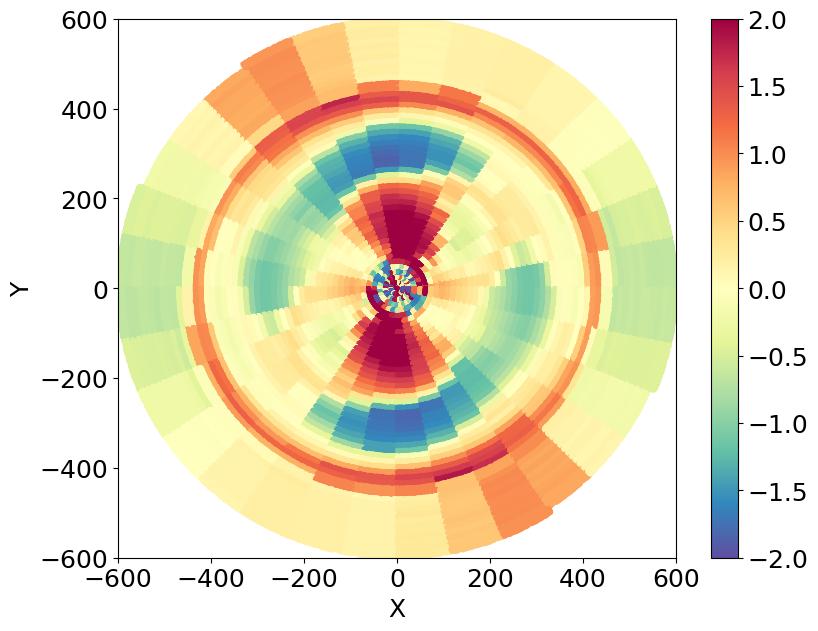}{0.3\textwidth}{(f)}}
\gridline{\fig{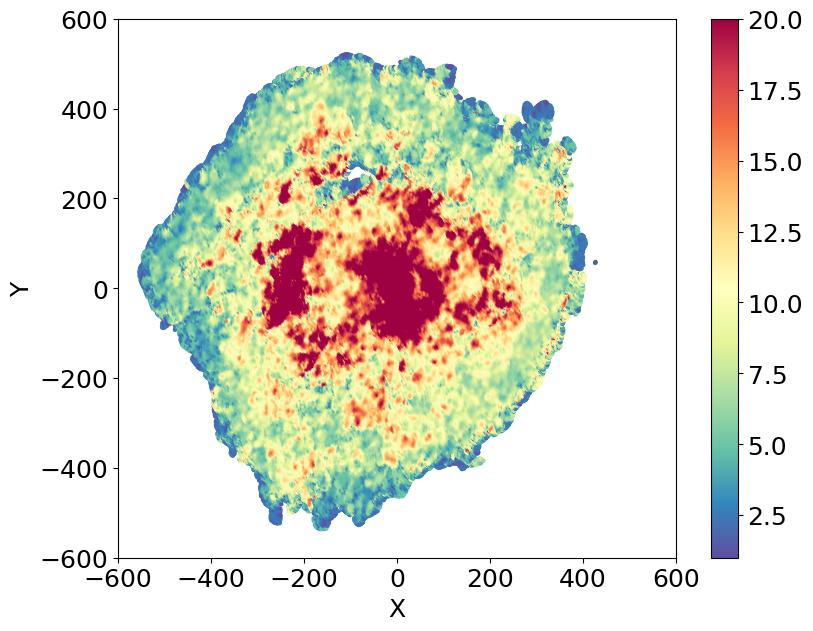}{0.3\textwidth}{(g)}
              \fig{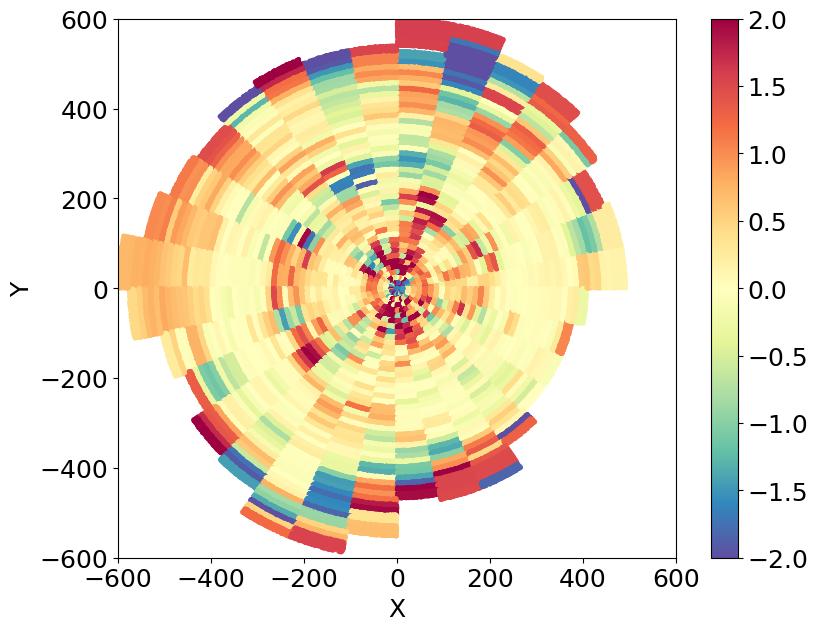}{0.3\textwidth}{(h)}
              \fig{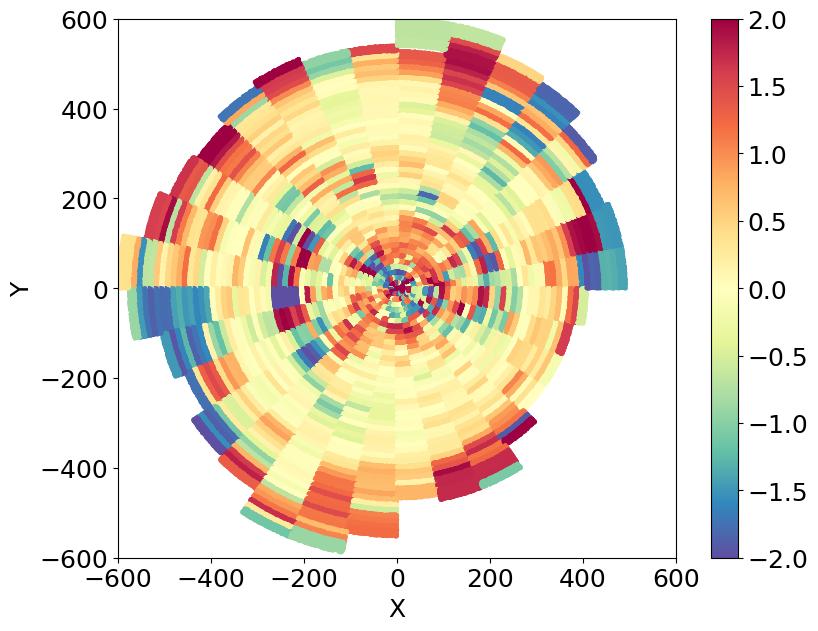}{0.3\textwidth}{(i)}}
\caption{As  Fig.\ref{fig:NGC2903-2} but for NGC 4736.} 
\label{fig:NGC4736-2} 
\end{figure*}
%


\clearpage


\subsection*{NGC 4826} 

In this case both the  inclination angle $i(R)$ and the P.A.  show  variations smaller than $10^\circ$ (Fig.\ref{fig:NGC4826-1}). The correlation coefficient is ${\cal C}$ = 0.08, suggesting that no warp is present. Intrinsic velocity perturbations dominate both velocity component maps. The 
rank   correlation coefficient  $r_{\sigma v_r}(R, \theta)$ shows a positive value in correspondence to the anisotropic pattern in the velocity dispersion map for $\theta \approx 180^\circ$   (Fig. \ref{fig:NGC4826-2}).
\begin{figure*}
\gridline{\fig{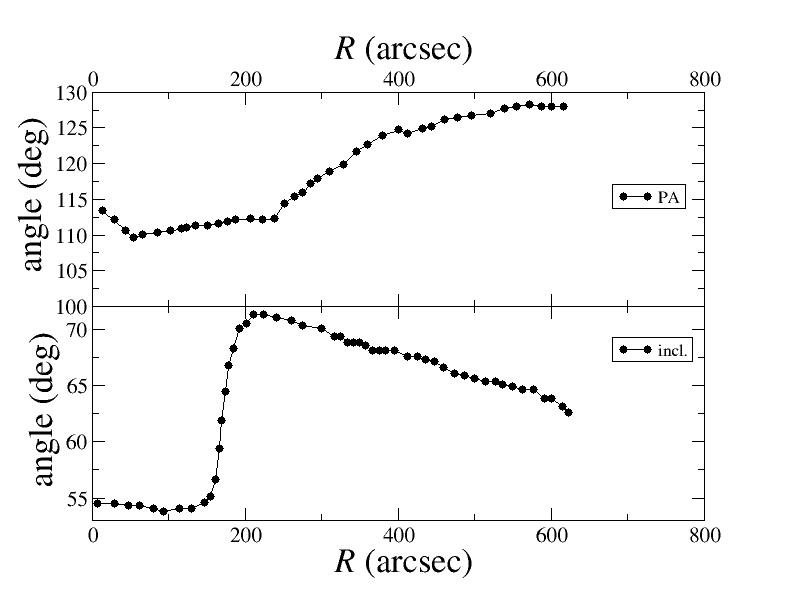}{0.45\textwidth}{(a)}
              \fig{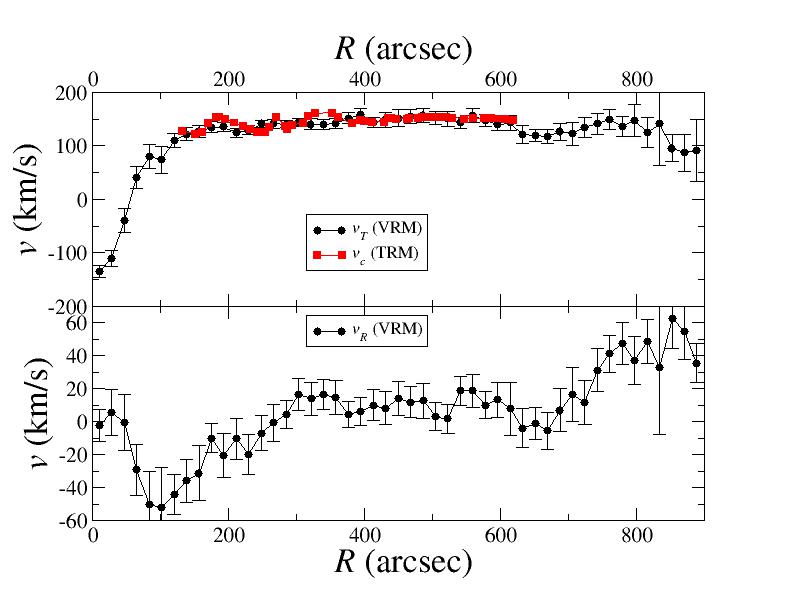}{0.45\textwidth}{(b)}
              }
  \gridline{
 \fig{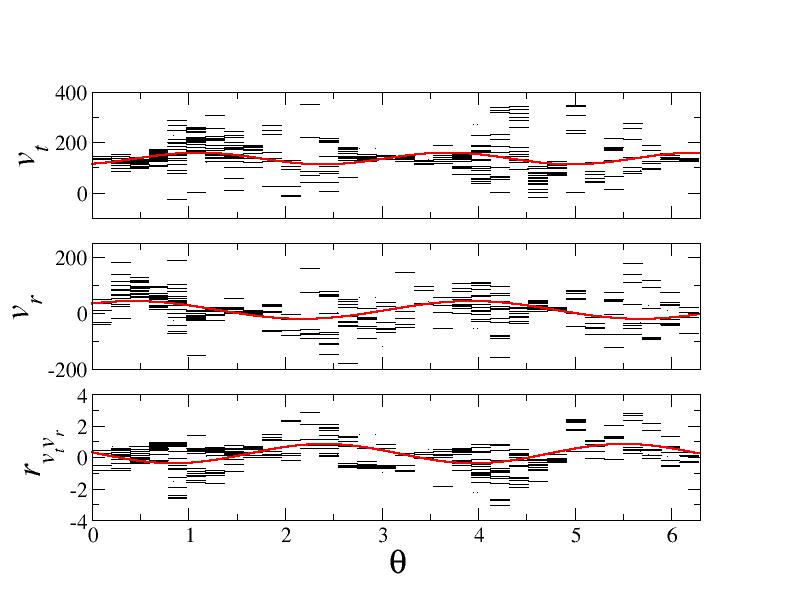}{0.45\textwidth}{(c)}
                \fig{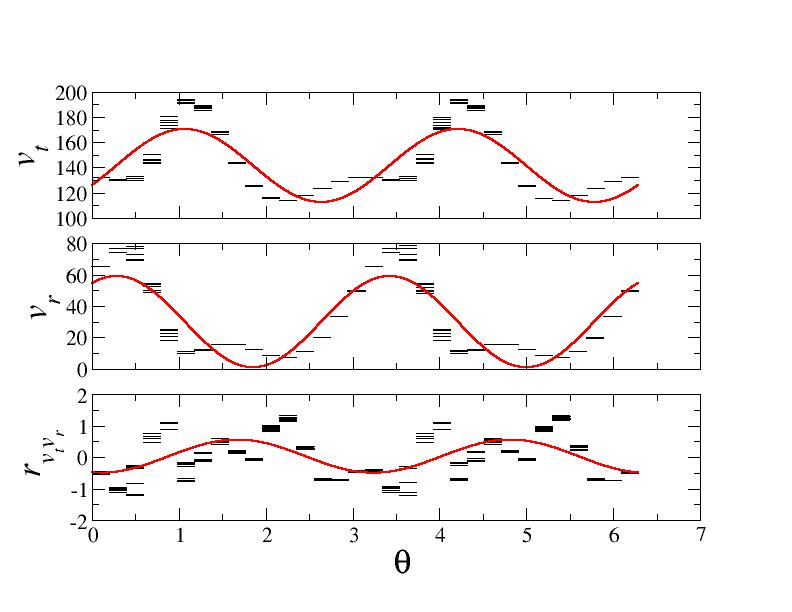}{0.45\textwidth}{(d)}}
     \caption{As  Fig.\ref{fig:NGC2903-1} but for NGC 4826.} 
\label{fig:NGC4826-1} 
\end{figure*}
\begin{figure*}
\gridline{\fig{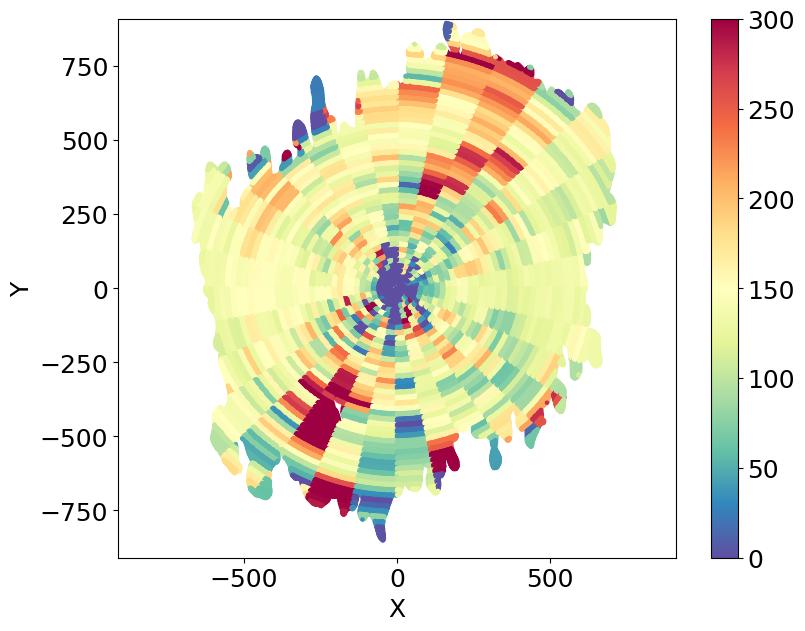}{0.3\textwidth}{(a)}
              \fig{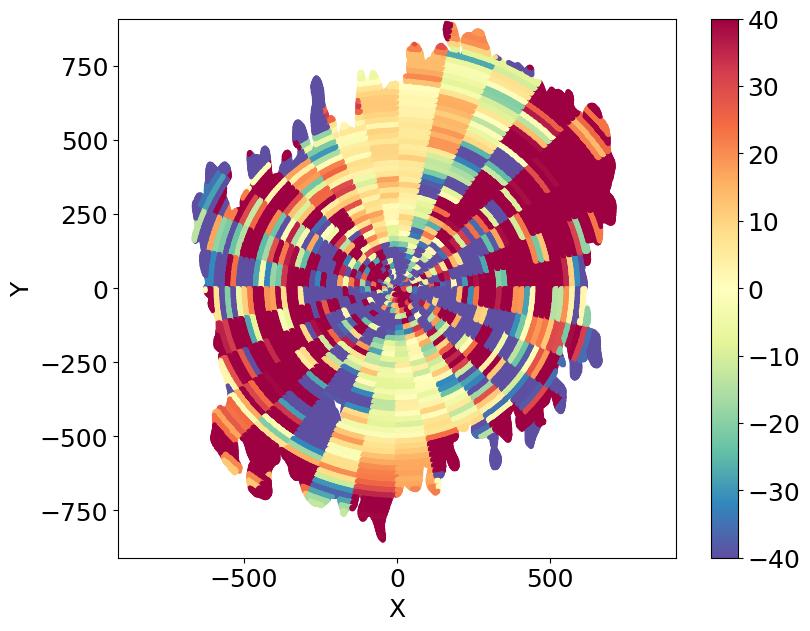}{0.3\textwidth}{(b)}
               \fig{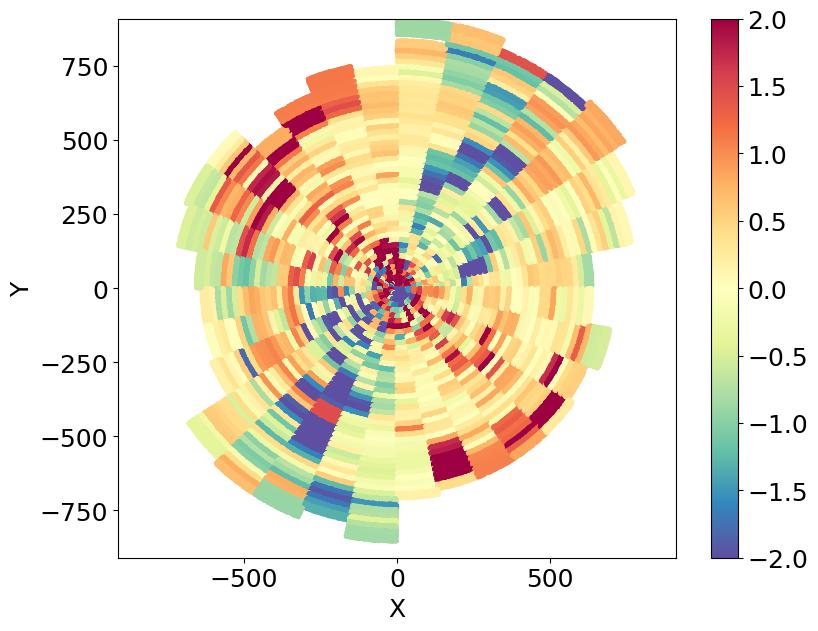}{0.3\textwidth}{(c)}}
\gridline{\fig{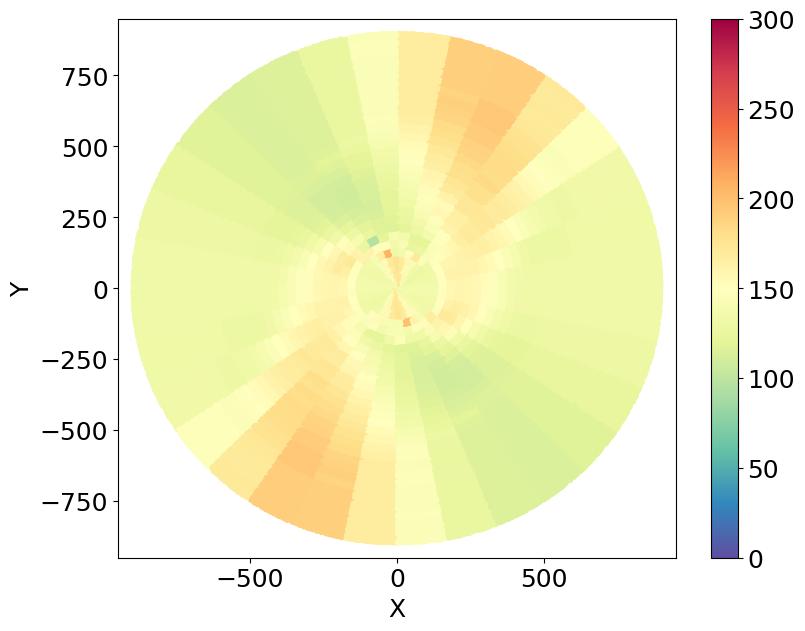}{0.3\textwidth}{(d)}
              \fig{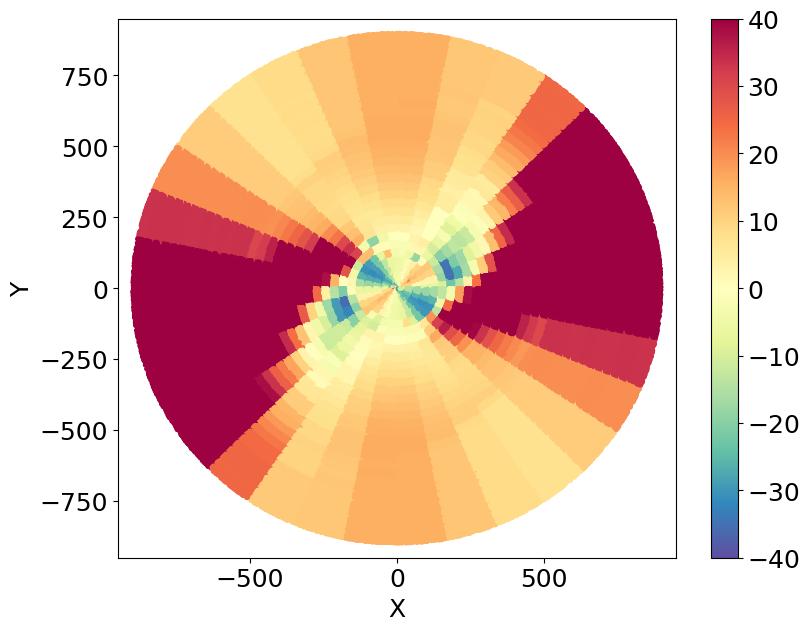}{0.3\textwidth}{(e)}
              \fig{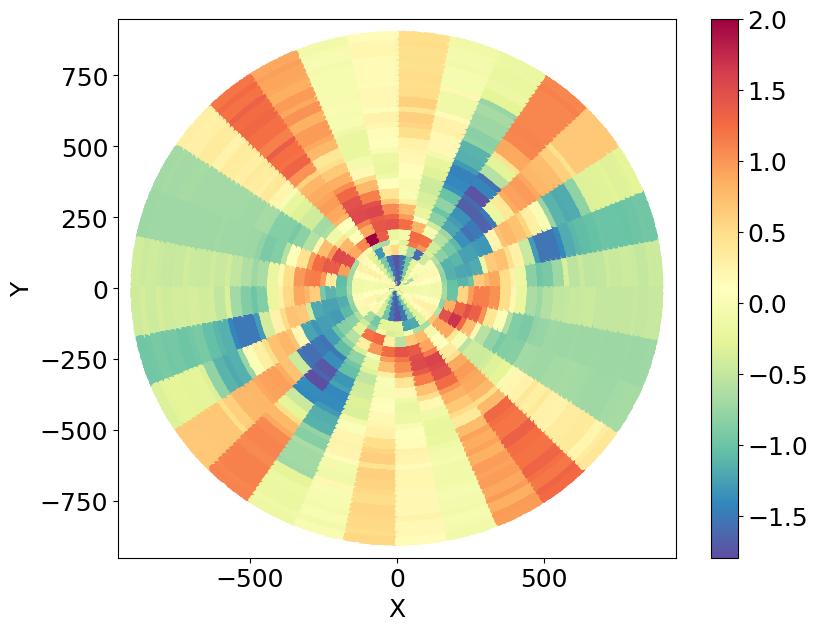}{0.3\textwidth}{(f)}}
\gridline{\fig{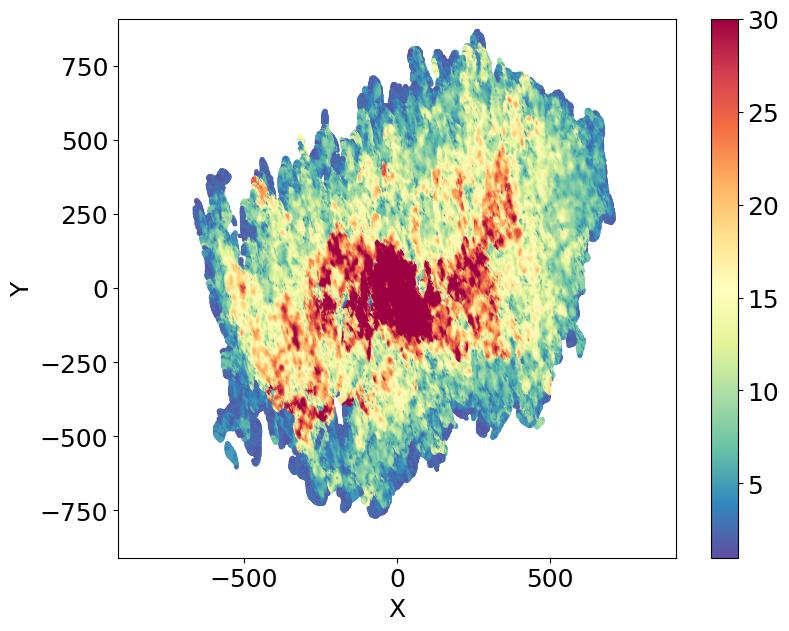}{0.3\textwidth}{(g)}
              \fig{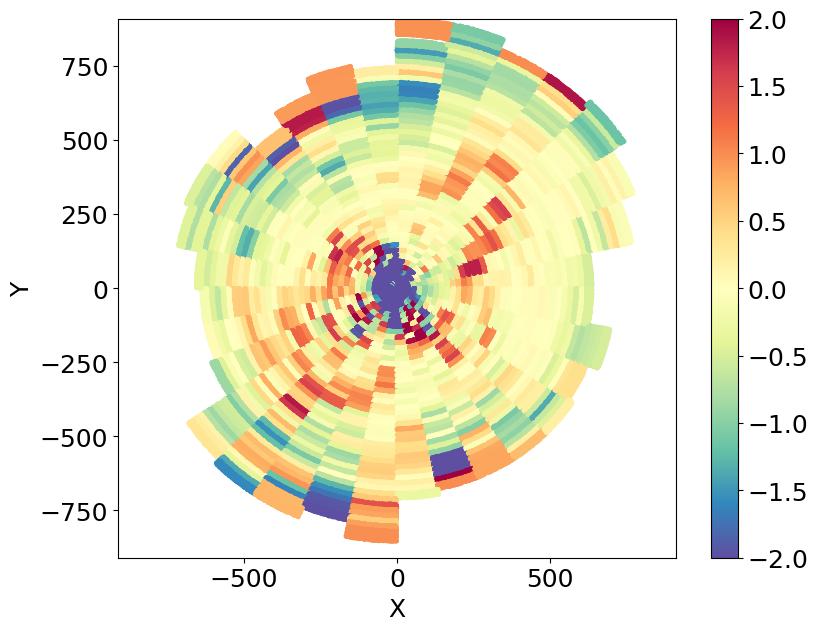}{0.3\textwidth}{(h)}
              \fig{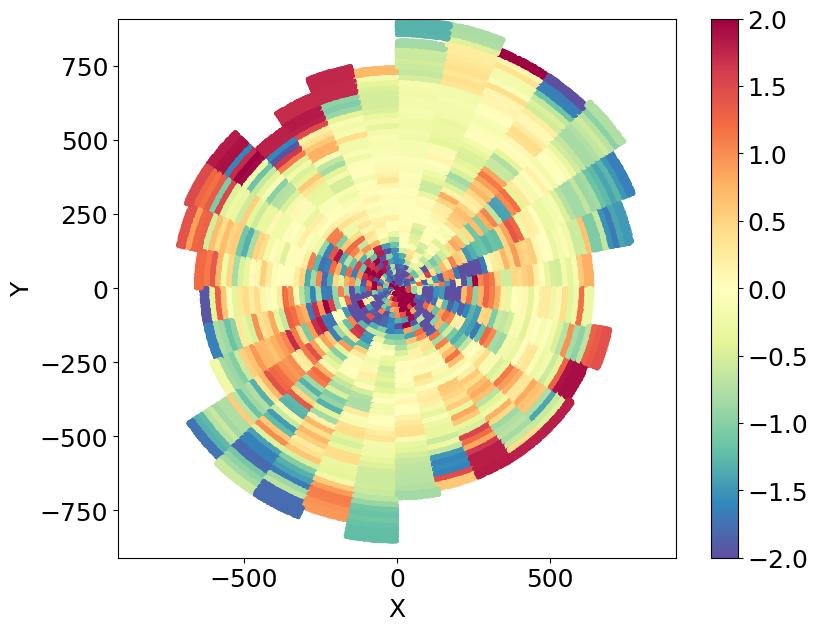}{0.3\textwidth}{(i)}}
\caption{As  Fig.\ref{fig:NGC2903-2} but for NGC 4826.} 
\label{fig:NGC4826-2} 
\end{figure*}
%

\subsection*{NGC 5055} 
The variation of the orientation angles detected by the TRM for the case of NGC 5055 (see Figure \ref{fig:NGC5055-1}) most probably corresponds to a warp. This is corroborated by the dipolar modulation of the velocity rank correlation coefficient and by the transversal velocity component. For the radial velocity component, the signal due to extrinsic perturbations is mixed with that due to intrinsic perturbations, and hence, the dipolar modulation is not well visible. 
{ The correlation coefficient is ${\cal C}$ = 0.36, indicating the presence of a  warp} .
The inner region seems to be characterized by the presence of a bar-like structure; this is present in all three rank correlation maps (see Figure \ref{fig:NGC5055-2}).

\begin{figure*}
\gridline{\fig{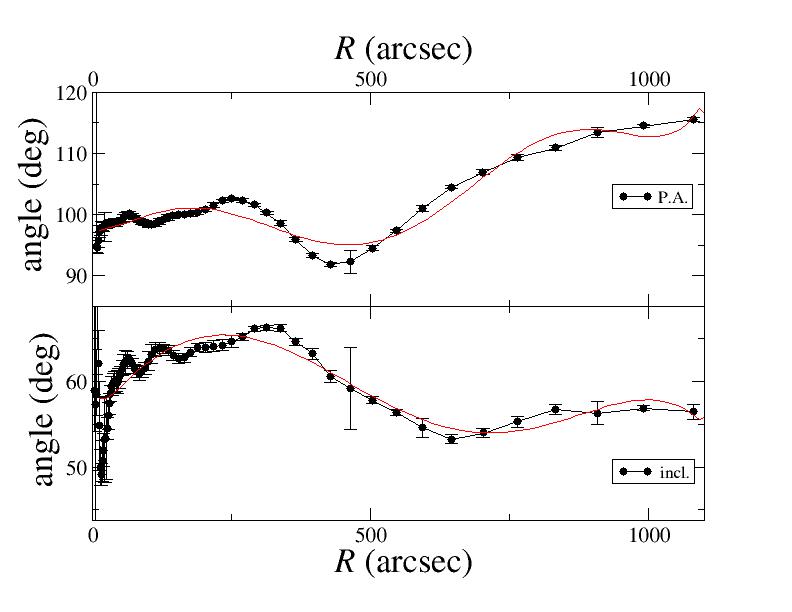}{0.45\textwidth}{(a)}
              \fig{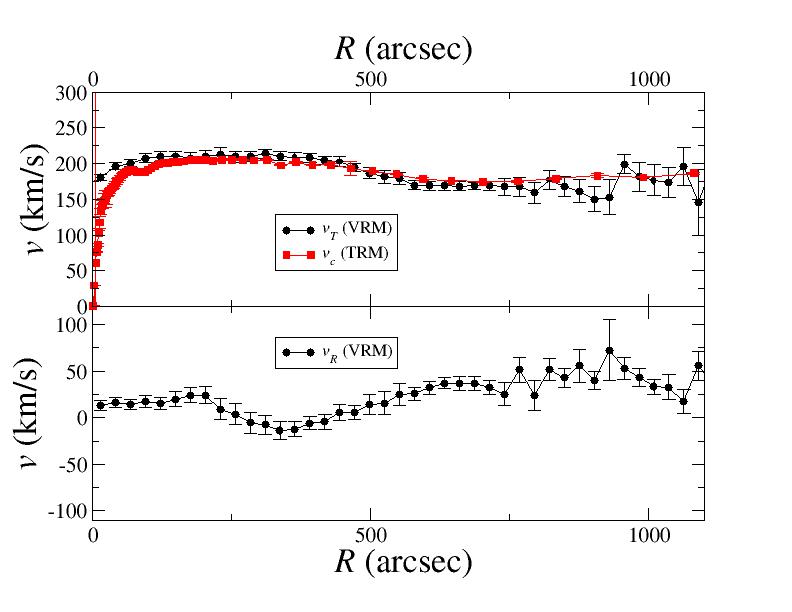}{0.45\textwidth}{(b)}
              }
  \gridline{
 \fig{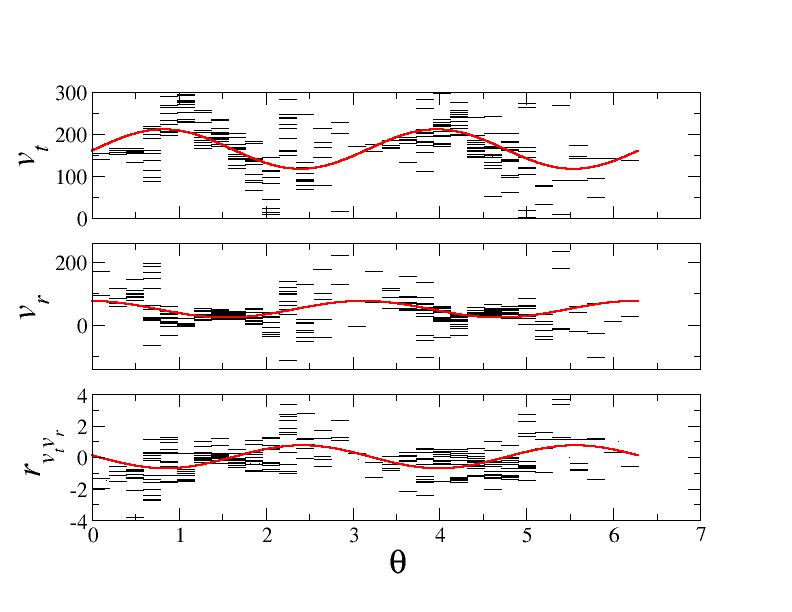}{0.45\textwidth}{(c)}
                \fig{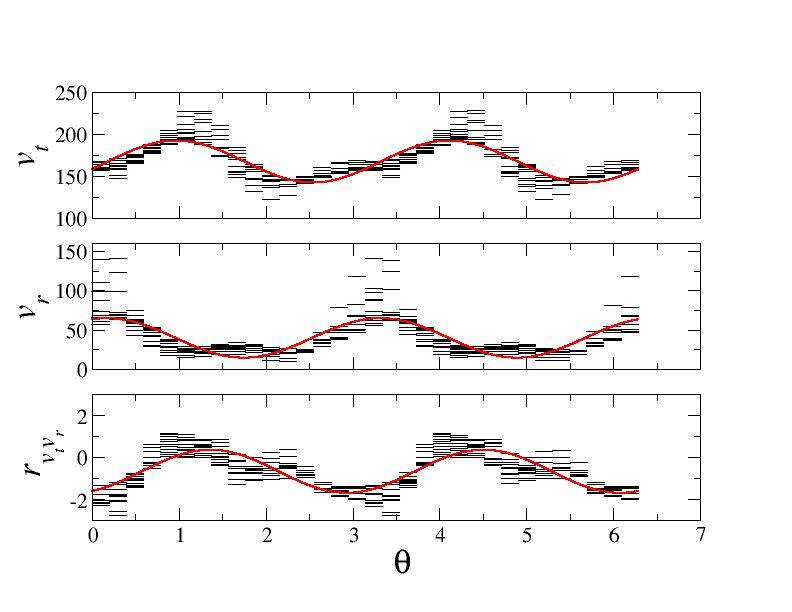}{0.45\textwidth}{(d)}}
     \caption{As  Fig.\ref{fig:NGC2903-1} but for NGC 5055.} 
\label{fig:NGC5055-1} 
\end{figure*}
%

\begin{figure*}
\gridline{\fig{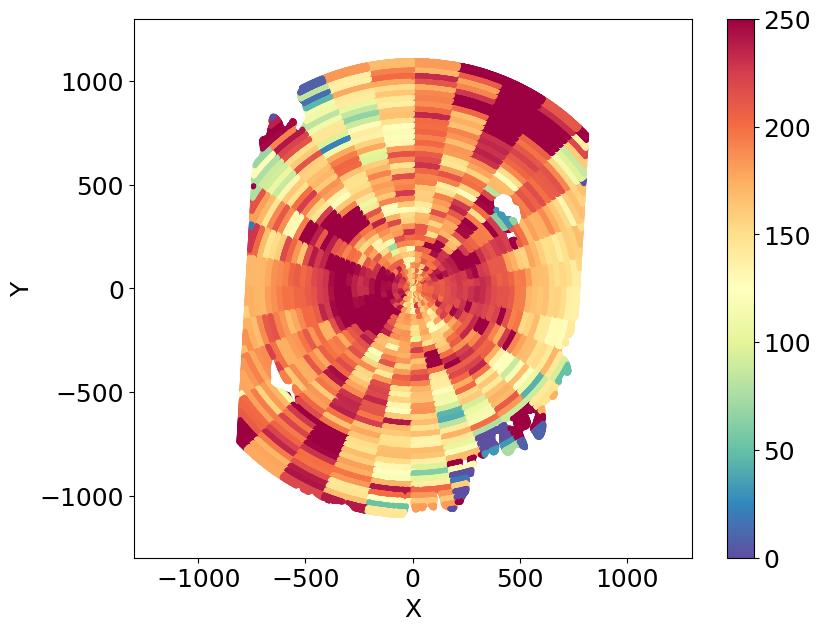}{0.3\textwidth}{(a)}
              \fig{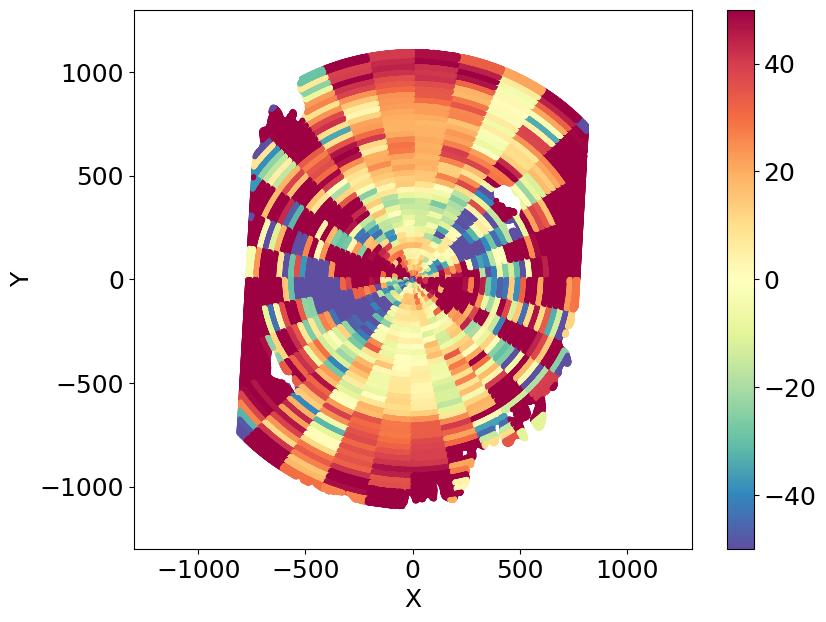}{0.3\textwidth}{(b)}
               \fig{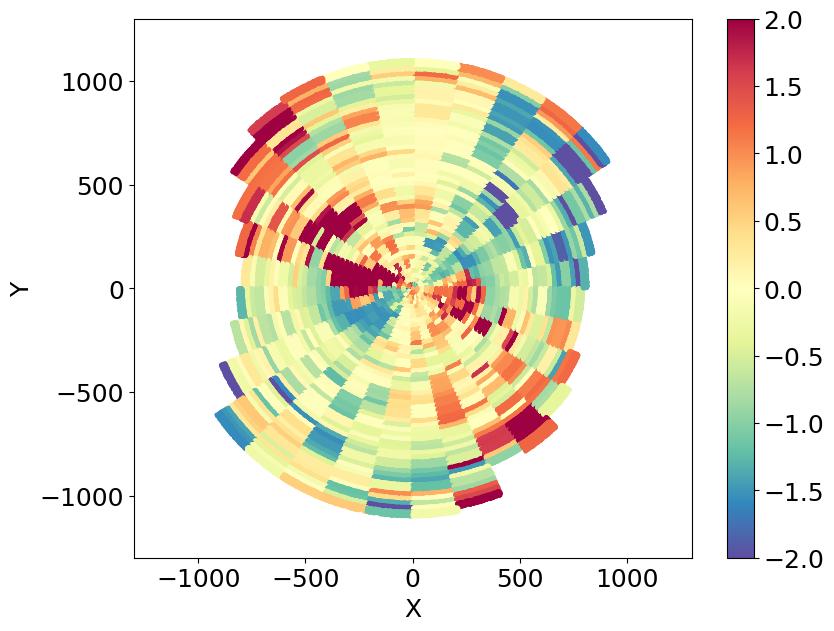}{0.3\textwidth}{(c)}}
\gridline{\fig{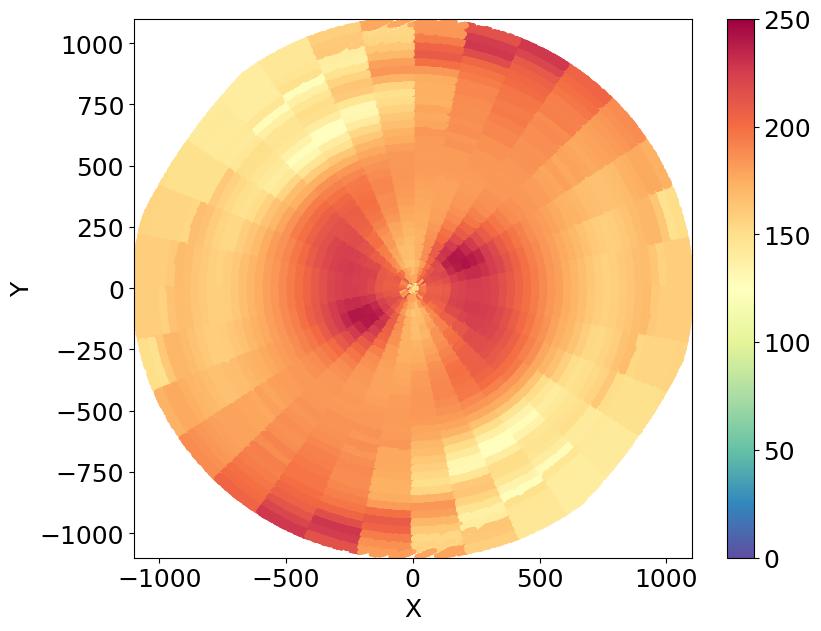}{0.3\textwidth}{(d)}
              \fig{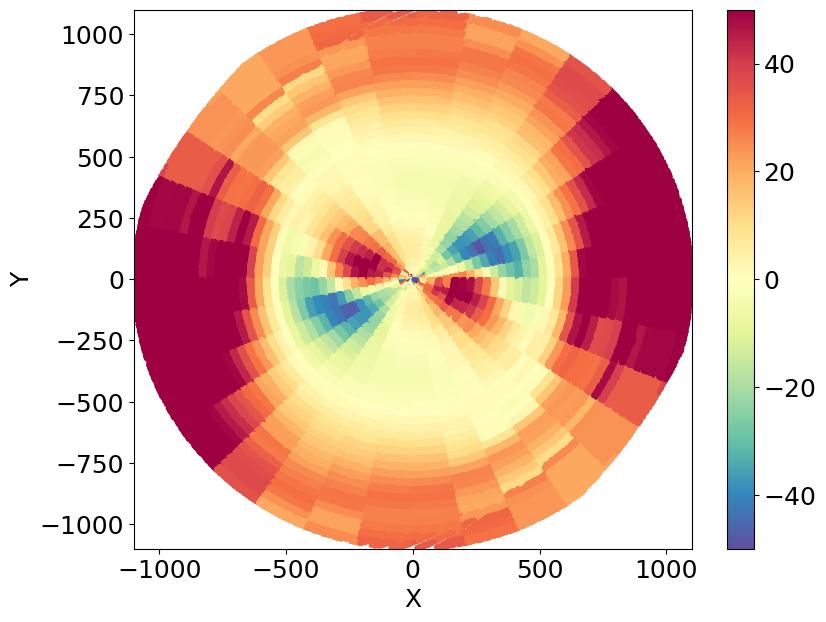}{0.3\textwidth}{(e)}
              \fig{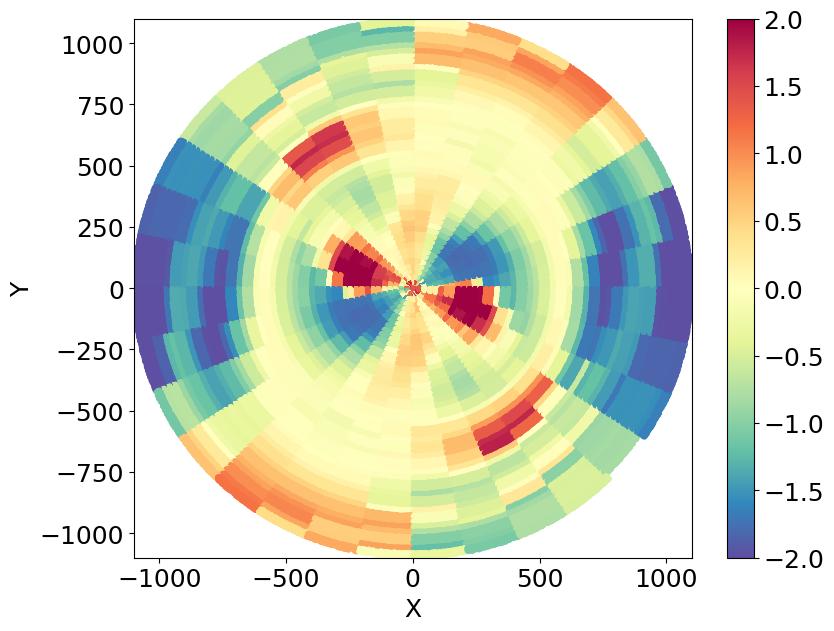}{0.3\textwidth}{(f)}}
\gridline{\fig{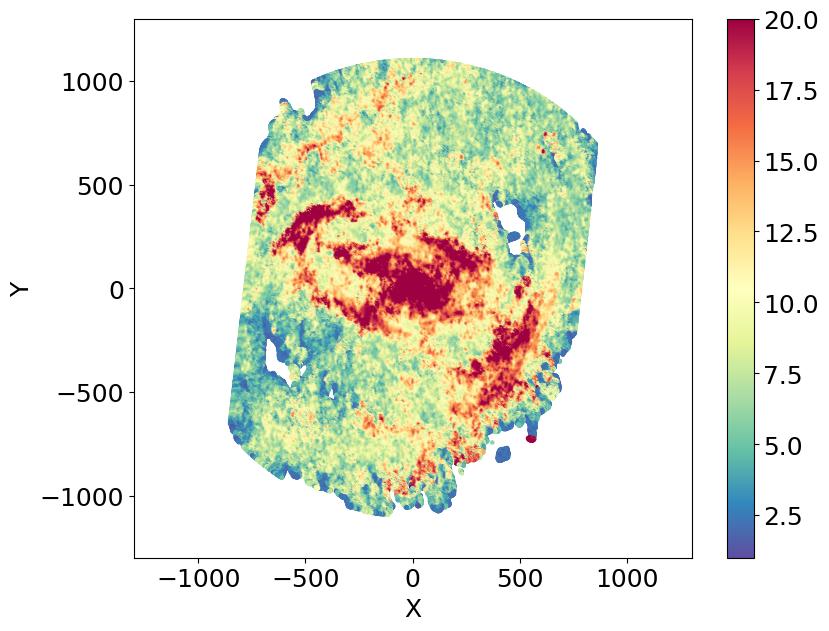}{0.3\textwidth}{(g)}
              \fig{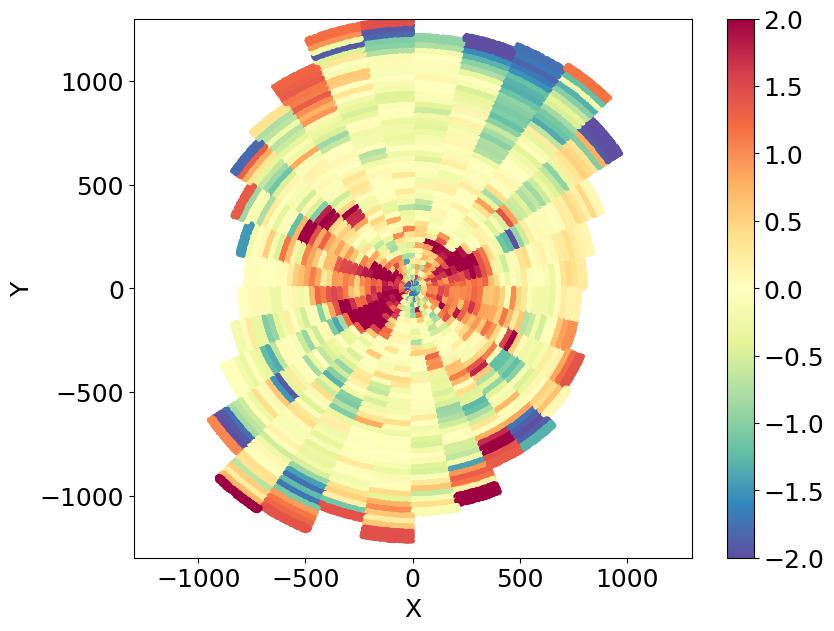}{0.3\textwidth}{(h)}
              \fig{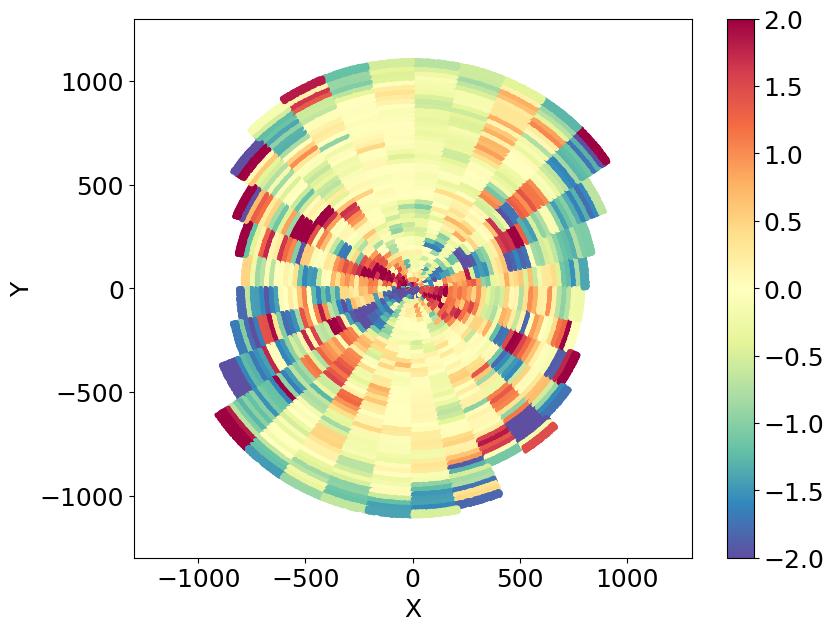}{0.3\textwidth}{(i)}}
\caption{As  Fig.\ref{fig:NGC2903-2} but for NGC 5055.} 
\label{fig:NGC5055-2} 
\end{figure*}
%
\clearpage


\subsection*{NGC 5236} 
The angular velocity component profiles of this galaxy are dominated by relatively large fluctuations, which do not allow the detection of a dipolar modulation corresponding to the warp, as seen by the TRM analysis  (Figure \ref{fig:NGC5236-1}). Furthermore, the velocity correlation coefficient does not show a dipolar modulation either, casting doubt on the very existence of a warp in this galaxy: { this is confirmed by a low value of the correlation coefficient is ${\cal C}$ = 0.19}. Not surprisingly, the reconstructed velocity component maps of the real galaxy are quite different from those of the toy disc model with the same orientation angles and circular velocity (Figure \ref{fig:NGC5236-2}).  The velocity dispersion field shows a local enhancement that corresponds to a positive correlation in the rank correlation coefficients $r_{\sigma v_t}(R, \theta)$ and $r_{\sigma v_r}(R, \theta)$.

\begin{figure*}
\gridline{\fig{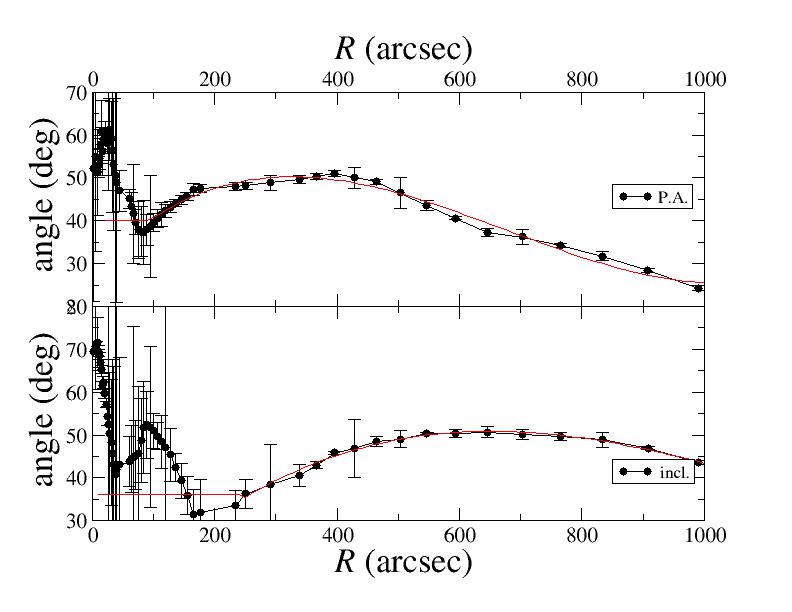}{0.45\textwidth}{(a)}
              \fig{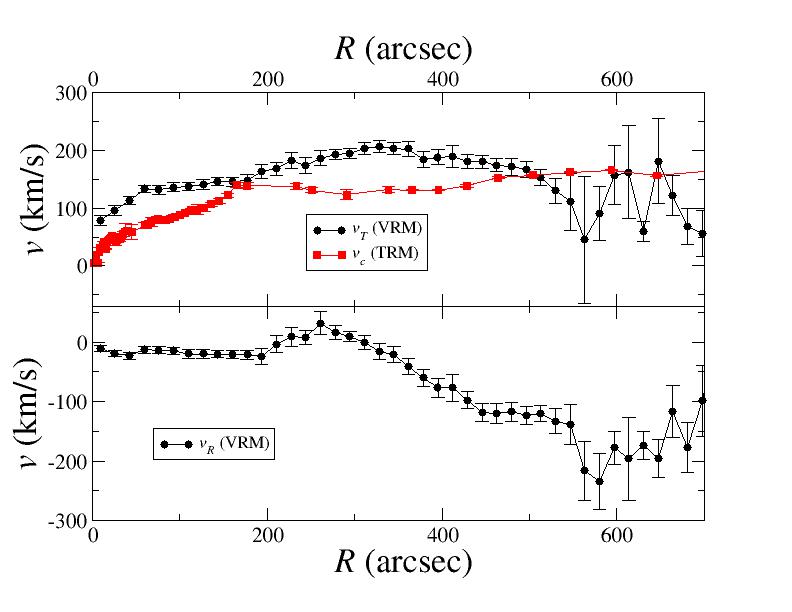}{0.45\textwidth}{(b)}
              }
  \gridline{
 	       \fig{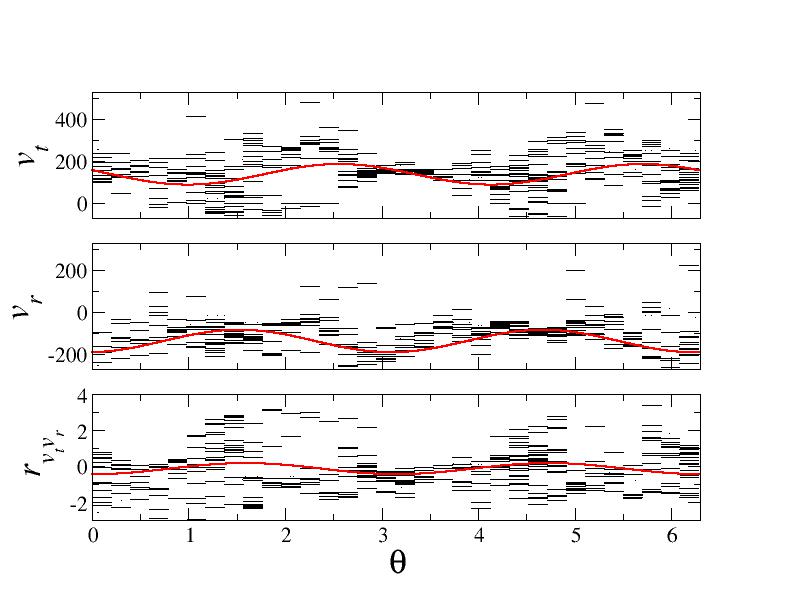}{0.45\textwidth}{(c)}
                \fig{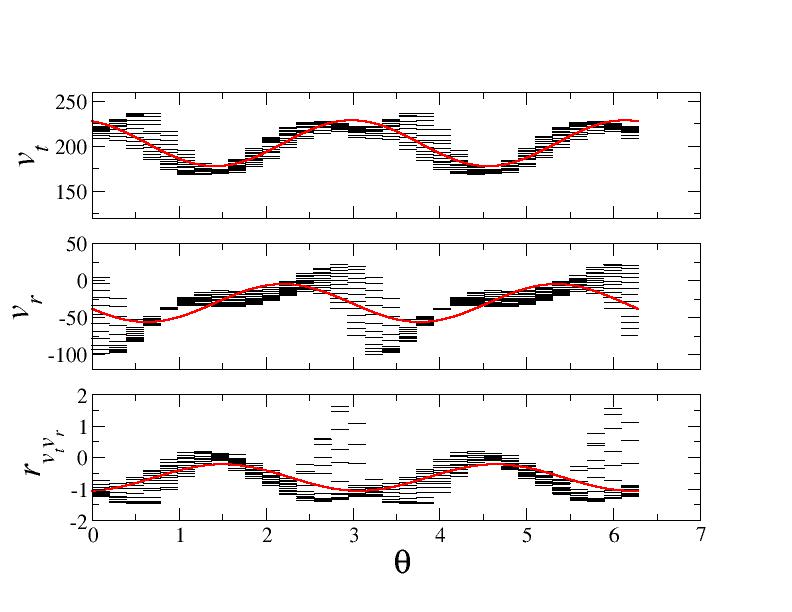}{0.45\textwidth}{(d)}}
     \caption{As  Fig.\ref{fig:NGC2903-1} but for NGC 5236.} 
\label{fig:NGC5236-1} 
\end{figure*}
%

\begin{figure*}
\gridline{\fig{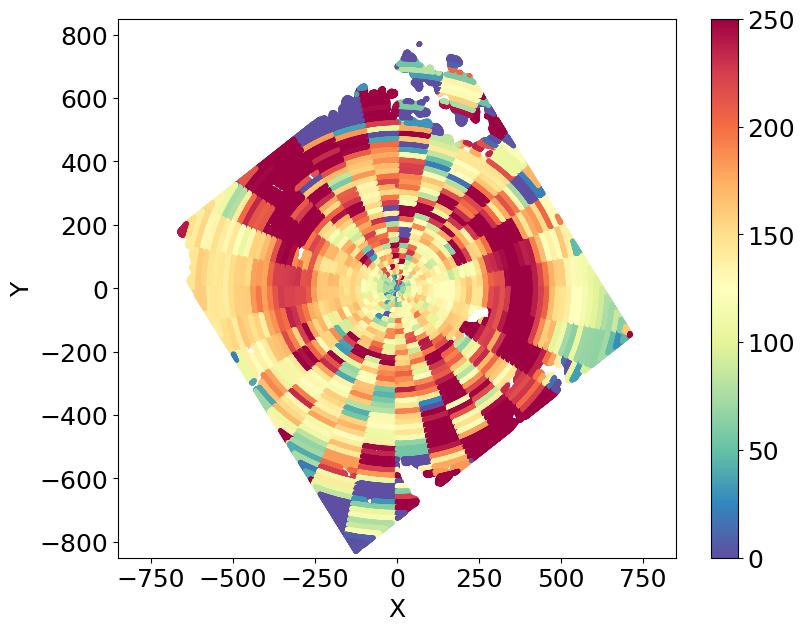}{0.3\textwidth}{(a)}
              \fig{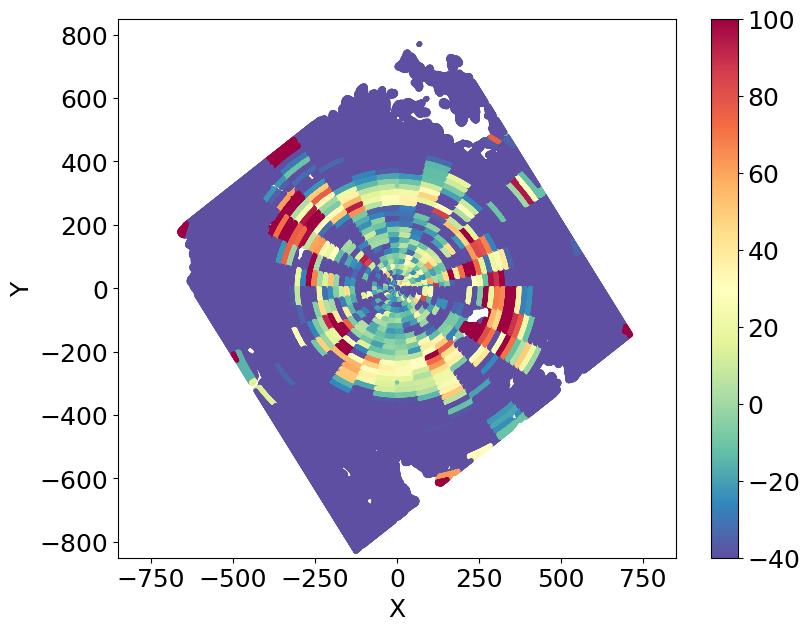}{0.3\textwidth}{(b)}
               \fig{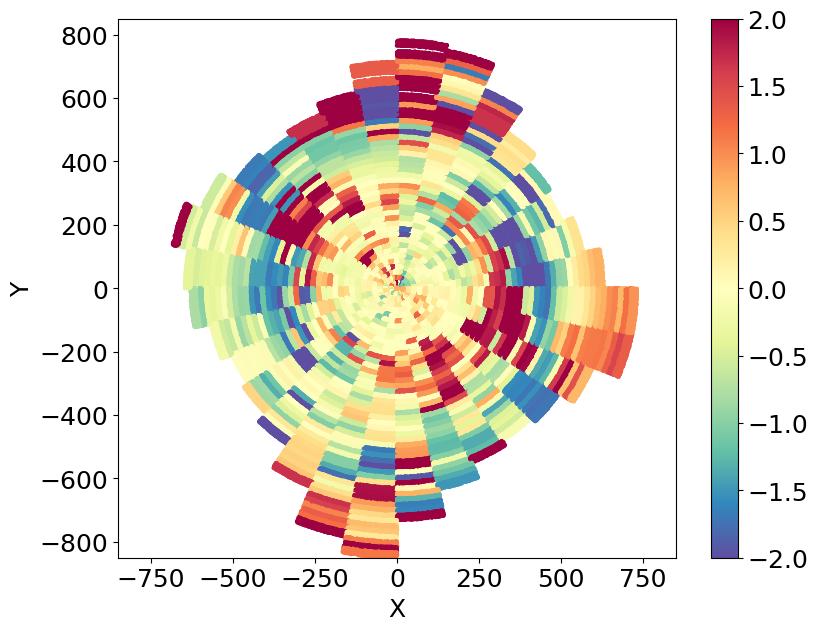}{0.3\textwidth}{(c)}}
\gridline{\fig{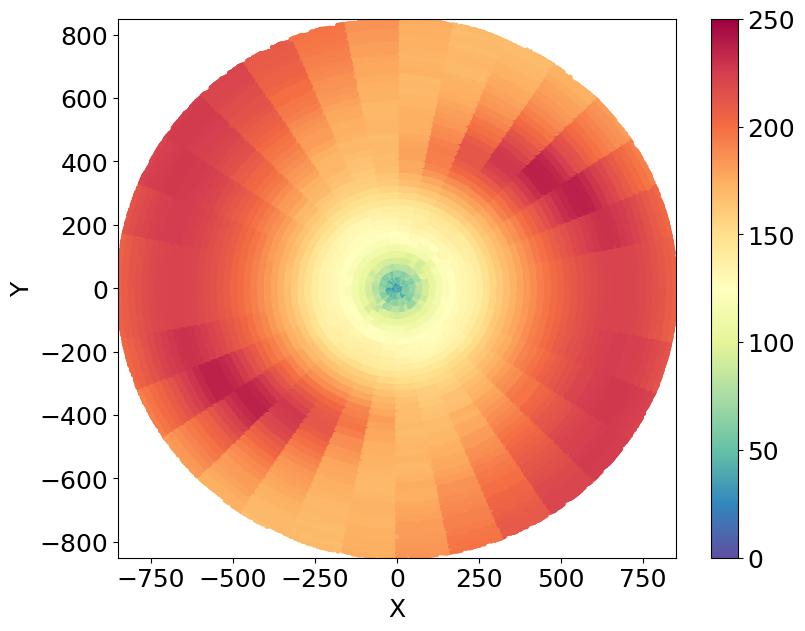}{0.3\textwidth}{(d)}
              \fig{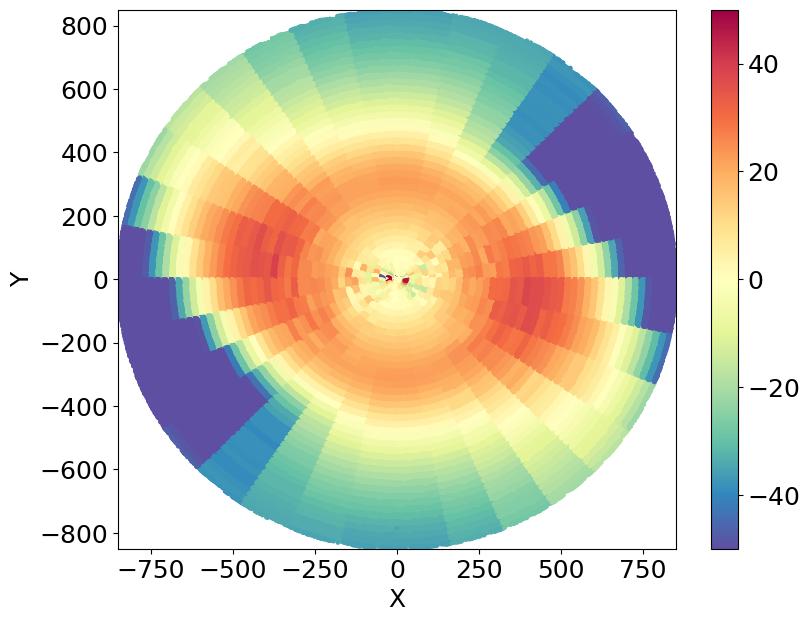}{0.3\textwidth}{(e)}
              \fig{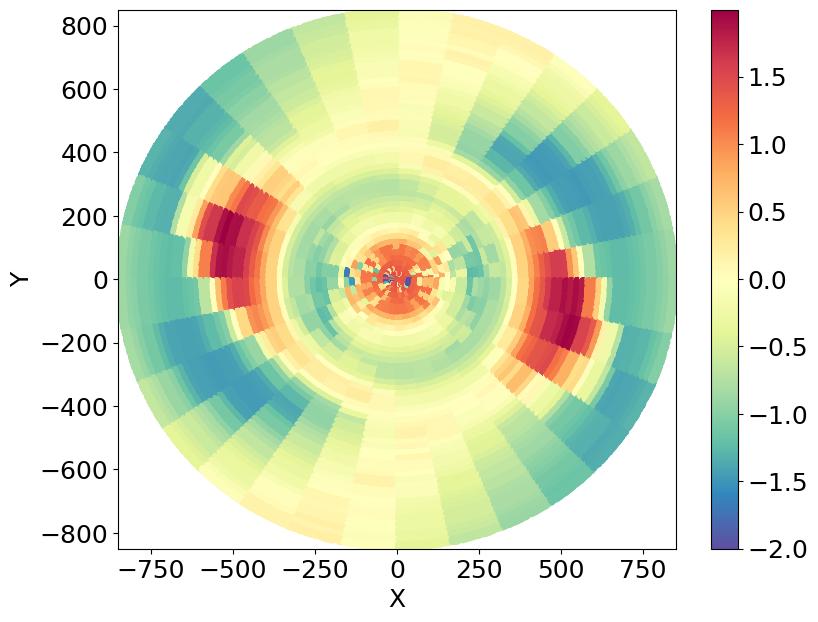}{0.3\textwidth}{(f)}}
\gridline{\fig{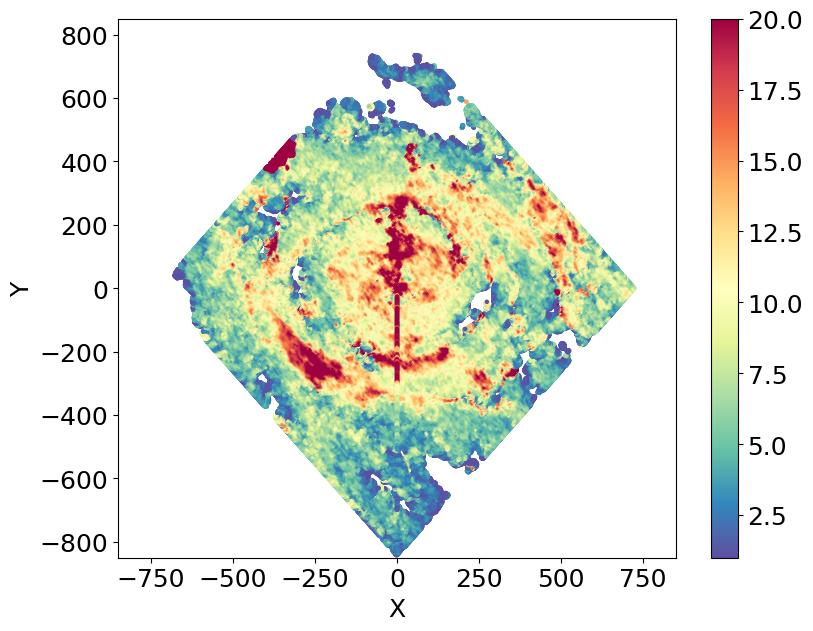}{0.3\textwidth}{(g)}
              \fig{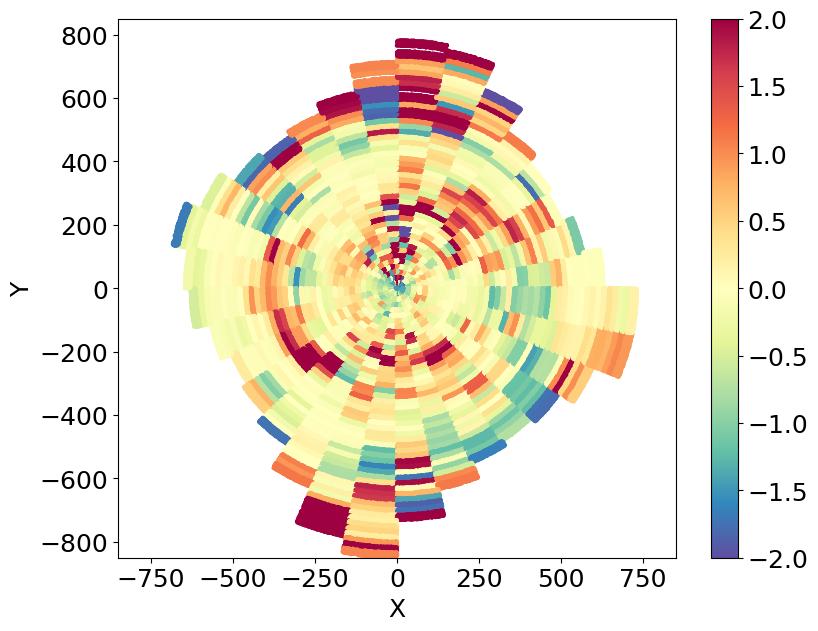}{0.3\textwidth}{(h)}
              \fig{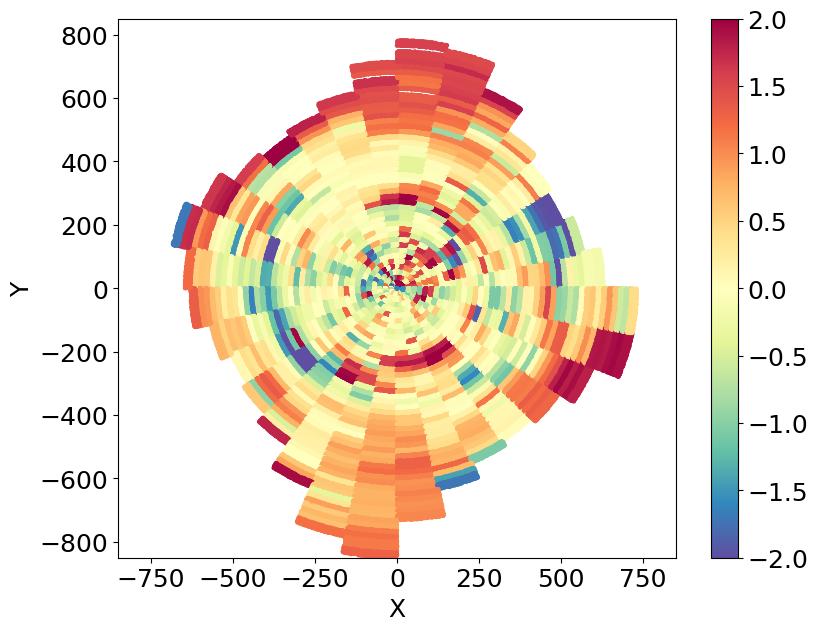}{0.3\textwidth}{(i)}}
\caption{As  Fig.\ref{fig:NGC2903-2} but for NGC 5236.} 
\label{fig:NGC5236-2} 
\end{figure*}

\clearpage


\subsection*{NGC 6946} 
The moderate variation of the orientation angles detected by the TRM for this galaxy (Figure \ref{fig:NGC6946-1}) seems to correspond to a real warp, as shown in our analysis by the dipolar modulation characterizing the velocity rank correlation coefficient (Figure \ref{fig:NGC6946-2}). 
{ Indeed, the correlation coefficient is ${\cal C}$ = 0.52.}
However, such a modulation is not well visible in the two velocity components, indicating the presence of relatively large intrinsic velocity perturbations. The velocity dispersion field is smooth and decays monotonically from the center.

\begin{figure*}
\gridline{\fig{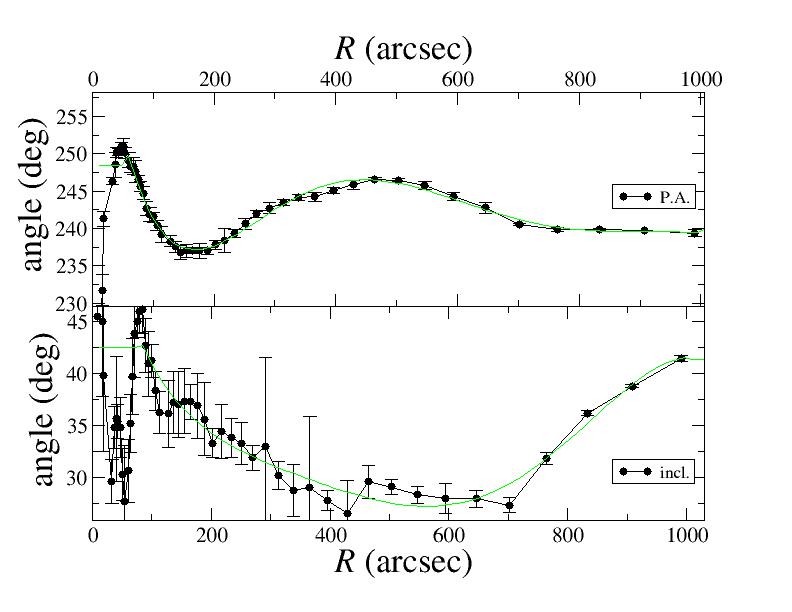}{0.45\textwidth}{(a)}
              \fig{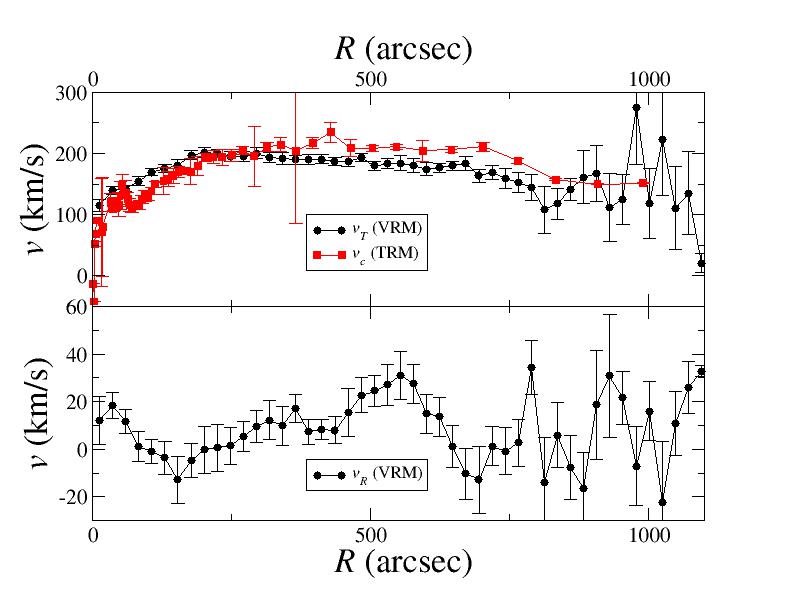}{0.45\textwidth}{(b)}
              }
  \gridline{
 	       \fig{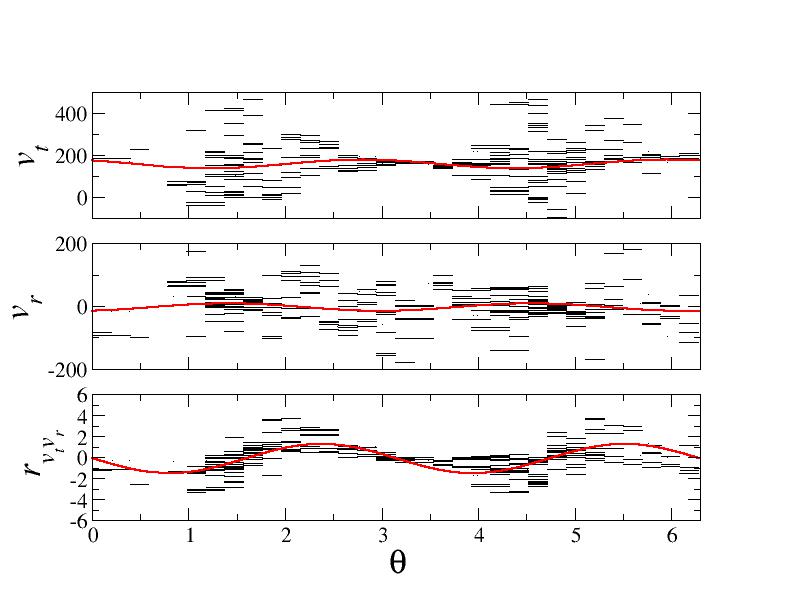}{0.45\textwidth}{(c)}
                \fig{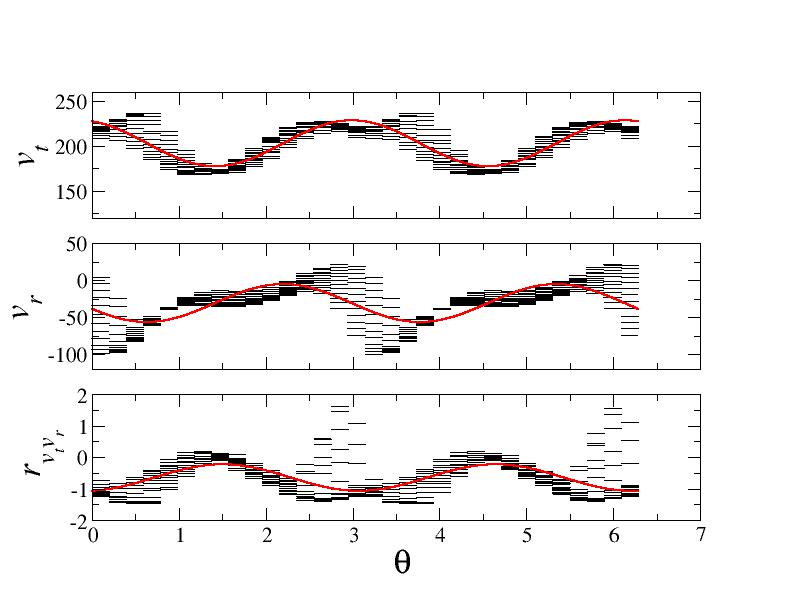}{0.45\textwidth}{(d)}}
     \caption{As  Fig.\ref{fig:NGC2903-1} but for NGC 6946.} 
\label{fig:NGC6946-1} 
\end{figure*}
%

\begin{figure*}
\gridline{\fig{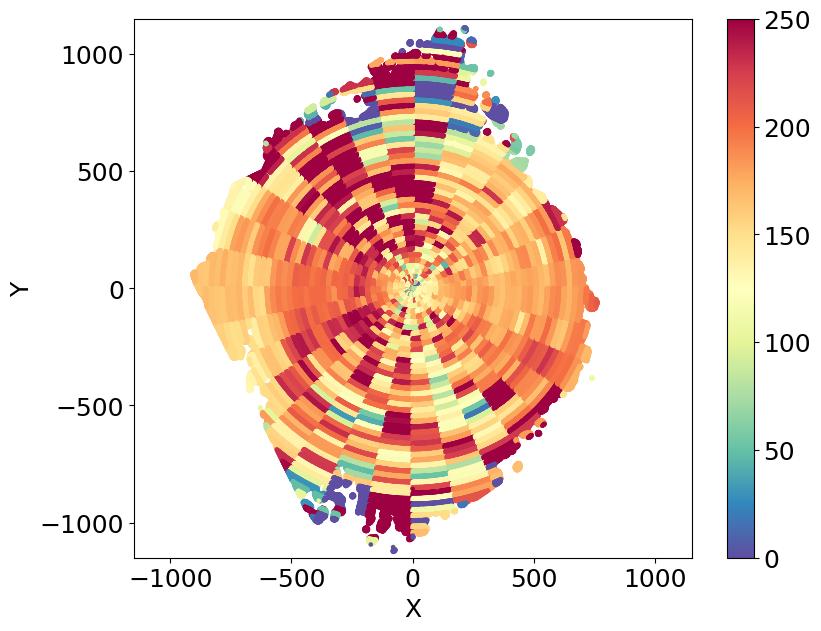}{0.3\textwidth}{(a)}
              \fig{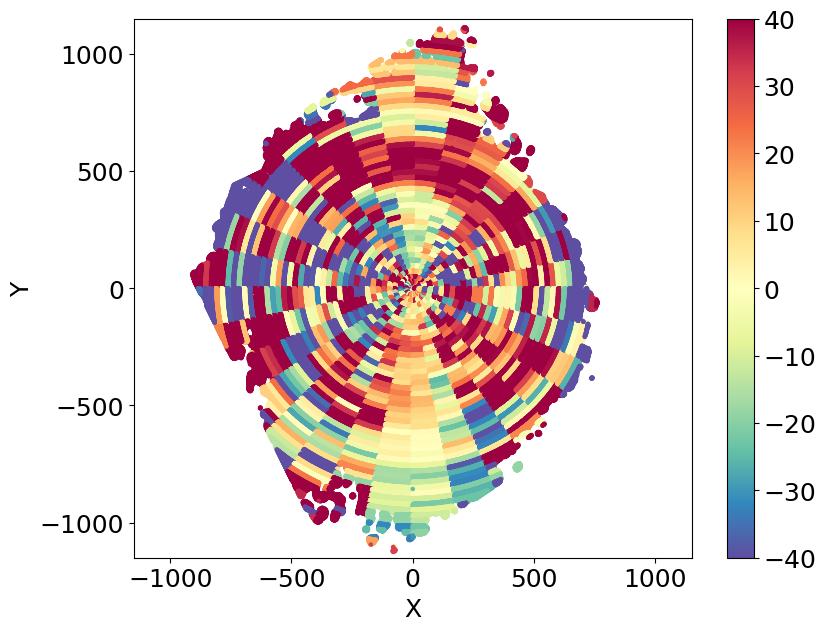}{0.3\textwidth}{(b)}
               \fig{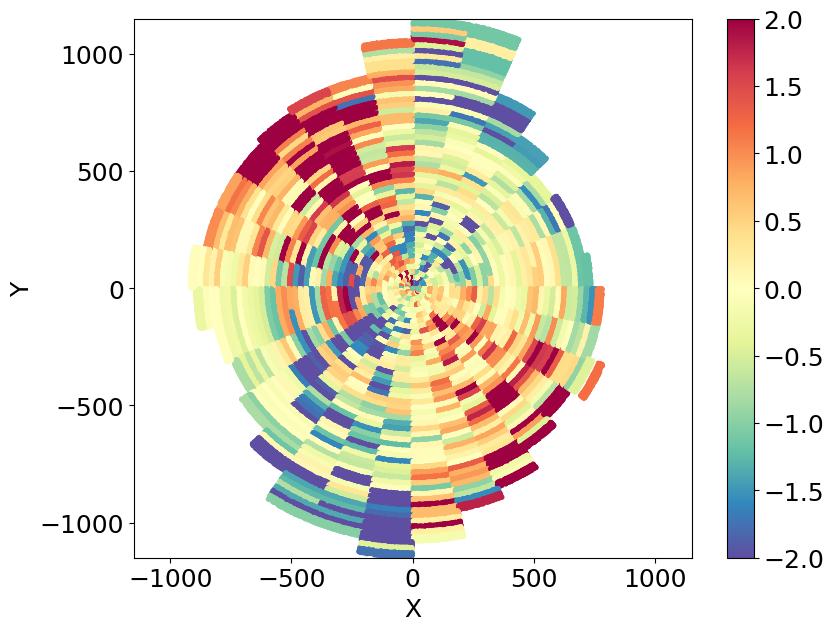}{0.3\textwidth}{(c)}}
\gridline{\fig{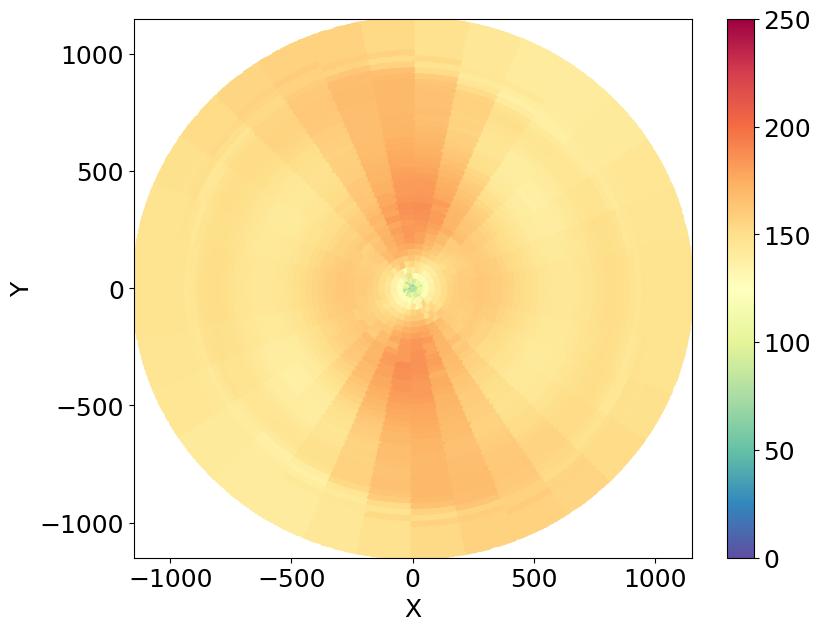}{0.3\textwidth}{(d)}
              \fig{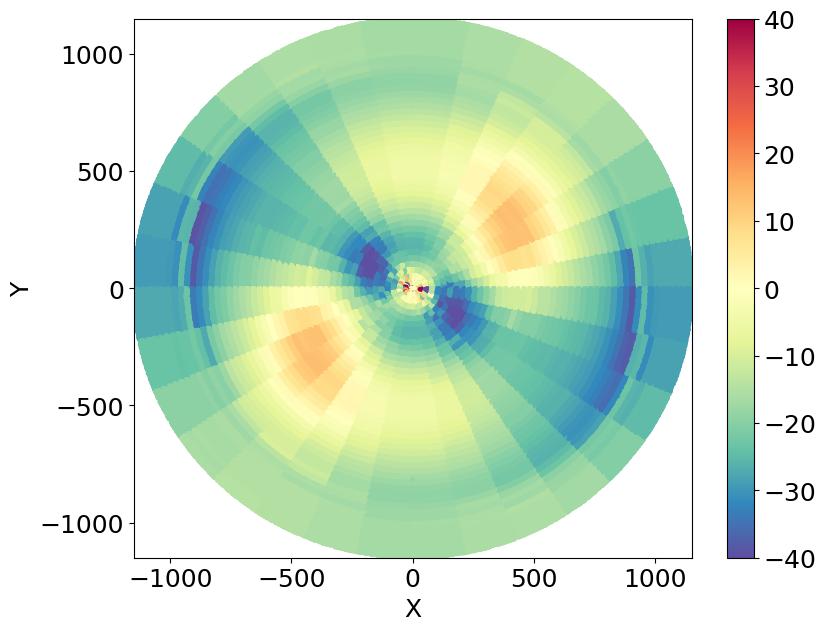}{0.3\textwidth}{(e)}
              \fig{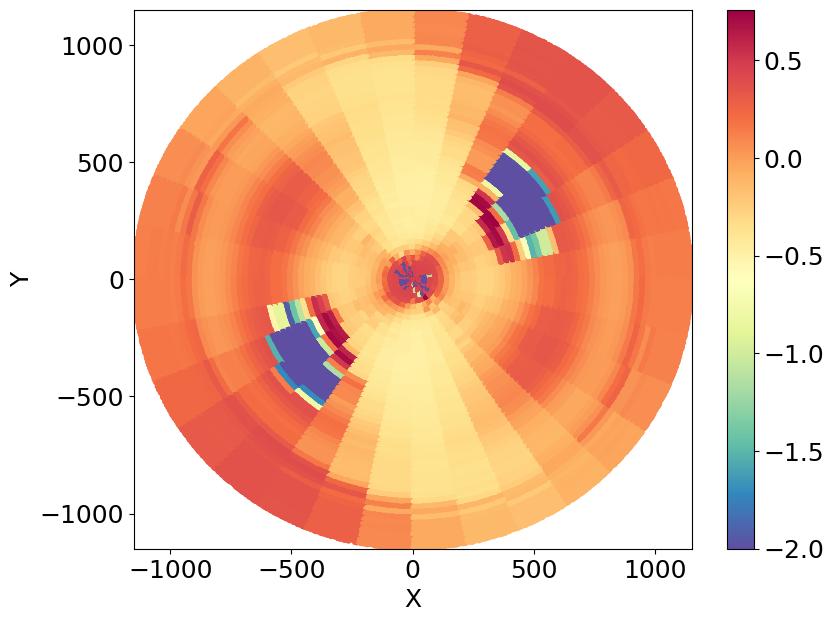}{0.3\textwidth}{(f)}}
\gridline{\fig{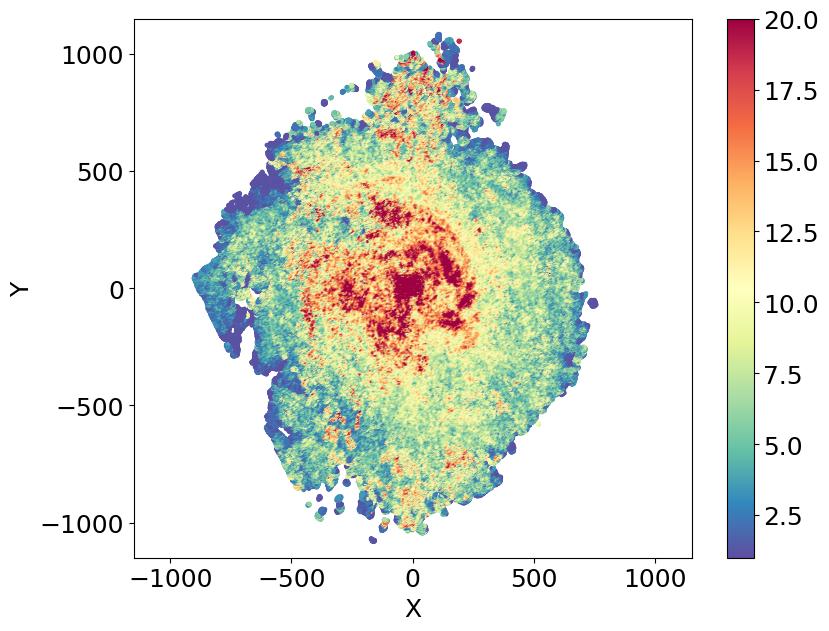}{0.3\textwidth}{(g)}
              \fig{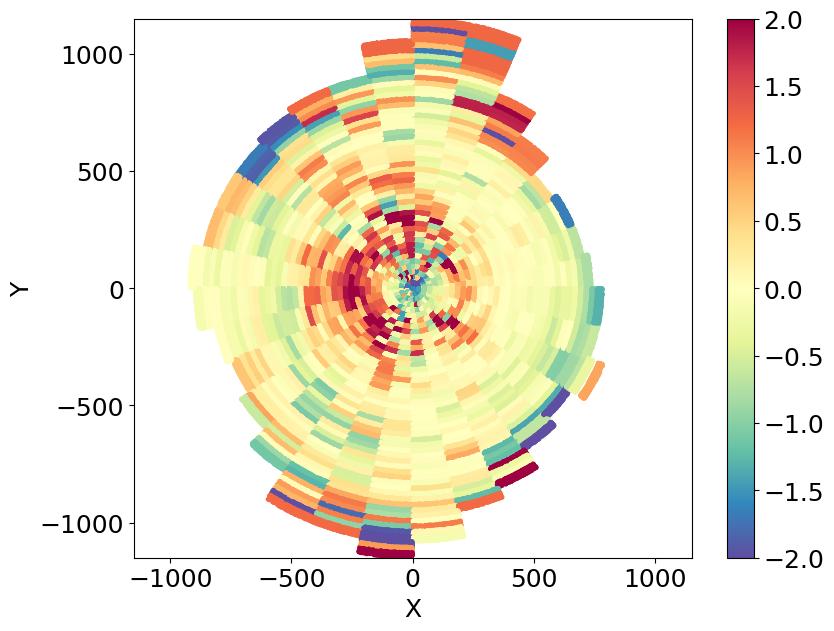}{0.3\textwidth}{(h)}
              \fig{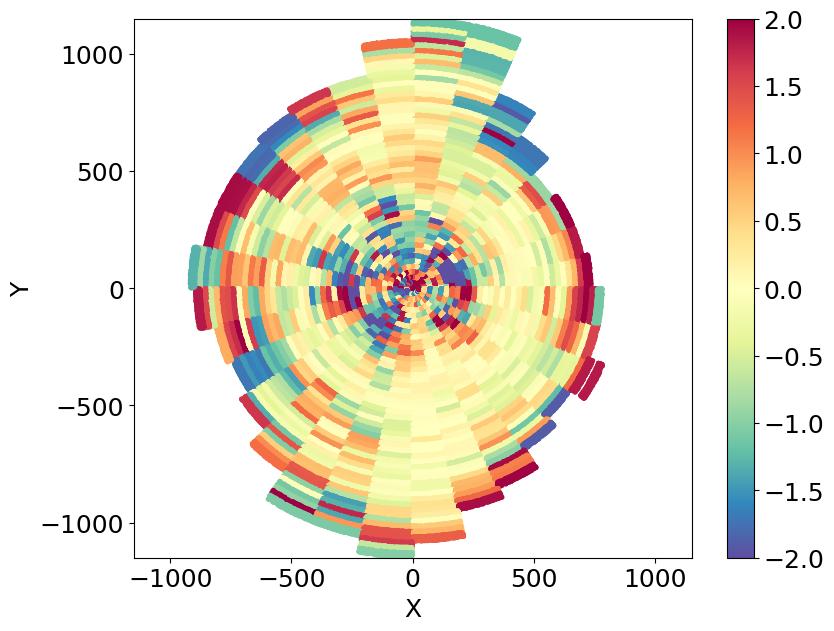}{0.3\textwidth}{(i)}}
\caption{As  Fig.\ref{fig:NGC2903-2} but for NGC 6946.} 
\label{fig:NGC6946-2} 
\end{figure*}


\subsection*{NGC 7331} 

A moderate warp, if present at all, is not the most evident characteristic of this galaxy. 
{ Indeed, the correlation coefficient is ${\cal C}$ = 0.16} .Instead, a perturbed and anisotropic velocity field is evidenced by the velocity components and velocity dispersion maps (see Fig.\ref{fig:NGC7331-1}-\ref{fig:NGC7331-2}).

\begin{figure*}
\gridline{\fig{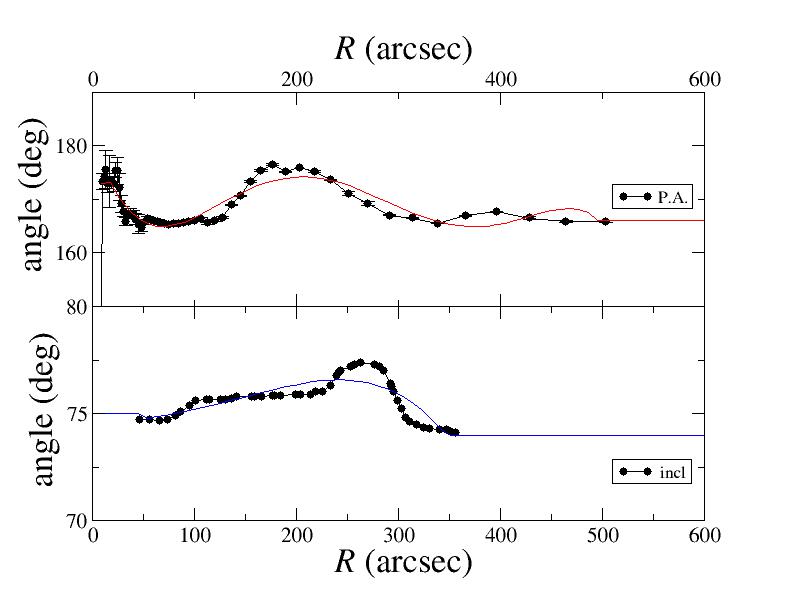}{0.45\textwidth}{(a)}
              \fig{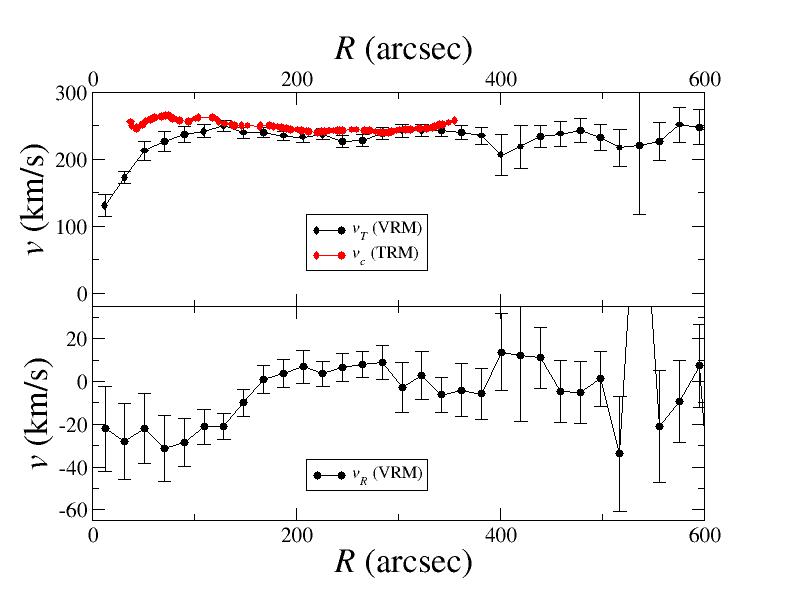}{0.45\textwidth}{(b)}
              }
  \gridline{
 	       \fig{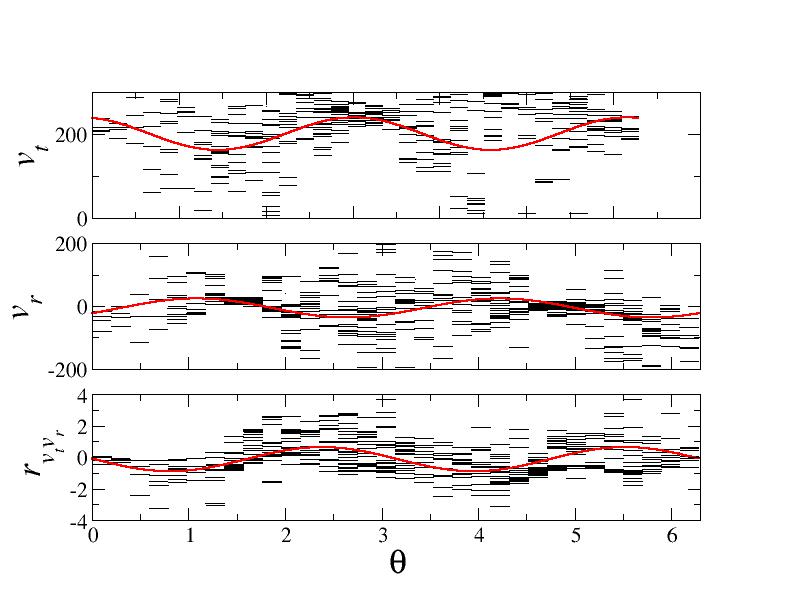}{0.45\textwidth}{(c)}
                \fig{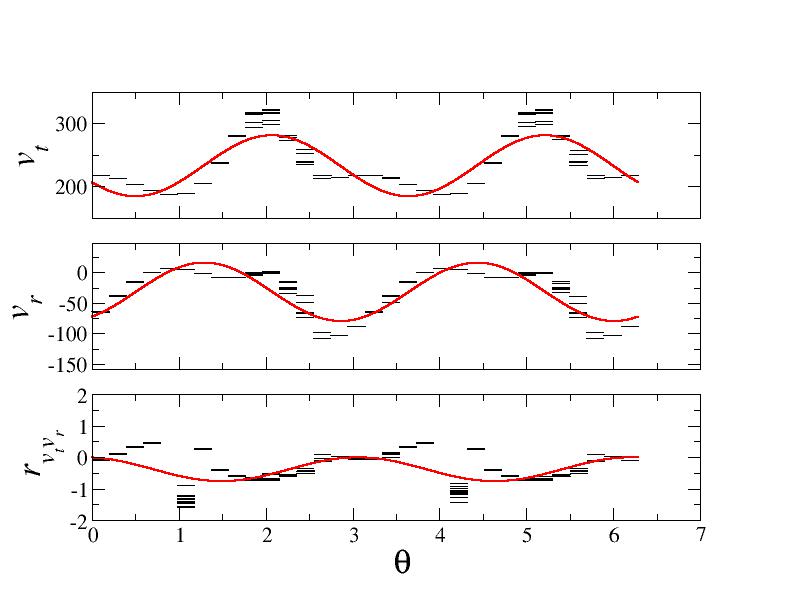}{0.45\textwidth}{(d)}}
     \caption{As  Fig.\ref{fig:NGC2903-1} but for NGC 7331.} 
\label{fig:NGC7331-1} 
\end{figure*}
%

\begin{figure*}
\gridline{\fig{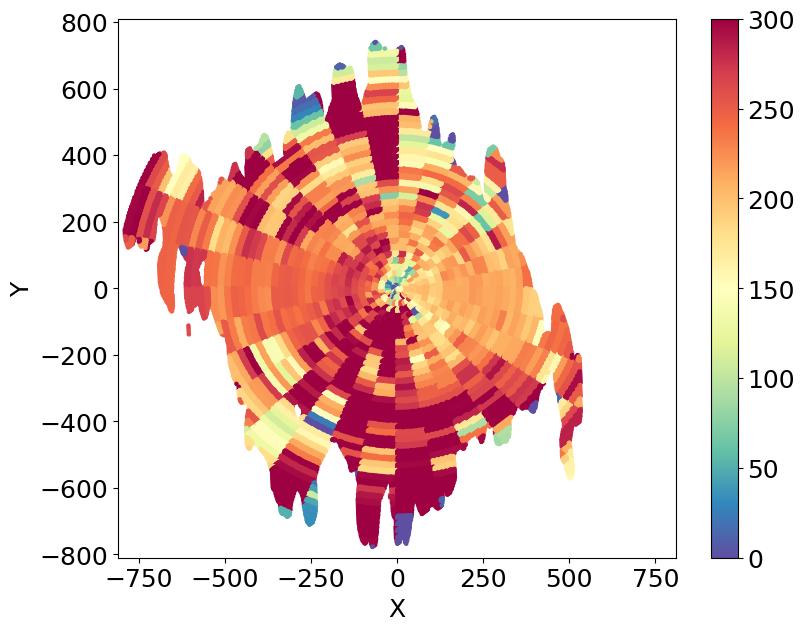}{0.3\textwidth}{(a)}
              \fig{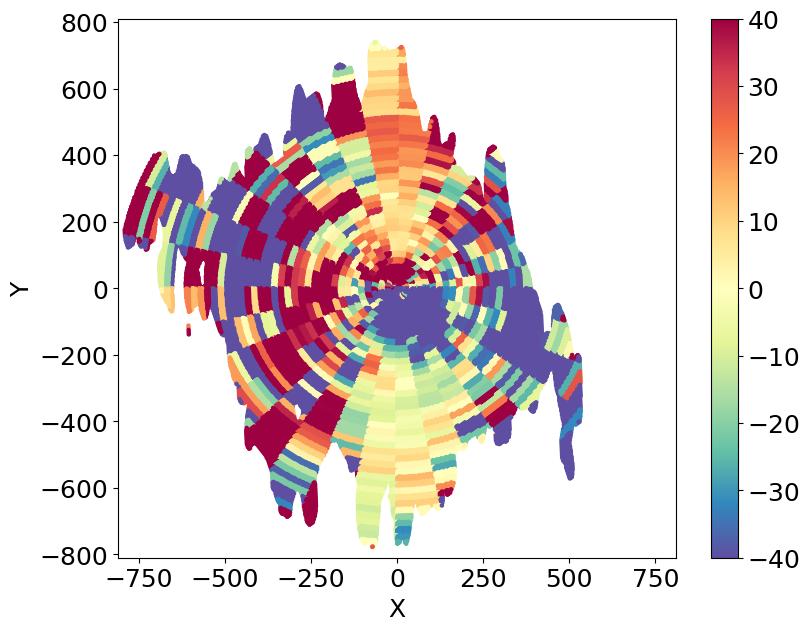}{0.3\textwidth}{(b)}
               \fig{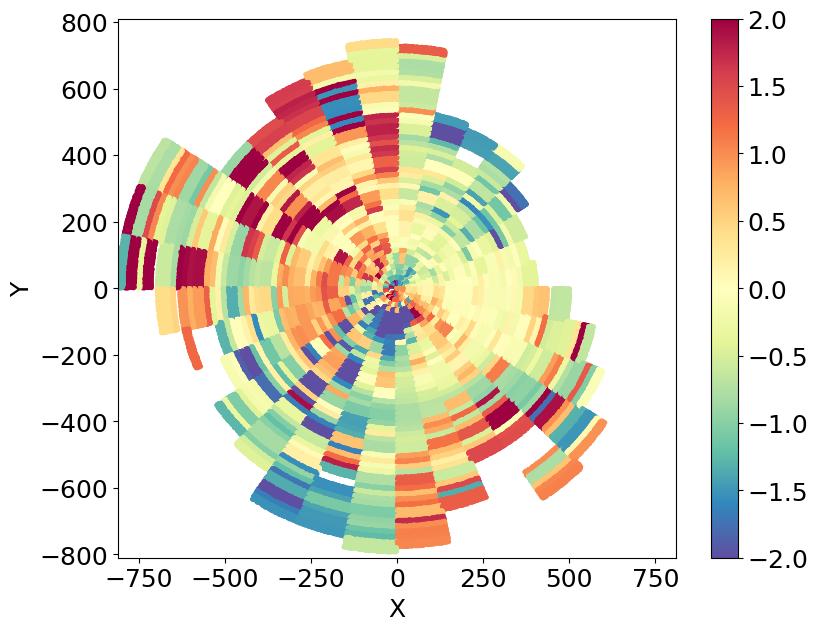}{0.3\textwidth}{(c)}}
\gridline{\fig{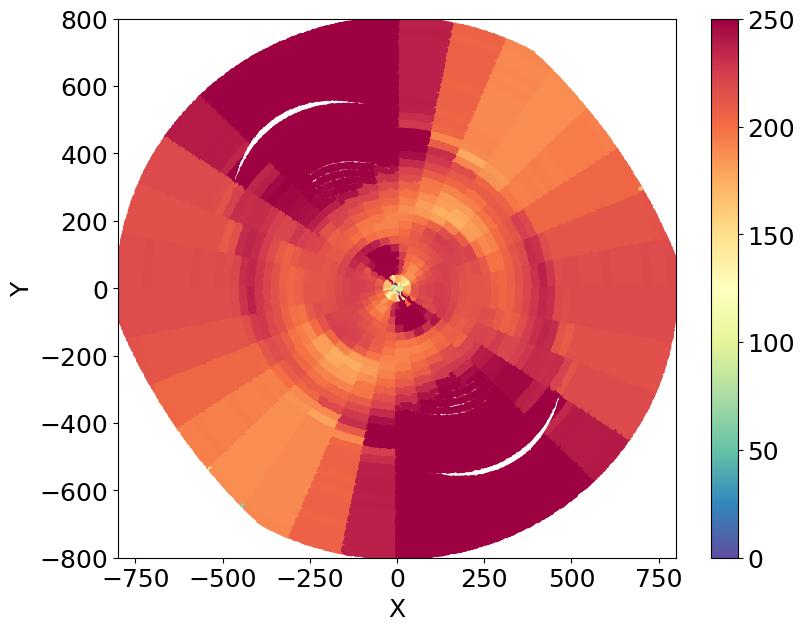}{0.3\textwidth}{(d)}
              \fig{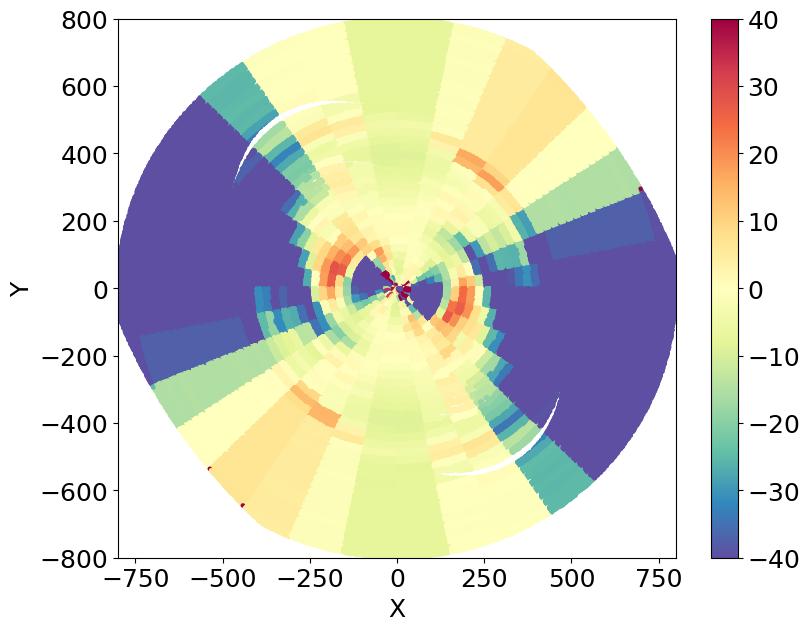}{0.3\textwidth}{(e)}
              \fig{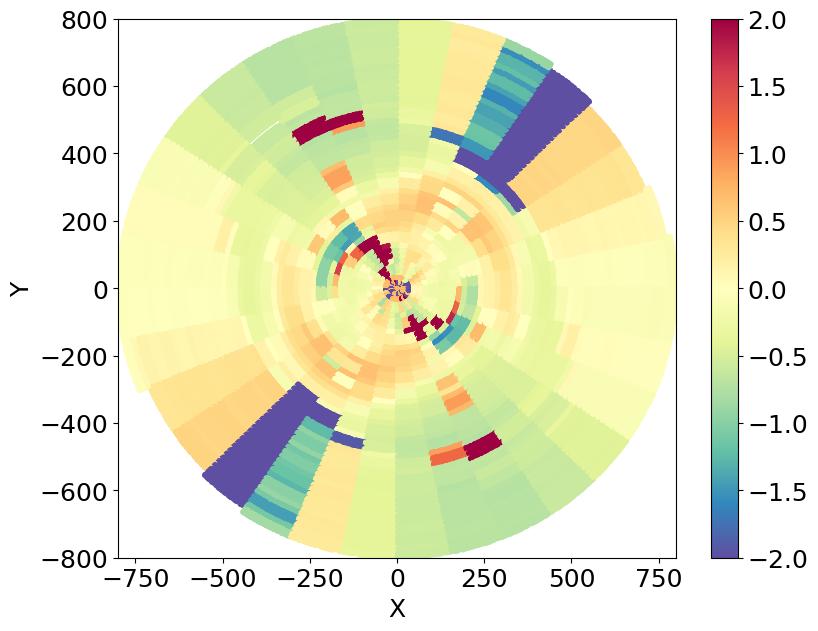}{0.3\textwidth}{(f)}}
\gridline{\fig{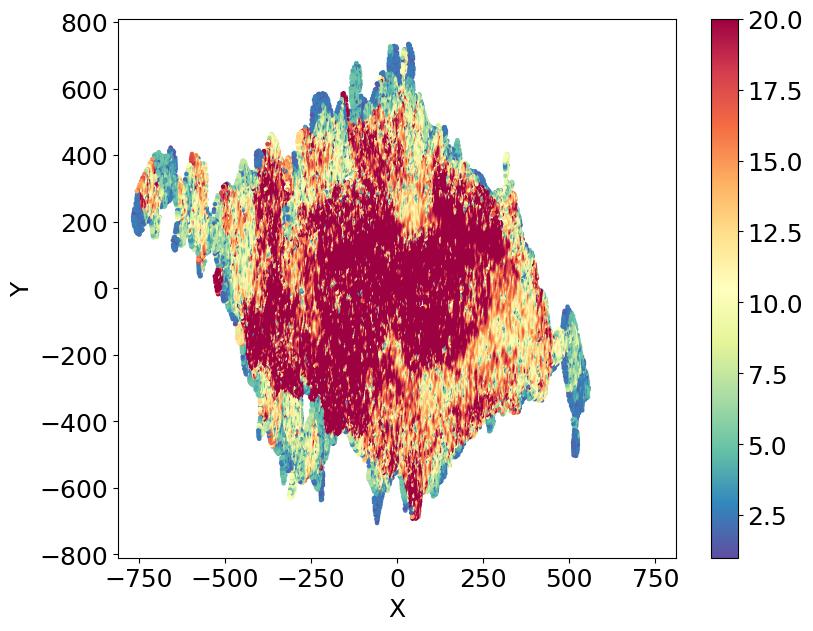}{0.3\textwidth}{(g)}
              \fig{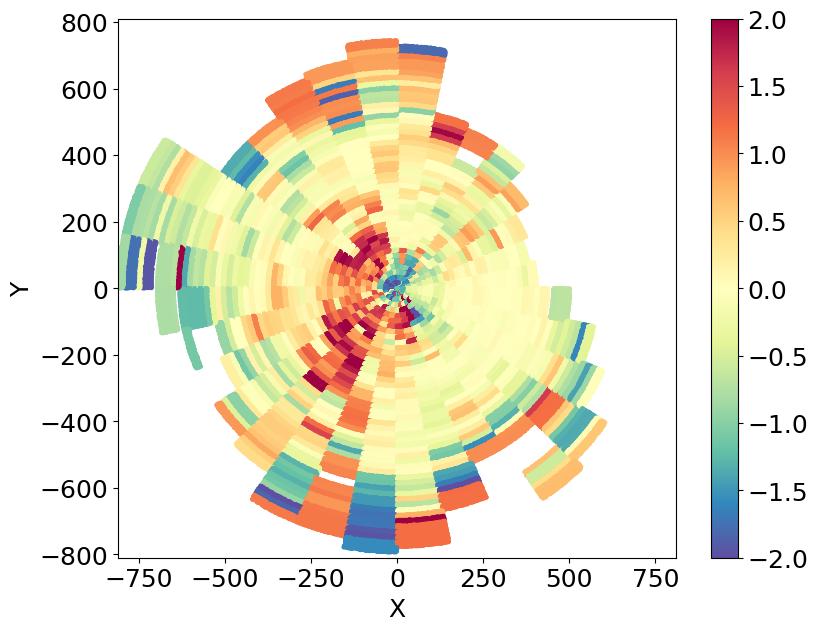}{0.3\textwidth}{(h)}
              \fig{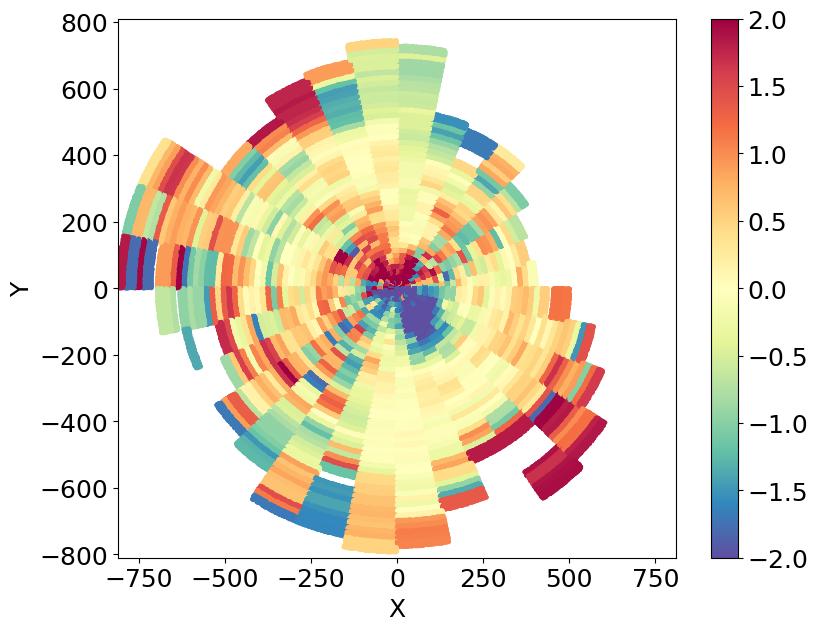}{0.3\textwidth}{(i)}}
\caption{As  Fig.\ref{fig:NGC2903-2} but for NGC 7331.} 
\label{fig:NGC7331-2} 
\end{figure*}
%


\subsection*{NGC 7793} 
The variation of the orientation angles, as detected by the TRM, seems to be compatible with the dipolar modulation of the rank correlation coefficient, pointing thus toward the reality of a  warp for this galaxy: { the correlation coefficient is ${\cal C}$ = 0.5.} However, the velocity component profiles are dominated by intrinsic perturbations, and a simple dipolar modulation is not visible (Figure \ref{fig:NGC7793-1}). The 2D maps reveal that the velocity field is quite smooth and quiet (Figure \ref{fig:NGC7793-1}).
%

\begin{figure*}
\gridline{\fig{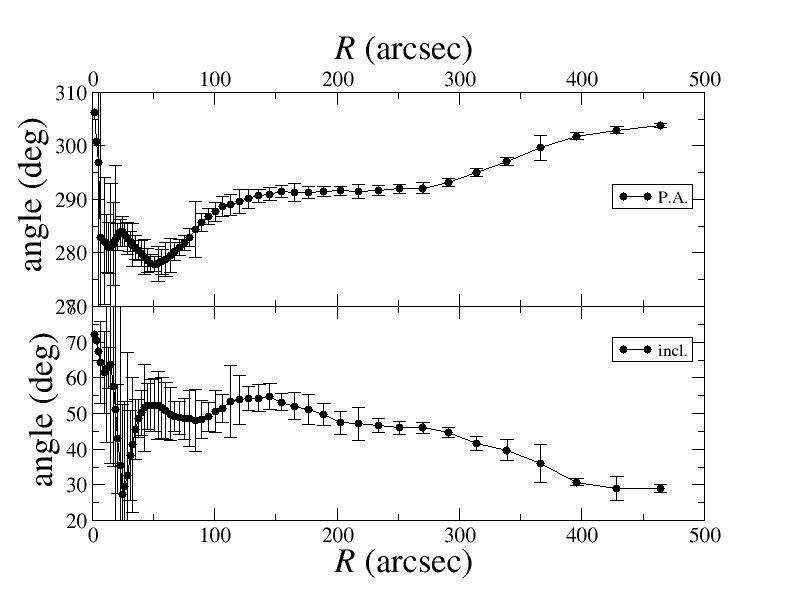}{0.45\textwidth}{(a)}
              \fig{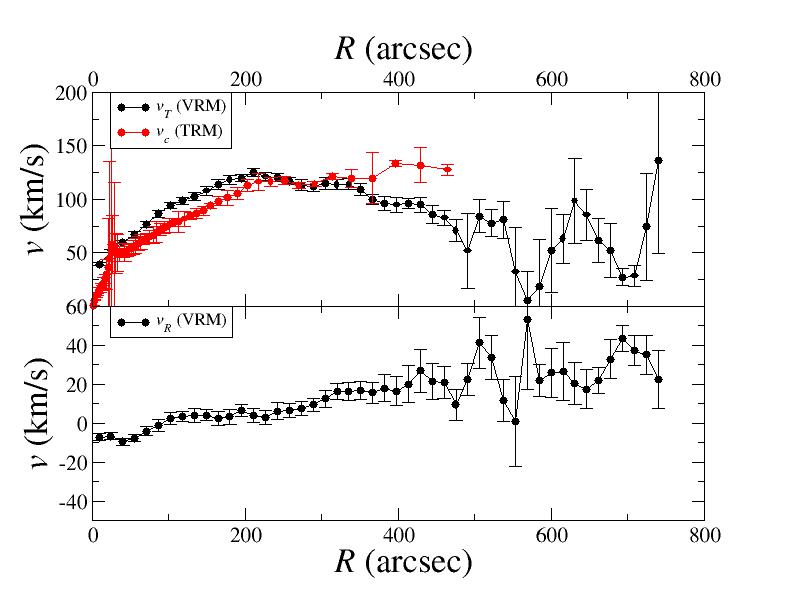}{0.45\textwidth}{(b)}
              }
  \gridline{
 	       \fig{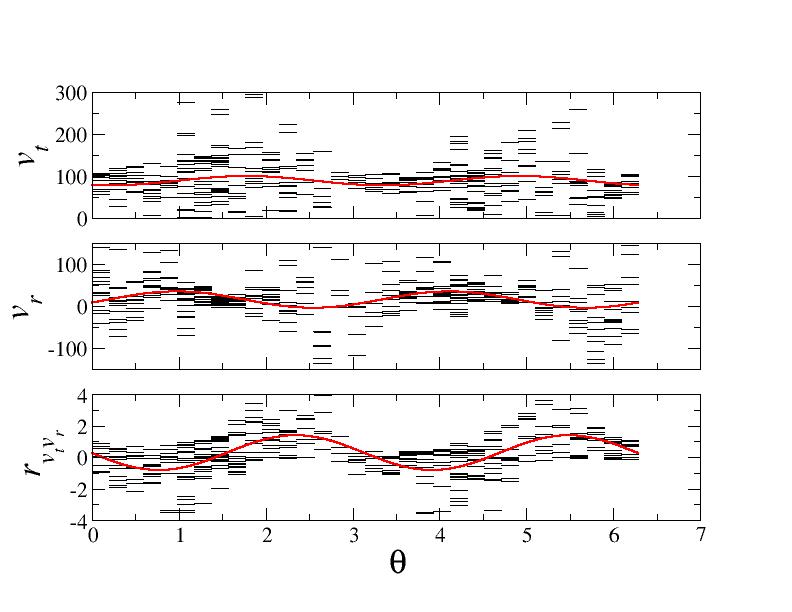}{0.45\textwidth}{(c)}
                \fig{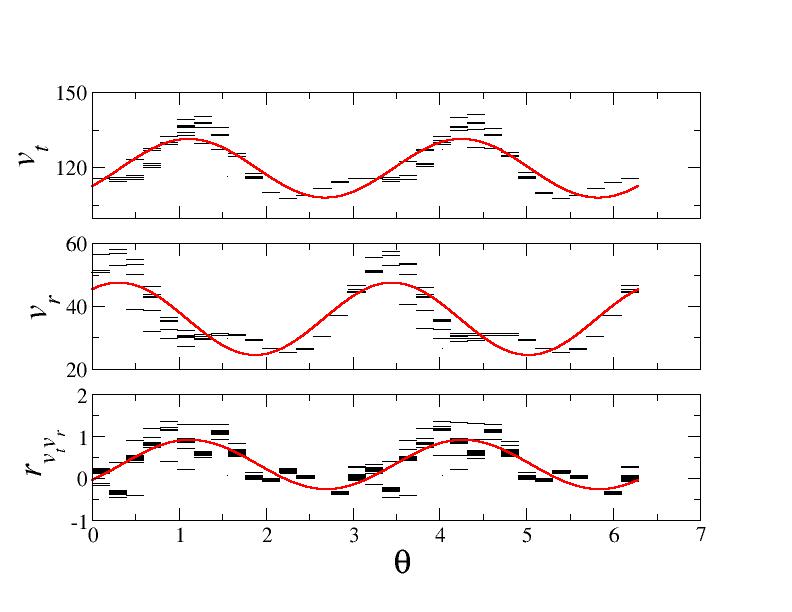}{0.45\textwidth}{(d)}}
     \caption{As  Fig.\ref{fig:NGC2903-1} but for NGC 7793.} 
\label{fig:NGC7793-1} 
\end{figure*}
%
 
\begin{figure*}
\gridline{\fig{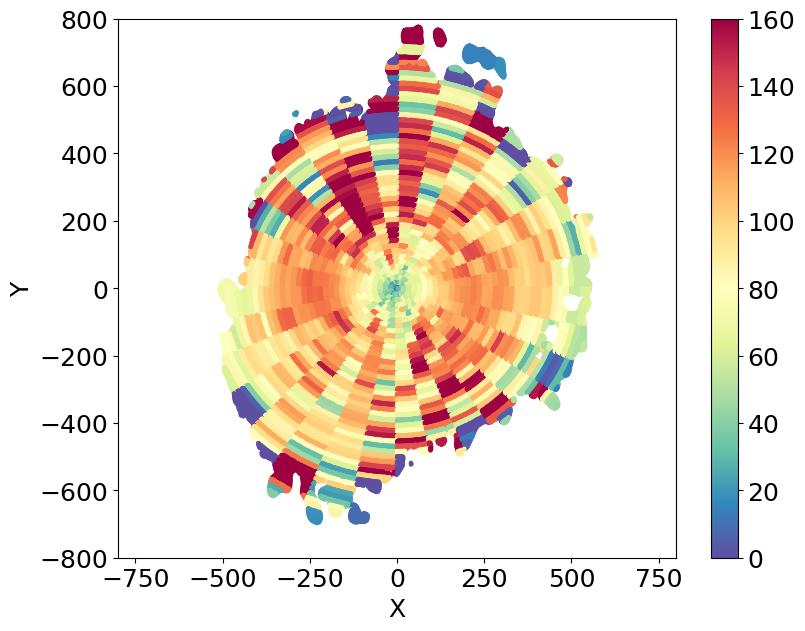}{0.3\textwidth}{(a)}
              \fig{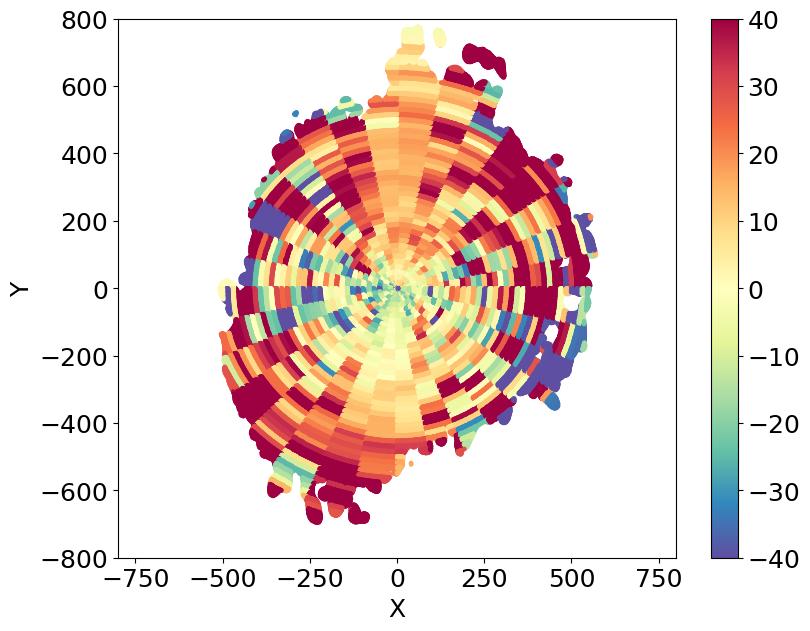}{0.3\textwidth}{(b)}
               \fig{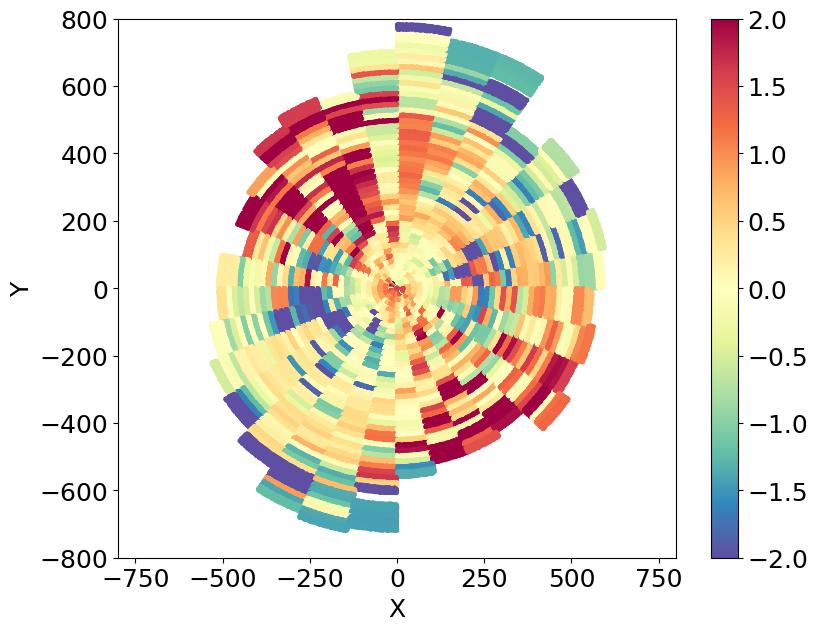}{0.3\textwidth}{(c)}}
\gridline{\fig{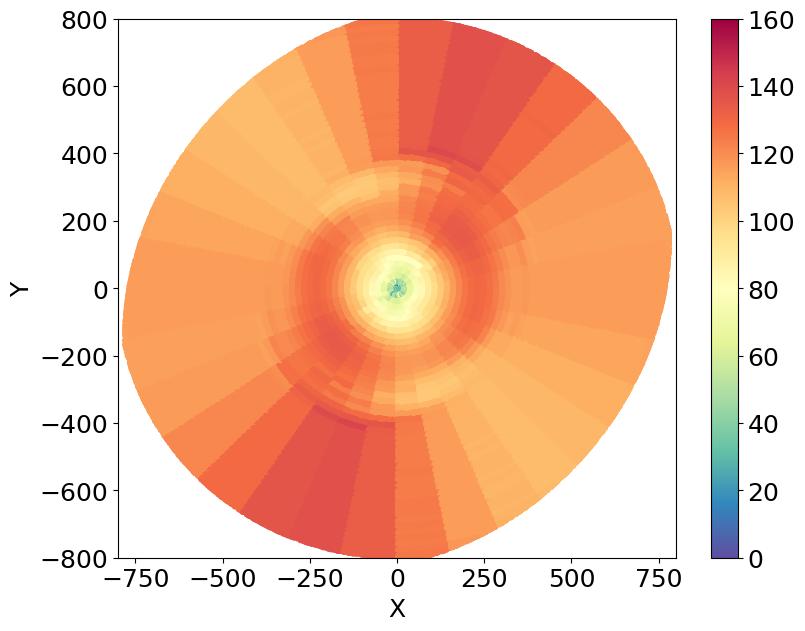}{0.3\textwidth}{(d)}
              \fig{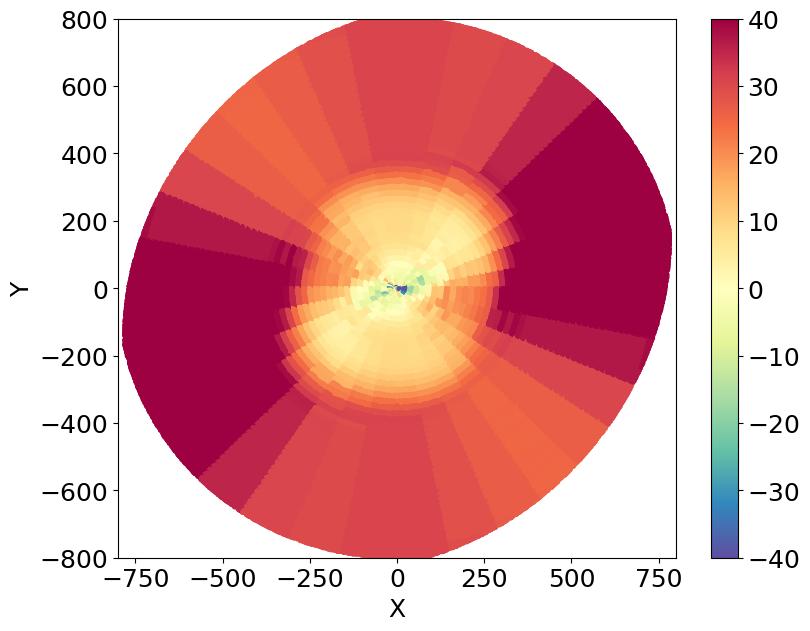}{0.3\textwidth}{(e)}
              \fig{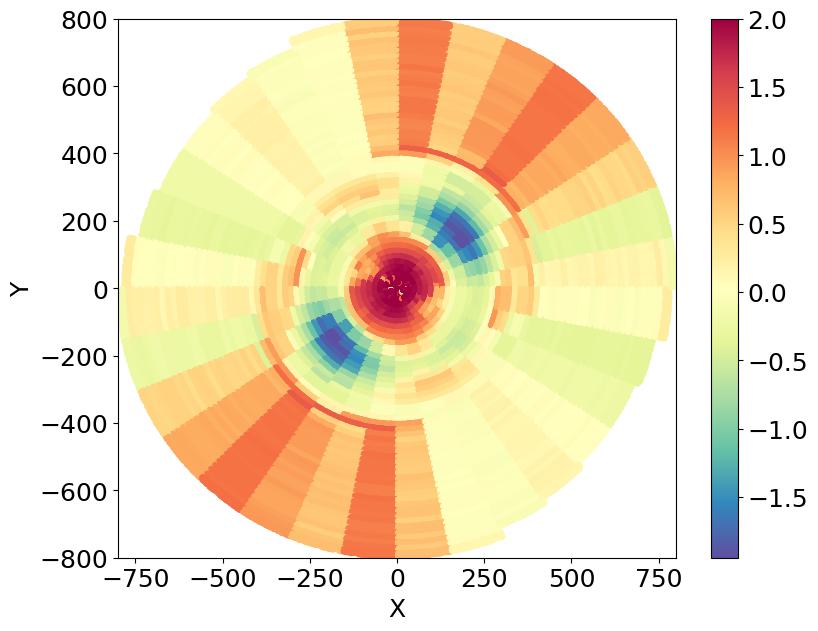}{0.3\textwidth}{(f)}}
\gridline{\fig{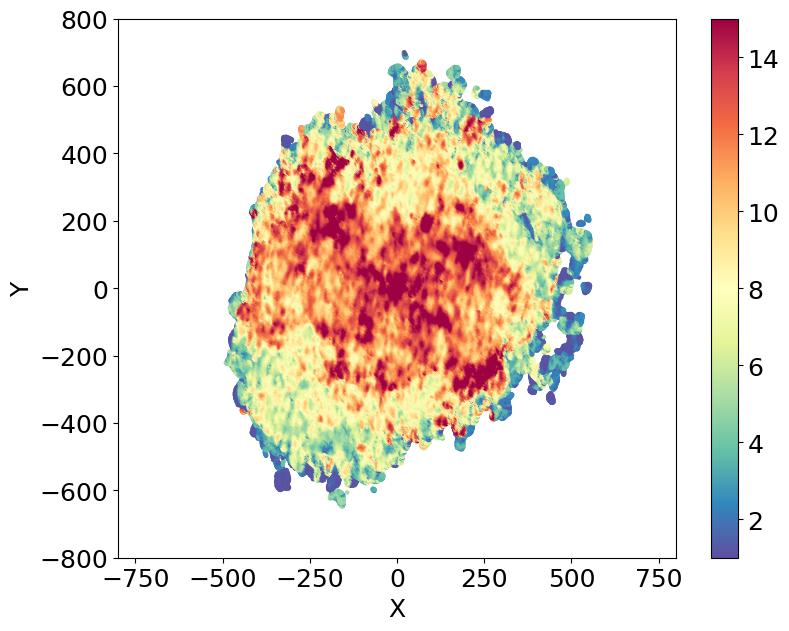}{0.3\textwidth}{(g)}
              \fig{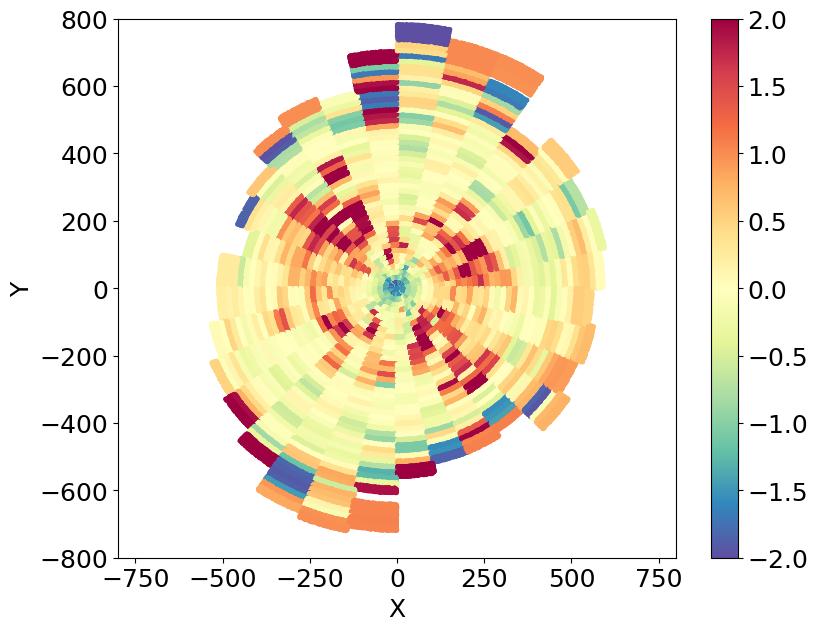}{0.3\textwidth}{(h)}
              \fig{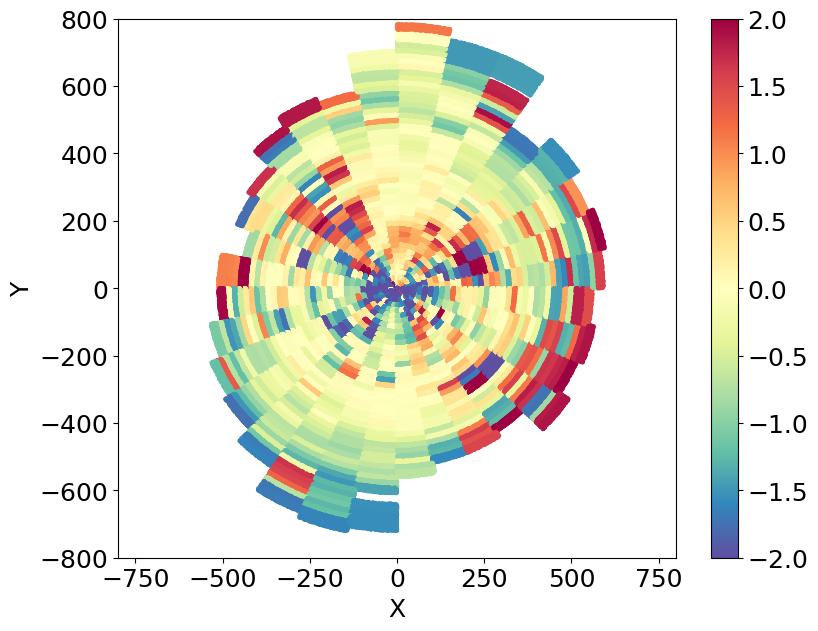}{0.3\textwidth}{(i)}}
\caption{As  Fig.\ref{fig:NGC2903-2} but for NGC 7793.} 
\label{fig:NGC7793-2} 
\end{figure*}
%


\subsection*{DDO 154} 

DDO 154 presents a weak evidence for the presence of a  warp as { the correlation coefficient is ${\cal C}$ = 0.09}, whereas the velocity field is weakly perturbed (Figs. \ref{fig:DDO154-1}- \ref{fig:DDO154-2}).
%

\begin{figure*}
\gridline{\fig{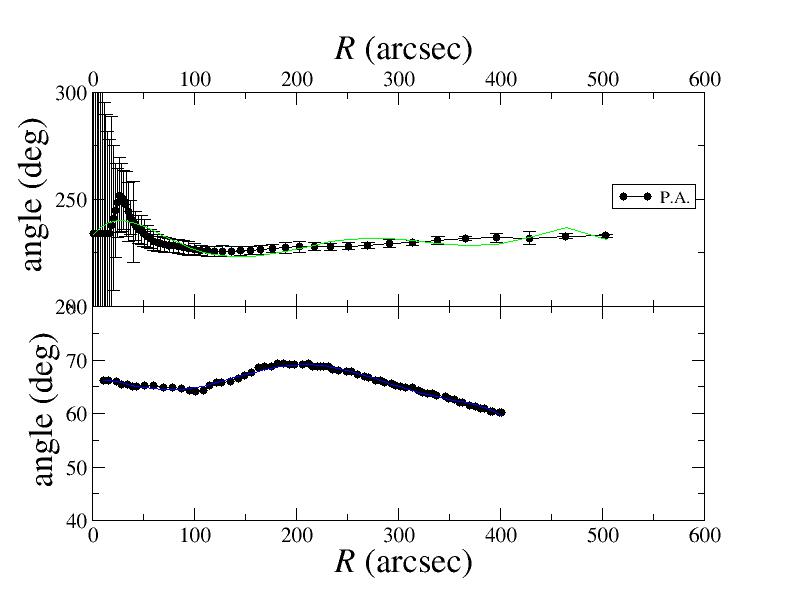}{0.45\textwidth}{(a)}
              \fig{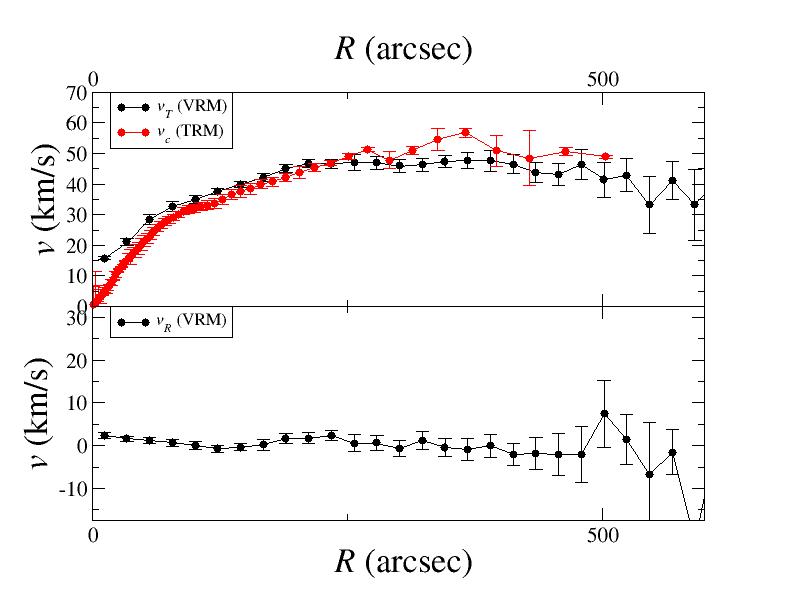}{0.45\textwidth}{(b)}
              }
  \gridline{
 	       \fig{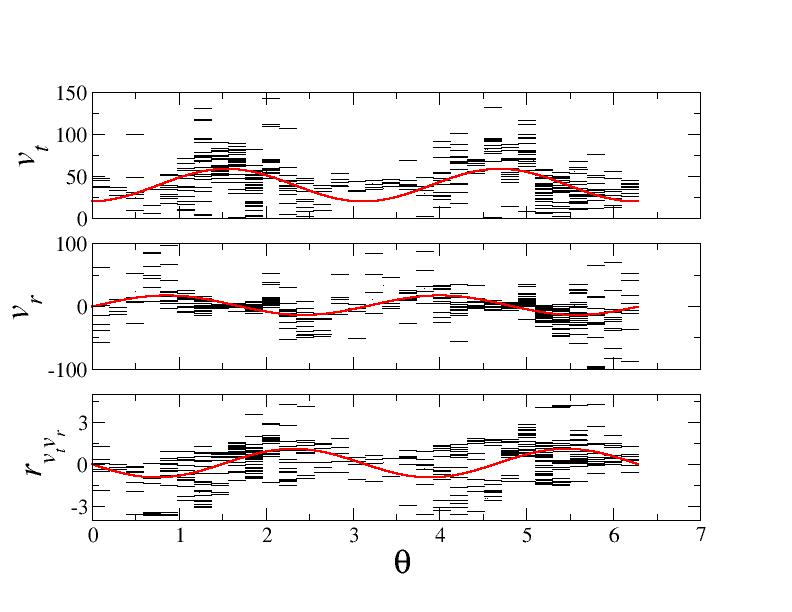}{0.45\textwidth}{(c)}
                \fig{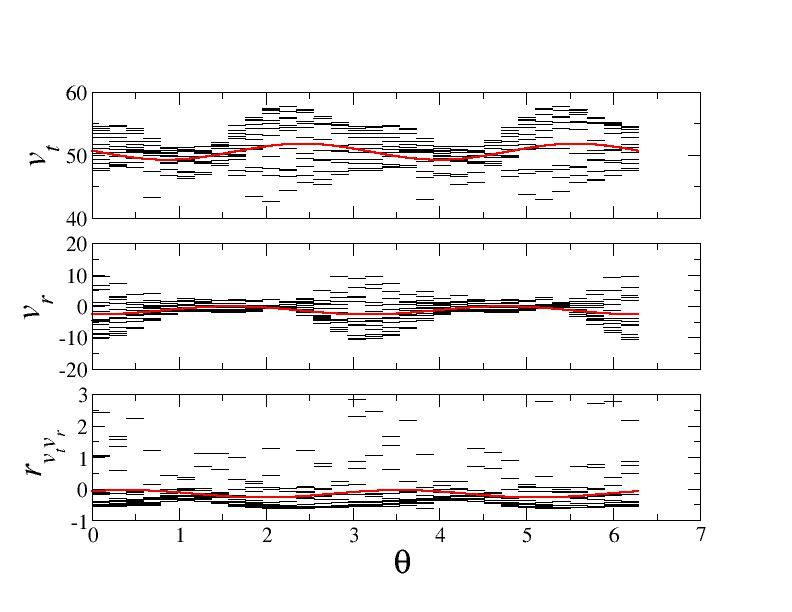}{0.45\textwidth}{(d)}}
     \caption{As  Fig.\ref{fig:NGC2903-1} but for NGC 5236.} 
\label{fig:DDO154-1} 
\end{figure*}
%

\begin{figure*}
\gridline{\fig{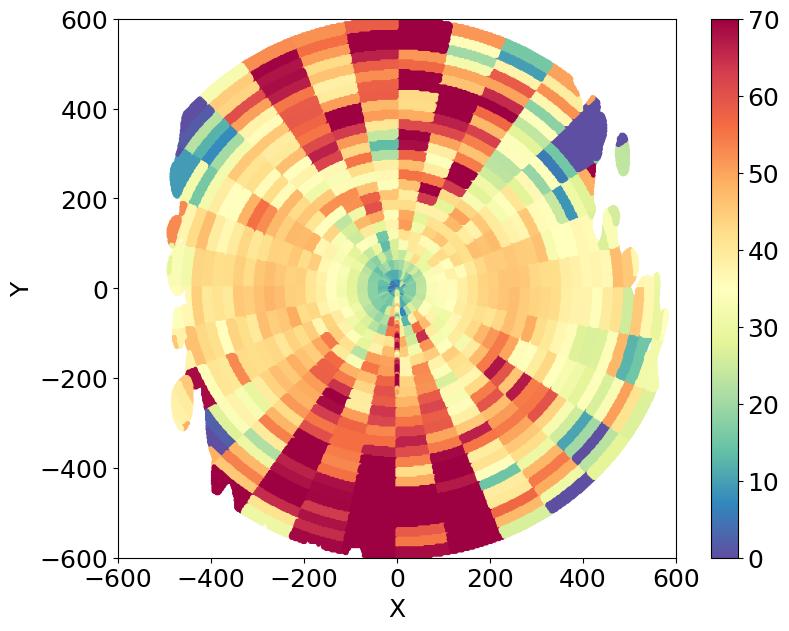}{0.3\textwidth}{(a)}
              \fig{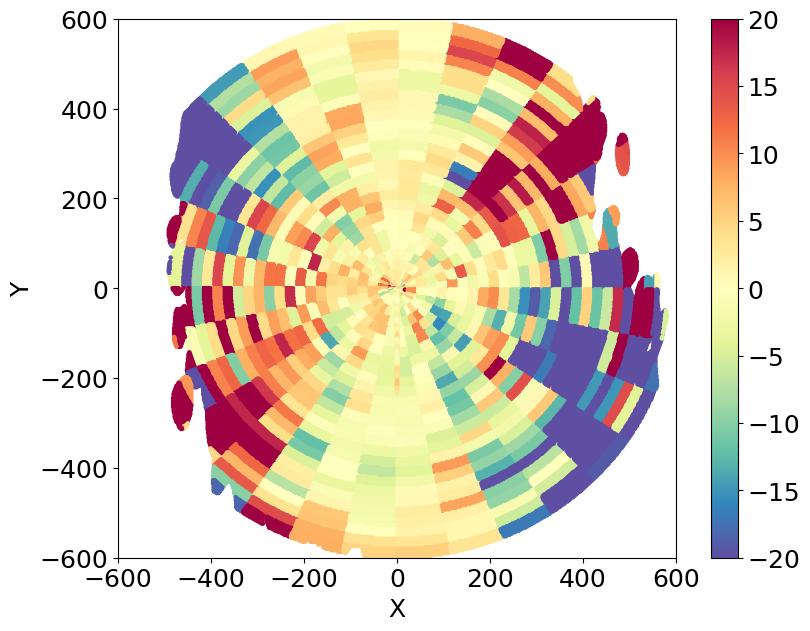}{0.3\textwidth}{(b)}
               \fig{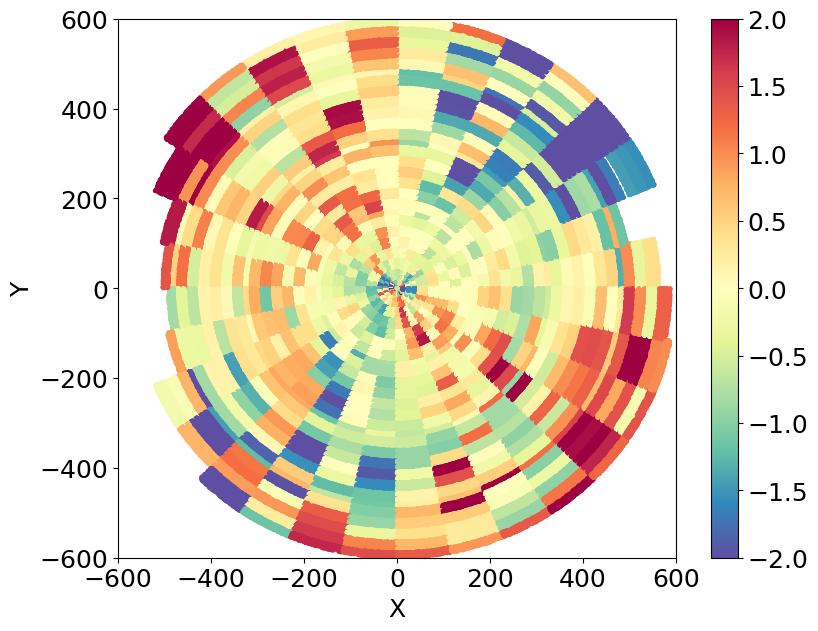}{0.3\textwidth}{(c)}}
\gridline{\fig{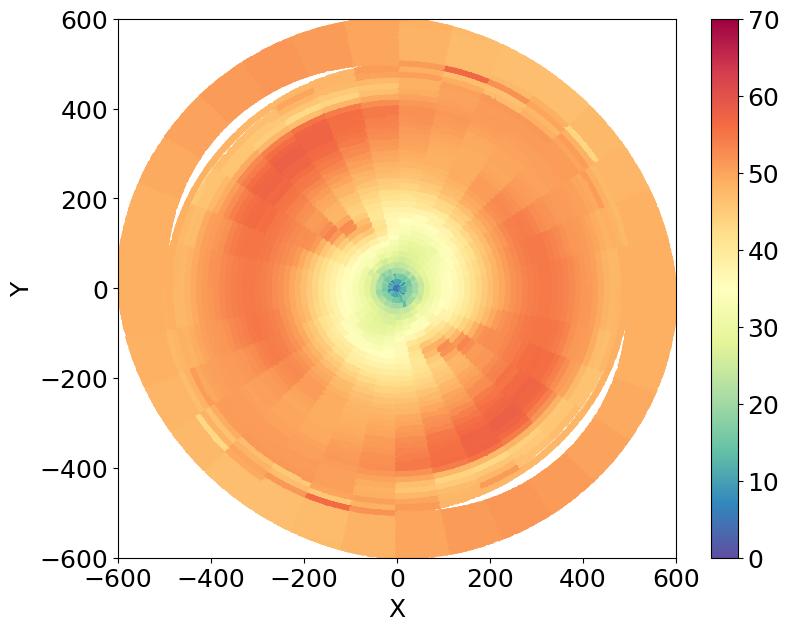}{0.3\textwidth}{(d)}
              \fig{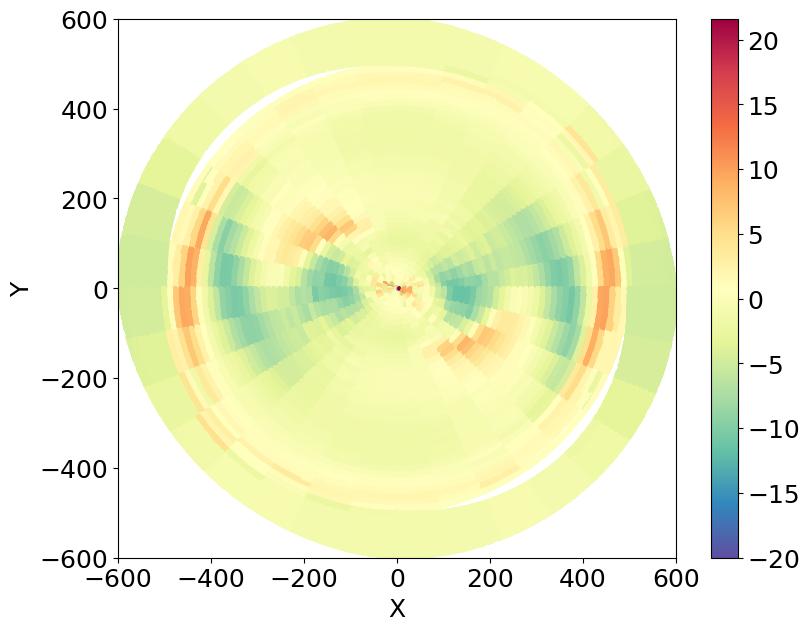}{0.3\textwidth}{(e)}
              \fig{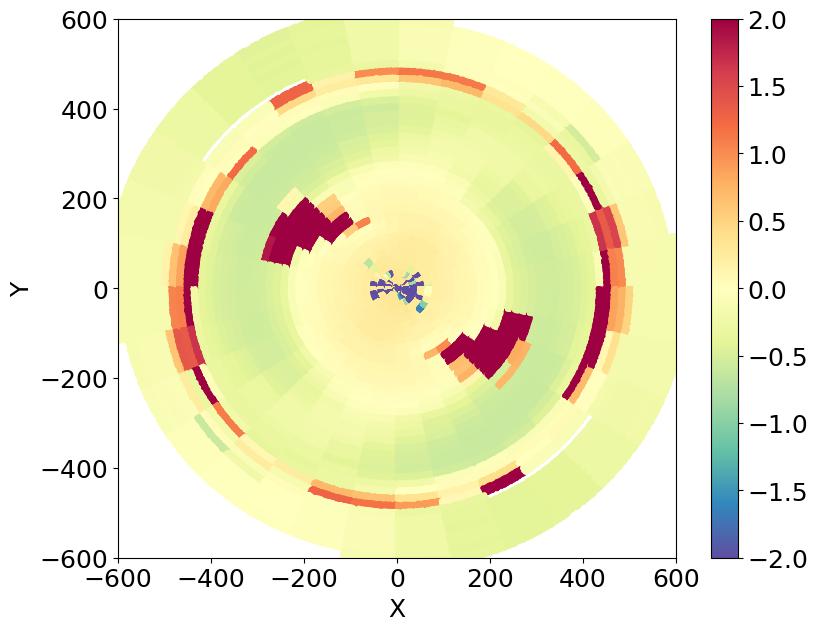}{0.3\textwidth}{(f)}}
\gridline{\fig{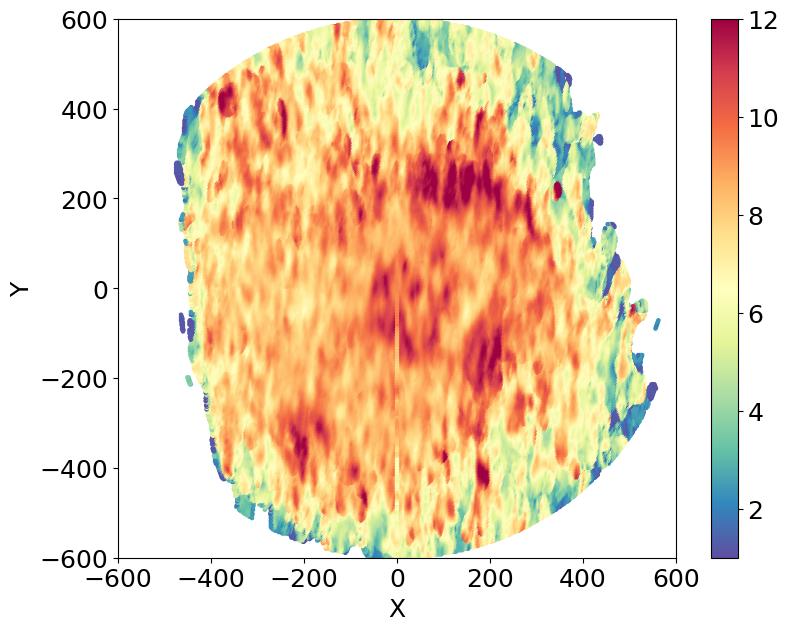}{0.3\textwidth}{(g)}
              \fig{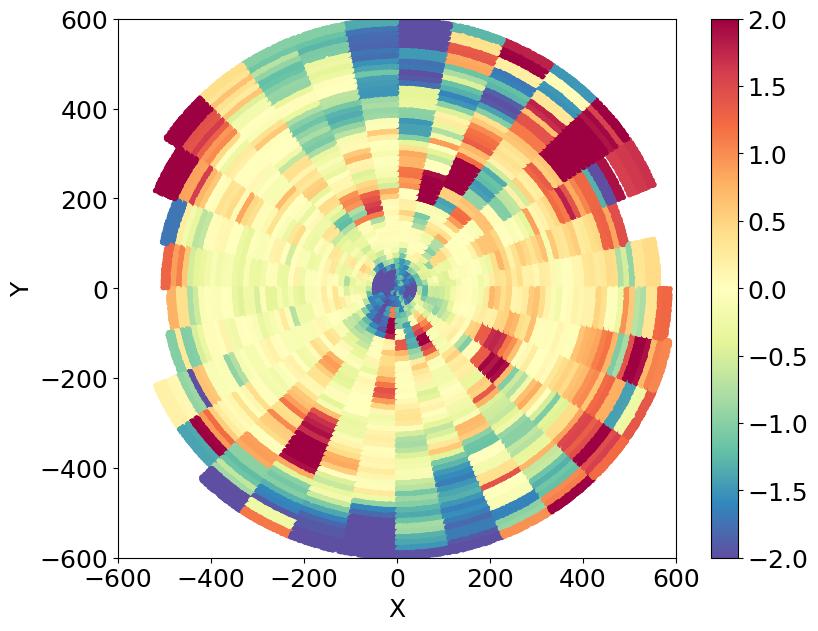}{0.3\textwidth}{(h)}
              \fig{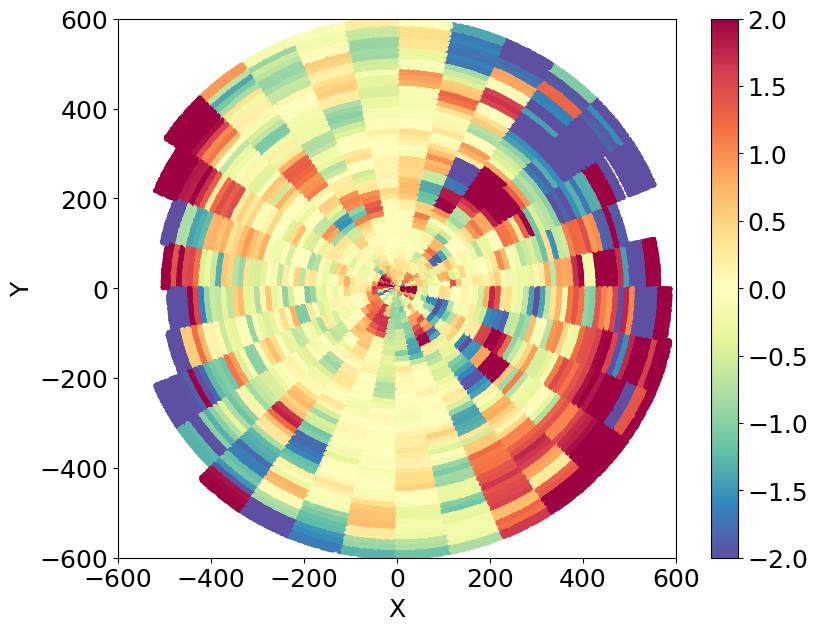}{0.3\textwidth}{(i)}}
\caption{As  Fig.\ref{fig:NGC2903-2} but for DDO 154.} 
\label{fig:DDO154-2} 
\end{figure*}
%

\clearpage
\subsection*{M 33}

In the analysis of the M33 galaxy, we observe that the inclination angle exhibits a relatively small and smooth variation when transitioning from the inner to the outer regions of the disk; on the other hand, the P.A. experiences a substantial change of around 40 degrees (refer to panel (a) of Fig. \ref{fig:M33-1}). Notably, for $R>3000$'', both velocity profiles display significant fluctuations. { The transverse velocity profile exhibits a behavior similar to the circular velocity detected by the TRM, but with larger fluctuations.}
Indeed, the velocity dispersion profile of M33, characterized by notable fluctuations and a slow decay as a function of radius, provides valuable insights into its dynamical state and kinematic properties. The slow decay indicates that the velocity dispersion remains relatively high even in the outer regions of the disk. This observation is compatible with the reality of the velocity perturbations detected by the VRM analysis. 

The toy disc model, constructed based on the measured inclination angle $i(R)$ and the P.A.  obtained from the TRM analysis, exhibits clear dipolar modulation in its external rings for both the transverse velocity component $v_t$ and the radial velocity component $v_r$. However, it is important to note that the dipolar modulation appears with different phases for these velocity components. Due to the differing phases, the velocity rank correlation coefficient also displays a dipolar modulation, albeit with a relatively low amplitude (as depicted in Figs. \ref{fig:M33-1}-\ref{fig:M33-2}). { By comparing the observed velocity field of M33 with that of the toy model,  we can conclude that a warp is present in galaxy's outermost regions in correspondance with the large change of the P.A.:   the correlation coefficient is  ${\cal C}$ = 0.18}. However,  significant fluctuations, on the order of 50 km/s or even higher, characterize both velocity components in these outermost parts of the galaxy. These fluctuations indicate the presence of complex velocity structures and dynamics in the outer skirts of M33. 

\begin{figure*}
\gridline{\fig{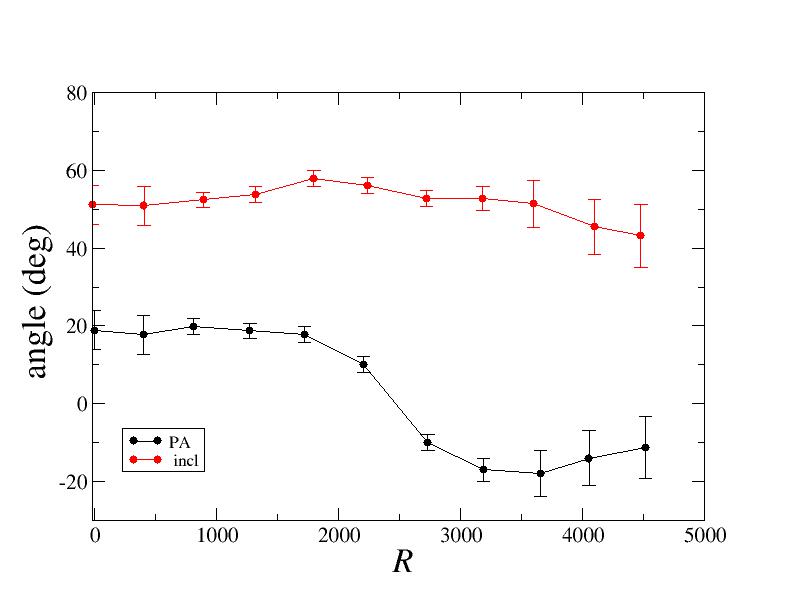}{0.45\textwidth}{(a)}
              \fig{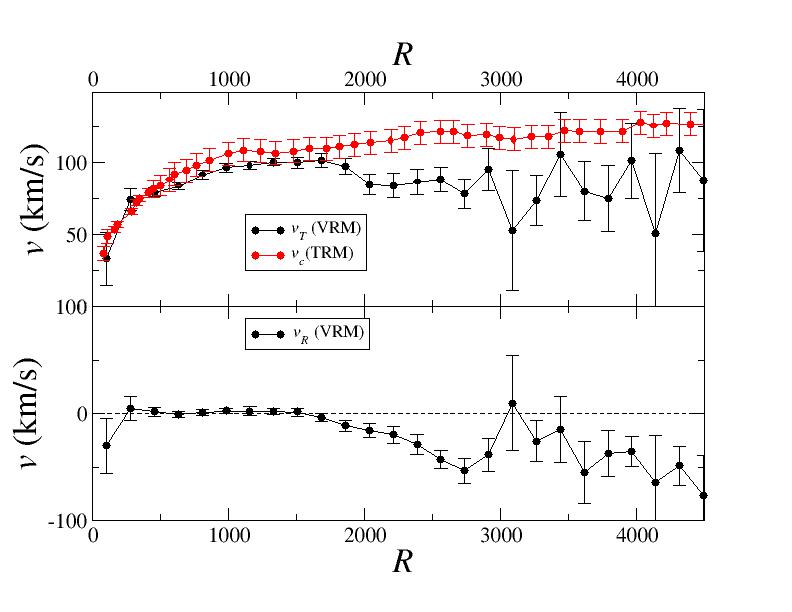}{0.45\textwidth}{(b)}
              }
  \gridline{
 	       \fig{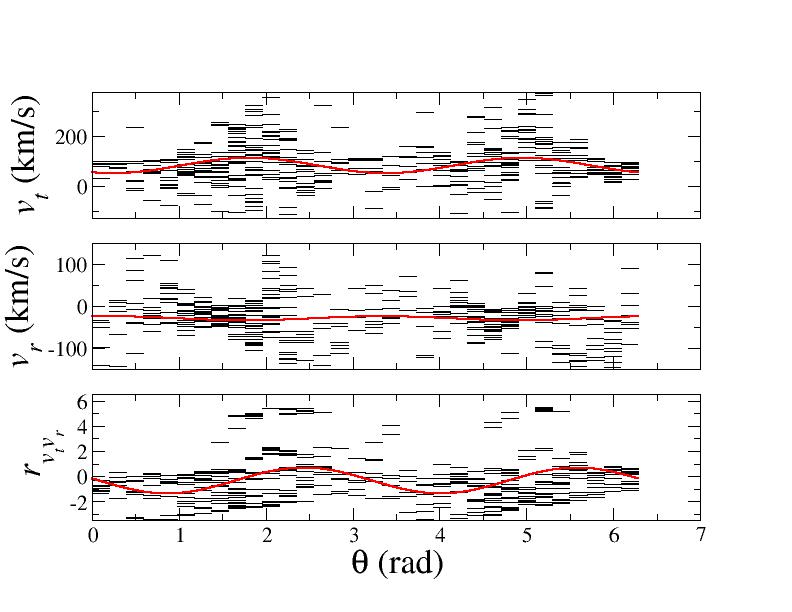}{0.45\textwidth}{(c)}
                \fig{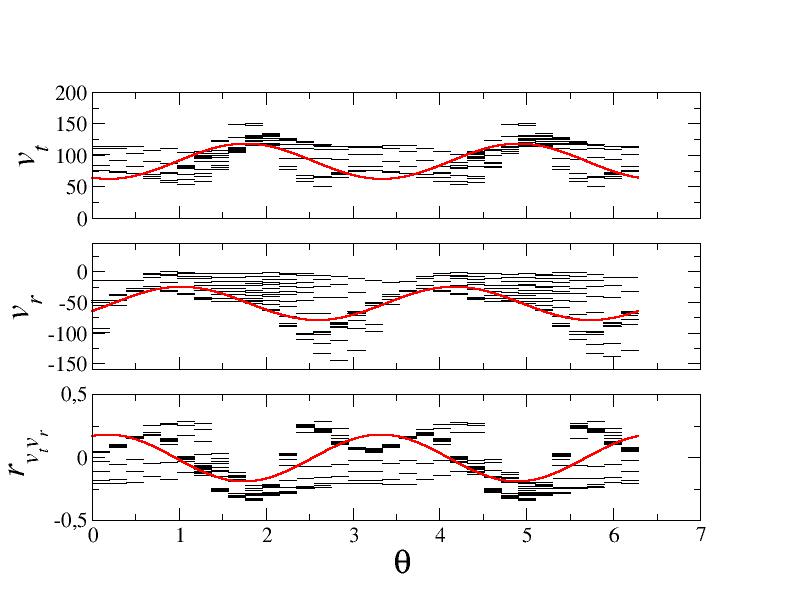}{0.45\textwidth}{(d)}}
     \caption{As  Fig.\ref{fig:NGC2903-1} but for M 33.} 
\label{fig:M33-1} 
\end{figure*}
%

\begin{figure*}
\gridline{\fig{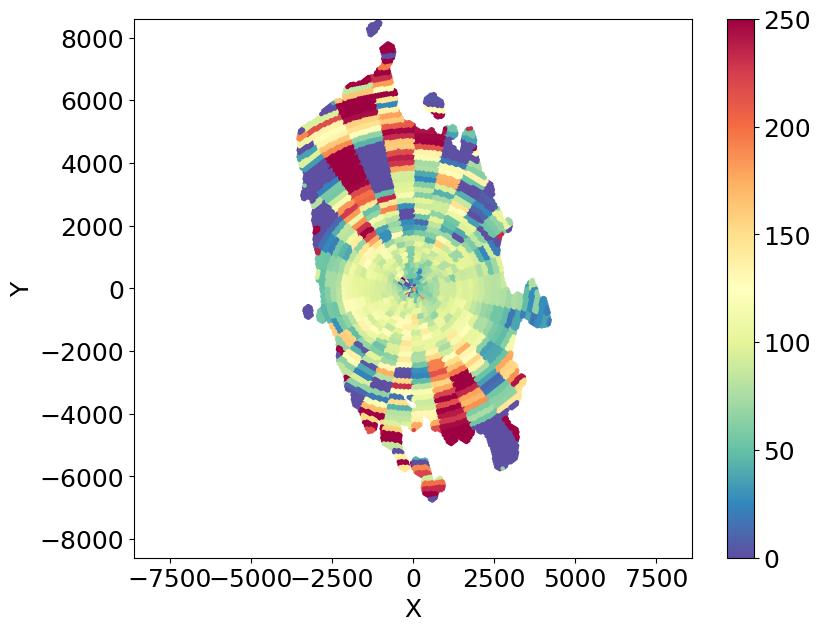}{0.3\textwidth}{(a)}
              \fig{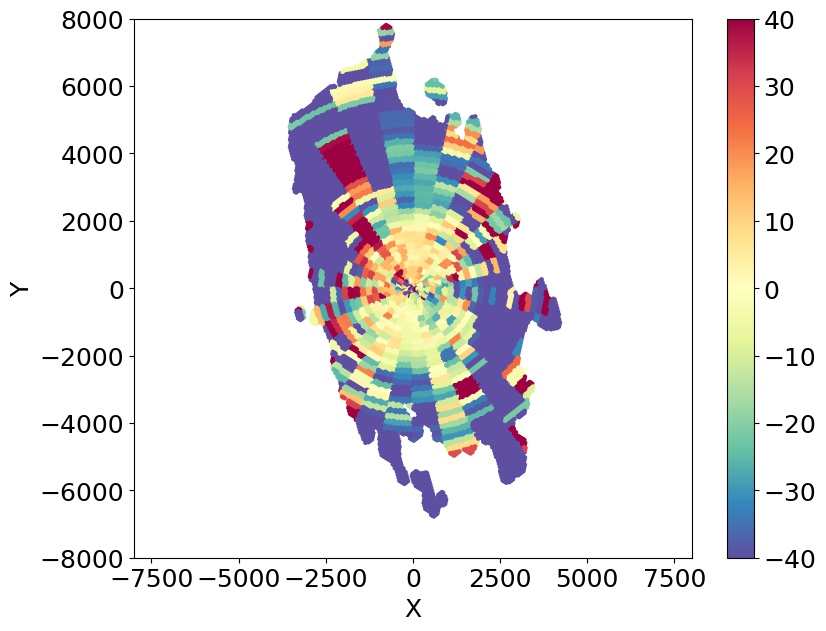}{0.3\textwidth}{(b)}
               \fig{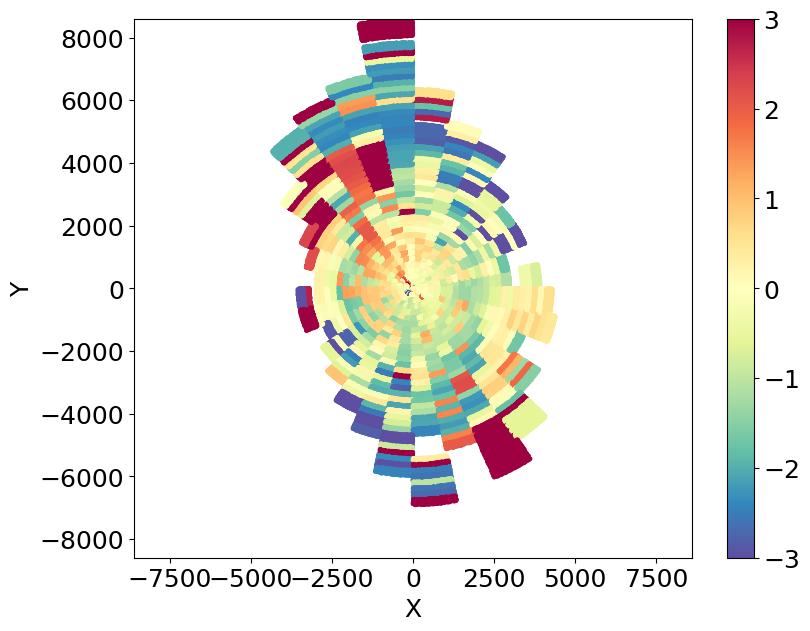}{0.3\textwidth}{(c)}}
\gridline{\fig{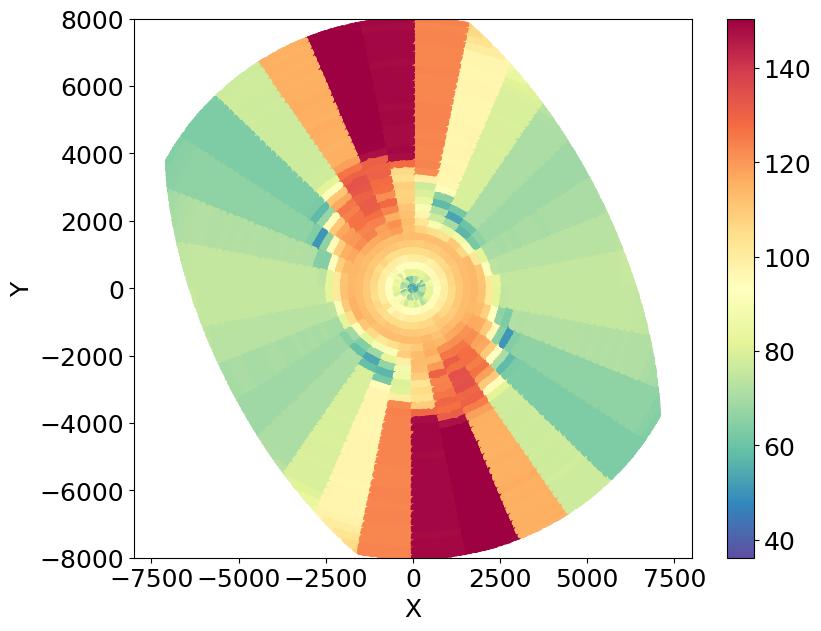}{0.3\textwidth}{(d)}
              \fig{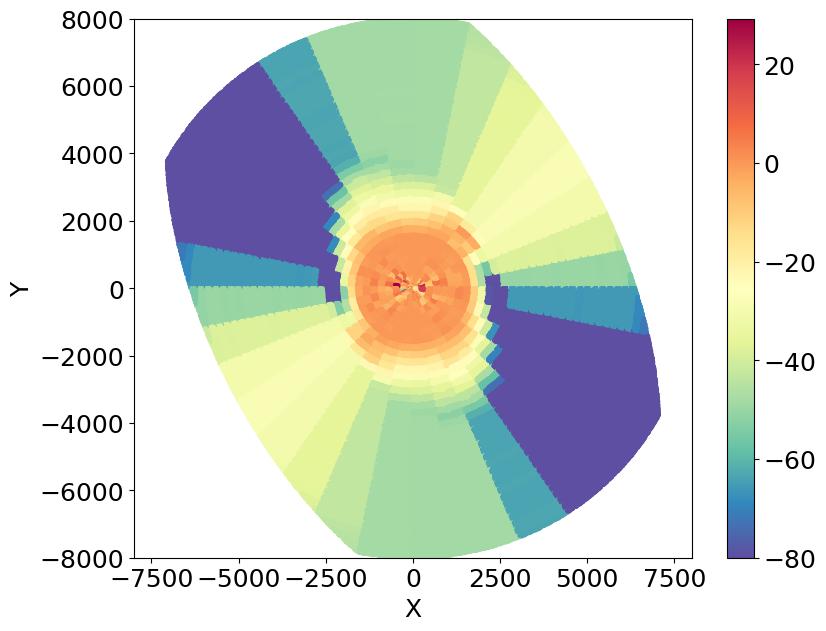}{0.3\textwidth}{(e)}
              \fig{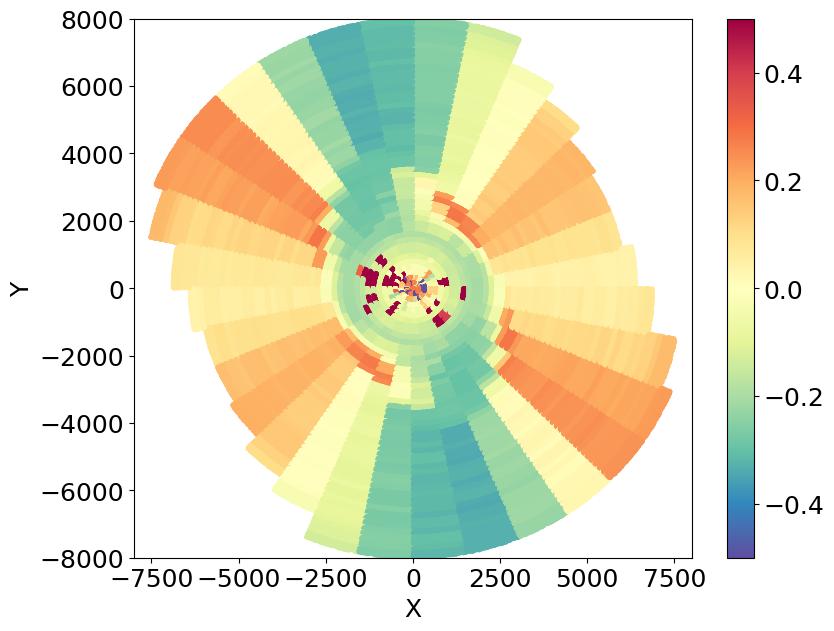}{0.3\textwidth}{(f)}}
\gridline{\fig{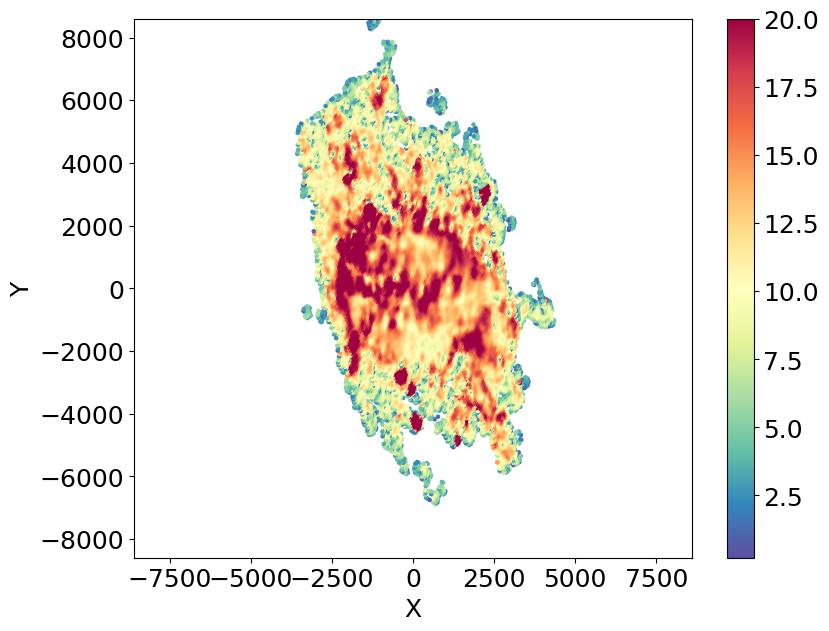}{0.3\textwidth}{(g)}
              \fig{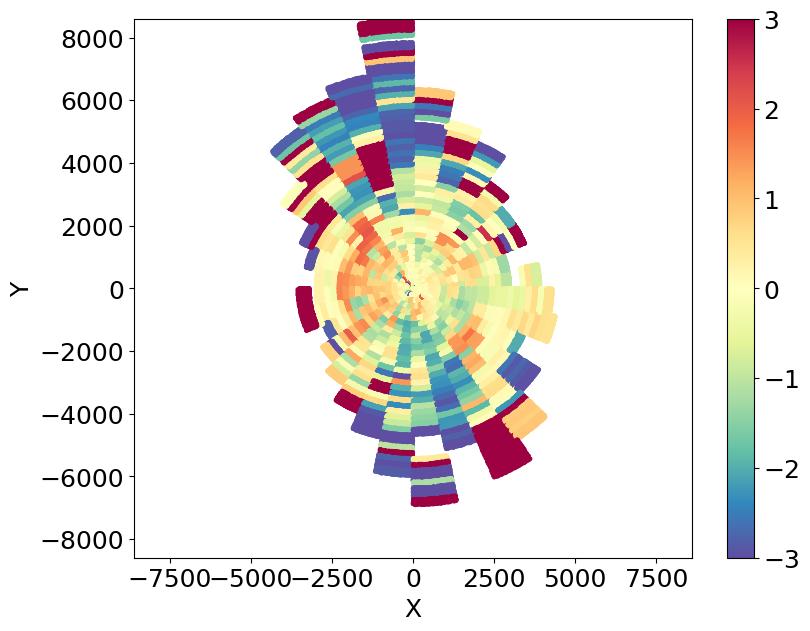}{0.3\textwidth}{(h)}
              \fig{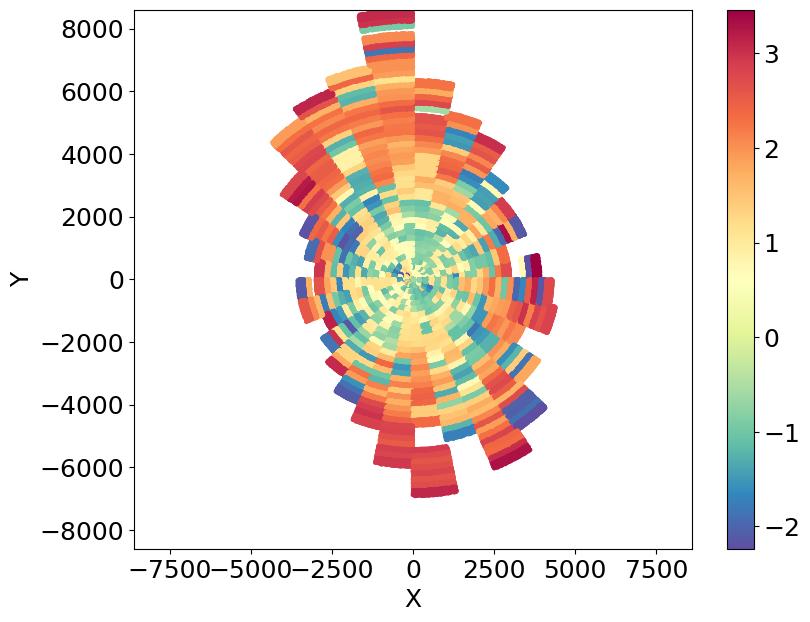}{0.3\textwidth}{(i)}}
\caption{As  Fig.\ref{fig:NGC2903-2} but for M33.} 
\label{fig:M33-2} 
\end{figure*}
%


\end{document}